\documentclass[12pt]{article}

\newlength{\abstractwidth}
\usepackage{amsmath, nccmath}
\usepackage{amsfonts}
\usepackage{amssymb}
\usepackage{latexsym}
\usepackage{shuffle}
\usepackage{enumerate}

\usepackage{color}
\usepackage{graphicx}
\usepackage{tikz}
\usepackage{dsfont}

\usepackage{tikz}
\usetikzlibrary{calc} 
\usetikzlibrary{patterns,snakes} 
\usetikzlibrary{decorations.pathreplacing} 
\usetikzlibrary{decorations.markings} 
\usetikzlibrary{decorations.pathmorphing} 
\usetikzlibrary{positioning}
\usetikzlibrary{arrows.meta}

\usetikzlibrary{arrows,shapes,positioning}
\usetikzlibrary{decorations.markings}
\usepackage[rightcaption]{sidecap}
\tikzstyle arrowstyle=[scale=1]
\tikzstyle directed=[postaction={decorate,decoration={markings,
    mark=at position .65 with {\arrow[arrowstyle]{stealth}}}}]
\tikzstyle reverse directed=[postaction={decorate,decoration={markings,
    mark=at position .65 with {\arrowreversed[arrowstyle]{stealth};}}}]
\usetikzlibrary{positioning}

\usepackage{hyperref}
\definecolor{darkred}{rgb}{0.8,0.1,0.1}
\hypersetup{colorlinks=true, linkcolor=darkred, citecolor=blue, linktoc=page}

\renewcommand{\thefootnote}{\fnsymbol{footnote}}
\renewcommand{\thanks}[1]{\footnote{#1}}
\newcommand{\starttext}{
\setcounter{footnote}{0}
\setcounter{section}{0}
\renewcommand{\thefootnote}{\arabic{footnote}}}

\newcommand{\bea}{\begin{eqnarray}}
\newcommand{\eea}{\end{eqnarray}}
\newcommand{\be}{\begin{eqnarray}}
\newcommand{\ee}{\end{eqnarray}}
\newcommand{\bma}{\begin{matrix}}
\newcommand{\ema}{\cr\end{matrix}}

\newtheorem{thm}{Theorem}[section]
\newtheorem{lem}[thm]{Lemma}

\newtheorem{cor}[thm]{Corollary}
\newtheorem{conj}[thm]{Conjecture}

\newcommand{\tGamma}[4]{\tilde \Gamma\Big(   \smallmatrix#1 \\  #2 \\ #3
 \endsmallmatrix ;#4\Big)}

\def\beq{\begin{equation}}
\def\eeq{\end{equation}}

\newcommand{\vecb}{\left(\begin{array}{c}}
\newcommand{\vece}{\end{array}\right)}
\newcommand{\ccb}{\left(\begin{array}{cc}}
\newcommand{\cce}{\end{array}\right)}
\newcommand{\cccb}{\left(\begin{array}{ccc}}
\newcommand{\ccce}{\end{array}\right)}
\newcommand{\ccccb}{\left(\begin{array}{cccc}}
\newcommand{\cccce}{\end{array}\right)}
\newcommand{\cccccb}{\left(\begin{array}{ccccc}}
\newcommand{\ccccce}{\end{array}\right)}

\def\cA{{\cal A}}
\def\cB{{\cal B}}
\def\cC{{\cal C}}

\def\cF{{\cal F}}
\def\cG{{\cal G}}
\def\cH{{\cal H}}
\def\cI{{\cal I}}
\def\cJ{{\cal J}}

\def\cO{{\cal O}}

\def\cR{{\cal R}}
\def\cS{{\cal S}}
\def\cT{{\cal T}}
\def\cU{{\cal U}}

\def\cW{{\cal W}}
\def\cX{{\cal X}}
\def\cY{{\cal Y}}
\def\cZ{{\cal Z}}

\def\mA{\mathfrak{A}}
\def\mB{\mathfrak{B}}
\def\mC{\mathfrak{C}}

\def\mJ{\mathfrak{J}}

\def\mN{\mathfrak{N}}

\def\mP{\mathfrak{P}}
\def\mQ{\mathfrak{Q}}

\def\mg{\mathfrak{g}}

\def\mw{\mathfrak{w}}

\def\ZZ{{\mathbb Z}}
\def\RR{{\mathbb R}}

\def\CC{{\mathbb C}}

\def\Im{{\rm Im \,}}

\def\det{{\rm det \,}}

\def\half{{1\over 2}}

\def\p{\partial}

\newcommand{\mcG}{\chi}
\newcommand{\Nf}{\widehat{\mathfrak{N}}}

\def\a{\alpha}

\def\tet{\vartheta}

\def\om{\omega}

\def\vI{{\overrightarrow{I}}}
\def\vP{{\overrightarrow{P}}}

\def\vPt{{\overleftarrow{P}}}
\def\tvarpi{\tilde\varpi}

\def\no{\nonumber}
\def\sm{\smallskip}

\def\pbx{\p_{\bar x}}
\def\pby{\p_{\bar y}}
\def\pbz{\p _{\bar z}}

\def\bom{\boldsymbol{\om}}

\begin{document}
\starttext
\setcounter{footnote}{0}

\begin{flushright}
2024 July 16  \\
revised May 2026  \\
UUITP--17/24
\end{flushright}

\bigskip

\begin{center}

{\Large \bf Fay identities for polylogarithms on}

\vskip 0.1in

{\Large \bf higher-genus Riemann surfaces} 

\vskip 0.4in

{\large Eric D'Hoker$^{(a)}$ and Oliver Schlotterer$^{(b,c)}$}

\vskip 0.1in

 ${}^{(a)}$ {\sl Mani L. Bhaumik Institute for Theoretical Physics}\\
 { \sl Department of Physics and Astronomy }\\
{\sl University of California, Los Angeles, CA 90095, USA}\\

\vskip 0.1in

 ${}^{(b)}$ { \sl Department of Physics and Astronomy,} 
 \\ {\sl Uppsala University, 75120 Uppsala, Sweden}
 
 \vskip 0.1in

 ${}^{(c)}$ 
 { \sl Department of Mathematics,} \\
  { \sl Centre for Geometry and Physics,} \\ 
  {\sl Uppsala University, 75106 Uppsala, Sweden}

 \vskip 0.1in
 
{\tt dhoker@physics.ucla.edu},  {\tt oliver.schlotterer@physics.uu.se}

\hskip 0.7in

\begin{abstract}
A recent construction of polylogarithms on Riemann surfaces of arbitrary genus in arXiv:2306.08644 is based on a flat connection assembled from single-valued non-holomorphic integration kernels that depend on two points on the Riemann surface. In this work, we construct and prove infinite families of bilinear relations among these integration kernels that are necessary for the closure of the space of higher-genus polylogarithms under integration over the points on the surface. Our bilinear relations generalize the Fay identities among the genus-one Kronecker-Eisenstein kernels to arbitrary genus. The multiple-valued meromorphic kernels in the flat connection of Enriquez are conjectured to obey higher-genus Fay identities of exactly the same form as their single-valued non-holomorphic counterparts. We initiate the applications of Fay identities to derive functional relations among higher-genus polylogarithms involving either single-valued or meromorphic integration kernels.
\end{abstract}

\end{center}

\newpage

\baselineskip=14.4pt
\setcounter{tocdepth}{2} 
\tableofcontents

\newpage

\baselineskip=16pt
\numberwithin{equation}{section}
\setcounter{equation}{0}
\setcounter{footnote}{0}

\newpage

\section{Introduction}
\label{sec:intro}

A variety of cutting-edge challenges in high-energy physics and different areas of mathematics
evolve around the treatment of iterated integrals on increasingly complex geometries. 
Different flavours of polylogarithm functions have become a common theme of 
Feynman-integral computations in quantum field theory 
\cite{Bourjaily:2022bwx, Abreu:2022mfk, Blumlein:2022qci, Weinzierl:2022} 
and the moduli-space integrals over punctured Riemann surfaces in string amplitudes  
\cite{Berkovits:2022ivl, Dorigoni:2022iem, DHoker:2022dxx, Mafra:2022wml}. 
At the same time, special values, Hopf-algebra  structures and related properties
of such polylogarithms connect deep questions in number theory and algebraic
geometry with computational advances at the interface of mathematics and physics
\cite{Duhr:2012fh, Schlotterer:2012ny, Schnetz:2013hqa, Brown:2015ylf}. 
A central objective in this field of research is to construct function spaces of polylogarithms  
that close under taking primitives and to exhibit effective algorithms
to determine these primitives.

\sm

The space of polylogarithms on a (compact) Riemann surface $\Sigma$ strongly depends 
on the genus of the surface $\Sigma$. On the sphere (genus zero),  polylogarithms arise from iterated integrals of rational functions \cite{Goncharov:1995, Goncharov:1998kja, Goncharov:2001iea}. Their closure under taking primitives can be traced back to standard partial  fraction identities \cite{Brown:2009qja}.

\sm 

On the torus, namely on a compact Riemann surface of genus one, elliptic polylogarithms were introduced in \cite{Beilinson:1994, Levin:2007, BrownLevin}, further developed in \cite{Enriquez:2023} and organized according to applications to superstring amplitudes and Feynman integrals \cite{Broedel:2014vla} and \cite{Broedel:2017kkb}, respectively. Elliptic polylogarithms may be obtained as iterated integrals on the torus of Kronecker-Eisenstein kernels. Essential to their closure under integration are certain identities among theta functions \cite{BrownLevin, Enriquez:2023} that are closely related to the Fay trisecant identity of \cite{Fay:1973} and will be referred to as \textit{Fay identities} in the sequel. Fay identities at genus one play the role of partial fraction decompositions at genus zero and rearrange  bilinear combinations in Kronecker-Eisenstein in a form that permits the evaluation of the primitives of all combinations of integration kernels and elliptic polylogarithms in any number of variables \cite{Broedel:2014vla, Broedel:2018iwv}. Fay identities also underly the algebraic and differential relations that elliptic polylogarithms satisfy at  special values of their arguments \cite{Broedel:2015hia, Gerken:2020aju},  known as elliptic multiple zeta values \cite{Enriquez:Emzv, Matthes:thesis} and, in the real-analytic case, modular graph functions and forms \cite{DHoker:2015gmr, DHoker:2015wxz, DHoker:2016mwo}.

\sm

The literature on integration kernels, associated flat connections and polylogarithms
on Riemann surfaces of higher genus $h\geq 2$ goes back to the study of correlators 
in Wess-Zumino-Witten models in \cite{Bernard:1988} and more recently 
features a broad bandwidth of approaches
\cite{Enriquez:2011, Enriquez:2021, Ichikawa:2022qfx, Enriquez:2022, DHS:2023, Baune:2024}.
In view of the growing relevance of higher-genus polylogarithms for
Feynman integrals  \cite{Huang:2013kh, Georgoudis:2015hca, Doran:2023yzu, Marzucca:2023gto, Jockers:2024tpc} and string amplitudes \cite{DHoker:2017pvk, DHoker:2018mys, DHoker:2020tcq, Berkovits:2022ivl, DHoker:2023khh}, the quest for  conceptual and computational control of the functional identities they obey is clearly a timely endeavor. Even so, while the Fay trisecant identity and its applications to bosonization are well-known for arbitrary genus, the generalization of the Fay identities that is required to promote the space of  higher genus polylogarithms into an algebra of functions  that closes under differentiation and integration has  remained part of largely uncharted territory. 

\sm

In this work, we close this gap by constructing and proving Fay identities for  integration kernels on compact Riemann surfaces $\Sigma$ of arbitrary genus $h$. We will mostly follow the explicit  approach to higher-genus polylogarithms in \cite{DHS:2023} where the integration kernels are  given by iterated convolutions of  the Arakelov Green function  
\cite{Faltings,Alvarez-Gaume:1986nqf, DHoker:2017pvk} as well as holomorphic Abelian differentials and their complex conjugates. As a consequence, the integration kernels of \cite{DHS:2023} are single-valued but non-meromorphic  functions of two points on $\Sigma$ and transform as tensors under the modular  group $Sp(2h,\mathbb Z)$ \cite{vdG3,Kawazumi:lecture,DHoker:2020uid}.

\sm 

Our main results include infinite families of \textit{tensor-valued Fay identities} among bilinears in the single-valued higher-genus integration kernels of \cite{DHS:2023} which are complete in the following sense. The  dependence of these bilinears on three points $x,y,z \in \Sigma$ can always be rearranged to avoid a repeated dependence on any of the points $x,y$ or $z$ in more than one integration kernel factor. This rewriting of higher-genus integration kernels in terms of products with at most one $x,y$ or $z$-dependent factor is essential for integration over the respective point in terms of the higher-genus polylogarithms of \cite{DHS:2023}. Throughout this work, the bilinear identities among integration kernels and their corollaries for higher-genus 
polylogarithms are considered on a fixed surface $\Sigma$, without varying its complex-structure~moduli.

\sm

We also investigate the three-point Fay identities in the limit of two coincident points and encounter modular tensors that solely depend on the moduli of $\Sigma$ and generalize (almost) holomorphic Eisenstein series to higher genus. 
As a simple subclass of two-point Fay identities, we recover the so-called \textit{interchange identities} which were presented  in \cite{DHoker:2020tcq, DHoker:2020uid, DHS:2023} relating integration kernels and Abelian differentials. The web of relations that is found to descend from the Fay identities in this work paves the way for deriving functional identities among higher-genus polylogarithms and proving their closure under taking primitives. In sections \ref{sec:clo} and \ref{sec:7.9}, we illustrate the role played by the interchange and Fay identities in the concrete construction of primitives involving different types of higher-genus polylogarithms.

\sm

At genus one, our understanding of elliptic polylogarithms and their special values
benefitted from the interplay between two types of Kronecker-Eisenstein integration
kernels: single-valued but non-meromorphic $f^{(r)}$ and meromorphic but multi-valued $g^{(r)}$ with $r \geq 0$.\footnote{While the single-valued $f^{(r)}$ entered the Brown-Levin formulation of
elliptic polylogarithms \cite{BrownLevin} as well as their string-theory applications in \cite{Broedel:2014vla, Broedel:2017jdo}, the alternative formulation of elliptic polylogarithms \cite{Broedel:2017kkb}
in terms of the meromorphic $g^{(r)}$ is predominantly used in Feynman-integral applications \cite{Bourjaily:2022bwx}.} Numerous main results of this paper are derived and proven 
through the properties of
single-valued but non-meromorphic modular tensors $f^{I_1 \cdots I_r}{}_J(x,y)$ 
which generalize the $f^{(r)}(x{-}y)$ to higher genus \cite{DHS:2023}. In particular,
the ubiquitous integration-by-parts identities in our computations crucially rely
on the single-valuedness of the $f$-tensors. Meromorphic but multiple-valued
higher-genus generalization of the $g^{(r)}(x{-}y)$ kernels,  to be denoted by 
$g^{I_1 \cdots I_r}{}_J(x,y)$,\footnote{We depart from the normalization
conventions of the meromorphic integration kernels 
$\om^{I_1 \cdots I_r}{}_J(x,y)$ introduced in Enriquez's work \cite{Enriquez:2011} 
by $ g^{I_1 \cdots I_r}{}_J(x,y)= (-2\pi i)^r \om^{I_1 \cdots I_r}{}_J(x,y)$ to attain
a smooth genus-one limit $ g^{I_1 \cdots I_r}{}_J(x,y) |_{h=1}= g^{(r)}(x{-}y)$ and
simple poles $g^{I}{}_J(x,y) = \delta^I_J/(x{-}y) + {\rm reg}$ with unit residue.} were introduced by Enriquez through their functional properties \cite{Enriquez:2011}.

\sm

Another main result of this work is a generalization of the interchange identities, which we prove, and a generalization of the Fay identities, which we conjecture, to the case of the meromorphic Enriquez kernels $g^{I_1 \cdots I_r}{}_J(x,y)$ at arbitrary genus and tensor rank. While explicit representations for the Enriquez kernels are somewhat cumbersome to exhibit,\footnote{See \cite{Baune:2024} for a recent proposal to express the Enriquez kernels $\omega^{I_1 \cdots I_r}{}_J(x,y)$ in  terms of Poincar\'e series and Schottky variables in the restricted subset of moduli space where the Poincar\'e series converges.} 
their defining properties provide sufficient guidance for anticipating and proving the conjectural identities in this work. Finally, the coincident limits $y\rightarrow x$ of Enriquez kernels $g^{I_1 \cdots I_r}{}_J(x,y)$ are conjectured to introduce meromorphic versions of the solely moduli-dependent modular tensors encountered in the analogous coincident limits of $f^{I_1 \cdots I_r}{}_J(x,y)$.


\subsection{Outline}

This work is organized as follows: We start by motivating the quest for higher-genus Fay identities in section \ref{sec:4zero} by highlighting the significance of partial fractions and genus-one Fay identities for iterated integrals on the sphere and the torus, respectively.  In section \ref{sec:2}, we review the protagonists of the Fay identities of this work, namely,  the single-valued but non-meromorphic integration kernels of \cite{DHS:2023} and the associated higher-genus polylogarithms. Section~\ref{sec:4} introduces a simple subclass of Fay identities on compact  Riemann surfaces of arbitrary genus $h$  that transform as scalars under the modular group $Sp(2h,\ZZ)$.  We then proceed to the general case of bilinear identities among higher-genus integration kernels with tensorial transformation law under $Sp(2h,\ZZ)$: interchange identities involving two points in section~\ref{sec:3} and Fay identities involving three or more points in section~\ref{sec:fay}.  In section~\ref{sec:clo} we illustrate the role of the interchange and the Fay identities in the closure under taking primitives of multivariable higher-genus polylogarithms. The coincident limits of higher-genus integration kernels and Fay identities featuring tensorial generalizations of (almost) holomorphic Eisenstein series  are discussed in section~\ref{sec:5}. Finally, in section~\ref{sec:7}, we gather counterparts of the results obtained in earlier sections for the meromorphic but multiple-valued integration kernels in the Enriquez connection \cite{Enriquez:2011}, to prove a meromorphic version of the interchange identities and conjecture a meromorphic version of the  Fay identities.

\sm

The appendices complement the discussion in the main text with additional background material on the prime form and the Arakelov Green function (Appendix \ref{appdefs}), an alternative approach to multi-variable Fay identities (Appendix \ref{sec:6.3}), proofs of the main lemmas and theorems (Appendix \ref{appprfs}) and a construction of higher-weight Fay identities from convolutions of lower-weight ones (Appendix~\ref{app.cons}). 

\sm

Pointers to the main  Theorems, and Conjectures of this work are as follows:
\begin{itemize}
\itemsep=-0.05in
\item For the single-valued but non-meromorphic integration kernels of \cite{DHS:2023}:
\begin{itemize}
\itemsep=0in
\item interchange identities in Theorem \ref{intlemma}, 
\item three-point Fay identities in Theorems \ref{3.thm:7} and \ref{3.thm:8}, 
\item their coincident limits in Theorems \ref{4.thm:1} and \ref{finthm},
\end{itemize}
such that readers with primary interest in applications of our results 
may directly jump to the key equations (\ref{genint}), (\ref{piszero}), (\ref{exfay.15}), (\ref{4.ff.99}), (\ref{3.coin.0}) and (\ref{coin.75}).
\item For the meromorphic but multiple-valued Enriquez kernels of \cite{Enriquez:2011}:
\begin{itemize}
\itemsep=0in
\item interchange identities in Theorem \ref{mintlemma},
\item three-point Fay identities in Conjectures \ref{mfaysscal} and \ref{mfaysform},
\item their coincident limits in Conjectures \ref{me.coin} and \ref{me.coin2},
\end{itemize}
see (\ref{ome.08}), (\ref{pisarezero}), (\ref{ome.13}), (\ref{ome.14}), (\ref{holo.15}) and (\ref{holo.22}) for the key equations.
\end{itemize}

\subsection{Results obtained after the first archive version of this paper}
\label{sec:1.c}

Several questions and conjectures that were stated in the earlier versions of this paper have since then been addressed or solved.
\begin{itemize}
\itemsep 0in
\item A proof of the Conjectures \ref{mfaysscal} and \ref{mfaysform} was advanced in \cite{Baune:2024ber};
\item Relations between the single-valued $f$-kernels and the meromorphic $g$-kernels,
and their consequences for the associated classes of polylogarithms 
can be found in joint work with Enriquez and Zerbini~\cite{DESZ:progress};
\item An alternative proof of the Conjecture \ref{mfaysscal}  is obtained in \cite{paper.II} 
by demonstrating the equivalence between interchange and Fay identities and
flatness of DHS or Enriquez connections in multiple variables.
\end{itemize}


\subsection{Acknowledgments}

We sincerely thank Benjamin Enriquez and Federico Zerbini for a variety of inspiring discussions and collaboration on related topics. We are grateful to Martijn Hidding for participation in early stages of this project and to Carlos Mafra, Martin Raum and Kaiwen Sun for valuable discussions and correspondence.

\sm

 The research of ED is supported in part by NSF grant PHY-22-09700. The research of OS is supported by the European Research Council under ERC-STG-804286 UNISCAMP as well as the strength area ``Universe and mathematical physics'' which is funded by the Faculty of Science and Technology at Uppsala University. This research was supported in part by grant NSF PHY-2309135 to the Kavli Institute for Theoretical Physics (KITP), and we thank the organizers of the program ``What is string theory'' for creating a stimulating atmosphere. OS is grateful to the Hausdorff Research Institute for Mathematics in Bonn 
 for their hospitality and the organizers and participants of the Follow-Up Workshop ``Periods in Physics, Number Theory and Algebraic Geometry'' for valuable discussions. Moreover, OS thanks the Galileo Galilei Institute for Theoretical Physics in Florence for the hospitality and the INFN as well as the Simons foundation for financial support during advanced stages of this work. Finally, OS thanks the Banff International Research Station in Oaxaca for their hospitality and the organizers \& participants of the workshop
 ``Beyond Elliptic Polylogarithms'' for inspiring discussions during final stages of this work.

\newpage

\section{Motivation: Fay identities at genus zero and one}
\label{sec:4zero}

The space of polynomials forms a ring under the operations of addition and multiplication and closes under differentiation and integration, namely the derivative and the primitive of a polynomial is again a polynomial. While rational functions also form a ring under addition and multiplication (they actually form a field) and close under differentiation, the primitive of a rational function is not necessarily again a rational function. Instead, the logarithm arises as the primitive of a simple pole, and polylogarithms \cite{Goncharov:1995, Goncharov:1998kja, Goncharov:2001iea} arise from further operations of multiplication by rational functions and integration.  The resulting space that combines rational functions and polylogarithms on the complex plane or on the Riemann sphere is closed under addition, multiplication, differentiation and integration \cite{Brown:2009qja}. While the study of polylogarithms has a long history \cite{Poincare:1884, Lappo:1953}, their significance for perturbative quantum field theory \cite{Remiddi:1999ew, Vollinga:2004sn, Goncharov:2010jf, Duhr:2012fh} and string theory \cite{Broedel:2013tta, Schlotterer:2018zce, Vanhove:2018elu} has been recognized only over the past few decades. 

\sm

On the torus, elliptic functions again close under addition, multiplication and differentiation, but integration again produces new functions, which are referred to as \textit{elliptic polylogarithms}. Double periodicity and meromorphicity are not always compatible with one another on the torus and this conflict leads to different formulations of the function spaces of elliptic polylogarithms. The standard choices are based on either single-valued but non-meromorphic flat connections or alternatively meromorphic but multiple-valued ones \cite{Beilinson:1994, Levin:2007, BrownLevin, Broedel:2017kkb, Enriquez:2023}.  Similar to their genus-zero counterparts, elliptic polylogarithms have become a common theme of perturbative computations in quantum field theory \cite{Bloch:2013tra, Adams:2017ejb, Ablinger:2017bjx, Remiddi:2017har, Broedel:2017siw} and string theory \cite{Broedel:2014vla, DHoker:2015wxz, Broedel:2017jdo, Mafra:2019xms, Broedel:2019gba}.

\sm

Further generalization to polylogarithms to a higher-genus Riemann surface $\Sigma$ have also been introduced recently.  The construction of polylogarithms is greatly facilitated by the introduction of a flat connection whose associated path-ordered exponential integral, or holonomy, between two points $x,y \in \Sigma$ depends only on the homotopy class of paths between the points $x$ and $y$ but not on the specific representative chosen to represent each class. The conflict between meromorphicity and single-valuedness, that existed already on the torus, persists for  higher-genus Riemann surfaces $\Sigma$ and again leads one to make choices. Formulations in terms of meromorphic flat connections on a punctured Riemann surface of arbitrary genus feature either multiple-valued integration kernels with simple poles \cite{Enriquez:2011, Baune:2024} or single-valued ones with higher poles \cite{Enriquez:2021, Enriquez:2022}. Their disadvantages are that modular invariance is obscured, and that the basic integration kernels are somewhat cumbersome to exhibit explicitly (though the Schottky parametrization has recently been used to evaluate   genus-two polylogarithms numerically \cite{Baune:2024}).

\sm

In a recent paper \cite{DHS:2023} a construction of polylogarithms was developed based on a non-meromorphic but single-valued and modular invariant flat connection with at most simple poles. More specifically, the $Sp(2h,\mathbb Z)$ invariance of the connection of \cite{DHS:2023} on a Riemann surface $\Sigma$ of arbitrary  genus $h$ is explicitly realized in terms of integration kernels that transform as modular tensors.

\sm
 
Closure under integration of the function space of a certain class of meromorphic hyperlogarithms was proven recently  in \cite{Enriquez:2022}. It has remained a challenge, however, to obtain effective algorithms for the explicit determination of primitives from that approach and any other.  It is the purpose of this paper to investigate the closure under integration of the polylogarithms introduced in \cite{DHS:2023} multiplied by the integration kernels in their  underlying flat connection. We shall prove bilinear identities among the higher-genus kernels that 
implement this closure as detailed in section \ref{sec:clo} and generalize the genus-one Fay identities
among the Kronecker-Eisenstein kernels in \cite{BrownLevin}.  We also propose concrete conjectures for certain relations that are needed to show the existence and determine the explicit form of primitives of the meromorphic polylogarithms derived from the kernels of \cite{Enriquez:2011, Baune:2024}.

\sm

In this section, we present brief reviews of the polylogarithms at genus zero, namely on the Riemann sphere, and for elliptic polylogarithms at genus one, namely on the torus. The remainder of the paper will be devoted to higher genus.

\subsection{Partial fraction decomposition at genus zero}
\label{sec:4.0}

On the Riemann sphere, any rational function of $x$ may be expressed using partial fraction decomposition. The primitive of every term in this decomposition, except for simple poles,  is again a rational function. Therefore, the extension beyond rational functions required to obtain closure under integration is generated by differentials $dx/(x-a_i)$. The corresponding polylogarithms $G(a_1, \cdots , a_n;x)$ with $a_i,x \in \mathbb C$ are defined recursively by $G(\emptyset; x)=1$ and,
\beq
G(a_1,\cdots,a_n;x) = \int^x_0 \frac{dz}{z{-}a_1} \, G(a_2,\cdots,a_n;z)
\label{g0poly}
\eeq
We assign regularized values to the divergent cases with 
$a_n=0$ and $a_1=x$ by defining $G(0;x)= \log(x)$ and $G(x;x)= - \log(x)$ and
imposing regularized $G(a_1,\cdots,a_n;x)$ with ${n\geq 2}$ to obey the
shuffle relations of convergent iterated integrals. By shuffle relations, the product of two
polylogarithms $G(\cdots;x)$ with the same endpoints $0$ and $x$ of the path is
a linear combination of the same type of integrals. Hence, in the discussion of
closure under integration over $x$, it is sufficient to consider expressions with at most
one factor of polylogarithms (\ref{g0poly}). In order to integrate over the labels $a_i$ 
of $G(a_1,\cdots,a_n;x)$, the differential equations of polylogarithms can be used
to move the integration variable into the endpoint of the path $G(\cdots;a_i)$ \cite{Brown:2009qja, Broedel:2013tta, Panzer:2014caa, Panzer:2015ida}. As detailed in section \ref{sec:clo}, these types of functional identities are known as a \textit{change of fibration basis}, and we will lay the ground for their generalizations to arbitrary genus.

\sm

The primitive of the product of a polylogarithm $G(a_1, \cdots ,a_n;z)$ and a rational function $\phi(z)$ may be decomposed into a sum of rational functions and polylogarithms. To show this, we decompose $\phi(z)$ into partial fractions. The primitive of any simple pole term $1/(z{-}b)$ in $\phi(z)$ clearly produces a new polylogarithm $G(b,a_1, \cdots, a_n;x)$ while polynomial and higher order poles may be integrated by parts and again recursively decomposed onto the polylogarithms of (\ref{g0poly}). For example, choosing $\phi(z) = \frac{1}{(z-b_1)(z-b_2)}$ with two linear factors in the denominator, the partial fraction decomposition needed to integrate the product 
$\phi(z)G(a_1,\cdots,a_n;z)$ via (\ref{g0poly}) is given by, 
\begin{align}
\int^x_0 dz \, \frac{G(a_1,\cdots,a_n;z)}{(z{-}b_1)(z{-}b_2) }
&= \frac{1}{b_1{-}b_2}  \,  \int^x_0 dz \, \bigg( \frac{ G(a_1,\cdots,a_n;z) }{z{-}b_1} -  \frac{ G(a_1,\cdots,a_n;z) }{z{-}b_2} \bigg) \notag \\
&= \frac{1}{b_1{-}b_2}  \, \Big ( G(b_1,a_1,\cdots,a_n;x) - G(b_2,a_1,\cdots,a_n;x) \Big )
\label{polylogint0}
\end{align}
Partial fraction decomposition is a property of rational functions and will not be available, as such, for genus one and beyond. Instead, what will be available at genus $h \geq 1$ are multi-periodic generalizations of the elementary partial fraction relation,
\beq
 \frac{1}{(z{-}x)(x{-}y)} + \frac{1}{(x{-}y)(y{-}z)} + \frac{1}{(y{-}z)(z{-}x)}= 0
\label{pfs}
\eeq
among three points on the sphere which implies partial fraction decompositions involving an arbitrary number of points.
More specifically, recursive application of (\ref{pfs}) to products of several simple poles will lead to standard partial fraction decomposition, such as in, 
\beq
\prod_{j=1}^r\frac{1}{z{-}x_j} = \frac{1}{(z{-}x_1)(x_1{-}x_2) \cdots (x_{r-1}{-}x_r)}+ {\rm perm}(x_1,x_2,\cdots,x_r)
\label{starg0}
\eeq
or in,
\beq
\frac{1}{(x_1{-}x_2) (x_2{-}x_3 ) \cdots (x_{r-1}{-}x_r)} + {\rm cycl}(x_1,x_2,\cdots,x_r) =0
\label{cycl0}
\eeq
As will be motivated further below, the multi-periodic generalizations of the elementary partial fraction relation (\ref{pfs}) to genus $h\geq 1$ will be referred to as \textit{Fay identities}. Similar to the situation on the sphere, elementary Fay identities among three points on a Riemann surface of genus $h$ will be sufficient to simplify functions of an arbitrary number of points. The desired simplifications to be achieved via higher-genus Fay identities are set by the closure of the genus-$h$ polylogarithms of \cite{DHS:2023} under integration in the same way as partial fraction enables the genus-zero integration in (\ref{polylogint0}).

\subsection{Kronecker-Eisenstein series at genus one}
\label{sec:4.1}

As mentioned in the introductory paragraphs to this section, function theory on the torus reflects the conflict between meromorphicity and single-valuedness, and leads to two natural but different generalizations of rational functions on the sphere. They are referred to as the \textit{Kronecker-Eisenstein coefficients}
$g^{(r)}(x)$ and $f^{(r)}(x)$ and are given by the following generating series,\footnote{The  $\tet$-function is  given by 
$\vartheta_1(x) = 2 q^{1/8} \sin(\pi x) \prod_{n=1}^{\infty} (1-q^n) (1-e^{2\pi i x} q^n) (1-e^{-2\pi i x} q^n) 
$ for $q= e^{2\pi i \tau}$.}
\begin{align}
 \sum_{r=0}^{\infty} \alpha^{r-1} g^{(r)}(x) & = 
 \frac{ \vartheta_1'(0)  \vartheta_1(x{+}\alpha) }{ \vartheta_1(x)  \vartheta_1(\alpha)  } 
 \label{exfay.22} \\
 \sum_{r=0}^{\infty} \alpha^{r-1} f^{(r)}(x) & = 
 \frac{ \vartheta_1'(0)  \vartheta_1(x{+}\alpha) }{ \vartheta_1(x)  \vartheta_1(\alpha)  } \,
 \exp\bigg( 2\pi i \alpha \, \frac{\Im x}{\Im \tau} \bigg)
 \notag
\end{align}
where $\a \in \CC$ plays the role of a bookkeeping device. Throughout this work, the
modulus $\tau$ of the torus $\Sigma = \CC/(\ZZ {+} \tau \ZZ)$ and the moduli of higher-genus
Riemann surfaces in later sections will be considered as fixed, and we therefore do not exhibit
the $\tau$-dependence in (\ref{exfay.22}). The functions $g^{(r)}(x)$ defined by (\ref{exfay.22}) are  meromorphic in~${x\in \Sigma}$ but multiple-valued, while the functions $f^{(r)}(x)$ are single-valued but not meromorphic. One has $g^{(0)} (x) = f^{(0)}(x) =1$ and the first non-trivial functions are,
\begin{align}
g^{(1)} (x) & =  \p_x \ln \tet_1(x) 
\label{gen1.02} \\
f^{(1)}(x) & =  \p_x \ln \tet_1(x) + 2 \pi i { \Im x \over \Im \tau}\notag
\end{align}
Both of $g^{(1)}(x)$ and $f^{(1)}(x)$ have simple poles for any $x\in (\ZZ {+} \tau \ZZ)$.
While all the single-valued $f^{(r)}(x)$ are regular on $\Sigma $ for $r \not=1$, the meromorphic functions
$g^{(r)}(x)$ for $r \geq 2$ on the universal cover $\mathbb C$ of the torus have
simple poles at $x\in (\ZZ {+} \tau \ZZ)\setminus \mathbb Z$. 

\sm

The single-valued functions $f^{(r)}(x{-}y)$ will generalize at higher genus to the modular tensors $f^{I_1 \cdots I_r}{}_J(x,y)$ introduced in \cite{DHS:2023} while the meromorphic functions $g^{(r)}(x{-}y)$ will generalize to the differential forms $g^{I_1 \cdots I_r}{}_J(x,y)$ introduced in \cite{Enriquez:2011}. Translation invariance on the torus admits the simple parity properties $f^{(r)}(x{-}y)= (-1)^r f^{(r)}(y{-}x)$ and  $g^{(r)}(x{-}y)= (-1)^r g^{(r)}(y{-}x)$ of the genus-one functions. However, their higher-genus generalizations obey more involved identities dubbed \textit{interchange identities}, see section \ref{sec:3} and section \ref{sec:7.2} below.
 
\sm 
 
 Recall that the scalar Green function $\cG(x,y)$ on the torus is defined by \cite{DHoker:2022dxx}, 
 \bea
\cG(x,y) = - \log \bigg| \frac{\vartheta_1(x{-}y)}{\eta} \bigg|^2+ 2\pi \frac{ \Im(x{-}y)^2}{\Im \tau} 
\label{agft}
\eea
where the Dedekind $\eta$ function satisfies $\tet_1'(0) = 2 \pi \eta^3$. Note that $\cG(x,y)$ is single-valued, symmetric under swapping $x$ and $y$, and depends only on the difference $x{-}y$ in view of translation invariance on the torus. The Kronecker-Eisenstein coefficients $f^{(r)}(x{-}y)$  can be naturally obtained from $\cG$ via differentiation and convolutions, as follows, \cite{Gerken:2018},
\begin{align}
f^{(1)}(x{-}y) &=  - \partial_x \cG(x,y)
\label{gen1.99} \\
f^{(r)}(x{-}y) &= \int_{\Sigma} d^2z \, \partial_x \cG(x,z) f^{(r-1)}(z{-}y) \, , \ \ \ \ \ \ r\geq 2
\notag
\end{align}
The meromorphic analogues of these formulae are given by (\ref{gen1.02}) for $g^{(1)}(x{-}y)$ and
its convolution integrals over homology cycles (as opposed to surface integrals) for $g^{(r)}(x{-}y)$ 
with $r\geq 2$ in equation (21) of \cite{DHoker:2025dhv}.

\sm 

The Kronecker-Eisenstein coefficients, either meromorphic or single-valued,  play a role for iterated integrals on the torus \cite{BrownLevin, Broedel:2014vla, Broedel:2017kkb} that is analogous to the role played by  the differentials $dx/(x{-}a_i)$ for the polylogarithms (\ref{g0poly}) on the sphere. Accordingly, both of $g^{(r)}(x)$ and $f^{(r)}(x)$ are referred to as \textit{integration kernels} or \textit{Kronecker-Eisenstein kernels}.  Both $g^{(r)}(x)$ and $f^{(r)}(x)$ are  said to have  \textit{weight} $r$ which in both cases equals the transcendental weight of the Fourier coefficient and in the case of $f^{(r)}(x)$ equals the \textit{modular weight} (though the $g^{(r)}(x)$ do not transform as Jacobi forms under $SL(2,\mathbb Z)$ \cite{Zagier:1991}).

\subsection{Three-point Fay identities at genus one}
\label{sec:4.1.1}

In this subsection, we  review the genus-one Fay identities in terms of both types of integration kernels $f^{(r)}$ and $g^{(r)}$ and provide a definition of the term \textit{$z$-reduced} for the genus-one case.

\subsubsection{Three-point Fay identities in terms of $f^{(r)}$}
\label{sec:4.1.12}

The genus-one analogue of the partial-fraction identity (\ref{pfs}) on the sphere is readily 
formulated in terms of the Kronecker-Eisenstein kernels $f^{(1)}$ in (\ref{gen1.02}) and $f^{(2)}$,
\cite{BrownLevin, Broedel:2014vla}
\begin{align}
& f^{(1)}(z{-}x) f^{(1)}(x{-}y) + f^{(1)}(x{-}y) f^{(1)}(y{-}z) + f^{(1)}(y{-}z) f^{(1)}(z{-}x)
\notag \\
&\quad + f^{(2)}(x{-}y) + f^{(2)}(y{-}z) + f^{(2)}(z{-}x)  = 0
\label{gen1.03}
\end{align}
  The factors $f^{(1)}$ in (\ref{gen1.03})  account for the pole terms in (\ref{gen1.02}), while 
the non-singular $ f^{(2)}$ terms  in the second line compensate for the non-holomorphicity  of the first line. The relation in (\ref{gen1.03}) for $f^{(1)}$ and $f^{(2)}$  and its meromorphic counterpart for $g^{(1)}$ and $g^{(2)}$ are the simplest examples of \textit{Fay identities} \cite{Fay:1973}  for the special case of  genus one.

\sm

At genus one, the Fay trisecant identity relating the meromorphic functions $g^{(1)}$ and $g^{(2)}$ may be derived   via Riemann identities for $\tet$-functions at arbitrary points in the Poincar\'e upper half plane. Similarly, for arbitrary genus, the Riemann identities hold at arbitrary points in the Siegel upper half space. By contrast, the Fay trisecant identity for arbitrary genus \cite{Fay:1973} hold only on the subset of the Siegel upper half space that corresponds to the period matrices of compact Riemann surfaces, referred to as \textit{Torelli space}. The identities derived here similarly hold only on Torelli space, whence we refer to them also as \textit{Fay identities}.

\sm

The Fay identity (\ref{gen1.03}) plays a crucial role in reducing the integrals of elliptic polylogarithms against products of $f^{(1)}$-functions to elliptic polylogarithms again, similarly to  the discussion following (\ref{g0poly}) for the sphere. For example, in an integral over the variable~$z$, the Fay identity (\ref{gen1.03}) allows one to reduce the product $f^{(1)}(y{-}z) f^{(1)}(z{-}x)$, both of whose factors involve $z$, to a sum of terms in which only a single factor is $z$-dependent. We shall refer to this process as \textit{$z$-reduction} and the final expression thus obtained as \textit{$z$-reduced}. In this \textit{$z$-reduced} form, the $z$-integral may now be carried out and produces again elliptic polylogarithms, possibly multiplied by factors $ f^{(r)}(x{-}y)$ with $r \geq 1$.

\sm

Analogous manipulations are needed to \textit{$z$-reduce} more general products $ f^{(r)}(y{-}z) f^{(s)}(z{-}x)$ for arbitrary values of $r,s\geq 1$, namely to express them in terms of a sum of products of Kronecker-Eisenstein kernels with at most one $z$-dependent factor.  This is accomplished by the following generalization of the Fay identity (\ref{gen1.03}) to arbitrary weight \cite{Broedel:2014vla},\footnote{The generating series of (\ref{exfay.16}) and its meromorphic counterpart, namely
\[
\Omega(x{-}z,\alpha_1) \Omega(y{-}z,\alpha_2) =  \Omega(x{-}y,\alpha_1) \Omega(y{-}z,\alpha_1{+}\alpha_2)
+  \Omega(y{-}x,\alpha_2) \Omega(x{-}z,\alpha_1{+}\alpha_2)
\]
and the same identity with $\Omega \rightarrow F$, follow from the Fay trisecant identity for the odd $\vartheta_1$ function via (\ref{exfay.22}). By a  slight abuse of terminology, we shall also refer to the coefficient identities (\ref{exfay.16}) themselves, and their higher-genus generalizations below, as Fay identities.}
\begin{align}
f^{(s)}(x{-}z)f^{(r)}(y{-}z)
&= - (-1)^s f^{(r+s)}(y{-}x)
+ \sum_{\ell=0}^{s} \mbinom{\ell{+}r{-}1}{ \ell}  f^{(s-\ell)}(x{-}y) f^{(r+\ell)}(y{-}z)
\notag \\
&\quad + \sum_{\ell=0}^{r} \mbinom{\ell{+}s{-}1}{ \ell}  f^{(r-\ell)}(y{-}x) f^{(s+\ell)}(x{-}z)
\label{exfay.16}
\end{align}
The \textit{$z$-reduction}  process, which was introduced and illustrated above for the case of genus one, will play a central role throughout this paper and will be defined more generally and more formally for arbitrary genus in section \ref{sec:3-x}. Generalizations of the Fay identities (\ref{exfay.16}) which implement the \textit{$z$-reduction} at arbitrary genus can be found in Theorems \ref{3.thm:7} and \ref{3.thm:8}.

\sm

A generating series for the Fay identities (\ref{exfay.16}) crucially enters the proof that the elliptic polylogarithms of Brown and Levin  are closed under taking primitives \cite{BrownLevin}. The Fay identity (\ref{exfay.16}) drives integration algorithms for the variants of the Brown-Levin elliptic polylogarithms used for genus-one string amplitudes \cite{Broedel:2014vla, Broedel:2017jdo}. First, integrating products of $f^{(s)}(x{-}z) f^{(r)}(y{-}z)$ and elliptic polylogarithms over $z$ necessitates a \textit{$z$-reduction} of the Kronecker-Eisenstein kernels via (\ref{exfay.16}). Second,  preparing these primitives with respect to $z$ for integration over $x$ or $y$ in a later step requires a \textit{change of fibration basis} of the elliptic polylogarithms which is performed through the 
differential equations they satisfy and the Fay identities of their integration kernels \cite{Broedel:2014vla}.
A detailed discussion of  changing fibration bases and explicit results on its implementation at higher genus can be found in section \ref{sec:clo} (also see section \ref{sec:7.9} for a formulation in terms of meromorphic polylogarithms).

\subsubsection{Three-point Fay identities in terms of $g^{(r)}$}
\label{sec:4.1.11}

The Kronecker-Eisenstein kernels $g^{(r)}$ literally satisfy the same Fay identities (\ref{exfay.16}) upon replacing  $f^{(r)}$ by $ g^{(r)}$ in all terms. The definition of \textit{$z$-reduction} straightforwardly carries over from $f^{(r)}$ to $g^{(r)}$. Accordingly, the meromorphic versions of the Fay identities (\ref{gen1.03}) and (\ref{exfay.16}) obtained from replacing $f^{(r)}$ by $ g^{(r)}$  are said to \textit{$z$-reduce} the product $g^{(s)}(x{-}z)g^{(r)}(y{-}z)$.

\sm

 In fact, these algorithms carry over to the meromorphic formulation of elliptic polylogarithms \cite{Broedel:2017kkb, Broedel:2018iwv} (see \cite{Enriquez:2023} for recent work on their closure under taking primitives) upon replacing the single-valued kernels $f^{(r)}$ by their meromorphic counterparts $g^{(r)}$ in (\ref{exfay.22}).

\subsection{Higher-point Fay identities at genus one}
\label{sec:4.1.2}

Similar to the identity (\ref{starg0}) among rational functions of multiple points $x_1,\cdots,x_r$ on the sphere, one can iterate the genus-one Fay identity (\ref{exfay.16}) to rewrite products $\prod_{j=1}^r f^{(k_j)}(z{-}x_j)$ in terms
of \textit{$z$-reduced} combinations of $f^{(r)}$. The genus-one uplift of the cyclic identity (\ref{cycl0}) among 
$((x_1{-}x_2) (x_2{-}x_3) \cdots (x_{r-1}{-}x_r))^{-1}$ may be expressed
in terms of the following elliptic (i.e.\ meromorphic and doubly-periodic) 
functions of $n$ points on the torus \cite{Dolan:2007eh, Tsuchiya:2012nf, Broedel:2014vla},
\beq
V_{w}(1,\cdots,n) = \! \! \! \! \!  \sum_{k_1+k_2+\cdots+k_n=w} \! \! \! \! \!   f^{(k_1)}(x_1{-}x_2)
f^{(k_2)}(x_2{-}x_3) \cdots f^{(k_{r-1})}(x_{r-1}{-}x_r) f^{(k_r)}(x_r{-}x_1)
\label{vids.01}
\eeq
Their special cases with $n=w{+}1$ vanish,
\beq
V_{w}(1,2,\cdots,w{+}1) = 0
\label{vids.02}
\eeq
as one can conveniently check from their generating series \cite{Dolan:2007eh}
or the following inductive argument: The elliptic functions
$V_{w}(1,2,\cdots,n)$ in (\ref{vids.01}) have simple poles in $(x_{j}{-}x_{j+1})$ with
residue $V_{w-1}(1,\cdots,j{-}1,j{+}1,\cdots n)$. 
Hence, the $V_{w}(1,2,\cdots,w{+}1)$ in (\ref{vids.02}) are non-singular
if their lower-weight counterparts $V_{w-1}(1,2,\cdots,w)$ vanish.
With the base case $V_{1}(1,2) = f^{(1)}(x_1{-}x_2) + f^{(1)}(x_2{-}x_1)=0$ 
of (\ref{vids.02}) and the fact that all the $V_{w}(1,2,\cdots,n)$ with $w<n$ vanish
upon integrating $x_1,\cdots,x_n$ over the torus, this leads to an inductive proof of  
(\ref{vids.02}). 

The second non-trivial example $V_{2}(1,2,3) = 0$ of
(\ref{vids.02}) is literally the weight-two Fay identity (\ref{gen1.03}).
At general $w\geq 3$ in turn, (\ref{vids.02}) realizes multiple instances of
higher-weight Fay identities (\ref{exfay.16}) applied to different triplets of
points. From the contribution with $w$ factors of $f^{(1)}(x_{j}{-}x_{j+1})$
to $V_{w}(1,2,\cdots,w{+}1)$, the pole structure of (\ref{vids.02}) is identical to
the genus-zero identity (\ref{cycl0}), namely given by the cyclic orbit of
$((x_1{-}x_2)\cdots (x_{r-1}{-}x_r))^{-1}$ under $x_j \rightarrow x_{j+1}$ with $x_{w+1}=x_1$.

\newpage

\section{The Arakelov Green function and polylogarithms}
\label{sec:2}

In this section, we review some basic ingredients that will enter  the formulation and proof of interchange and Fay identities, including the homology of Riemann surfaces for arbitrary genus $h$, modular transformations, Abelian differentials, the Arakelov Green function, integration kernels, and the construction of polylogarithms in \cite{DHS:2023} from flat connections. Additional details on the construction of the Arakleov Green function via the prime form may be found in Appendix \ref{appdefs} and in \cite{DHoker:2022dxx}.

\subsection{Homology, cohomology and $Sp(2h,\mathbb Z)$ basics}
\label{sec:2.1}

We follow the notation and conventions of \cite{DHS:2023} for the basic ingredients for integration on compact Riemann surfaces $\Sigma$ of arbitrary genus $h$. A canonical basis of 
$H_1(\Sigma,\mathbb Z) \cong \mathbb Z^{2h}$ is spanned by homology cycles $\mathfrak{A}^I$ and 
$\mathfrak{B}_J$ with $I,J=1,2,\cdots,h$ subject to a symplectic intersection pairing 
$\mathfrak{I}(\mathfrak{A}^I , \mathfrak{B}_J) = - \mathfrak{I}(\mathfrak{B}_J , \mathfrak{A}^I)=\delta^I_J$
and $\mathfrak{I}(\mathfrak{A}^I , \mathfrak{A}^J) = \mathfrak{I}(\mathfrak{B}_I , \mathfrak{B}_J)=0$.

\sm

The $h$ Abelian differentials $\bom_I\in H^1(\Sigma,\mathbb Z)$ are normalized on the $\mA$-cycles while the $\mB$-cycles give rise to the components $\Omega_{IJ} = \Omega_{JI}$ of the period matrix $\Omega$, 
\beq
\oint_{\mA^I} \bom_J = \delta^I_J \, ,
\hskip 1in
\oint_{\mB_I} \bom_J = \Omega_{IJ}
\label{bsec.01}
\eeq
The positive definite imaginary part of $\Omega_{IJ} $ and its matrix inverse will be denoted by,
\beq
Y_{IJ}=\Im \Omega_{IJ} \, , \ \ \ \ \ \
Y^{IJ}= \big( (\Im \Omega)^{-1} \big)^{IJ}
\label{bsec.02}
\eeq
and used to raise and lower indices, for instance,\footnote{Here and throughout this work, repeated indices are understood to be summed over unless indicated otherwise, i.e.\ $ Y^{IJ} \bom_J =  \sum_{J=1}^h Y^{IJ} \bom_J$ 
and $Y_{IJ}  \bom^J=   \sum_{J=1}^h Y_{IJ}  \bom^J$. Unless stated otherwise, the dependence on the period matrix of the Abelian differentials, and other functions in the sequel, will be suppressed.} 
\beq
\bom^I = Y^{IJ} \bom_J \, , \ \ \ \ \ \ 
\bom_I = Y_{IJ}  \bom^J
\label{bsec.03}
\eeq
In local complex coordinates $z,\bar z$ on $\Sigma$, we will frequently peel the differential $dz$ off
the Abelian differentials $\bom_I$ and denote the component functions $\omega_I(z)$ in normal font,
\beq
\bom_I = 
\om_I(z) dz \, , \ \ \ \ \ \
\bar \bom^I = 
\bar \om^I(z) d\bar z 
\label{bsec.04}
\eeq
Since the moduli of $\Sigma$ are kept fixed throughout this work,
we do not display the moduli dependence of $\om_I(z)$ and various
later quantities.
With the notation $d^2z = \frac{i}{2} dz \wedge d\bar z$ for the coordinate volume form
the Riemann bilinear relations take the following form,
\beq
\frac{i}{2}\int_\Sigma \bom_I \wedge \bar \bom^J =  \int_\Sigma d^2 z\,  \omega_I(z)  \bar \omega^J(z) = \delta^J_I 
\label{bsec.05}
\eeq
Modular transformations $M \in Sp(2h,\mathbb Z)$ implement changes of canonical $H_1(\Sigma,\mathbb Z)$ bases that preserve the intersection pairing, i.e.\ $M^t \mathfrak{I} M = \mathfrak{I}$ as $2h\times 2h$ matrices. In the notation $A=A_I{}^J, \, B=B_{IJ},\, C=C^{IJ}$ and $D= D^I{}_J$ for the $h\times h$ blocks of $M = (\begin{smallmatrix} A &B \\ C &D\end{smallmatrix})$, the modular transformation of the homology cycles is given by,
\begin{align}
\tilde \mB_I & = A_I{}^J \mB_J + B_{IJ}\mA^J \cr 
\tilde \mA^I & = C^{IJ} \mB_J +  D^I{}_J \mA^J 
\label{bsec.06}
\end{align}
The holomorphic Abelian differentials $\bom$ and their complex conjugates $\bar \bom$, the period matrix $\Omega$, its  imaginary part $Y$, and the inverse of $Y$  transform as follows under $Sp(2h,\ZZ)$,
\begin{align}
\label{bsec.08}
\tilde \bom_I & =  \bom_J \, R(\Omega)^J{}_I  &  
     \tilde \Omega_{IJ} & = (A\Omega+B)_{IK} R(\Omega)^K{}_J 
\no \\  
\tilde {\bar \bom} ^I  & = Q(\Omega) ^I{}_J \bar  \bom^J & 
    \tilde Y_{IJ} & = Y_{KL} \, R(\Omega)^K{}_I \, \overline{ R(\Omega)}^L{}_J
\no \\ &&
\tilde Y^{IJ} & = Y^{KL} \, Q(\Omega)^I{}_K \, \overline{ Q(\Omega)}^J{}_L
\end{align}
where we use the following shorthand for the ubiquitous combination $C \Omega +D$ and its inverse,
\beq
Q(\Omega) = C \Omega+D \, ,
\hskip 1in
R(\Omega) =(C \Omega+D)^{-1}
\label{bsec.07}
\eeq
By raising  and/or lowering indices via contraction with $Y^{IJ}$ and/or $Y_{IJ}$, one can trade transformations via anti-holomorphic factors  $ \overline{ Q(\Omega)}$ and  $\overline{ R(\Omega)}$ for transformations via holomorphic factors $R(\Omega)$ and $ Q(\Omega)$, respectively. For instance, while the anti-holomorphic form $\bar \bom_I$ with lower index transforms by a factor of $ \overline{R(\Omega)}$, its counterpart $\bar \bom^I$ transforms via a factor of $Q(R)$ as shown in the second line on the left of (\ref{bsec.08}). It will be convenient to convert all indices in such a way that their modular transformations are either under $Q(\Omega)$ or $R(\Omega)$, i.e.\ not under their complex conjugates.  A function $ {\cal T}^{I_1\cdots I_r}_{J_1\cdots J_s}$ that depends on  $\Omega$ and  possibly on a number of points on $\Sigma$ and transforms as follows under $Sp(2h,\ZZ)$, 
\beq
\tilde {\cal T}^{I_1\cdots I_r}_{J_1\cdots J_s}(\tilde \Omega)
= Q(\Omega)^{I_1}{}_{K_1}\cdots Q(\Omega)^{I_r}{}_{K_r} \, 
{\cal T}^{K_1\cdots K_r}_{L_1\cdots L_s}(\Omega) \, 
R(\Omega)^{L_1}{}_{J_1} \cdots R(\Omega)^{L_s}{}_{J_s}
\label{bsec.09}
\eeq
is referred to as a \textit{modular tensor}. A tensor of vanishing rank, namely with   $r=s=0$, will be referred to as a \textit{modular scalar}. Siegel modular forms constitute a special case of (\ref{bsec.09}) for which suitable anti-symmetrization of the indices reduces the transformation to multiplication by a power of the determinant $\det(C\Omega+D)$. Modular tensors may be viewed as sections of holomorphic vector bundles on Torelli space $\cT_h$, namely the moduli space of Riemann surfaces with a specified canonical homology basis (see also Appendix \ref{appdefs}).

\subsection{Higher-genus integration kernels}
\label{sec:2.2}

Polylogarithms on higher-genus Riemann surfaces were constructed in \cite{DHS:2023} in terms of 
complex-valued integration kernels $f^{I_1\cdots I_r}{}_J(x,y) $ that depend on the period matrix $\Omega$ and on two points $x,y \in \Sigma$, and transform as modular tensors under $Sp(2h,\mathbb Z)$ in the sense of (\ref{bsec.09}). Their explicit  construction may be carried out in terms of convolutions of Abelian differentials and the Arakelov Green function ${\cal G}(x,z)$ \cite{DHS:2023}, and starts off with the following modular tensor, introduced by Kawazumi in \cite{Kawazumi:lecture, Kawazumi:seminar}, and exploited further in \cite{DHoker:2020uid},
\bea
\label{3.Phi}
\Phi ^I {}_J(x) = \int _\Sigma d^2 z \, \cG(x,z) \, \bar \om^I(z) \om_J(z)
\eea
Modular tensors of higher rank are defined via the following iterated integrals, 
\begin{align}
\Phi^{I_1\cdots I_r}{}_J(x) &=  
\int_{\Sigma} d^2 z \, \cG(x,z) \, \bar \omega^{I_1}(z) \, \p_z \Phi^{I_2\cdots I_r}{}_J(z)
\no \\
\cG^{I_1\cdots I_r}(x,y) & = 
\int_{\Sigma} d^2 z \, \cG(x,z) \, \bar \omega^{I_1}(z) \, \p_z \cG^{I_2\cdots I_r}(z,y) 
\label{fphig.2} 
   \end{align}
where we define $\cG^{I_2 \cdots I_r}(z,y){=} \cG(z,y)$ for $r{=}1$ such that $\cG^{I}(x,y) {=} 
\int_{\Sigma} d^2 z  \cG(x,z)  \bar \omega^{I}(z)  \p_z \cG(z,y)$. Both of $\Phi^{I_1\cdots I_r}{}_J(x) $ and
$\cG^{I_1\cdots I_r}(x,y)$ are complex-valued scalar functions of $x,y \in \Sigma$ and obey the following trace and symmetry relations,
\bea
\Phi ^{I_1 \cdots I_{r-1} J } {}_J(x) & = & 0
\no \\
\cG^{I_1\cdots I_r}(x,y) & = & (-)^r \cG^{I_r \cdots I_1}(y,x)
\label{Gtrcless}
\eea
where the former implies the vanishing of the genus-one restriction $\Phi ^{I_1 \cdots I_{r} } {}_J(x) |_{h=1}$ and the latter is established by successive integrations by parts. The integration kernels $f^{I_1\cdots I_r}{}_J(x,y) $ are defined as follows,
\beq
f^{I_1\cdots I_r}{}_J(x,y) = \p_x \Phi^{I_1\cdots I_r}{}_J(x) -  \p_x \cG^{I_1\cdots I_{r-1}}(x,y) \, \delta^{I_r}_J
\label{fphig.1}
\eeq
They are $(1,0)$ forms in $x$ and $(0,0)$ forms in $y$ and transform as follows under $Sp(2h,\ZZ)$, 
\bea
\tilde f^{I_1\cdots I_r}{}_J(x,y)  
= Q(\Omega)^{I_1}{}_{K_1} \cdots Q(\Omega)^{I_r}{}_{K_r} f^{K_1\cdots K_r}{}_L (x,y) R^L{}_J(\Omega)
\label{ftentrf}
\eea
Combining the definition of (\ref{fphig.1}) with the convolutions of (\ref{3.Phi}) and (\ref{fphig.2}), we get the following formula directly for $f^I {}_J(x,y)$ and the convolution formulas for $f^{I_1\cdots I_r}{}_J(x,y) $ with $r\geq 2$,
\bea
f^I{}_J(x,y) & = & \int _\Sigma d^2 z \, \p_x \cG(x,z) \Big ( \bar \om^I(z) \om_J(z) - \delta (z,y) \delta ^I _J \Big )
\no \\
f^{I_1\cdots I_r}{}_J(x,y) & = & 
\int_{\Sigma} d^2 z \, \partial_x \cG(x,z) \, \bar \omega^{I_1}(z) \, f^{I_2\cdots I_r}{}_J(z,y)
\label{ften.06}
\eea 
Note that the trace $f^{I_1\cdots I_{r-1} J }{}_J(x,y) $  gives $- h \p_x \cG^{I_1\cdots I_{r-1}}(x,y)$ while the traceless part in the two rightmost indices gives $\p_x \Phi^{I_1\cdots I_r}{}_J(x)$, i.e.\ no information is lost in taking the sum~(\ref{fphig.1}). 

\sm

While the Arakelov Green function $\cG(x,y)$  is a conformal scalar in  $x,y$, string theory calculations often make use of the string Green function $G(x,y)$ defined in (\ref{stringgf}) of Appendix~\ref{appdefs} which is \textit{not} a proper conformal scalar in $x,y$ but admits a simple representation in terms of the prime form and Abelian integrals. Using the relation of (\ref{A.GG}), one readily verifies that $\cG(x,y)$ used in the iterative definition of $f^I{}_J(x,y) $ and $f^{I_1\cdots I_r}{}_J(x,y) $ may equivalently be replaced by  $G(x,y)$, as all dependence on their difference cancels out. 

\sm

Finally, we define the \textit{weight} $r$ of a modular tensor to be the minimal number of Green functions $\cG(x,y)$ required to define the tensor. Thus, by this counting, all of $f^{I_1\cdots I_r}{}_J(x,y)$, $\p_x \cG^{I_1\cdots I_{r-1}}(x,y)$ and $ \p_x \Phi^{I_1\cdots I_r}{}_J(x) $  have weight $r$.

\subsection{Anti-holomorphic derivatives}
\label{sec:2.3}

The proofs of the main results in this work will be based on the anti-holomorphic derivatives of the integration kernels $f^{I_1\cdots I_r}{}_J(x,y)$ in (\ref{fphig.1}) and (\ref{ften.06}).
Their $\partial_{\bar x}$ and $\partial_{\bar y}$ derivatives can be traced back to the
Laplace equation of the Arakelov Green function and the $\Phi$ tensor,
\begin{align}
\partial_{\bar x} \partial_x  {\cal G}(x,y) &= 
- \pi \, \delta(x,y)  + \pi \, \kappa(x) 
\notag \\
\partial_{\bar y} \partial_x  {\cal G}(x,y) &=  
\pi \, \delta(x,y)  - \pi \, \omega_I(x) \, \bar \omega^I(y) 
\notag \\
\partial_{\bar x} \partial_x  \Phi^I{}_J(x) &= 
- \pi \, \bar \omega^I(x) \, \omega_J(x) +  \pi \, \delta^I_J \, \kappa(x)
\label{AG.05}
\end{align}
where $\kappa(x)= \om_I(x) \bar \om^I(x)/h$ is the normalized modular and conformally invariant volume form on $\Sigma$ discussed more extensively in Appendix \ref{appdefs}. The above relations  readily imply the following formulas for the derivatives of $f^I{}_J(x,y)$, 
\begin{align}
\partial_{\bar x} f^I{}_J(x,y) &= - \pi \, \bar \omega^I(x) \, \omega_J(x) + \pi \, \delta^I_J \, \delta(x,y) 
\notag \\
\partial_{\bar y} f^I{}_J(x,y) &=  \pi \, \delta^I_J \, \bar \omega^K(y) \, \omega_K(x) - \pi \, \delta^I_J \, \delta(x,y) 
\label{ften.02}
\end{align}
The delta function is normalized by $\int_{\Sigma} d^2 x\, \delta(x,y) = 1$ and reflects the singular behavior,
\beq
\partial_x {\cal G}(x,y) =  - \frac{1}{x{-}y} + \hbox{reg}  \, ,
\hskip 1in
f^I{}_J(x,y) = \frac{\delta^I_J}{x{-}y} + \hbox{reg}
\label{singf1}
\eeq
The analogous anti-holomorphic derivatives at higher weight $r \geq 2$ are given by,
\begin{align}
\partial_{\bar x} f^{I_1\cdots I_r}{}_J(x,y) &= - \pi \bar \omega^{I_1}(x) f^{I_2\cdots I_r}{}_J(x,y)  
 \notag \\
\partial_{\bar y} f^{I_1\cdots I_r}{}_J(x,y) &=  \pi \delta^{I_r}_J
f^{I_1 \cdots I_{r-1}}{}_K(x,y)
 \bar \omega^K(y) 
\label{ften.03}
\end{align}
or equivalently ($s\geq 1$ and $r \geq 2$),
\begin{align}
\partial_{\bar x} \partial_x  {\cal G}^{I_1 \cdots I_s}(x,y) &= 
- \pi \, \bar \omega^{I_1}(x) \, {\cal G}^{I_2 \cdots I_s}(x,y) 
 \notag \\
\partial_{\bar y} \partial_x   {\cal G}^{I_1 \cdots I_s}(x,y) &= 
 \pi  \, \partial_x {\cal G}^{I_1 \cdots I_{s-1}}(x,y)  \, \bar \omega^{I_s}(y)
 - \pi \, \partial_x \Phi^{I_1 \cdots I_s}{}_J(x) \, \bar \omega^J(y)
\notag \\
\partial_{\bar x} \partial_x   \Phi^{I_1 \cdots I_r}{}_J(x) &= 
- \pi \, \bar \omega^{I_1}(x) \, \partial_x \Phi^{I_2 \cdots I_r}{}_J(x)\label{ften.93}
\end{align}
At various intermediate stages in the sequel  another family of modular functions, defined by iterated convolutions, will occasionally enter,  
\bea
\cG_n (x,y) = \int _\Sigma d^2 z \,  \cG(x,z) \, \kappa (z) \, \cG_{n-1} (z,y)
\label{defgns}
\eea
where we set $\cG_1(x,y) = \cG(x,y)$ and $\kappa(z)$ was defined below (\ref{AG.05}). One readily verifies that for $n \geq 2$, they are symmetric $\cG_n (x,y) = \cG_n(y,x)$ and satisfy the following Laplace equation, 
\bea
\label{ften.77}
\pbx \p_x \cG_n(x,y) = - \pi \, \kappa (x) \, \cG_{n-1}(x,y)
\eea
As we will see in (\ref{papag2}), the $\p_x  \p_ y$ derivatives of the $\cG_2(x,y)$ function are ultimately expressible in terms of $f$-tensors and Abelian differentials of total weight $2$ with all of their indices contracted.

\subsection{Polylogarithms  via a flat connection}
\label{sec:3.DHS}

The modular tensors $f^{I_1 \cdots I_r}{}_J(x,y)$ may be used to construct a flat connection and associated polylogarithms on a compact Riemann surface $\Sigma$ of arbitrary genus $h \geq 1$, which generalize the genus-one non-holomorphic polylogarithms of Brown and Levin in \cite{BrownLevin}.

\subsubsection{The flat connection $\cJ_\text{DHS}$}

    To do so, we introduce a Lie algebra $\mg$ that is freely generated by $2h$ non-commutative elements denoted by $a^I$ and $b_I$ for $I=1, \cdots, h$. In addition, we construct a $\mg$-valued  connection $\cJ_{\text{DHS}}(x,p)$ on the punctured Riemann surface $\Sigma _p = \Sigma \setminus \{ p \}$, given by \cite{DHS:2023},\footnote{The generators denoted by $a^I$ below were denoted by $\hat a^I = a^I + \pi Y^{IJ} b_J$ in \cite{DHS:2023}.}
\bea
\label{3.conn}
\cJ_{\text{DHS}} (x,p) = - \pi \, \bar \bom ^I (x) b_I + \bom_J(x) a^J + \sum _{r=1}^\infty dx \, f^{I_1 \cdots I_r}{}_J(x,p) B_{I_1} \cdots B_{I_r} a^J
\eea
where $B_I$ is a derivation in $\mg$ that generates the adjoint action $B_I X = [b_I, X]$ for any $X \in \mg$. 
The connection $\cJ_{\text{DHS}} (x,p) $ is a differential form of type $(1,0) \oplus (0,1)$ in $x$ and a scalar in~$p$, the $(0,1)$ part being generated solely by the first term in (\ref{3.conn}). Using the closure of the forms  $\bom_J$ and $\bar \bom^I$, and the anti-holomorphic derivatives of $f$ given in (\ref{ften.02}) and (\ref{ften.03}), one readily shows that the connection $\cJ_{\text{DHS}} (x,p)$ satisfies the Maurer-Cartan equation, 
\bea
d_x \cJ_{\text{DHS}} (x,p) -  \cJ_{\text{DHS}} (x,p) \wedge \cJ_{\text{DHS}} (x,p) = 
\pi \, d\bar x  \wedge dx \, \delta(x,p) \, [b_I,a^I] 
\label{flatDHS}
\eea
and is therefore a flat connection, away from the singular point $p$. The $\delta(x,p)$
distribution in (\ref{flatDHS}) reflects the singular terms
\bea
\cJ_{\text{DHS}} (x,p) = { dx \over (x{-}p)} \, [b_I,a^I] + {\rm reg}
\eea 
following from the simple
poles of $f^I{}_J(x,p)$ in (\ref{singf1}) (regularity of $f^{I_1\cdots I_r}{}_J(x,p)$ with $r\geq 2$ 
throughout $\Sigma$ is implicit in (\ref{ften.03})).
In view of the modular transformation laws under $Sp(2h,\ZZ)$ of $\om_I, \bar \om^J$ given in (\ref{bsec.08}), and $f^{I_1 \cdots I_r}{}_J(x,p)$ given in (\ref{ftentrf}), the connection $\cJ_\text{DHS}$ will be invariant under $Sp(2h,\ZZ)$ provided the generators $a^I$ and $b_I$ transform as follows (see (\ref{bsec.07}) for $Q(\Omega) $ and $R(\Omega) $),
\bea
\label{3.abtrans}
\tilde a^I = Q(\Omega)^I{}_J \, a ^J\, , \hskip 1in \tilde b_I = b_J \, R(\Omega) ^J{}_I
\eea
Restricting to genus one and redefining $a \rightarrow a+\pi b / \Im \tau$ produces the non-holomorphic connection of \cite{BrownLevin} valued in a Lie algebra freely generated by two elements $a,b$.

\subsubsection{Polylogarithms from the flat connection $\cJ_\text{DHS}$}

Flatness of $\cJ_{\text{DHS}} (x,p) $ guarantees that the differential equation, 
\bea
d_x \, \mathbf{\Gamma} (x,y;p)  = \cJ_{\text{DHS}} (x,p) \,   \mathbf{\Gamma} (x,y;p)
\eea
is integrable. Its solution, subject to the initial condition $\mathbf{\Gamma} (y,y;p)  = I$, is valued in the Lie group of $\mg$ and may be represented by the path-ordered exponential, 
\bea
\label{3.int}
\mathbf{\Gamma} (x,y; p) = \text{P} \exp \int _y^x \cJ_{\text{DHS}} (z,p)
\eea 
which is considered here for $p\neq x,y$ (it would otherwise require 
a regularization prescription for endpoint divergences, see the preamble of section \ref{sec:clo}) and satisfies the composition law,
\bea
\label{3.comp}
\mathbf{\Gamma} (x,y; p) = \mathbf{\Gamma} (x,z; p) \, \mathbf{\Gamma} (z,y; p) 
\eea
The multiplication on the right side is understood to be that of the Lie group of $\mg$.  Flatness of $\cJ_{\text{DHS}} (x,p)$ also guarantees that $\mathbf{\Gamma} (x,y; p)$ is \textit{homotopy invariant}, namely that  its value only depends on the homotopy class of paths used to integrate from $y$ to $x$ but is independent of the representative path chosen within a given homotopy class.

\sm

Higher-genus polylogarithms are obtained by expanding $\mathbf{\Gamma} (x,y; p)  $ in \textit{words} $\mw$ consisting of a finite number of letters in the alphabet made up of the letters $a^J$ and $b_I$ for $I,J = 1 , \cdots , h$. 
The set $\cW_{ab}$ of all words in the alphabet of letters $a^I, b_I$ closes under the associative concatenation product, has the empty word $\emptyset$ as its neutral element, and is thereby a monoid.  This expansion requires working in the enveloping algebra of $\mg$ and takes the form,
\bea
\mathbf{\Gamma} (x,y; p) =  \sum_{\mw \in \cW_{ab} } \mw \,  \Gamma (\mw; x,y;p)
\label{Gacoeff}
\eea 
where the sum is over all different words $\mw$ in $\cW_{ab}$,  including the empty word with $\Gamma (\emptyset; x,y;p)=1$. For a given word $\mw$, the function $\Gamma (\mw; x,y;p)$ is homotopy invariant and referred to as a \textit{higher-genus polylogarithm}. Since $\cJ_\text{DHS}(x,p)$ is modular invariant, so is $\boldsymbol{\Gamma}(x,y;p)$, and the  polylogarithms $\Gamma (\mw; x,y;p)$ are modular tensors.  In section \ref{sec:Jmv} below, we shall generalize these polylogarithms to depend on an arbitrary number of variables.

\sm

The integral representation of their generating series (\ref{3.int}) implies  the following shuffle product rule (see section \ref{sec:fay0} for properties of the shuffle product) on polylogarithms, 
\bea
\Gamma (\mw_1; x,y;p) \, \Gamma (\mw_2; x,y;p) = \sum_{\mw \in \mw_1 \shuffle \, \mw_2} \Gamma (\mw; x,y;p) 
\label{Gashuff}
\eea
The polylogarithm $\Gamma (\mw; x,y;p)$ for a word $\mw$ of length $\ell$, may be calculated by expanding the path-ordered integral of (\ref{3.int}) in powers of $\cJ_\text{DHS} $ (with $t_0=x$ in the $n=1$ term),
\beq
\label{3.expandP}
\boldsymbol{\Gamma} (x,y;p) = 1 
+ \sum _{n=1}^\infty \int _y ^x   \cJ_\text{DHS} (t_1,p) \int _y ^{t_1}   \cJ_\text{DHS} (t_2,p)
\cdots \int _y ^{t_{n-1}}   \cJ_\text{DHS} (t_n,p)
\eeq
retaining only the terms with $n \leq \ell$, and projecting onto the contributions for the word $\mw$. Note that  the polylogarithm $\Gamma (\mw; x,y;p)$ for a word $\mw$ of length $\ell$ will generically receive contributions from all $n \leq \ell$. 

\subsubsection{Examples}

The simplest examples correspond to words composed of the letters~$a^J$ only, or of the letters $b_I$ only. They admit the following expressions,\footnote{In the conventions of (\ref{3.conn}) for $\cJ_{\text{DHS}} (z,p)$, the coefficients $\Gamma (\mw; x,y;p)$ of words in $a^J$ and $b_I$ defined by (\ref{Gacoeff}) were denoted by $\hat \Gamma (\mw; x,y;p)$ in \cite{DHS:2023}.}
\bea
\Gamma ( a^{J_1}  a^{J_2} \cdots a^{J_r}; x,y;p) & = & 
\int ^x _y dt_1 \, \om_{J_1}(t_1)  \int ^{t_1} _y dt_2 \, \om_{J_2}(t_2)  \cdots \int ^{t_{r-1}} _y dt_r \, \om_{J_r}(t_r)
\label{abints} \\
\Gamma ( b_{I_1} b_{I_2} \cdots b_{I_r}; x,y;p) & = & 
(- \pi)^r \int ^x _y d \bar t_1 \, \bar \om^{I_1} (t_1)  \int ^{t_1} _y d\bar t_2 \, \bar \om^{I_2} (t_2)  \cdots 
\int ^{t_{r-1}} _y d\bar t_r \, \bar \om^{I_r} (t_r)
\no
\eea
Both are homotopy-invariant, independent of $p$ and multiple-valued in $x,y$. The first is holomorphic in $x,y$ while the second is anti-holomorphic in $x,y$. 

\sm

Polylogarithms corresponding to words that involve both letters $a^J$ and $b_I$, however,  feature sums of iterated integrals, each of which generically fails to be homotopy-invariant by itself. Thus, carrying out the expansion in (\ref{3.expandP})  requires one to define all integrals to be evaluated along the same path.
Only when all contributions to the polylogarithm are combined will the dependence on the
 choice of representative for a given homotopy class of paths cancel out.
We illustrate this mechanism for the simplest non-trivial case where the word is $\mw = b_I a^J$, and we obtain, 
\bea
\Gamma ( b_I a^J; x,y;p) = \int ^x _y dt \, f^I{}_J(t,p) - \pi \int ^x _y d \bar t \, \bar \om^I(t) \int ^t _y dt' \, \om_J(t')
\label{gamba}
\eea 
Neither integral on the right side is homotopy-invariant, and their separate evaluation requires  specifying a path of integration from $y$ to $x$ along which the point $t'$ also takes values. To see that $\Gamma ( b_I a^J; x,y;p) $ is path-independent within a given homotopy class of paths on the punctured surface $\Sigma_p$\footnote{The simple pole $f^I{}_J(t,p)= \delta^I_J/(t{-}p)+{\rm reg}$ causes $\Gamma ( b_I a^J; x,y;p)$ to change by integer multiples of $2\pi i \delta^I_J$ once the homotopy class of the path from $x$ to $y$ is modified by loops around the singular point $p$.} we recast the integral in the following form, 
\bea
\Gamma ( b_I a^J; x,y;p) = \int ^x _y  \nu^I{}_J(t,p)  
\eea 
where the $(1,0) \oplus (0,1)$ form $\nu ^I{}_J(t,p)$ is given by,
\bea
\nu^I{}_J(t,p) = dt \, f^I{}_J(t,p) - \pi d \bar t \, \bar \om^I(t) \int ^t _y dt' \, \om_J(t')
\eea
The integral defining $\Gamma ( b_I a^J; x,y;p)$ is homotopy-invariant because the form $\nu^I{}_J(t,p)$ is closed with respect to $t$ and satisfies $d_t \nu^I{}_J(t,p)=0$ for $t \not=p$ in view of the first equation in (\ref{ften.02}). Note that special values $p=x$ or $p=y$ give rise to
endpoint divergence whose regularizations can for instance be 
approached via tangential base points \cite{Deligne:1989, Panzer:2015ida, Abreu:2022mfk}.

\sm

Similarly, polylogarithms $\Gamma (\mw;x,y;p)$ for longer words containing both letters of type $a^J$ and $b_I$ may be obtained by expanding the path-ordered exponential of (\ref{3.int}) to higher order, and collecting all contributions with the same word $\mw$. Individual iterated integrals in the expansion are of the form, 
\begin{align}
 \int ^x _y dt_1 \, f^{I_1\cdots I_r}{}_{J}(t_1,p)
   \int ^{t_1} _y dt_2 \, f^{K_1\cdots K_s}{}_{L}(t_2,p) \cdots
    \int ^{t_{m-1}}_y dt_m \, f^{P_1\cdots P_u}{}_{Q}(t_m,p) 
\label{nhomotop}
\end{align}
multiplied by the coefficient,
\beq
[ b_{I_1},[ \cdots ,[b_{I_r} ,a^{J}] \cdots ]] 
 \, [b_{K_1},[ \cdots ,[b_{K_s}, a^{L} ] \cdots ]] \cdots 
 [ b_{P_1} ,[ \cdots ,[b_{P_u}, a^{Q} ] \cdots]]
\eeq
Each of these individual iterated integrals in (\ref{nhomotop}) fails to be homotopy-invariant, but the flatness of the connection (\ref{flatDHS}) guarantees  that  (\ref{nhomotop}) is always accompanied by a tail of additional
path-dependent integrals involving lower-weight $f$-tensors, that eventually render a polylogarithm such as
$\Gamma ( b_{I_1} \cdots b_{I_r} a^{J} b_{K_1} \cdots b_{K_s} a^{L} \cdots  b_{P_1} \cdots b_{P_u} a^{Q} ; x,y;p)$ 
and all the other  higher-genus polylogarithms homotopy invariant.

\subsection{Polylogarithms in multiple variables via a flat connection}
\label{sec:Jmv}

Applications of polylogarithms to quantum field theory and string theory necessitate generalizations of the polylogarithms discussed in the previous subsection to multiple variables, namely dependent on several points $p_i \in \Sigma$ for $i =1, \cdots, n$.  An explicit construction of such polylogarithms at arbitrary genus as provided in \cite{DHS:2023} will now be reviewed. 
Their construction requires enlarging the Lie algebra $\mg$ to a Lie algebra $\mg_c$ which is freely generated by $a^I, b_I$ for $I=1,\cdots , h$ and one extra generator $c_i$ for $i=1,\cdots, n$ per additional point $p_i$.  The corresponding multi-variable connection $\cJ_\text{mv} (x,p; p_1, \cdots, p_n )$ introduced in  \cite{DHS:2023} is,
\bea
\cJ_\text{mv} (x,p; p_1, \cdots, p_n) = 
\cJ_\text{DHS} (x,p) + \sum_{i=1}^n dx \Big (\cH(x,p;B)- \cH(x,p_i;B)  \Big ) c_i
\eea
where $\cH $ is given by,
\bea
\cH(x,y;B) = 
 \p_x\cG(x,y) + \sum_{r=1}^\infty \p_x \cG^{I_1 \cdots I_r} (x,y) B_{I_1} \cdots B_{I_r}
\eea
or may alternatively be expressed solely in terms of the integration kernels $f$, 
\bea
\cH(x,y;B)  = - { 1 \over h} f^J{}_J(x,y) - {1 \over h} \sum _{r=1}^\infty f^{I_1 \cdots I_r J}{}_J(x,y) B_{I_1} \cdots B_{I_r}
\eea
The connection $\cJ_\text{mv} (x,p; p_1, \cdots, p_n)$ is flat away from the points $p$ and $p_i$, as may be verified by evaluating its curvature form,
\bea
d_x \cJ_\text{mv} - \cJ_\text{mv} \wedge \cJ_\text{mv} 
=
\pi d \bar x \wedge dx \bigg  ( \delta (x,p) [b_I, a^I] + \sum_{i=1}^n c_i \Big ( \delta (x,p_i) - \delta (x,p) \Big ) \bigg  )
\eea
The connection $\cJ_\text{mv}$ is modular invariant under $Sp(2h,\ZZ)$, provided that $\om_I$ and $ \bar \om^J$ transform as in (\ref{bsec.08}), $f^{I_1 \cdots I_r}{}_J(x,p)$ as in (\ref{ftentrf}), the generators $a^I$ and $ b_I$ as in (\ref{3.abtrans}), and  the generators $c_i$ as scalars. The connection $\cJ_\text{mv}$ reduces to the multi-variable Brown-Levin connection \cite{BrownLevin} upon restricting to genus $h=1$. 

\sm

Higher-genus polylogarithms in multiple variables may now be defined in analogy with the case of polylogarithms of a single variable, where the connection $\cJ_\text{DHS}$ and the Lie algebra $\mg$ 
in the expansion of the path-ordered exponential in (\ref{3.int}) and (\ref{Gacoeff})
are now adapted to $\cJ_\text{mv}$ and $\mg_c$, respectively, 
\bea
\boldsymbol{\Gamma} (x,y;p;p_1, \cdots , p_n) & = & 
\text{P} \exp \int _y^x  \cJ_\text{mv} (t,p; p_1, \cdots, p_n) 
\no \\ & = &
 \sum_{\mw \in \cW_{abc} } \mw \,  \Gamma (\mw; x,y;p; p_1,\cdots,p_n)
 \label{expmv}
\eea
This expansion assigns a multi-variable polylogarithm $ \Gamma (\mw; x,y;p; p_1,\cdots,p_n)$ to each $\mw$ in 
$\cW_{abc}$ composed of all possible letters in the alphabet $\{ a^1, \cdots , a^h, b_1 , \cdots b_h, c_1 \cdots c_n \}$.  The resulting multi-variable polylogarithms are homotopy-invariant upon complete assembly of all contributions to a given word $\mw$ and depend only on the homotopy class of the path taken from $x$ to $y$ on the punctured surface $\Sigma \setminus \{ p, p_1, \cdots , p_n\}$. Moreover, products of multi-variable polylogarithms with the same endpoints $x,y$ of their integration path satisfy the same shuffle  relations (\ref{Gashuff}) noted in the single-variable case, implying their closure under multiplication.
 
 \sm 
 Simple examples of multi-variable polylogarithms which depend non-trivially on an extra point $p_1$ include,
 \begin{align}
 \Gamma (c_1; x,y;p; p_1) &= \int^x_y dt \, \big(  \p_t \cG(t,p ) -  \p_t \cG(t,p_1 )\big)
 \label{examv} \\
  \Gamma(a^K c_1;x,y;p;p_1) &=  \int^x_y d  t \, \omega_K(t)
 \int^{t}_y dt' \, \big(  \p_{t'} \cG(t',p ) -  \p_{t'} \cG(t',p_1 )\big)
  \notag 
 \\
  \Gamma (b_I c_1; x,y;p; p_1) &= \int^x_y dt \, \big(  \p_t \cG^I(t,p ) -  \p_t \cG^I(t,p_1 )\big) \notag 
 \\
  &\quad
  - \pi  \int^x_y d\bar t \, \bar \omega^I(t) \int^{t}_y dt' \, \big(  \p_{t'} \cG(t',p ) -  \p_{t'} \cG(t',p_1 )\big)
 \notag
\end{align}
 where the homotopy invariance of the third example does not hold for the individual terms
 and is tied to their special linear combination selected by the expansion of (\ref{expmv}), also see
the discussion below (\ref{gamba}) for a single-variable analogue.

\subsection{Definition of $z$-reduced}
\label{sec:3-x}

Besides their intrinsic interest, the Fay identities will serve to carry out fundamental reductions in the construction of polylogarithms that lead to their closure under addition, multiplication, and taking primitives. To organize these reductions, we generalize the notion of  \textit{$z$-reduced}, introduced informally for genus one in section \ref{sec:4.1.1}, to arbitrary genus.

\sm

We shall present the definition here in the non-meromorphic context, and defer the minor modifications needed for its adaptation to the meromorphic case to section \ref{sec:7}. Informally, a sum of products of tensors $f$ is \textit{$z$-reduced} if it can be expressed as a linear combination of tensors $f(z,y)$ or $f(y,z)$ with coefficients that are independent of $z$.

\sm

More formally, the building blocks of the connection $\cJ_\text{DHS}$ of section \ref{sec:3.DHS} and its generalization $\cJ_\text{mv}$  to multiple points $z_1, \cdots , z_N$ are given by  the differential forms,\footnote{In this subsection, we shall denote the points $x,p$ and $p_1, \cdots, p_n$ involved in the connections $\cJ_\text{DHS}$ and $\cJ_\text{mv}$ by $z_1, \cdots , z_N$ with $N=n{+}2$ in order to stress the generality of the definition of \textit{$z$-reduction}.}
\bea
\label{3.forms}
\bom_I(z_i)\,, \qquad \bar \bom^I(z_i)\,, \qquad 
\boldsymbol{f}^{I_1 \cdots I_r}{}_J(z_i, z_j)= f^{I_1 \cdots I_r}{}_J(z_i, z_j) dz_i
\eea
for $i,j=1,\cdots , N$ and all possible values of $r\geq 0 $ and $I, I_1, \cdots I_r, J=1,\cdots ,h$ (setting $\boldsymbol{f}^{\emptyset}{}_J(z_i, z_j)= \bom_J(z_i)$). They generalize the forms $dz_i/(z_i{-}z_j)$ at genus zero and the forms $dz_i$, $d\bar z_i$,  $f^{(r)}(z_i{-}z_j)dz_i$ at genus one. Here and below, we are assuming that the points are non-coincident, namely $z_i \not = z_j$ for $i \not= j$. The differential forms of (\ref{3.forms}),  together with the two-forms obtained by applying the  total differential $d_j = dz_j \p_{z_j} + d\bar z_j \p _{\bar z_j}$ to $\boldsymbol{f}$, 
\bea
d_j \boldsymbol{f}^{I_1 \cdots I_r}{}_J(z_i, z_j) = - \p_{z_j} f^{I_1 \cdots I_r}{}_J(z_i, z_j) dz_i \wedge dz_j
- \pi \delta ^{I_r}_J  \boldsymbol{f}^{I_1 \cdots I_{r-1} }{}_K(z_i, z_j) \wedge \bar \bom^K(z_j)
\quad 
\eea
generate an algebra $\cA_N$ of differential forms in $N$ variables whose multiplication is the exterior product of  differential forms.   By construction, the algebra $\cA_N$ is closed under addition,  under exterior product multiplication and under total differentiation by $d_j$. This is clear for $\bom_I(z_i)$ and $\bar \bom^I(z_i)$ and holds true for the forms $\boldsymbol{f}$ and $d_j \boldsymbol{f}$  thanks to the relations (\ref{ften.02}), which we recast here in terms of differentials, 
\bea
d_i \boldsymbol{f}^{I_1 \cdots I_r}{}_J(z_i, z_j) & = & 
- \pi \bar \bom^{I_1} (z_i) \wedge \boldsymbol{f} ^{I_2 \cdots I_r}{}_J(z_i,z_j)
\no \\
d_i \big ( d_j \boldsymbol{f}^{I_1 \cdots I_r}{}_J(z_i, z_j) ) & = & 
- \pi \bar \bom^{I_1} (z_i) \wedge d_j \boldsymbol{f} ^{I_2 \cdots I_r}{}_J(z_i,z_j)
\eea
Note that wedge products of the form $\boldsymbol{f}^{I_1 \cdots I_r}{}_J(z_i, z_j)  \wedge \boldsymbol{f}^{K_1 \cdots K_s}{}_L(z_i, z_k)$ that share their first point $z_i$ vanish identically.

\sm

An arbitrary element $\boldsymbol{\phi}(z_1, \cdots, z_N) \in \cA_N$ is defined to be \textit{$z_i$-reduced}, for a given value of $i \in \{1,\cdots, N\}$,  if it is a linear combination of $z_i$-independent terms and those generators of the algebra $\cA_N$  that depend on $z_i$,  with coefficients that are independent of $z_i$. More explicitly, $\boldsymbol{\phi}(z_1, \cdots, z_N)$ is \textit{$z_i$-reduced} if its $z_i$-dependent parts are a linear combination of the  differential forms $\bom_I(z_i)$, $\bar \bom^I(z_i)$, $\boldsymbol{f}^{I_1 \cdots I_r}{}_J(z_i,z_j)$, $\boldsymbol{f}^{I_1 \cdots I_r}{}_J(z_j,z_i)$, $d_j \boldsymbol{f}^{I_1 \cdots I_r}{}_J(z_i, z_j)$ and $d_i \boldsymbol{f}^{I_1 \cdots I_r}{}_J(z_j, z_i) $ with $z_i$-independent coefficients, and arbitrary assignments of the indices $I,J, I_1, \cdots I_r$. The process of obtaining the \textit{$z_i$-reduced} form of an element in $\cA_N$ will be referred to as \textit{$z_i$-reducing} or \textit{$z_i$-reduction}. 

\sm 

The Fay identities in section \ref{sec:fay} will perform the \textit{$z_i$-reduction} for coefficients
$f^{I_1 \cdots I_r}{}_J(z_i,z_j)$ of the above differentials $dz_i$. For instance, Theorem \ref{3.thm:7} provides
the \textit{$z$-reduced} form of products $f^{P_1\cdots P_s M}{}_J(y,z) f^{I_1\cdots I_r J}{}_K(x,z)$, written in terms of bilinears of the schematic form $f(y,z) f(x,y)$ and $f(y,x) f(x,z)$ with no more than one $z$-dependent factor.
Given that the tensors $\p \Phi$ and $\p \cG$ may be obtained from the trace and traceless part of the kernels $f^{I_1\cdots I_r}{}_J(x,y)$ via (\ref{fphig.1}), the definitions of  \textit{$z$-reducing} apply to the products involving $\p \Phi$ and $\p \cG$ as well.

\newpage

\section{Scalar prototypes of higher-genus Fay identities}
\label{sec:4}

The simplest higher-genus Fay identities involving three points $x,y,z$ will be modeled on the relation between rational functions  in (\ref{pfs}) and doubly periodic  functions  in (\ref{gen1.03}) for genus zero and genus one, respectively. In both cases, the points $x,y,z$ enter on an equal footing, as the relations may be viewed as scalars in $x,y,z$, and invariant under cyclic permutations of $x,y,z$. On a Riemann surface of higher genus, however, it is the derivative of the Arakelov Green function $\p_x \cG(x,y)$ that exhibits a simple pole, as shown in (\ref{singf1}). The fact that $\p_x \cG(x,y)$ is a $(1,0)$ form in $x$ and a $(0,0)$ form in $y$ creates an asymmetry between the dependences on $x$ and $y$.  It is not hard to see that  the generalization of the Fay identity for three points to higher genus  cannot be cyclically symmetric in the points $x,y,z$, but rather must be a $(1,0)$ form in two of the points and a $(0,0)$ form in the other point. 

\sm

To exhibit this structure and its implications in the simplest possible setting first, we  begin with a discussion of the higher-genus Fay identity in three points for modular scalars. An immediate extension to scalar Fay identities in an arbitrary number of points can be found in section \ref{sec:4.3.1}, and the more comprehensive generalizations to tensorial Fay identities at arbitrary rank and weight are discussed in section \ref{sec:fay}.

\subsection{The modular scalar Fay identity in three points}

A natural Ansatz for a sum of products  of the derivative of the Arakelov Green function that contains the pole terms of (\ref{pfs}) is provided by the following combination,
\bea
\p_x \cG(x,y) \p_y \cG(y,z) + \p_y \cG(y,x) \p_x \cG(x,z) -  \p_x \cG(x,z) \p_y \cG(y,z)
\eea
which we choose to be a $(1,0)$ form in $x$ and $y$ and a $(0,0)$ form in $z$. Applying the $\bar \p$ operator to this combination in $x,y,z$ using (\ref{AG.05}) reveals that it is not holomorphic and therefore cannot vanish. This situation is familiar from the corresponding identity in (\ref{gen1.03}) for doubly periodic functions at genus one in which contributions from weight-two functions $f^{(2)}$ were required. Similar contributions are required also here, and the result may be summarized by the following theorem.
{\thm
\label{4.thm:7}
The three-point Fay identity that is a scalar under modular transformations states that the following combination, 
 which is a $(1,0)$ for in $x,y$ and a $(0,0)$ form in $z$,
\begin{align}
F_3(x,y,z) & =  \p_x \cG(x,y) \, \p_y \cG(y,z) + \p_y \cG(y,x) \, \p_x \cG(x,z) -  \p_x \cG(x,z) \, \p_y \cG(y,z)
\notag \\
&\quad - \om_I (x) \, \p_y \cG^I(y,z) - \om_I(y) \, \p_x \cG^I(x,z) + \p_x \p_y \cG_2(x,y)
\label{3.a.2}
 \end{align}
vanishes identically on a Riemann surface $\Sigma$ of arbitrary genus,  
 \beq 
F_3(x,y,z) =0
\eeq 
Recall that the ingredients of (\ref{3.a.2}) were defined in section \ref{sec:2}, and we will see in section \ref{sec:3} that the last term $ \p_x \p_y \cG_2(x,y) $ may equivalently be expressed solely in terms of $f$ and~$\Phi$.}

\subsection{Method of proof}
\label{prfsec.1}

The proof of Theorem \ref{4.thm:7}  follows the same method that will be used throughout this work to demonstrate the vanishing of certain single-valued modular tensors. For this reason the method of proof presented below is structured so that it applies to the proof of Theorem \ref{4.thm:7} as well as to the proofs of many results in the sequel. For simplicity, we consider the case where the identity involves three points $x,y,z$ on an arbitrary compact Riemann surface $\Sigma$, the case of additional points being  a straightforward generalization of the three-point case.

\sm

We consider a sequence of modular scalars or modular tensors ${\cal T}_{(n)}(x,y;z)$  (tensor indices will be suppressed throughout this subsection) labeled by a non-negative integer $n$ indicating their weight in the sense of section \ref{sec:2.2}. The sequence may have a finite or an infinite number of elements, and each element ${\cal T}_{(n)}(x,y;z)$ is a polynomial in the integration kernels $f$, single-valued in $x,y,z$, and assumed to be a $(1,0)$ form in $x$ and $y$ and a scalar in $z$. We shall assume that the relation $\cT_{(0)}(x,y;z)=0$ has been established to hold. The proof of a sequence of identities for $n \geq 1$ of the form,
\bea
\cT_{(n)} (x,y;z)=0
\eea
proceeds via the following two steps. 
\begin{itemize}
\item[1.] 
First, one  proves that the anti-holomorphic derivatives of $\cT_{(n)} (x,y;z)$ in $x,y$ and $z$ all vanish when 
$\cT_{(m)} (x,y;z)=0$ for all $m$ in the range $0 \leq m < n$,
\bea
\left. \begin{array}{r} \pbx \cT_{(n)} (x,y;z) \equiv 0 
\\  \pby \cT _{(n)} (x,y;z) \equiv 0 
\\ \pbz \cT_{(n)} (x,y;z) \equiv 0 \end{array}\ \right\} \hskip 1in {\rm mod} ~ \big \{ \cT_{(m)} =0 , \, 0 \leq m<n \big \}
\eea
using the differential equations in section \ref{sec:2.3}. Holomorphicity in $x,y,z$ implies that  ${\cal T}_{(n)}(x,y;z)$ is independent of $z$ (since it is a scalar in $z$) and can be expanded in a basis of holomorphic $(1,0)$ forms in $x$ and $y$ as follows, 
\bea
\cT_{(n)} (x,y;z) =  \omega_K(x) \, \omega_L(y) \, T^{KL}_{(n)}
\eea
for an $x,y$ independent  modular tensor $T^{KL}_{(n)}$.
\item[2.] 
Second, one proceeds to verify that $\cT_{(n)} (x,y;z)$ integrates to zero against a basis of antiholomorphic $(0,1)$ forms in $x$ and $y$, 
\bea
\int_{\Sigma} d^2x\, \bar \omega^K(x) \int_{\Sigma} d^2y \,\bar \omega^L(y)  {\cal T}_{(n)} (x,y;z)
&= &  T^{KL }_{(n)} =0
 \label{prfstrat}
\eea
Establishing the vanishing of these integrals is greatly facilitated by the fact that many terms in ${\cal T}_{(n)} (x,y;z)$ are total derivatives of a single-valued function in $x$ or $y$, or both. Note in particular that, by virtue of (\ref{ften.06}), the tensors $f^{I_1\cdots I_r}{}_J(x,y)$ and therefore also $\p_x \Phi^{I_1\cdots I_r}{}_J(x) $ and $\p_x \cG^{I_1\cdots I_{r-1}}(x,y) $ are total derivatives in $x$ at arbitrary rank $r\geq 1$.
\end{itemize}

\subsection{Proof of Theorem \ref{4.thm:7}}
\label{prfsec.2}

Let us now apply the two steps of the previous section to prove the
vanishing of $F_3$ in (\ref{3.a.2}).
\begin{itemize}
\item[1.] One first verifies that the $\partial_{\bar x},\partial_{\bar y},\partial_{\bar z}$ derivatives vanish. Holomorphicity in $x,y$  follows outright from (\ref{AG.05}),  (\ref{ften.93}) and (\ref{ften.77}). However, the $\pbz$ derivative of $F_3$,
\begin{align}
\partial_{\bar z}F_3(x,y,z) = \pi \bar \omega^K(z)
\Big(
\omega_I(x) \partial_y \Phi^I{}_K(y) - \omega_K(x) \partial_y {\cal G}(y,x) + (x\leftrightarrow y)
\Big)
\end{align}
gives rise to a particular weight-one combination in the parenthesis which vanishes by the
interchange identity of  (\ref{preften.08}). The latter was already demonstrated in \cite{DHoker:2020tcq, DHoker:2020uid} and is reviewed in more detail in section \ref{sec:3}.
Therefore, $F_3(x,y,z)$ must be independent on $z$, a holomorphic $(1,0)$ form in $x,y$, and admit an expansion $ F_3(x,y,z)=  \omega_K(x) \omega_L(y) F_3^{KL}$ with a modular tensor $F_3^{KL}$ independent on $x,y$.
\item[2.] 
Second, one verifies that $F_3$ integrates to zero against
$\int_{\Sigma} d^2x\, \bar \omega^K(x) \int_{\Sigma} d^2y \,\bar \omega^L(y) $ to show that
 $F_3^{KL}=0$. The vanishing of the integral over $x$ is manifest from the first, third, fifth and sixth term on the right side of (\ref{3.a.2}) since each one of these terms is a total derivative of single-valued functions in $x$.
 Similarly, the second, third, fourth and sixth terms in (\ref{3.a.2}) are total derivatives of single-valued functions in $y$ and integrate to zero against $ \int_{\Sigma} d^2y \,\bar \omega^L(y) $.
\end{itemize}

\subsubsection{Comments on Theorem \ref{4.thm:7}}
\label{prfsec.3}

Although the Fay identity (\ref{3.a.2}) is a $(1,0)$-form in $x,y$ and a scalar in $z$  and thus fails to be cyclically symmetric in $x,y,z$ for higher genus, its restriction to genus one is  cyclically symmetric and reduces to (\ref{gen1.03})  in view of the following restrictions to genus one, 
\begin{align}
\om_I (x) \big|_{h=1} &= 1 \, ,
 &\p_x \cG(x,y) \big|_{h=1} &= - f^{(1)}(x{-}y)
\notag \\
 \p_x \cG^I(x,y) \big|_{h=1} &=- f^{(2)}(x{-}y)
  \, , 
 &\p_x \p_y \cG_2(x,y) \big|_{h=1} &= f^{(2)}(x{-}y)
 \label{htoone}
\end{align}
Similarly, the genus-zero counterpart (\ref{pfs}) is also cyclically symmetric in $x,y,z$.

\subsubsection{Application of \textit{$z$-reduction} to arbitrary genus}
\label{4.deff}

The scalar Fay identity in Theorem \ref{4.thm:7} provides a first example that motivates the generalization of the notion of \textit{$z$-reduction} to arbitrary genus $h$ given in section \ref{sec:3-x}.

\sm

While the scalar Fay identity (\ref{3.a.2}) is symmetric in $x,y$, it has no further symmetry involving  the variable $z$. As a result, (\ref{3.a.2}) may be used in two inequivalent ways towards the calculation of iterated integrals.  As a $(1,0)$ form in $x,y$ and a scalar in $z$, it may be rearranged  either in a \textit{$z$-reduced} or in an \textit{$x$-reduced} form.  More explicitly, 
\begin{itemize}
\itemsep=0in
\item \textit{$z$-reduce} the product $ \p_x \cG(x,z) \p_y \cG(y,z)$ in the third term of (\ref{3.a.2}), which is a $(0,0)$ form in $z$,   to a sum of  terms in which at most one factor is $z$-dependent; or
\item \textit{$x$-reduce} the product  $\p_y \cG(y,x) \p_x \cG(x,z) $ in the second term of (\ref{3.a.2}), which is a $(1,0)$ form in $x$, to a sum of  terms in which one factor is $x$-dependent
\end{itemize}
In neither case will the terms in (\ref{3.a.2})  yield homotopy-invariant integrals over $x$ or $z$ all by themselves. Still, the generating-series construction of higher-genus polylogarithms in (\ref{3.int}), (\ref{Gacoeff}) and (\ref{expmv}) provides a complete prescription for how to arrange individual  iterated integrals over $f$-tensors to produce homotopy invariant combinations in a fully constructive manner. Our main results in later sections are tensorial Fay identities 
among $(1,0)$-forms  in $x,y$ and scalars in $z$ which bring arbitrary products of  $\omega_I$ and higher-weight tensors $f$, $\p\cG$,  $\p\Phi$ into either a \textit{$z$-reduced} or a \textit{$x$-reduced} form.

\subsection{Higher-point modular scalar Fay identities}
\label{sec:4.3.1}

On the sphere, the identity (\ref{pfs}) for three points suffices to carry out a partial fraction decomposition for an arbitrary rational function of an arbitrary number of points, as exhibited for the denominators in (\ref{starg0}) and (\ref{cycl0}). We shall establish here that an analogous strategy essentially also works for arbitrary genus. Here, we shall again focus on the simplest higher-genus identities that share the pole structure of (\ref{cycl0}) and are modular scalars.
 
\sm
 
It will be convenient to denote the various points by $x_i$ and use the standard abbreviations for the arguments of functions such as in $ \cG(i,j) = \cG(x_i,x_j)$, and derivatives $\p_i = \p_{x_i}$. In particular, we introduce the following notation for the vanishing expression (\ref{3.a.2}),
\begin{align}
F_3(1,2,3) & =  \p_1 \cG(1,2) \p_2 \cG(2,3) + \p_2 \cG(2,1) \p_1 \cG(1,3) -  \p_1 \cG(1,3) \p_2 \cG(2,3)
\no \\ 
&\quad
- \om_I (1) \p_2 \cG^I(2,3) - \om_I(2) \p_1 \cG^I(1,3) + \p_1 \p_2 \cG_2(1,2)
\label{4.a.1}
\end{align}
The combination $F_3(1,2,3)$ is a $(1,0)$ form in $x_1,x_2$ and a scalar in $x_3$ with the manifest symmetry $F_3(1,2,3)= F_3(2,1,3)$.  

\sm

One readily  engineers an expression with the pole structure of (\ref{cycl0}) for four points which is a $(1,0)$ form in $x_1, x_2, x_3$ and a $(0,0)$ form in $x_4$, given by, 
\begin{align}
\label{6.a.1}
F_4 (1,2,3,4) &=  
 \p_1 \cG(1,2) \p_2 \cG(2,3)  \p_3 \cG(3,4)
 - \p_2 \cG(2,3)  \p_3 \cG(3,4) \p_1 \cG(1,4)
 \\ 
&\quad +  \p_3 \cG(3,4) \p_1 \cG(1,4) \p_2 \cG(2,1)
-  \p_1 \cG(1,4) \p_2 \cG(2,1)  \p_3 \cG(3,2) \notag\\
&\quad
+ \big (  \p_1 \p_2 \cG_2(1,2) - \om_I(2) \p_1 \cG^I(1,4) - \om_I(1) \p_2 \cG^I(2,4) \big ) 
\big [ \p_3 \cG(3,4) - \p_3 \cG(3,2) \big ] \notag\\
&\quad
+  \big (\p_2 \p_3 \cG_2(2,3) - \om_I(2) \p_3 \cG^I(3,4) - \om_I(3) \p_2 \cG^I(2,4)   \big ) 
\big [   \p_1 \cG(1,2) - \p_1 \cG(1,4)\big ]
\no
\end{align}
The first two lines on the right side capture the pole structure of the decomposition of (\ref{cycl0}) in a minimal manner, namely with the smallest number of terms. The third and fourth lines consist of terms required to make the full expression holomorphic in $x_1, \cdots, x_4$. As a result, $F_4$ is independent of $x_4$ and is a holomorphic $(1,0)$ form in $x_1, x_2, x_3$. Finally, as in the case of three points in Theorem \ref{3.a.2}, one readily shows that the integral of $F_4(1,2,3,4)$ against $\bar \om^A(1) \bar \om^B(2) \bar \om^C(3)$ vanishes so that we must have $F_4(1,2,3,4)=0$.  

\sm

One may construct an analogous combination for five points, 
\begin{align}
\label{4.c.1}
F_5 (1,2,\cdots,5) & =  
 \p_1 \cG(1,2)  \p_2 \cG(2,3)  \p_3 \cG(3,4)  \p_4 \cG(4,5)
{-} \p_2 \cG(2,3)  \p_3 \cG(3,4)  \p_4 \cG(4,5)  \p_1 \cG(1,5)
\no \\ 
&\quad
{+} \p_3 \cG(3,4)  \p_4 \cG(4,5)  \p_1 \cG(1,5)  \p_2 \cG(2,1) 
{-} \p_4 \cG(4,5)  \p_1 \cG(1,5)  \p_2 \cG(2,1)  \p_3 \cG(3,2)
\no \\
&\quad 
{+} \p_1 \cG(1,5)  \p_2 \cG(2,1)  \p_3 \cG(3,2)  \p_4 \cG(4,3) + \cdots
\end{align} 
where the ellipses stand for another 45 terms that are required for $F_5(1,2,3,4,5)=0$. These terms may be constructed as we did for the cases of three and four points.

\sm

Instead of the above expressions for $F_4$ and $F_5$ in terms of individual monomials in the derivatives of the Green functions and related functions, one may re-organize their expressions recursively, as given for the four and five points functions in the following Theorem. 
{\thm 
\label{4.thm:5}
The modular scalar Fay identities for four and five points may be recursively expressed in terms of,
\bea
\label{5ptFFay} 
F_4 (1,2,3,4) &=  &
\big ( \p_1 \cG(1,2) - \p_1 \cG(1,4) \big ) F_3(2,3,4) 
+ \big ( \p_3 \cG(3,4) - \p_3 \cG(3,2) \big ) F_3(1,2,4) 
\no \\
F_5 (1,2,3,4,5) & =  &
\big ( \p_1 \cG(1,2) - \p_1 \cG(1,5) \big )  F_4 (2,3,4,5) 
+ \Big [  \p_3 \cG(3,4)  \p_4 \cG(4,5) 
\no \\ && \hskip 0.5in
- \p_4 \cG(4,5)  \p_3 \cG(3,2)
+  \p_3 \cG(3,2) \p_4 \cG(4,3) \Big ] F_3(1,2,5)  
\eea
which both vanish identically on a Riemann surface $\Sigma$ of arbitrary genus,
\bea
F_4 (1,2,3,4) & = & 0
\no \\
 F_5 (1,2,3,4,5) & = & 0
\eea
In the expression for $F_5$ the function $F_4$ may be eliminated in terms of $F_3$ functions using the first equation, so that both $F_4$ and $F_5$ are linear combinations of $F_3$ functions only.}

\sm

Theorem \ref{4.thm:5} may be proven in two different ways. Either one may algebraically rearrange the explicit expressions for $F_4$ found in (\ref{6.a.1}) and for $F_5$ found in (\ref{4.c.1}) into the above forms. Or one may show that the expressions for $F_4$ and $F_5$ given in Theorem \ref{4.thm:5} precisely contain the corresponding minimal pole parts, and no other poles. In particular, one argues that all poles  between non-adjacent points, which arise from individual terms in (\ref{5ptFFay}), cancel in the sums that make up $F_4$ and $F_5$. Specifically, the pole term $\p_2 \cG(2,4)$ cancels in $F_4$ while the pole term $\p_2 \cG(2,5)$ cancels in $F_5$. All other pole terms are between adjacent points.   Since $F_3$ was already shown to vanish, it then follows straightforwardly that also $F_4$ and $F_5$ vanish.   

\sm

The generalization of Theorems \ref{4.thm:5} and \ref{4.thm:7} to the case of an arbitrary number of points $x_1 , \cdots, x_n$ is most easily provided by following the second argument above. 
{\thm
\label{4.thm:9}
The modular scalar Fay identity for an arbitrary number of points $n$, characterized by the following minimal pole structure, 
\bea
\label{4.c.2}
F_n(1,2,\cdots,n) & =  &
 \p_1 \cG(1,2)  \p_2 \cG(2,3)\cdots  \p_{n-1} \cG(n{-}1,n) 
 \\ &&
+  \p_1 \cG(1,n) 
 \sum_{j=1}^{n-1} (-1)^j  \bigg( \prod_{i=j+1}^{n-1} \partial_i \cG(i,i{+}1) \bigg)
\bigg( \prod_{k=2}^{j} \partial_k \cG(k,k{-}1) \bigg) +\cdots
\notag
\eea
may be recursively related to $F_m$ for $m <n$ as follows,
\bea
\label{nptFFay} 
F_n(1,2,\cdots,n) & =  &
\big ( \p_1 \cG(1,2) - \p_1 \cG(1,n) \big )  F_{n-1} (2,\cdots,n) 
 \\ &&
+ \sum_{j=2}^{n-1} (-1)^j  \bigg( \prod_{i=j+1}^{r-1} \partial_i \cG(i,i{+}1) \bigg)
\bigg( \prod_{k=3}^{j} \partial_k \cG(k,k{-}1) \bigg) F_3(1,2,n) 
\no
\eea
and therefore vanishes on a Riemann surface $\Sigma$ of arbitrary genus,
\bea
F_n(1,2,\cdots,n) =0
\eea}

The proof of this theorem may be carried out with the help of the second approach followed above for $F_4$ and $F_5$. The contributions with $n{-}1$ factors of $\p_i \cG(i,j)$ involving adjacent points $i,j$, spelt out in (\ref{4.c.2})  have exactly the pole structure of the genus-zero identity (\ref{cycl0}). Some of the factors $ \partial_i \cG^I(i,j)$ in the ellipsis of  (\ref{nptFFay}) involve non-adjacent points $i,j$. The cancellation of terms $\partial_i \cG(i,j)$ involving non-adjacent $i,j$, already established for $F_4$ and $F_5$,  can be recursively generalized to any number $n$ of points. In fact, imposing the cancellation of the poles $\partial_2 \cG(2,n)$ in individual terms of $F_{n-1} (2,\cdots,n) $ fixes the form of the second line in (\ref{nptFFay}). Accordingly, the vanishing of $F_n(1,\cdots,n) $ given by (\ref{nptFFay})  can be viewed as the higher-genus uplift of the identity (\ref{cycl0}) on the sphere. In the same way as the higher-point identities (\ref{starg0}) and (\ref{cycl0}) among rational functions boil down to iterations of the three-point partial-fraction identity (\ref{pfs}), the recursion (\ref{nptFFay}) reduces $n$-point modular scalar Fay identities
at arbitrary genus to the elementary three-point identity $F_3(i,j,k)=0$.

\subsubsection{Comments on Theorem \ref{4.thm:9}}

As a genus-one counterpart of the $n$-point identity (\ref{cycl0}) among rational functions, we reviewed the vanishing of elliptic functions $V_{n-1}(1,\cdots,n)$ in section \ref{sec:4.1.2}. While the expression (\ref{vids.01}) for arbitrary $V_w$ functions only involves Kronecker-Eisenstein kernels $f^{(r)}(x_i{-}x_{j})$ with adjacent $j=i{\pm}1\, {\rm mod} \, n$, the recursion (\ref{nptFFay}) for  higher-genus $F_n(1,\cdots,n)$ at $n\geq 4$ introduces $\cG^I(i,j)$ with non-adjacent $i,j$ , see for instance (\ref{6.a.1}). In Appendix~\ref{sec:6.3}, we present an alternative construction of vanishing $n$-point combinations of $f$-tensors with the pole structure of (\ref{cycl0}) which furnish a more direct generalization of $V_{n-1}(1,\cdots,n)=0$ to arbitrary genus.

\newpage

\section{Interchange identities}
\label{sec:3}

The goal of this section is to formulate and prove interchange identities that relate products of the form 
$\omega_M(x)f^{I_1\cdots I_r}{}_J(y,x)$, with two $x$-dependent factors, to their counterparts 
$\omega_M(y)f^{I_1\cdots I_r}{}_J(x,y)$ with $x$ and $y$ swapped plus a sum of products in which no more than one factor depends on $x$.\footnote{One can view interchange identities as simpler versions of Fay identities that only involve two instead of three points and trivialize at genus one by translation invariance
 on the torus and the parity $f^{(r)}(x{-}y)= (-1)^r f^{(r)}(y{-}x)$ of Kronecker-Eisenstein kernels.}  
In the spirit of the definition \textit{$x$-reduction} given in section \ref{sec:3-x} and illustrated for scalar Fay identities in section \ref{4.deff} for arbitrary genus, interchange identities will produce \textit{$x$-reductions} of  $\omega_M(x)f^{I_1\cdots I_r}{}_J(y,x)$  necessary to express their  primitives 
with respect to $x$ in terms of the higher-genus polylogarithms 
reviewed in sections \ref{sec:3.DHS} and \ref{sec:Jmv}.
{\lem 
\label{5.lem:9}
The basic  interchange identity {\rm \cite{DHoker:2020tcq, DHoker:2020uid}} for lowest weight reads as follows, 
\bea
\omega_M(x) \p_y \Phi^M{}_J(y) + \omega_M(y) \p_x \Phi^M{}_J(x)    
 -   \omega_J(x) \p_y \cG(y,x)  -   \omega_J(y) \p_x \cG(x,y)   = 0
  \label{preften.08}
\eea
or equivalently as follows in terms of $f$-tensors,
\bea
 \omega_M(x) f^M{}_J(y,x) + \omega_M(y) f^M{}_J(x,y)    = 0
  \label{ften.08a}
\eea
}
The role of the tensor $\Phi$, which was defined in (\ref{3.Phi}), may be viewed as compensating for the lack of translation invariance of the Arakelov Green function $\cG(x,y)$ on a Riemann surface of higher genus $h\geq 2$. The equivalence between (\ref{preften.08}) and (\ref{ften.08a}) is readily established using the decomposition of (\ref{fphig.1}).
The proof of Lemma  \ref{5.lem:9} in  \cite{DHoker:2020tcq} follows the two steps explained in detail for Theorem~\ref{4.thm:7} in section \ref{prfsec.1}: The left sides of (\ref{preften.08}) and (\ref{ften.08a})
\begin{itemize}
\itemsep=0in
\item[1.] are easily verified to be holomorphic in $x,y$ via (\ref{AG.05}) and (\ref{ften.02}), respectively,
\item[2.] integrate to zero since all of $\p_x \Phi^M{}_J(x),\p_x \cG(x,y)$ and $f^M{}_J(x,y)$
are total derivatives of single-valued functions in $x$ (and the remaining terms are similarly total $y$-derivatives).
\end{itemize}

\subsection{Interchange identities at higher weight}
\label{sec:3.1}

Convolutions of the basic interchange identity (\ref{preften.08}) or (\ref{ften.08a}) with $\partial_z {\cal G}(z,x)$ lead to higher-weight analogues \cite{DHS:2023}. At weight two, the compact formulation in terms of $f$-tensors is,
\beq
 \omega_M(x) f^{IM}{}_J(y,x)- \omega_M(y) f^{IM}{}_J(x,y)  
 +f^I{}_M(y,a)  f^M{}_J (x,b)  - f^I{}_M(x,b) f^M{}_J(y,a) = 0
   \label{ften.09}
\eeq
This relation may be derived either from convolutions of the weight-one interchange identity with $\partial_z \cG(z,x) \bar \omega^I(x)$\footnote{The derivation of (\ref{ften.09}) by integrating (\ref{ften.08a}) against $d^2x \, \partial_z \cG(z,x) \bar \omega^I(x)$ requires an additional application of the weight-one interchange identity (\ref{preften.08}) to the term $ \partial_z \cG(z,x) \bar \omega^I(x)\omega_M(x) f^M{}_J(y,x) $ in the integrand to \textit{$x$-reduce} the product $\partial_z \cG(z,x)\omega_M(x) $.} or by following the steps in the proof of the basic interchange identity in  (\ref{ften.08a}). The combination of the last two terms may be viewed as a matrix commutator which is actually independent of the points  $a,b \in \Sigma$, and may be re-expressed as follows, 
\begin{align}
\label{ften.10} 
 f^I{}_M(y,a) & f^M{}_J(x,b) - f^I{}_M(x,b) f^M{}_J(y,a) 
 \no \\ &
= \partial_y \Phi^I{}_M(y) \partial_x \Phi^M{}_J(x) - \partial_x \Phi^I{}_M(x) \partial_y \Phi^M{}_J(y)
\quad
\end{align}
The interchange identities (\ref{preften.08}) and (\ref{ften.09}) at weight one and two
allow us to express derivatives of the ${\cal G}_2(x,y)$ function (\ref{defgns})
entering the Fay identity of Theorem \ref{4.thm:7} in terms of $f$-tensors: 
Integrating the weight-one lemma (\ref{preften.08}) against the product
$ \partial_x {\cal G}(x,z)  \om_I(z)$ gives, 
\beq
 \int_{\Sigma} d^2 z \, \partial_x {\cal G}(x,z)  \om_I(z)
\bar \om^J(z)  \partial_y {\cal G}(y,z) 
=  \omega_M(x)f^{JM}{}_I(y,x)
+\partial_x \Phi^M{}_I(x) \partial_y \Phi^J{}_M(y)  
\label{w2convol}
\eeq
Then, upon contraction in $I,J$ and using the contracted version of the weight-two identity (\ref{ften.09}),
$ \omega_M(x) f^{IM}{}_I(y,x)= \omega_M(y) f^{IM}{}_I(x,y)$, we arrive at the two equivalent representations,
\begin{align}
h \, \p_x \p_y \cG_2(x,y) &=
 \om_M(x) f^{IM}{}_I(y,x) +  \p_x \Phi ^M{}_I(x) \p_y \Phi^I{}_M(y) \notag \\
  &=
 \om_M(y) f^{IM}{}_I(x,y) +  \p_x \Phi ^M{}_I(x) \p_y \Phi^I{}_M(y)
 \label{papag2}
 \end{align}
One may further rewrite $\p_x \Phi ^J{}_I(x)$ as the traceless
part of $f^J{}_I(x,a)$ for an arbitrary point $a$.

\sm

The generalization of (\ref{ften.09}) to arbitrary weight $r{+}1$ is provided by the following theorem.
{\thm 
\label{intlemma}
The modular tensors $\mP^{I_1 \cdots I_r}{}_J(x,y) $ defined by,
\begin{align}
& \mP ^{I_1 \cdots I_r}{}_J(x,y) = \omega_M(x) f^{I_1 \cdots I_r M}{}_J(y,x)
+ (-1)^r \omega_M(y) f^{I_r \cdots I_1 M}{}_J (x,y)
 \label{genint}\\
 &\quad + \sum_{k = 1}^r (-1)^{k+r}
 \Big[
f^{I_1 \cdots I_k}{}_M(y,a_k) f^{I_r  \cdots I_{k + 1} M}{}_J(x,b_k)
- f^{I_1 \cdots I_{k-1} M}{}_J(y,a_k) f^{I_r \cdots I_{k} }{}_M(x,b_k)
 \Big]
 \notag
\end{align}
with arbitrary points $a_1,\cdots, a_r, b_1,\cdots, b_r \in \Sigma$ vanish for all $r\geq 0$,
\beq
\mP^{I_1 \cdots I_r}{}_J(x,y) = 0
\label{piszero}
\eeq
}
The proof of the theorem is carried out by repeating the two steps in section \ref{prfsec.1} just as we did  in the above proof of (\ref{ften.08a}).
\begin{itemize}
\itemsep=0in
\item[1.] 
Holomorphicity in $x$ is most conveniently proven by induction in $r$ by noting that,
\beq
\pbx \, \mP^{I_1 \cdots I_r}{}_J (x,y)   = \pi \bar \omega ^{I_r} (x) \, \mP^{I_1 \cdots I_{r-1}}{}_J (x,y) 
\eeq
and that the base case $\mP^{\emptyset}{}_J (x,y) =  \omega_M(x) f^{ M}{}_J(y,x)
+ \omega_M(y) f^{M}{}_J(x,y)$ at $r=0$ vanishes by (\ref{ften.08a}). Holomorphicity
in $y$ follows from the previous result $\pbx \, \mP^{I_1 \cdots I_r}{}_J (x,y)=0$ through
the symmetry property $\mP^{I_1 \cdots I_r}{}_J(x,y) = (-)^r \, \mP^{I_r \cdots I_1}{}_J(y,x) $
under simultaneous exchange $x\leftrightarrow y$ and reversal $I_1 \cdots I_r \rightarrow 
I_r \cdots I_1$ of the indices.
\item[2.] The integral $\int_{\Sigma} d^2 x \, \bar \omega^P(x)
\int_{\Sigma} d^2 y \, \bar \omega^Q(y) \, \mP^{I_1 \cdots I_r}{}_J(x,y) $
 vanishes since each term in (\ref{genint}) is a total derivative in $x$ or $y$ of a single-valued function on $\Sigma \times \Sigma$.
\end{itemize}
Note that the second line of (\ref{genint}) can be alternatively rewritten as,
\bea
 \sum_{k = 1}^r (-1)^{k+r}
 \Big[
\partial_y \Phi^{I_1 \cdots I_k}{}_M(y) \partial_x \Phi^{I_r \cdots I_{k + 1} M}{}_J(x)
- \partial_y \Phi^{I_1 \cdots I_{k-1} M}{}_J(y) \partial_x \Phi^{I_r \cdots I_{k } }{}_M(x)
 \Big]
 \eea 
in terms of the higher-weight $\Phi$-tensors in (\ref{fphig.2}) since the ${\cal G}$-tensors in the decomposition (\ref{fphig.1}) cancel separately at each value of $k$. In this way, we recover the formulation of higher-weight interchange identities in section 4.6.1 of \cite{DHS:2023} that manifests the independence on 
the arbitrary points $a_i, b_i$ of (\ref{genint}).

\subsection{Uncontracted interchange identities}
\label{sec:3.2}

The interchange identities of Theorem \ref{intlemma} may be used to obtain the \textit{$x$-reduced} form of the specific contraction  $\omega_M(x) f^{I_1 \cdots I_r M}{}_J(y,x)$ over $M$. In a more general situation, however, 
one may wish to \textit{$x$-reduce} a product $\omega_J(x) f^{I_1 \cdots I_r L}{}_K(y,x)$ 
with free indices $I_1,\cdots,I_r,J,L$ and $K$ in preparation for integration over $x$ in terms of the higher-genus polylogarithms  of~\cite{DHS:2023}. In this section, the product $\omega_J(x) f^{I_1 \cdots I_r L}{}_K(y,x)$ with ``uncontracted'' indices will be \textit{$x$-reduced} by means of the ``contracted'' interchange identities of Theorem \ref{intlemma} at general weight and the elementary identity,
\begin{align}
 \omega_J(x)  \big[f^{\overrightarrow{ I} L}{}_K(y,x) -  f^{\overrightarrow{ I} L}{}_K(y,a)  \big] 
 &= \omega_J(x) \delta^L_K \big[ \p_y \cG^{\overrightarrow{ I} }(y,a) - \p_y \cG^{\overrightarrow{ I} }(y,x) \big] 
 \label{rewr.1}  \\
 &= \delta^L_K \omega_M(x) \big[f^{\overrightarrow{ I} M}{}_J(y,x) -  f^{\overrightarrow{ I} M}{}_J(y,a)  \big] \notag
 \end{align}
 valid for arbitrary $a,x,y \in \Sigma$. Here and below, we use multi-index 
 notation $\overrightarrow{ I}= I_1 I_2\cdots I_r$ for ordered sets of $r\geq 0$ indices $I_j$
 and denote the reversal $\overleftarrow{ I}= I_r \cdots I_2I_1$ through a flipped arrow.
 The rearrangement (\ref{rewr.1}) is a straightforward consequence of the decomposition
 (\ref{fphig.1}) since the $\partial_y \Phi$ contributions to $ f^{\overrightarrow{ I} L}{}_K(y,\cdot)$
and $f^{\overrightarrow{ I} M}{}_J(y,\cdot)$ clearly cancel from both lines. This takes advantage of the fact that all the dependence of   $f^{I_1 \cdots I_r}{}_J(x,y)$ on the second point $y$ is concentrated in the trace $\delta^{I_r}_J$
 with respect to the last two indices. In other words, when $f^{I_1 \cdots I_r}{}_J(x,y)$ is viewed as an $h\times h$ matrix indexed by $I_r,J$, the decomposition  (\ref{fphig.1}) implies that each term is either proportional to the unit matrix
 or independent on $y$.
 
 \sm
 
 The rearrangement (\ref{rewr.1}) paves the way for the following uncontracted version of the interchange identities of Theorem \ref{intlemma}:
 {\cor
 The modular tensor $ \omega_J(x)  f^{\overrightarrow{I}L}{}_K(y,x)$ with multi-index 
 $\overrightarrow{I}=I_1\cdots I_r$ and weight $r{+}1$ may be \textit{$x$-reduced} as follows,
\begin{align}
 \omega_J(x)  f^{\overrightarrow{I} L}{}_K(y,x) 
&=  -  (-1)^r \omega_J(y)  f^{ \overleftarrow{I}  L}{}_K(x,y) +  \omega_J(x)  f^{\overrightarrow{I} L}{}_K(y,a) - \delta^L_K \omega_M(x)    f^{\overrightarrow{I} M}{}_J(y,a) 
\notag \\
&\quad
+ (-1)^r \omega_J(y)  f^{  \overleftarrow{I} L}{}_K(x,b) 
 - (-1)^r  \delta^L_K \omega_M(y)    f^{ \overleftarrow{I} M}{}_J(x,b)  \notag \\
 &\quad+ \delta^L_K
\sum_{\ell = 1}^r (-1)^{\ell+r}
 \Big [
 f^{I_1 \cdots I_{\ell-1} M}{}_J(y,a_\ell) f^{I_r \cdots I_{\ell } }{}_M(x,b_\ell) \notag \\
 &\quad\quad\quad\quad\quad\quad\quad\ 
 - f^{I_1 \cdots I_\ell}{}_M(y,a_\ell) f^{I_r \cdots I_{\ell + 1} M}{}_J(x,b_\ell)
 \Big ]
 \label{rewr.3} 
\end{align}
\label{unclemma}
}
This corollary is readily proven by applying (\ref{rewr.1}) to both terms on the left side of
\begin{align}
& \omega_J(x)  f^{\overrightarrow{I} L}{}_K(y,x) + (-1)^r \omega_J(y)  f^{\overleftarrow{I}L}{}_K(x,y) 
 \label{rewr.2}  \\
 &
=  \omega_J(x)  f^{\overrightarrow{I} L}{}_K(y,a) - \delta^L_K \omega_M(x)    f^{\overrightarrow{I} M}{}_J(y,a)  \notag \\
&\quad
+ (-1)^r \omega_J(y)  f^{\overleftarrow{I} L}{}_K(x,b) 
 - (-1)^r  \delta^L_K \omega_M(y)    f^{\overleftarrow{I} M}{}_J(x,b)  \notag \\
 &\quad+ \delta^L_K \big[
  \omega_M(x)    f^{\overrightarrow{I} M}{}_J(y,x)
  + (-1)^r   \omega_M(y)    f^{\overleftarrow{I} M}{}_J(x,y)
  \big]
  \notag
\end{align}
and eliminating the coefficient of $\delta^L_K$ in the last line through the
contracted interchange identities of Theorem \ref{intlemma}.

\newpage

\section{Tensorial Fay identities}
\label{sec:fay}

This section is dedicated to the systematic construction and proof of higher-genus Fay identities among bilinears in the tensors  $f^{I_1\cdots I_r}{}_J(x,y) $ of section \ref{sec:2.2} involving three points. In section \ref{sec:fay.1} we shall extend the three-point identity in (\ref{3.a.2}) among bilinears in the modular scalar $\p_i\cG(i,j)$ to a tensor-valued identity. Such an identity is needed already to  obtain an \textit{$x$-reduced} form (in the spirit of the definition given in section \ref{sec:3-x} and its illustration in section \ref{4.deff}) of products $\p_y \cG(y,x) \p_x \Phi^M{}_K(x)$ in the same way as (\ref{3.a.2}) provides the \textit{$x$-reduced} form of the product $\p_y \cG(y,x) \p_x \cG(x,z)$.

\sm

The shuffle product will greatly facilitate and shorten the formulation and proof of  tensor-valued Fay identities of  higher rank and higher weight, and will be briefly reviewed in section \ref{sec:fay0}. The fundamental Lemma \ref{5.lem:1} of section \ref{sec:fay.1.5} will underly many of the subsequent results in this section.   In section \ref{sec:fay.2}, we will construct explicit all-weight formulas for tensorial Fay identities that \textit{$z$-reduce} the expression $f^{\overrightarrow{P}M}{}_J(x,z) f^{\overrightarrow{I}J}{}_K(y,z)$ which is a scalar in $z$ and a $(1,0)$-form in both $x$ and $y$ and where we use  the multi-index notation,
\bea
\overrightarrow{I} = \left\{ \begin{array}{cl} \emptyset &:\ r=0 \cr I_1 \cdots I_r &:\ r \geq 1 \end{array} \right.
\hskip 1in 
\overrightarrow{P}=  \left\{ \begin{array}{cl} \emptyset &:\ s=0 \cr P_1\cdots P_s &:\ s \geq 1 \end{array} \right.
\eea
 introduced already informally in section \ref{sec:3.2}. In section \ref{sec:fay.3} we shall rearrange the Fay identities of section \ref{sec:fay.2} in order to obtain the \textit{$x$-reduced} expression for a product of the type  $f^{\overrightarrow{I}} {}_J(x,z)   f^{\overrightarrow{P} J}{}_K(y,x)$, which is a $(1,0)$-form in~$x$.

\sm

The contraction of one index $J$ in the Fay identities of sections \ref{sec:fay.2} and \ref{sec:fay.3} is convenient to formulate compact expressions. In section \ref{sec:4.9}, we deduce Fay identities for expressions of the form $f^{\overrightarrow{P}Q}{}_L(x,z)  f^{\overrightarrow{I}M}{}_K(y,z)$ and
$f^{\overrightarrow{I}}{}_K(x,z)  f^{\overrightarrow{P}Q}{}_L(y,x) $ from their counterparts with one index contraction, using the same techniques that allowed us to deduce the uncontracted interchange identities in section \ref{sec:3.2}. 
Most importantly, iterative use of these uncontracted Fay identities produces \textit{$z$-reduced} expressions for higher products of $f$-tensors with an $(N\geq 3)$-fold appearance of a given point $z$.

\subsection{Tensorial Fay identity at weight two}
\label{sec:fay.1}

The simplest tensorial Fay identity has weight two and is given by
\begin{align}
& f^M{}_J(x,y) f^J{}_K(y,z) + f^M{}_J(y,x) f^J{}_K(x,z) - f^M{}_J(x,z) f^J{}_K(y,z) \notag  \\
&\quad + \omega_J(x) f^{MJ}{}_K(y,x) 
+ \omega_J(y) f^{JM}{}_K(x,z) +  \omega_J(x) f^{JM}{}_K(y,z) = 0
\label{hf.22}
\end{align}
It comprises $h^2$ components from the values $M,K=1,2,\cdots,h$ of the free indices. The left side of (\ref{hf.22}) is symmetric in $x\leftrightarrow y$ which is manifest for the first two terms and the last two terms. Verifying the
$x\leftrightarrow y$ symmetry of the remaining two terms $\omega_J(x) f^{MJ}{}_K(y,x) - f^M{}_J(x,z) f^J{}_K(y,z)$ requires the weight-two interchange identity (\ref{ften.09}).

\sm

We shall discuss the following two alternative proofs of (\ref{hf.22}):
\begin{itemize}
\item 
Following the two-step procedure of section \ref{prfsec.1}, one first verifies that the left side of (\ref{hf.22}) has vanishing anti-holomorphic derivatives in $x,y,z$,  which relies on the weight-one interchange identity (\ref{ften.08a}). The integral of the left side of (\ref{hf.22}) against $\int_\Sigma \bar \omega^I(x) \int_\Sigma \bar \omega^J(y)$ vanishes, since each term on the left side of (\ref{hf.22}) is a total derivative in $x$ or in $y$ of a single-valued function. 
\item 
Alternatively, one applies the arguments of the previous paragraph to prove that,
\beq
{\cal V}_I^{(2)}(x,y,z) = \omega_J(y) \omega_K(z) f^{KJ}{}_I(x,y) 
+ \omega_J(y) f^J{}_K(z,x) f^K{}_I(x,y)+ {\rm cycl}(x,y,z)
\label{calv2}
\eeq
vanishes, thereby generalizing the vanishing of the elliptic $V_2(1,2,3)$ function  in (\ref{vids.01}) to arbitrary genus. The identity ${\cal V}_I^{(2)}(x,y,z)=0$ used in section 4.6.2 of \cite{DHS:2023} is a $(1,0)$-form in all of $x,y,z$
as opposed to the left side of (\ref{hf.22}) which is a $(1,0)$-form in $x,y$ and a scalar in $z$. Even though only three out of six terms in the cyclic sum (\ref{calv2}) have an exposed factor of $\omega_M(z)$, one can apply (contracted and uncontracted)  interchange identities to rewrite ${\cal V}_K^{(2)}(x,y,z)=  \omega_M(z) \Xi^{M}{}_K(x,y,z)$. The tensor $\Xi^{M}{}_K(x,y,z)$ turns out to exactly reproduce the left side of (\ref{hf.22}).

\sm

In Appendix \ref{sec:6.3}, we conjecture a construction of identities ${\cal V}_I^{(w)}(x_1,\cdots,x_{w+1})=0$ at arbitrary genus and arbitrary multiplicity which generalize the genus-one identity $V_w(1,2,\ldots,w{+}1)=0$ of section \ref{sec:4.1.2}.
\end{itemize}
The scalar three-point identity in (\ref{3.a.2}) may be recovered 
from (\ref{hf.22}), up to a factor of~$h$, via contraction with $\delta^K_M$ which for instance
reduces the last two terms to $ - \om_I (x) \p_y \cG^I(y,z) - \om_I(y) \p_x \cG^I(x,z) $
by the tracelessness condition $\partial_x \Phi^{\overrightarrow{I}M}{}_K(x)\delta^K_M = 0$.
The remaining $h^2{-}1$ components of (\ref{hf.22}) are captured by the
traceless part in $M,K$,
\begin{align}
\p_y \cG(y,x) \p_x \Phi^M{}_K(x) &= - \p_x \cG(x,y) \p_y \Phi^M{}_K(y)
+ \omega_J(x) \partial_y \Phi^{JM}{}_K(y)
+ \omega_J(y) \partial_x \Phi^{JM}{}_K(x) \notag\\
&\quad + \omega_J(x) f^{MJ}{}_K(y,x) 
- \tfrac{1}{h} \delta^M_K  \omega_J(x) f^{LJ}{}_L(y,x) \notag\\
&\quad +  \partial_y \Phi^M{}_J(y) \partial_x \Phi^J{}_K(x)
- \tfrac{1}{h} \delta^M_K \partial_y \Phi^L{}_J(y) \partial_x \Phi^J{}_L(x)
\label{1stGphi}
\end{align}
which \textit{$x$-reduces the left side.} Hence, the added value of the tensorial Fay identity (\ref{hf.22}) beyond the trace component in (\ref{3.a.2}) is an \textit{$x$-reduced expression for $\p_y \cG(y,x) \p_x \Phi^M{}_K(x) $.} Note that the last two lines are, up to renaming of indices, the  traceless projection of the tensorial weight-two
convolution in (\ref{w2convol}).

\subsection{The shuffle product}
\label{sec:fay0}

The shuffle product provides an efficient tool in terms of which to organize and prove various tensor-valued Fay identities for higher rank and higher weight. Here, we review the essentials of the shuffle product and shuffle algebra that will be needed in the subsequent developments (for a standard reference see for example \cite{Reutenauer}).

\sm

The \textit{shuffle product} $\overrightarrow{X} \shuffle \overrightarrow{Y}$ is a binary operation on two words $\overrightarrow{X}$ and $\overrightarrow{Y}$ formed out of a given alphabet of letters and is given by the sum of all possible ways of interlacing the letters of $\overrightarrow{X}$ and $\overrightarrow{Y}$ such that the order of the letters  in each word is preserved in $\overrightarrow{X} \shuffle \overrightarrow{Y}$. The shuffle product  has the following properties that make the set of words equipped with addition and the shuffle product into a \textit{shuffle algebra}: 
\begin{enumerate}
\itemsep =-0.04in
\item associativity $(\overrightarrow{X} \shuffle \overrightarrow{Y}) \shuffle \overrightarrow{Z} = \overrightarrow{X} \shuffle (\overrightarrow{Y} \shuffle \overrightarrow{Z}) = \overrightarrow{X} \shuffle \overrightarrow{Y} \shuffle \overrightarrow{Z}$;
\item commutativity $\overrightarrow{X} \shuffle \overrightarrow{Y} = \overrightarrow{Y} \shuffle \overrightarrow{X}$;
\item neutral element provided by the empty set $\emptyset$ such that $\overrightarrow{X} \shuffle \emptyset = \overrightarrow{X}$;
\item recursive decomposition for non-empty words $\overrightarrow{X}=X_1 \cdots X_r$ and $\overrightarrow{Y}=Y_1 \cdots Y_s$,
\bea
\overrightarrow{X} \shuffle \overrightarrow{Y} & = & X_1 (X_2 \cdots X_r \shuffle \overrightarrow{Y}) + Y_1 (\overrightarrow{X} \shuffle Y_2 \cdots Y_s)
\no \\
& = & (X_1 \cdots X_{r-1}  \shuffle \overrightarrow{Y}) X_r  + (\overrightarrow{X} \shuffle Y_1 \cdots Y_{s-1}) Y_s
\label{recshf}
\eea
\end{enumerate}
The shuffle products considered here will be on words formed out of multi-indices denoted $\vI=I_1 \cdots I_r$ and $\vP=P_1 \cdots P_s$ containing letters  in the alphabet $\{ 1,\cdots,  h \}$. The representations of the shuffle algebra on the tensors encountered here is obtained by implementing the recursive decomposition of item 4.\ above on tensors as follows,
\beq
f^{\cdots (\overrightarrow{I} \shuffle \overrightarrow{P}) \cdots}{}_K(x,y) 
= f^{\cdots I_1 (I_2\cdots I_r  \shuffle \overrightarrow{P}) \cdots}{}_K(x,y) + f^{\cdots P_1 (\overrightarrow{I} \shuffle P_2 \cdots P_s ) \cdots}{}_K(x,y) 
\label{fsh}
\eeq
Accordingly, the anti-holomorphic derivatives (\ref{ften.03}) of
$f$ tensors generalize to shuffles via,
\begin{align}
 \partial_{\bar x} f^{(\overrightarrow{I} \shuffle \overrightarrow{P})\overrightarrow{M}}{}_K(x,y) 
 &= 
 - \pi  \bar \omega^{I_1}(x) f^{(I_2\cdots I_r \shuffle \overrightarrow{P}) \overrightarrow{M}}{}_K(x,y) 
 -  \pi  \bar \omega^{P_1}(x) f^{( P_2 \cdots P_s \shuffle \overrightarrow{I} ) \overrightarrow{M}}{}_K(x,y) 
 \label{prf.02}\\
 \partial_{\bar y} f^{\overrightarrow{M} (\overrightarrow{I} \shuffle \overrightarrow{P})}{}_K(x,y) 
 &=  
 \pi \Big [ \delta^{I_r}_K  f^{\overrightarrow{M}(I_1\cdots I_{r-1} \shuffle \overrightarrow{P})  }{}_R(x,y) 
 + \delta^{P_s}_K  f^{\overrightarrow{M}( P_1 \cdots P_{s-1} \shuffle \overrightarrow{I} ) }{}_R(x,y) \Big ] \bar \omega^R(y) 
 \notag
\end{align}
for arbitrary $\overrightarrow{M}= M_1\cdots M_t$ with $t\geq 0$ and $\overrightarrow{I},\overrightarrow{P} \neq \emptyset$. Many of the subsequent formulas simplify by writing $f^{\emptyset}{}_J(x,y)= \omega_J(x)$, in analogy with the kernel $f^{(0)}=1$ at genus one.

\subsection{A fundamental lemma}
\label{sec:fay.1.5}

In subsequent subsections, we shall derive two different types of Fay identities. Suppressing all index structure, they may be schematically represented as follows: 
\begin{itemize}
\itemsep=0in
\item in section \ref{sec:fay.2} to \textit{$z$-reduce} the product $f(x,z) f(y,z)$, namely where the repeated point $z$ corresponds to the scalar on both factors;
\item in section \ref{sec:fay.3}  to \textit{$x$-reduce} the product $f(x,z) f(y,x)$, namely where the repeated point $x$ corresponds to a $(1,0)$ form on one factor and a scalar on the other factor. 
\end{itemize}
On a genus-one Riemann surface these two cases are equivalent to one another, but for genus $h \geq 2$ they are inequivalent and require separate treatments. The case of $f(x,y) f(x,z)$ will not be considered in this work
by its form degree $(2,0)$ in $x$ and the vanishing of $dx \wedge dx$ as its wedge product vanishes $\boldsymbol{f}(x,y) \wedge \boldsymbol{f}(x,z)=0$ and its $(2,0)$-form tensor product $\boldsymbol{f}(x,y) \otimes \boldsymbol{f}(x,z)$ never occurs as an integration kernel under the multiplication
of the algebra ${\cal A}_N$ introduced in section \ref{sec:3-x}. The two cases of $f(x,z) f(y,z)$ and $f(x,z) f(y,x)$
will be built on a single lemma, valid for arbitrary rank, weight and genus, which we now state. 
{\lem
\label{5.lem:1}
The following combination, defined for $\vI = I_1 \cdots I_r$ and $\vP=P_1 \cdots P_s$ via,
\bea
\label{5.lem:1a}
\cS ^{\vI | \vP  }{}_K(x,y,z) & = & 
 f^{\vI} {}_J(x,z)  \Big ( f^{\vP J}{}_K(y,x) - f^{\vP J}{}_K(y,z) \Big )
\no \\ &&
+ f^{\vP}{}_J(y,z) \Big ( f^{\vI J}{}_K(x,y) - f^{\vI J}{}_K(x,z) \Big ) 
\no \\ &&
+ \sum_{k=0}^r f^{I_1 \cdots I_k } {}_J(x,y) f^{\vP \shuffle JI_{k+1} \cdots I_r }{}_K(y,z) 
\no \\ &&
+ \sum_{\ell=0}^s f^{P_1 \cdots P_\ell}{}_J(y,x) f^{\vI \shuffle J P_{\ell+1} \cdots P_s} {}_K(x,z)
\eea
vanishes identically for arbitrary $r,s\geq 0$,
\beq
\cS ^{\vI | \vP  }{}_K(x,y,z) = 0
\eeq
Here and throughout we use $f^\emptyset {}_J(x,y) = \om_J(x)$.}

\sm

The lemma is proven in Appendix \ref{appB.0}. The right side of (\ref{5.lem:1a}) exposes the symmetry,
\bea
\cS ^{\vI | \vP  }{}_K(x,y,z)=\cS ^{\vP | \vI  }{}_K(y,x,z)
\eea 
under simultaneous exchange of $\vI \leftrightarrow \vP$ and $x\leftrightarrow y$. For $\vI = \vP=\emptyset$, all dependence on $z$ cancels, and the remaining terms reduce to the basic interchange identity  given in (\ref{ften.08a}).

\subsection{Eliminating repeated scalar points at all weights}
\label{sec:fay.2}

In this section, we extend the tensorial weight-two identity (\ref{hf.22}) to arbitrary weight which is one of the main results of this work. A first variant of all-weight Fay identities among three points $x,y,z \in \Sigma$ 
is stated in the following theorem.
{\thm
\label{3.thm:7}
The contracted product $f^{\vP M}{}_J(y,z) f^{\vI J}{}_K(x,z)$, which is a scalar in the 
repeated point $z$, may be \textit{$z$-reduced} as follows,
\begin{align}
\label{exfay.15}
&f^{ \vP M}{}_J(y,z) f^{ \vI J}{}_K(x,z) = (-1)^{s} \, \omega_J(y) \, f^{\vI M \vPt J}{}_K(x,y)  
  \\
&\quad 
+ f^{\vI M}{}_J(x,y) f^{\vP J}{}_K(y,z) 
+\sum_{k = 0}^r   f^{I_1 \cdots I_k}{}_J(x,y) f^{(\vP  \shuffle J I_{k+1}\cdots I_r )M}{}_K(y,z) 
 \notag \\
&\quad 
+ f^{\vP M}{}_J(y,x) f^{\vI J}{}_K(x,z)
+\sum_{\ell=0}^s f^{P_1 \cdots P_\ell }{}_J(y,x) f^{(\vI  \shuffle J  P_{\ell+1} \cdots P_s)M}{}_K(x,z)
\notag \\
&\quad 
+ \sum_{\ell =1}^s (-1)^{s-\ell} 
\Big [ f^{P_1 \cdots   P_\ell}{}_J(y,b_\ell) f^{\vI  M P_s\cdots P_{\ell+1} J}{}_K(x,a_\ell)
-  f^{P_1 \cdots P_{\ell-1} J}{}_K(y,b_\ell ) f^{\vI M P_s\cdots  P_\ell}{}_J(x,a_\ell ) \Big ]
\notag
\end{align}
where $\vI= I_1\cdots I_r$, $\vP=P_1 \cdots P_s$ and $\vPt= P_s\cdots P_2 P_1$. The points $a_1,\cdots ,a_s$ and $b_1,\cdots ,b_s$ in the last line are arbitrary and actually  drop out of the combination on the last line.}

\sm

The proof of Theorem \ref{3.thm:7} is given in Appendix \ref{appB.1} and relies on
Lemma \ref{5.lem:1}. In view of our convention $f^{\emptyset}{}_J(x,y)= \omega_J(x)$, the $k=0$ and $\ell=0$ summands in the second and third line of (\ref{exfay.15}) are given by
$ \omega_J(x) f^{( \overrightarrow{P}\shuffle J\overrightarrow{I}) M}{}_K(y,z) $ and
$ \omega_J(y) f^{( \overrightarrow{I} \shuffle J \overrightarrow{P}) M}{}_K(x,z)$, respectively. 
The trace component of (\ref{exfay.15}) with respect to $M,K$ expresses $\p_y \cG^{\overrightarrow{P}}(y,z) 
\p_x \cG^{\overrightarrow{I}}(x,z)$ for arbitrary pairs $\overrightarrow{P}, \overrightarrow{I}$ of multi-indices
in terms of $\cG$ and $\Phi$-tensors without any repeated appearance of~$z$.
Inserting the decomposition (\ref{fphig.1}) into the last line of (\ref{exfay.15}) cancels 
all $\p \cG$ tensors and one is left with manifestly $a_i,b_i$-independent
bilinears of $\p_x \Phi$ and $\p_y \Phi$ tensors.

\subsubsection{Comments on Theorem \ref{3.thm:7}}

Since the proof of Theorem \ref{3.thm:7} in Appendix \ref{appB.1}
is not constructive, we sketch two constructive algorithms in Appendix \ref{app.cons} that may be used to 
generate higher-weight Fay identities from convolutions of
lower-weight ones. The examples at weight $3\leq w \leq 6$ obtained from
the methods of Appendix \ref{app.cons} led to anticipating (\ref{exfay.15}), initially as
a conjecture, which is now underpinned by the proof in Appendix \ref{appB.1}.

\sm

In the specialization of (\ref{exfay.15}) to genus one, the last line cancels, and the
shuffle products lead to Kronecker-Eisenstein kernels (\ref{exfay.22}) multiplied by
 combinatorial factors according to 
\beq
f^{ C_1\cdots C_p (A_1\cdots A_m \shuffle B_1\cdots B_n)  D_1\cdots D_q}{}_K(x,y) \big|_{h=1} =
{m{+}n\choose m} f^{(m+n+p+q)}(x{-}y)
\label{binom}
\eeq
In this way, one recovers the binomial coefficients in (\ref{exfay.16}) and can
readily verify consistency with the genus-one Fay identities at arbitrary weight.

\subsubsection{Examples at weight three and four}

The simplest example of (\ref{exfay.15}) at $\vI =\vP =\emptyset$ is the weight-two identity (\ref{hf.22}). For choices of $\vI$ and $\vP$ with a total of one and two letters, we obtain the following Fay identities at weight three and four, 
\begin{align} 
 f^{M}{}_J(y,z) f^{IJ}{}_K(x,z) &= 
f^{M}{}_{ J}(y,x) f^{I J}{}_{ K}(x,z) + 
 f^{I}{}_{ J}(x,y) f^{J  M}{}_{ K}(y,z)  \notag \\
 &\quad + 
f^{I  M}{}_{ J}(x,y)  f^{J}{}_{ K}(y,z) + 
\omega_J(x)  f^{J I M}{}_{ K}(y,z)  \notag \\
 &\quad + 
\omega_J(y) f^{I M J}{}_{ K}(x,y) 
+ \omega_J(y) f^{ (J\shuffle I ) M}{}_{ K}(x,z) 
\notag \\
 f^{M}{}_J(y,z) f^{I_1 I_2 J}{}_K(x,z) &= 
 f^{M}{}_{ J}(y,x) f^{I_1 I_2 J}{}_{ K}(x,z) + f^{I_1 I_2}{}_{ J}(x,y) f^{J M}{}_{ K}(y,z)  
  \notag \\
 &\quad 
+  f^{I_1}{}_{ J}(x,y) f^{J I_2 M}{}_{ K}(y,z)  
+ f^{I_1 I_2 M}{}_{ J}(x,y)  f^{J}{}_{ K}(y,z)  \notag \\
 &\quad + 
 \omega_J(x) f^{J I_1 I_2 M}{}_{ K}(y,z) + 
\omega_J(y) f^{I_1 I_2 M J}{}_{ K}(x,y)  \notag \\  \displaybreak[3]
 &\quad 
+ \omega_J(y) f^{(J\shuffle I_1  I_2) M}{}_{ K}(x,z)
\notag \\ 
f^{PM}{}_J(y,z) f^{IJ}{}_K(x,z) &= f^{PM}{}_J(y,x) f^{IJ}{}_K(x,z) 
+ f^{P}{}_J(y,x) f^{(I \shuffle J)M}{}_K(x,z)
\notag \\
&\quad
+ \omega_J(y) f^{( I \shuffle JP)M}{}_K(x,z)
+ f^{IM}{}_J(x,y) f^{PJ}{}_K(y,z)\notag \\
&\quad
+ f^{I}{}_J(x,y) f^{(P\shuffle J)M}{}_K(y,z)
+ \omega_J(x) f^{(P \shuffle JI)M}{}_K(y,z)
\notag \\
&\quad
+ f^{P}{}_J(y,b) f^{IMJ}{}_K(x,a) -  f^{J}{}_K(y,b) f^{IMP}{}_J(x,a)
\notag \\
&\quad
- \omega_J(y) f^{IMPJ}{}_K(x,y)
\label{explf34}
\end{align}

\subsubsection{Examples involving weight-one factors}
\label{sec:obs}

The all-weight family of Fay identities (\ref{exfay.15}) with $\vP = \emptyset$ 
takes the simple form 
\begin{align}
\label{Pempty}
&f^{M}{}_J(y,z) f^{ \vI J}{}_K(x,z) = 
f^{M}{}_J(y,x) f^{\vI J}{}_K(x,z) 
+ f^{\vI M}{}_J(x,y) f^{ J}{}_K(y,z) 
  \\
&\quad 
+\sum_{k = 0}^r   f^{I_1 \cdots I_k}{}_J(x,y) f^{  J I_{k+1}\cdots I_r M}{}_K(y,z) 
+ \omega_J(y) \, f^{\vI M J}{}_K(x,y) 
+\omega_J(y) f^{(\vI  \shuffle J  )M}{}_K(x,z)
\notag
\end{align}
where the bilinears of $\partial_x \Phi$ and $\partial_y \Phi$ tensors
in the last line of (\ref{exfay.15}) are absent. More importantly, the right side of
(\ref{Pempty}) features just a single repeatedly $x$-dependent term $f^{M}{}_J(y,x) f^{\vI J}{}_K(x,z) $
even though the Fay identities (\ref{exfay.15}) are engineered to eliminate repeated points $z$ rather
than $x$. Hence, as exploited in Appendix \ref{s:fay.3.2.b},
the Fay identities (\ref{Pempty}) can also be solved to \textit{$x$-reduce} $f^{M}{}_J(y,x) f^{\vI J}{}_K(x,z) $ on the right side instead of \textit{$z$-reducing} the left side.
In other words, (\ref{Pempty}) intersects with the Fay identities of the next section
which are dedicated to the removal of repeated one-form points. This is a peculiarity of
having $\vP = \emptyset$ in (\ref{exfay.15}) and will no longer be the case for non-empty $\vP $.

\subsection{Eliminating repeated one-form points at all weights}
\label{sec:fay.3}

We shall now proceed to another main result of this work which may be summarized in the following theorem.
{\thm
\label{3.thm:8}
The contracted product $f^{\overrightarrow{I}} {}_J(x,z)   f^{\overrightarrow{P} J}{}_K(y,x)$, 
which is a $(1,0)$-form in the repeated point $x$, may be \textit{$x$-reduced} as follows,
\begin{align}
&f^{\overrightarrow{I}} {}_J(x,z)   f^{\overrightarrow{P} J}{}_K(y,x) =
f^{\overrightarrow{I}} {}_J(x,z)   f^{\overrightarrow{P} J}{}_K(y,z)   \label{4.ff.99}  \\
&\quad
- \sum_{\ell=0}^s (-1)^{s-\ell} \sum_{k = 0}^r 
f^{P_s \cdots P_{\ell+1} \shuffle I_1 \cdots I_k} {}_J(x,y) 
f^{P_1\cdots  P_\ell J I_{k + 1} \cdots I_r}{}_K(y,z)
\notag\\
&\quad
-  \sum_{\ell=0}^s (-1)^{s-\ell} f^{P_1 \cdots P_\ell}{}_J(y,z) 
\Big[  f^{(P_s\cdots P_{\ell+1} \shuffle \vI )J} {}_K(x,y) 
+  f^{(P_s\cdots P_{\ell+1} J \shuffle  I_1 \cdots I_{r-1} ) I_r} {}_K(x,z) \Big] 
\notag
\end{align}
where $\vI = I_1\cdots I_r$ and $\vP= P_1\cdots P_s$ with $r\geq 1$ and $s\geq 0$.}

The proof is presented  in Appendix \ref{appB.2} and proceeds in two parts. In the first part we  prove  Lemma \ref{5.lem:6} below. In the second part the result of Lemma \ref{5.lem:6} is used to prove Theorem \ref{3.thm:8}.
{\lem
\label{5.lem:6}
The contracted product $f^{\vI} {}_J(x,z)   f^{\vP J}{}_K(y,x)$ may be expressed in terms of, 
\bea
 \label{recxx.1}
 f^{\vI } {}_J(x,z)   f^{\vP J}{}_K(y,x)  =  
 \sum_{\ell =0}^s (-)^{s-\ell} \, \Lambda ^{\vI  \shuffle P_s \cdots P_{\ell+1} | P_1 \cdots P_\ell}{}_K(x,y,z)
 \eea
where the right side is built from the following \textit{$x$-reduced} products,
\begin{align}
\label{3.obv9}
\Lambda ^{\vI | \vP }{}_K & (x,y,z)  = 
 f^{\vI } {}_J(x,z)  f^{\vP J}{}_K(y,z) 
- f^{\vP }{}_J(y,z) \Big ( f^{\vI J}{}_K(x,y) - f^{\vI J}{}_K(x,z) \Big ) 
 \\ 
&\quad 
- \om_J(y) f^{\vI \shuffle J \vP }{}_K(x,z) 
- \sum_{\ell=0}^r f^{I_1 \cdots I_\ell } {}_J(x,y) f^{\vP  \shuffle JI_{\ell+1} \cdots I_r }{}_K(y,z) 
\notag \\
&\quad
+ \sum_{\ell=1}^s \big[ f^{\vI \shuffle  P_{\ell} \cdots P_s} {}_J(x,z)  f ^{P_1 \cdots P_{\ell-1} J}{}_K (y,a_\ell) 
-  f^{\vI \shuffle J P_{\ell+1} \cdots P_s} {}_K(x,z) f ^{P_1 \cdots P_\ell}{}_J(y,a_\ell)  \big]
\notag
\end{align}
The points $a_1,\cdots,a_s \in \Sigma$ in the last line are arbitrary. The specific choice by which they are all identified with $z$ produces various cancellations in (\ref{recxx.1}) that lead to (\ref{4.ff.99}).}

\sm

Given the non-constructive proofs of Theorems \ref{3.thm:7} and \ref{3.thm:8} in Appendices \ref{appB.1} and \ref{appB.2}, respectively, we sketch two constructive algorithms in Appendix \ref{app.cons} that  were initially used to generate examples and played an essential role in proposing (\ref{exfay.15}) and (\ref{4.ff.99}).

\subsubsection{Examples}

The simplest example of (\ref{4.ff.99}) with $\vI= I$ and $\vP=\emptyset$ reproduces the weight-two identity (\ref{hf.22}) after applying the interchange identity (\ref{ften.09}). Specializing $\vI= I$ and $\vP=P$ to single letter words leads to the weight-three identity
\bea
\label{ipexpl}
f^{I} {}_J(x,z)   f^{P J}{}_K(y,x)
&=& 
-f^{I}{}_J(x, y) f^{P J}{}_K( y, z)
+f^{P}{}_J(x, y) f^{J I}{}_K(y, z)
\no \\ &&
-f^{P}{}_J(y, z) f^{I J}{}_K(x, y) 
{+} \, f^{J}{}_K(y, z) f^{(P \shuffle I)}{}_J(x, y)
\no \\ &&
\, {-} \, f^{P}{}_J(y, z) f^{J I}{}_K(x, z)
\, {+} \, f^{I}{}_J(x, z) f^{PJ}{}_K(y, z) 
\no \\ &&
{-} \,\omega_J(x) f^{PJI}{}_K(y, z)
\,{+}\, \omega_J(y) f^{PJ I}{}_K( x, z) 
\no \\ &&
\,{+} \, \omega_J(y)  f^{( I \shuffle P ) J}{}_K(x, y) \! \!
\eea
One can derive all instances of Theorem \ref{3.thm:8} 
for arbitrary $\overrightarrow{P}$ and $\overrightarrow{I}\neq \emptyset$
from suitable combinations of Theorem \ref{3.thm:7} with different 
choices of the multi-indices. The key idea is to solve (\ref{exfay.15}) for the 
term $f^{\overrightarrow{P} M}{}_J(y,x) f^{\overrightarrow{I}J}{}_K(x,z) $ on the right side
which has a repeated appearance of the $(1,0)$-form point $x$ and
where the factor of $f^{\overrightarrow{P} M}{}_J(y,x)$ carries a maximum number of indices.\footnote{The term $f^{\overrightarrow{P} M}{}_J(y,x) f^{\overrightarrow{I}J}{}_K(x,z) $ in (\ref{exfay.15})
can be lined up with the index structure of the left side $f^{\overrightarrow{I}} {}_J(x,z)   f^{\overrightarrow{P} J}{}_K(y,x)$ of (\ref{4.ff.99}) by means of the matrix commutator identity (with arbitrary $a,b\in \Sigma$),
\[
f^{\overrightarrow{P} M}{}_J(y,x) f^{\overrightarrow{I}J}{}_K(x,z)  
= f^{\overrightarrow{P} J}{}_K(y,x) f^{\overrightarrow{I}M}{}_J(x,z) 
- f^{\overrightarrow{P} J}{}_K(y,a) f^{\overrightarrow{I}M}{}_J(x,b) 
+ f^{\overrightarrow{P} M}{}_J(y,a) f^{\overrightarrow{I}J}{}_K(x,b) 
\]}
In this way, $f^{\overrightarrow{I} M}{}_J(y,x) f^{\overrightarrow{P}J}{}_K(x,z) $ can be iteratively
expressed via terms $  f^{P_1 \cdots P_\ell}{}_J(y,x) f^{(\overrightarrow{I} \shuffle J P_{\ell+1}\cdots P_s)M}{}_K(x,z)$ with fewer indices in the $f^{\cdots}{}_J(y,x)$-tensor and \textit{$x$-reduced} terms.
This recursion terminates in the base case
$\omega_J(y) f^{(\overrightarrow{I} \shuffle J   \overrightarrow{P} )M}{}_K(x,z)$ where 
$f^{\emptyset}{}_J(y,x)= \omega_J(y)$ only leaves a
single $x$-dependent factor.

\subsubsection{Comments on Theorem \ref{3.thm:8}}

In view of $f^{\emptyset}{}_J(x,y)= \omega_J(x)$, the summand with  $k=0$ and $\ell=s$
 in the second line of (\ref{4.ff.99}) is given by
$- \omega_J(x) f^{ \overrightarrow{P}  J\overrightarrow{I}}{}_K(y,z) $. 
Similarly, the $\ell=0$ term of the last line is
$(-1)^{s-1} \omega_J(y)$ multiplying $[  f^{( \overleftarrow{P}  \shuffle \overrightarrow{I} )J} {}_K(x,y) +  f^{(\overleftarrow{P} J \shuffle  I_1 \cdots I_{r-1} ) I_r} {}_K(x,z) ] $ with
$\overleftarrow{P} = P_s\cdots P_2 P_1$. 

\sm

The terms with a repeated appearance of $x$ on the
left side of (\ref{4.ff.99}) are
\beq
- \partial_y \cG^{\overrightarrow{P}}(y,x) f^{\overrightarrow{I}}{}_K(x,z)  = 
\partial_y \cG^{\overrightarrow{P}}(y,x) \big(  \delta^{I_r}_K \partial_x \cG^{I_1\cdots I_{r-1}}(x,z) 
- \p_x \Phi^{\overrightarrow{I}}{}_K(x) \big)
\label{rptx}
\eeq
Accordingly, the Fay identities needed to \textit{$x$-reduce} the products
$\partial_y \cG^{\overrightarrow{P}}(y,x) \partial_x \cG^{I_1\cdots I_{r-1}}(x,z) $
and $\partial_y \cG^{\overrightarrow{P}}(y,x)  \p_x \Phi^{\overrightarrow{I}}{}_K(x)$
are obtained from the trace and the traceless part of (\ref{4.ff.99}) with respect to $I_r, K$, respectively.

\subsection{Uncontracted and iterated Fay identities}
\label{sec:4.9}

Our main results for tensorial Fay identities at arbitrary weights in (\ref{exfay.15}) and (\ref{4.ff.99}) feature a contracted index $J$ in $f^{\overrightarrow{P}M}{}_J(y,z) f^{\overrightarrow{I}J}{}_K(x,z)$ and $f^{\overrightarrow{I}} {}_J(x,z)   f^{\overrightarrow{P} J}{}_K(y,x) $. In this section, we describe simple manipulations that generalize
the earlier Fay identities to situations where all indices are free, leading to ``uncontracted Fay identities''.
In this form, Fay identities can be iterated, and we provide an algorithm to  \textit{$z$-reduce}
 products $\prod_{j=1}^N f^{\overrightarrow{P_j}}{}_{K_j}(x_j,z) $
with arbitrary multi-indices $\overrightarrow{P_j}$ and possibly an extra factor of
$f^{\overrightarrow{R} }{}_{M}(z,y)$.

\subsubsection{Uncontracted Fay identities for repeated scalar points}
\label{s:fay.6.6}

The driving force for the derivation of uncontracted Fay identities is the
mild generalization of (\ref{rewr.1})
\begin{align}
f^{\overrightarrow{P}M}{}_K(y,z) \big[ f^{\overrightarrow{I}Q}{}_L(x,z) - f^{\overrightarrow{I}Q}{}_L(x,a) \big]
&= f^{\overrightarrow{P}M}{}_K(y,z) \delta^Q_L \big[ \p_x \cG^{\overrightarrow{I}}(x,a) - \p_x \cG^{\overrightarrow{I}}(x,z) \big]
\label{rewr.6} \\
&= \delta^Q_L f^{\overrightarrow{P}M}{}_J(y,z)  \big[ f^{\overrightarrow{I} J}{}_K(x,z) - f^{\overrightarrow{I} J}{}_K(x,a) \big]
\notag
\end{align}
valid for arbitrary $x,y,z,a \in \Sigma$ which makes use of the fact that all the dependence
of $f^{\overrightarrow{I}Q}{}_L(x,z)$ on the second point $z$ is concentrated in the trace $\delta^Q_L$ 
with respect to its last two indices. The same idea leads to the rearrangement
\begin{align}
\big[ f^{\overrightarrow{P}Q}{}_L(y,z) -  f^{\overrightarrow{P}Q}{}_L(y,a) \big] f^{\overrightarrow{I}M}{}_K(x,z)
&= \delta^Q_L \big[ \p_y \cG^{\overrightarrow{P}}(y,a) -  \p_y \cG^{\overrightarrow{P}}(y,z) \big] f^{\overrightarrow{I}M}{}_K(x,z) \label{rewr.7} \\
&= \delta^Q_L \big[ f^{\overrightarrow{P}M}{}_J(y,z) - f^{\overrightarrow{P}M}{}_J(y,a) \big] f^{\overrightarrow{I}J}{}_K(x,z) \notag
\end{align}
As a result, we can enforce index contractions in an uncontracted Fay identity either via~(\ref{rewr.6})
\begin{align}
& f^{\overrightarrow{P}M}{}_K(y,z)  f^{\overrightarrow{I}Q}{}_L(x,z) =
 \delta^Q_L f^{\overrightarrow{P}M}{}_J(y,z)  f^{\overrightarrow{I} J}{}_K(x,z) \label{rewr.8} \\
 &\quad
   +  f^{\overrightarrow{P}M}{}_K(y,z)  f^{\overrightarrow{I}Q}{}_L(x,a)
  -  \delta^Q_L f^{\overrightarrow{P}M}{}_J(y,z)    f^{\overrightarrow{I} J}{}_K(x,a)  
\notag
\end{align}
or via (\ref{rewr.7}) 
\begin{align}
&f^{\overrightarrow{P}Q}{}_L(y,z)  f^{\overrightarrow{I}M}{}_K(x,z) =
 \delta^Q_L  f^{\overrightarrow{P}M}{}_J(y,z)   f^{\overrightarrow{I}J}{}_K(x,z)
\label{rewr.9} \\
&\quad
+ f^{\overrightarrow{P}Q}{}_L(y,a)   f^{\overrightarrow{I}M}{}_K(x,z)
-  \delta^Q_L  f^{\overrightarrow{P}M}{}_J(y,a) f^{\overrightarrow{I}J}{}_K(x,z)
 \notag
\end{align}
In both of (\ref{rewr.8}) and (\ref{rewr.9}), the only term with a repeated point $z$ on the right side is $f^{\overrightarrow{P}M}{}_J(y,z)   f^{\overrightarrow{I}J}{}_K(x,z)$.
The latter has exactly the right index configuration (including the contraction of $J$) to apply the contracted Fay identity (\ref{exfay.15}), eliminating the repeated appearance of $z$. Hence, both of (\ref{rewr.8}) and (\ref{rewr.9}) can be viewed as uncontracted Fay identities that \textit{$z$-reduce} the product
$ f^{\overrightarrow{P}M}{}_K(y,z)  f^{\overrightarrow{I}Q}{}_L(x,z)$
by applying (\ref{exfay.15}) to $f^{\overrightarrow{P}M}{}_J(y,z)   f^{\overrightarrow{I}J}{}_K(x,z)$ on the right side.

\sm

The arbitrary points $a$ in (\ref{rewr.8}) and (\ref{rewr.9}) can be identified with $x$ or $y$ without altering
the desired simplification of the $z$ dependence. In presence of $f$-tensors depending on additional points,
however, different choices of $a$ might turn out to be even more opportune.

\subsubsection{Uncontracted Fay identities for repeated one-form points}
\label{s:fay.6.7}

The uncontracted version of the Fay identities (\ref{4.ff.99}) to eliminate the
repeated $(1,0)$-form point $x$ in $f^{\overrightarrow{I}} {}_J(x,z)   f^{\overrightarrow{P} J}{}_K(y,x) $ 
can be obtained from the same techniques. We shall only spell out one of the two
possible rearrangements analogous to (\ref{rewr.6}) and (\ref{rewr.7})
\begin{align}
f^{\overrightarrow{I}}{}_K(x,z) \big[ f^{\overrightarrow{P}Q}{}_L(y,x) - f^{\overrightarrow{P}Q}{}_L(y,a) \big]
&= f^{\overrightarrow{I}}{}_K(x,z) \delta^Q_L \big[ \p_y \cG^{\overrightarrow{P}}(y,a) - \p_y \cG^{\overrightarrow{P}}(y,x) \big]
\label{rewr.11} \\
&= \delta^Q_L f^{\overrightarrow{I}}{}_J(x,z)  \big[ f^{\overrightarrow{P} J}{}_K(y,x) - f^{\overrightarrow{P} J}{}_K(y,a) \big]
\notag
\end{align}
As a result, we are led to the uncontracted Fay identity
\begin{align}
&f^{\overrightarrow{I}}{}_K(x,z)  f^{\overrightarrow{P}Q}{}_L(y,x) 
= \delta^Q_L f^{\overrightarrow{I}}{}_J(x,z)   f^{\overrightarrow{P} J}{}_K(y,x)  
\label{rewr.12} \\
&\quad + f^{\overrightarrow{I}}{}_K(x,z)  f^{\overrightarrow{P}Q}{}_L(y,a) 
-  \delta^Q_L f^{\overrightarrow{I}}{}_J(x,z)    f^{\overrightarrow{P} J}{}_K(y,a)
\notag
\end{align}
where the only term $ f^{\overrightarrow{I}}{}_J(x,z)   f^{\overrightarrow{P} J}{}_K(y,x) $ with a repeated appearance of $x$ on the right side can be \textit{$x$-reduced}
by means of the contracted Fay identity (\ref{4.ff.99}). The arbitrary
point $a \in \Sigma$ can be identified with $z$ without impairing the
simplification of the $x$ dependence, though situations with additional
marked points may suggest different choices.

\subsubsection{Iterated Fay identities}
\label{s:fay.6.8}

The uncontracted Fay identities (\ref{rewr.8}) and (\ref{rewr.12}) allow for
an iterative reduction of higher products of $f$-tensors that share a given
point $z$ an arbitrary number of times. We shall consider products of the form
\beq
\tikzpicture[line width=0.3mm]
\draw(-4.65,0)node{$\displaystyle \prod_{j=1}^N f^{  \overrightarrow{P_j} }{}_{K_j}(x_j,z) \ \ \longleftrightarrow$};
\draw(0,0)node{$\bullet$}node[right]{$z$};
\draw(-2,0)node{$\bullet$}node[above]{$x_1$};
\draw(-1,1.732)node{$\bullet$}node[left]{$x_2$};
\draw(1,1.732)node{$\bullet$}node[right]{$x_3$};
\draw(-1,-1.732)node{$\bullet$}node[left]{$x_N$};
\draw(1,-1.732)node{$\bullet$}node[right]{$x_{N-1}$};
%
%
\draw[->](-2,0)--(-1,0);
\draw(-1,0)--(0,0);
\draw[->](-1,1.732)--(-0.5,0.866);
\draw(-0.5,0.866)--(0,0);
\draw[->](1,1.732)--(0.5,0.866);
\draw(0.5,0.866)--(0,0);
\draw[->](-1,-1.732)--(-0.5,-0.866);
\draw(-0.5,-0.866)--(0,0);
\draw[->](1,-1.732)--(0.5,-0.866);
\draw(0.5,-0.866)--(0,0);
\draw[dashed](1.13137, -1.13137) arc (-45:45:1.6) ;
\endtikzpicture
\label{rewr.31} 
\eeq
which can be viewed as the higher-genus uplift of the product of $(z{-}x_j)^{-1}$ 
in (\ref{starg0}). The visualization as a star graph is based on drawing a directed
edge between vertices $x_j$ and $z$ for each factor of $f^{\overrightarrow{P_j}}{}_{K_j}(x_j,z)$.

\sm

The following algorithm will eventually \textit{$z$-reduce} the product (\ref{rewr.31}). To see this, one starts by applying the uncontracted Fay identity (\ref{rewr.8}) to any two factors (\ref{rewr.31}) -- without loss of generality the first two -- resulting in a  single $z$-dependent factor in each term with a different index structure, 
\begin{align}
f^{\overrightarrow{P_1}}{}_{K_1}(x_1,z) f^{\overrightarrow{P_2}}{}_{K_2}(x_2,z)
&= \sum_{\overrightarrow{Q_2}} 
\Big [ C^{\overrightarrow{P_1} \overrightarrow{P_2}L_2}_{K_1 K_2 \overrightarrow{Q_2}}(x_2,x_1) 
f^{ \overrightarrow{Q_2}}{}_{L_2}(x_1,z)  
\label{rewr.32}  \\
& \hskip 0.5in 
+ D^{\overrightarrow{P_1} \overrightarrow{P_2}L_2}_{K_1 K_2 \overrightarrow{Q_2}}(x_1,x_2) 
f^{ \overrightarrow{Q_2}}{}_{L_2}(x_2,z) \Big ]
\notag
\end{align}
The modular tensors
 $C^{\overrightarrow{P_1} \overrightarrow{P_2} L_2}_{K_1 K_2 \overrightarrow{Q_2}}(x_2,x_1)$
and 
$D^{\overrightarrow{P_1} \overrightarrow{P_2}L_2}_{K_1 K_2 \overrightarrow{Q_2}}(x_1,x_2)$ 
are $(1,0)$-forms in their first arguments built from $f$-tensors and Kronecker-deltas  which can be made fully explicit by combining (\ref{rewr.8}) with the contracted Fay identity (\ref{exfay.15}).\footnote{In view of the arbitrary points 
$(a_\ell,b_\ell)$ in the last line of the contracted Fay identity  (\ref{exfay.15}), its individual terms only take the form of
(\ref{rewr.32}) upon setting $(a_\ell,b_\ell)$ to either $(y,z)$ or $(z,x)$. However, there is no need to make this choice since all of $a_\ell,b_\ell$ drop out separately for each value of $\ell$ in the last line of (\ref{exfay.15}).}
The sum over the multi-index $\overrightarrow{Q_2}$ in (\ref{rewr.32}) includes the case of 
$\overrightarrow{Q_2}= \emptyset$ and is finite since it preserves the weight of both sides. Upon multiplication
with the remaining factors in (\ref{rewr.31}) with $j\geq 3$, the maximal number of $z$-dependent factors is $N{-}1$.

\sm

In the next step, the factors of $ f^{ \overrightarrow{Q_2}}{}_{L_2}(x_1,z)$
and $ f^{ \overrightarrow{Q_2}}{}_{L_2}(x_2,z)$ on the right side of (\ref{rewr.32}) are
combined with another factor from (\ref{rewr.31}) -- without loss of generality 
$f^{\overrightarrow{P_3}}{}_{K_3}(x_3,z)$ -- and one applies
(\ref{rewr.32}) again to these products of two $z$-dependent factors. The result involves
additional modular tensors $C^{\overrightarrow{Q_2} \overrightarrow{P_3}L_3}_{L_2 K_3 \overrightarrow{Q_3}}$ and 
$D^{\overrightarrow{Q_2}\overrightarrow{P_3}L_3}_{L_2 K_3 \overrightarrow{Q_3}}$
that depend on $x_1,x_2,x_3$ but not on $z$. Together with the $j\geq 4$
contributions to (\ref{rewr.31}), the maximal number of $z$-dependent factors is now $N{-}2$.

\sm

By iteratively applying (\ref{rewr.32}) to the product of the residual $z$-dependent factors
of the previous step and the next $f^{\overrightarrow{P_j}}{}_{K_j}(x_j,z)$ from (\ref{rewr.31}),
the $(N{-}1)^{\rm th}$ step eventually results in at most one $z$-dependent factor per term, i.e.\ a \textit{$z$-reduced} expression.
These final $z$-dependent $f$-tensors will be accompanied by up to $N{-}1$ tensors
$C$ and $D$ in (\ref{rewr.32}) (with various index contractions among
different factors) whose explicit form is fully determined by (\ref{rewr.8}) and (\ref{exfay.15}).
The number of terms upon expanding these contracted $C$- and $D$-tensors
will grow drastically with $N$ and the length of the multi-indices $\overrightarrow{P_j}$ in (\ref{rewr.31}).

\sm

The above algorithm can be straightforwardly extended to the products (\ref{rewr.31}) multiplying 
an additional 
$(1,0)$-form $f^{\overrightarrow{R} }{}_{M}(z,y)$ in $z$: apply the uncontracted Fay identity
(\ref{rewr.12}) followed by its contracted counterpart (\ref{4.ff.99}) to the product of
$f^{\overrightarrow{R} }{}_{M}(z,y)$ and the final $z$-dependent factor from
the $N{-}1$ iterations of (\ref{rewr.32}). 

\sm

With the above reduction of (\ref{rewr.31}) and its extension to include additional factors of $f^{\overrightarrow{R} }{}_{M}(z,y)$ at hand, we have \textit{$z$-reduced}  the most general polynomial in $f$-tensors compatible with the $(1,0)$-form degree $\leq 1$ (see section \ref{sec:3-x}), and systematically eliminated a wide class of obstructions to the $z$-integration of the higher-genus polylogarithms in~\cite{DHS:2023}. The above argument was carried out for products with star-graph topology in (\ref{rewr.31}), but we will see in section~\ref{sec:5} that coincident limits $x_i \rightarrow x_j$ or $y \rightarrow x_i$ (introducing loops into the star graphs) do not alter the conclusion. The procedure of this section implies that Fay identities involving three  points $x,y,z\in \Sigma$ are sufficient to  eliminate the repeated appearance of any given point in functions of an arbitrary number of points $x_1,\cdots,x_N,y,z$ that are built from $f$-tensors and Abelian differentials.

\newpage

\section{Fay identities and polylogarithms}
\label{sec:clo}

In this section, we illustrate the role of the interchange and Fay identities of the previous sections for the closure of the higher-genus polylogarithms of sections \ref{sec:3.DHS} and \ref{sec:Jmv} under integration over all points they depend on. 

\sm

Primitives with respect to the endpoints $x,y$ of the path that defines the polylogarithms 
$\Gamma (\mw; x,y;p)$ in (\ref{Gacoeff}) and their multi-variable generalization in (\ref{expmv})
readily follow from their construction as iterated integrals. More specifically,
the differential equation, 
\beq
d_x \, 
\boldsymbol{\Gamma} (x,y;p_0;p_1, \cdots , p_n) 
= \cJ_\text{mv} (x,p_0; p_1, \cdots, p_n)  \,  
\boldsymbol{\Gamma} (x,y;p_0;p_1, \cdots , p_n) 
\label{Gsec.01}
\eeq
of the multi-variable path-ordered exponential (\ref{expmv}) determines the primitives of
any $(1,0)\oplus (0,1)$-form in $x$ occurring in the
expansion of the right side. This settles the closure under integration over $x$ for
products of $f^{I_1\cdots I_r}{}_J(x,p_i)$
with polylogarithms $\Gamma(\mw;x,y;p_0;p_1,\cdots, p_n)$ labeled by arbitrary words
$\mw$ in the letters $a^J,b_I$ and $c_i$ with $i=1,\cdots,n$. The \textit{$x$-reduction}
performed by the Fay identities of the previous section furthermore determines
the primitive in $x$ for $\Gamma(\mw;x,y;p_0;p_1,\cdots, p_n)$ multiplied by arbitrary products,
\bea
f^{I_1\cdots I_r}{}_J(x,z_i) \prod_{j=1}^N f^{\overrightarrow{P_j}}{}_{K_j}(x_j,x)
\eea
Recall that products of the type $f^{I_1 \cdots I_r}{}_J(z_i, z_j)  f^{K_1 \cdots K_s}{}_L(z_i, z_k)$ which share their first point $z_i$ never arise, since the corresponding wedge product $\boldsymbol{f}^{I_1 \cdots I_r}{}_J(z_i, z_j)  \wedge \boldsymbol{f}^{K_1 \cdots K_s}{}_L(z_i, z_k)$ of $(1,0)$ forms in the algebra $\cA_N$ of section \ref{sec:3-x} vanishes identically.

\sm 

While closure under integration in the variable $x$ clearly holds true in view of the discussion above, we also claim closure under integration  in all the other points $p_i$ of arbitrary products of multi-variable polylogarithms $\Gamma(\mw;x,y;p_0;p_1,\cdots, p_n)$ and $f^{I_1 \cdots I_r}{}_J(p_i,z)$.
The quest for primitives in the additional points $p_i$ ($i=0,1,\cdots,n$) of the flat connection in (\ref{Gsec.01})
is considerably more challenging since they enter the
defining representation of higher-genus polylogarithms through the second argument of $\p_t \cG^{I_1\cdots I_r}(t,p_i)$.
Our main strategy to prepare for integration over $p_i$ is to rewrite all polylogarithms in
the integrand such that their entire dependence on $p_i$ is moved to the integration limit.
These rewritings are said to \textit{change the fibration basis}.\footnote{The terminology stems from the fact that singling out a particular point $p_i$ amongst the points $p_0,p_1 , \cdots, p_n$ may be formalized in terms of a choice of  fibration of moduli spaces $\mathfrak{M}_{h,n}\rightarrow \mathfrak{M}_{h,n-1}$ for genus $h$ with $n$ and $n{-}1$ punctures, respectively, as advocated, for example, in \cite{Brown:2009qja, BrownLevin}.}
For a generic scalar or tensor-valued function $\Gamma(p_i)$ of the point $p_i$ to be integrated, changes of fibration bases are implemented through the
fundamental theorem of calculus,
\beq
\Gamma(p_i) = \Gamma(q) + \int^{p_i}_q d_\xi \Gamma(\xi)
\label{Gsec.02}
\eeq
as done for polylogarithms at genus zero \cite{Brown:2009qja, Broedel:2013tta, Panzer:2014caa, Panzer:2015ida} and genus one \cite{Broedel:2014vla, Broedel:2018iwv}.
In the cases of our interest, the placeholder $\Gamma(p_i)$ is identified with multi-variable 
higher-genus polylogarithms  $\Gamma(\mw;x,y;p_0;p_1,\cdots, p_n)$. The central task is then to express the
 $(1,0)\oplus (0,1)$-form $d_\xi \Gamma(\xi)$ in terms of the expansion coefficients 
 on the right side of (\ref{Gsec.01}) with $\xi$ in the place of~$x$ -- without any
 additional dependence on $p_i$ in the integrand of (\ref{Gsec.02}). Whenever
 this is accomplished, integrations of (\ref{Gsec.02}) over $p_i$ can be performed 
 with the same ease as the integral of $\Gamma(\mw;x,y;p_0;p_1,\cdots, p_n)$ over $x$.
 In particular, the primitives with respect to $p_i$ of arbitrary products  
 $\Gamma(\mw;x,y; p_0;p_1,\cdots, p_n)$
multiplying $f(p_i,\cdot)$ and possibly additional $f(\cdot,p_i)$-tensors 
become available through algorithmic methods in this case.

\sm

Instead of attempting a general proof that the differentials  $d_\xi \Gamma(\xi) $
can be brought into the desired form,
we shall present three non-trivial case studies in this section. The first one in section
\ref{sec:clo.1} shows through the computation of $d_\xi \Gamma ( b_I a^J; x,y;\xi)$
 that the polylogarithms generated from the connection $  \cJ_\text{DHS} (x,  p)$ of
(\ref{3.conn}) in two variables do not by themselves close under integration over $p$.
Instead, primitives with respect to $p$ automatically introduce the 
multi-variable polylogarithms of section \ref{sec:Jmv}.
The second case study in section \ref{sec:clo.2} necessitates an interchange identity
to attain the desired form of $d_\xi \Gamma (\mw; x,y;\xi)$
with $\mw=a^K a^J b_I$ and exemplifies
that changes of fibration bases of $ \Gamma(\mw;x,y;\xi;p_1,\cdots,p_n)$ are performed
recursively in the length of the words $\mw$. The
third case study in section \ref{sec:clo.3} illustrates the need for Fay identities
to integrate generic $\Gamma(\mw;x,y;p_0;p_1,\cdots, p_n)$ with words $\mw$
of length $\geq 4$ over $p_i$.
These three examples should incorporate the key features of a general integration
algorithm for the higher-genus polylogarithms of \cite{DHS:2023}, and it would be 
valuable to have computer implementations similar to those for polylogarithms \cite{Panzer:2014caa, Duhr:2019tlz, Schnetz:2024}.

\sm

Throughout this section, we assume $p_i \neq x,y$ for the singular points
of the connection in (\ref{Gsec.01}) for all of $i=0,1,\cdots,n$. Cases with
$p_i = x,y$ require a regularization prescription for endpoint divergences of higher-genus 
polylogarithms\footnote{See for instance \cite{Broedel:2014vla, Broedel:2017kkb, Broedel:2018iwv} for regularizations
of elliptic polylogarithms at genus $h=1$
and section 4 of \cite{Baune:2025sfy} for a detailed discussion of regularizing polylogarithms at arbitrary genus.},
e.g.\ via tangential base points \cite{Deligne:1989, Panzer:2015ida, Abreu:2022mfk},
and generate extra terms from partial derivatives in multiple arguments when for instance evaluating
$d_\xi \Gamma(\mw;\xi,y;\xi;p_1,\cdots,p_n)$.
Both of these features in case of $p_i =~x,y$ are straightforwardly reconciled with the 
methods of this section, that is why we keep $p_0,p_1,\cdots,p_n$ distinct from $x,y$ to streamline the discussion.

\subsection{The need for multi-variable polylogarithms}
\label{sec:clo.1}

The protagonist of the first case study is the polylogarithm,
\beq
\Gamma^I{}_J(x,y;p) = \Gamma(b_I a^J; x,y;p) =  \int ^x _y dt \, f^I{}_J(t,p) - \pi \int ^x _y d \bar t \, \bar \om^I(t) \int ^t _y dt' \, \om_J(t')
\label{Gsec.11}
\eeq
which was already discussed in (\ref{gamba}). The opening line (\ref{Gsec.02}) towards integration over $p=p_i$ then specializes to,
\begin{align}
\Gamma^I{}_J(x,y;p) &= \Gamma^I{}_J(x,y;q) + \int^p_q d_\xi \Gamma^I{}_J(x,y;\xi) 
\label{Gsec.12}
\end{align}
where the $p$-independent term $\Gamma^I{}_J(x,y;q)$ on the right side 
is straightforward to integrate over $p$ (upon multiplication with one-forms in $p$). 
The actual challenge is a rewriting of
the integral over $\xi$ in terms of polylogarithms $\Gamma(\mw; p,q;x_i) $ with
$p$-independent coefficients. The double integral in (\ref{Gsec.11})
and the $\partial_t \Phi^I{}_J(t)$ part of $ f^I{}_J(t,p) $ do not depend 
on $p$ and therefore do not contribute to,
\begin{align}
 \int^p_q d_\xi  \Gamma^I{}_J(x,y;\xi)  &= \int^x_y dt \,  \int^p_q d_\xi f^I{}_J(t,\xi) 
 \no \\ & 
= - \delta^I_J  \int^x_y dt  \int^p_q \big( d \xi \, \p_\xi  \p_t \cG(t,\xi)+ d \bar \xi \, \p_{\bar \xi}  \p_t \cG(t,\xi) \big) 
\notag \\
&=   \delta^I_J  \int^p_q d\xi  \int^x_y \big( {-} d_t \, \p_\xi  \cG(t,\xi) + d\bar t \, \p_\xi  \p_{\bar t} \cG(t,\xi) \big)
+  \pi  \delta^I_J \int^p_q   \bar \bom^M \int^x_y \bom_M \notag \\
&= \delta^I_J\, \bigg\{  \int^p_q d\xi \, \big( \p_\xi  \cG(\xi,y) - \p_\xi  \cG(\xi,x) \big)  
\notag \\
&\quad\quad\quad
- \pi  \int^p_q    \bom_M \int^x_y \bar \bom^M
+ \pi  \int^p_q   \bar \bom^M \int^x_y \bom_M \bigg\} \notag \\
&= \delta^I_J\, \big( \Gamma(c_x;p,q; y;x) - \pi \Gamma_M(p,q) \bar \Gamma^M(x,y) + \pi \Gamma_M(x,y) \bar \Gamma^M(p,q)
\big)
\label{Gsec.13}
\end{align}
In passing to the third and to the fourth line, we used $\p_{\bar \xi}  \p_t \cG(t,\xi) = -\pi \omega_M(t) \bar \om^M(\xi)$
and $\p_{ \xi}  \p_{\bar t} \cG(t,\xi) = -\pi \omega_M(\xi) \bar \om^M(t)$
away from the support of the delta function in (\ref{AG.05}).
In the last line, we have identified combinations of the
Abelian integrals (\ref{abints}) 
\begin{align}
\Gamma_{J_1\cdots J_r} ( x,y) &=
\Gamma ( a^{J_1} \cdots a^{J_r}; x,y;p)  = 
\int ^x _y   \bom_{J_1}(t_1)  \int ^{t_1} _y  \bom_{J_2}(t_2)  \cdots \int ^{t_{r-1}} _y  \bom_{J_r}(t_r)
\label{Gsec.14} \\
\bar \Gamma^{I_1\cdots I_r} ( x,y) &=  \frac{1}{(- \pi)^{r}} \Gamma ( b_{I_1} b_{I_2} \cdots b_{I_r}; x,y;p)
= \int ^x _y \bar \bom^{I_1} (t_1)  \int ^{t_1} _y  \bar \bom^{I_2} (t_2)  \cdots 
\int ^{t_{r-1}} _y \bar \bom^{I_r} (t_r)
\notag
\end{align}
and the simplest example (\ref{examv})
of multi-variable polylogarithms at higher genus.

\sm

The final form of (\ref{Gsec.13}) with all $p$-dependence as an integration limit
of some $\Gamma(\mw; p,q; \cdots ) $ is tailored to facilitate integration over $p$. As an
example, we compute the primitive of $dp\,\om_K(p) ( \Gamma^I{}_J(x,y;p) 
- \pi \delta^I_J  \Gamma_M(x,y) \bar \Gamma^M(p,q))$, where the subtraction of the
anti-holomorphic term $\sim \bar \Gamma^M(p,q)$ in $p$ ensures closure under $d_p$ and
homotopy invariance of
\begin{align}
&\int^z_q  dp\, \om_K(p) \big( \Gamma^I{}_J(x,y;p) 
- \pi \delta^I_J  \Gamma_M(x,y) \bar \Gamma^M(p,q) \big) \label{Gsec.15}  \\
&= \int^z_q  dp\, \om_K(p) \big(
\Gamma^I{}_J(x,y;q) + \delta^I_J  \Gamma(c_x;p,q; y;x) 
 - \pi \delta^I_J \Gamma_M(p,q) \bar \Gamma^M(x,y) 
\big) \notag \\
&= \Gamma_M(z,q) \Gamma^I{}_J(x,y;q) 
+ \delta^I_J   \big(
 \Gamma(a^K c_x;z,q; y;x)  -\pi  \Gamma_{KM}(z,q) \bar \Gamma^M(x,y) 
\big) \notag
\end{align}
The term $\Gamma(c_x;p,q; y;x)$ in the last line of the rewriting (\ref{Gsec.13}) of 
$\Gamma^I{}_J(x,y;p)$ and its contribution $ \Gamma(a^K c_x;z,q; y;x)$ to the primitive in
(\ref{Gsec.15}) illustrate an important property of the function spaces: Even though
the polylogarithm $\Gamma^I{}_J(x,y;p)$ is generated by the path-ordered exponential
(\ref{3.int}) of the connection $\cJ_{\text{DHS}} (z,p)$ in two variables,
its primitives with respect to the last point $p$ inevitably involve
multi-variable polylogarithms such as $ \Gamma(a^K c_x;z,q; y;x) $ in (\ref{examv}).

\subsection{Primitives from interchange identities}
\label{sec:clo.2}

While the $p$-integration of the example in the previous section did not require any functional
identities of the integration kernels other than $\p_\xi  \p_t \cG(t,\xi)= \p_\xi  \p_t \cG(\xi,t)$, we will now
demonstrate the necessity of interchange identities by means of the example,
\begin{align}
\Gamma_{KJ}{}^I(x,y;p) &= \Gamma(a^K a^J b_I ;x,y;p) 
 \label{Gsec.16} \\
 &= - \int^x_y dt_1 \, \omega_K(t_1) \int^{t_1}_y dt_2 \, f^I{}_J(t_2,p) \notag \\
 &\quad
 - \pi \int^x_y dt_1 \, \omega_K(t_1) \int^{t_1}_y dt_2\, \omega_J(t_2)
 \int^{t_2}_y d\bar t_3\, \bar \omega^I(t_3)  \notag
 \end{align}
Similar to the previous section, we follow the integration strategy of (\ref{Gsec.02}),
exposing that all the $p$-dependence concentrates in the diagonal $\delta^I_J$,
\begin{align}
&\Gamma_{KJ}{}^I(x,y;p) - \Gamma_{KJ}{}^I(x,y;q) = \int^p_q d_\xi \Gamma_{KJ}{}^I(x,y;\xi)
 \label{Gsec.17} \\
 &= \delta^I_J \int^x_y dt_1 \, \omega_K(t_1) \int^{t_1}_y dt_2 \, 
\int^p_q \big( d \xi \, \p_\xi  \p_{t_2} \cG(t_2,\xi)+ d \bar \xi \, \p_{\bar \xi}  \p_{t_2} \cG(t_2,\xi) \big) 
 \notag \\
  &= \delta^I_J \, \bigg\{
\int^p_q d\xi \int^x_y dt_1 \, \omega_K(t_1) \big(
\p_\xi \cG(\xi,t_1) - \p_\xi \cG(\xi,y)
\big)
 \notag \\
 &\quad + \Gamma_M(p,q) 
    \int^x_y dt_1 \, \omega_K(t_1) \int^{t_1}_y d\bar t_2 \, \bar \om^M(t_2)
- \pi \bar \Gamma^M(p,q) \Gamma_{KM}(x,y)    
 \bigg\}
 \notag
 \end{align}
 In passing to the last two lines, we have again used the Laplace equation of the Arakelov Green
 function, rewrote $dt_2   \p_\xi  \p_{t_2} \cG(t_2,\xi)= d_{t_2}   \p_\xi  \cG(t_2,\xi)
 - d\bar t_2   \p_\xi  \p_{\bar t_2} \cG(t_2,\xi)$ and identified the (anti)holomorphic
 polylogarithms via (\ref{Gsec.14}). Even though all the $p$-dependence on the right 
 side of (\ref{Gsec.17}) enters through an upper integration limit, the double integral in the first term
 $\int^p_q d\xi \int^x_y dt_1 \, \omega_K(t_1)\p_\xi \cG(\xi,t_1)$ is not yet of the
 right form to be identified with a polylogarithm in the multi-variable path-ordered exponential
 (\ref{Gsec.01}). In order to show consistency of $\Gamma_{KJ}{}^I(x,y;p)$ with the closure of
 higher-genus polylogarithms under integration over $p$, we need to further simplify this double integral.
 
 \sm
 
 The weight-one interchange identity (\ref{preften.08}) turns out to provide the desired rewriting
 \begin{align}
&\int^p_q d\xi \int^x_y dt_1 \, \omega_K(t_1)\p_\xi \cG(\xi,t_1)  \label{Gsec.18}\\
&= \int^p_q d\xi \int^x_y dt_1 \,\big( \omega_M(t_1)\p_\xi \Phi^M{}_K(\xi)
+ \omega_M(\xi)\p_{t_1} \Phi^M{}_K(t_1)
- \omega_K(\xi)\p_{t_1} \cG(t_1,\xi)
\big) \notag \\
&= \Gamma_M(x,y) \int^p_q d\xi \, \p_\xi \Phi^M{}_K(\xi)
+  \Gamma_M(p,q) \int^x_y dt \, \p_\xi \Phi^M{}_K(t)
- \int^p_q d\xi \, \omega_K(\xi) \int^x_y dt_1 \, \p_{t_1} \cG(t_1,\xi)
\notag
 \end{align}
 The last integral in the third line has been brought into a suitable fibration basis in section \ref{sec:clo.1},
e.g.\ the trace components of (\ref{Gsec.12}) and (\ref{Gsec.13}) are equivalent to
\begin{align}
\int^x_y dt_1 \, \p_{t_1} \cG(t_1,\xi) &= \int^x_y dt_1 \, \p_{t_1} \cG(t_1,q) - \Gamma(c_x;\xi,q;y;x)
\label{Gsec.19}\\
&\quad - \pi \Gamma_M(x,y) \bar \Gamma^M(\xi,q) + \pi \Gamma_M(\xi,q) \bar \Gamma^M(x,y) \notag
\end{align}
Upon insertion into (\ref{Gsec.18}), the challenging double integral
 $\int^p_q d\xi \int^x_y dt_1 \, \omega_K(t_1)\p_\xi \cG(\xi,t_1)$ is
 expressed in terms of multi-variable polylogarithms generated by
 (\ref{Gsec.01}), and we can bring the right side of (\ref{Gsec.17})
into the following final form,
\begin{align}
\Gamma_{KJ}{}^I(x,y;p) - &\Gamma_{KJ}{}^I(x,y;q) =
\delta^I_J \, \Big\{ {-}\Gamma_M(x,y)  \Gamma_K{}^M(p,q;y)
-\Gamma_M(p,q)  \Gamma_K{}^M(x,y;q)
 \label{Gsec.21}  \\
 &\quad\quad  +  \Gamma(a^K c_x;z,q; y;x)  - \pi \Gamma_{KM}(p,q)  \bar \Gamma^M(x,y)
 - \pi  \Gamma_{KM}(x,y) \bar \Gamma^M(p,q) \Big\}
 \notag
 \end{align}
using the notation $\Gamma(a^K c_x;z,q; y;x) $ for the multi-variable polylogarithm
in (\ref{examv}) and the following shorthand for the variant of the polylogarithm (\ref{Gsec.11})
\begin{align}
 \Gamma_K{}^M(x,y;q) = 
\Gamma(a^K b_M;x,y;q) = \int^x_y dt \, f^M{}_K(t,q) + \pi  \int^x_y dt \, \omega_K(t) \int^t_y d\bar t' 
\, \bar \omega^M(t')
 \label{Gsec.22}
\end{align} 
In summary, the quest for primitives of $\Gamma_{KJ}{}^I(x,y;p) $ with respect to $p$ necessitates
both the weight-one interchange identity (\ref{preften.08}) and the change of fibration basis
performed for the simpler polylogarithm $\Gamma^I{}_J(x,y;p) $ in section \ref{sec:clo.1}. This
illustrates the more general phenomenon that the changes of fibration bases for polylogarithms 
$\Gamma(\mw;x,y;p_0;p_1,\cdots, p_n)$ needed for closure under integration over any $p_i$ are
implemented recursively in the length of the word~$\mw$.

\subsection{Primitives from Fay identities}
\label{sec:clo.3}

Our last case study is dedicated to the simplest double integral involving two
non-trivial kernels $dt_1 f^L{}_K(t_1,z) dt_2 f^I{}_J(t_2,p)$ with two distinct points $z\neq p$ in their second arguments.
A convenient homotopy-invariant realization via multi-variable polylogarithms of (\ref{Gsec.01}) is given by,
\begin{align}
&\hat \Gamma^L{}_{KJ}{}^I(x,y;p,z) = \Gamma(b_L a^K a^J b_I;x,y;p) + \delta^L_K 
\Gamma(c_z a^J b_I;x,y;p;z)  
 \label{Gsec.26} \\
 &\ \ =  - \int^x_y dt_1 \, f^L{}_K(t_1,z) \int^{t_1}_y dt_2 \, f^I{}_J(t_2,p)
 + \pi^2     \int^x_y  \bar \bom^L(t_1) \int^{t_1}_y  \bom_K(t_2) 
\int^{t_2}_y  \bom_J(t_3) \int^{t_3}_y   \bar \bom^I(t_4) 
  \notag \\
  &\quad \ \
 + \pi     \int^x_y \bar \bom^L(t_1) \int^{t_1}_y   \bom_K(t_2) 
 \int^{t_2}_y dt_3\,  f^I{}_J(t_3,p)
   - \pi  \int^x_y dt_1\, f^L{}_K(t_1,z) \int^{t_1}_y \bom_J(t_2) \int^{t_2}_y  \bar \bom^I(t_3)
 \notag
 \end{align}
 %
Our general strategy (\ref{Gsec.02}) then brings the $p$-dependence into the form of,
 \begin{align}
& \hat \Gamma^L{}_{KJ}{}^I(x,y;p,z)  - \hat \Gamma^L{}_{KJ}{}^I(x,y;q,z)  
  \label{Gsec.27} \\
  &= - \int^x_y dt_1 \, f^L{}_K(t_1,z) \int^{t_1}_y dt_2 \,
  \int^p_q \big( d\xi \, \p_\xi f^I{}_J(t_2,\xi)+  d\bar \xi \, \p_{\bar \xi} f^I{}_J(t_2,\xi) \big) \notag \\
  &\quad  + \pi     \int^x_y \bar \bom^L(t_1) \int^{t_1}_y   \bom_K(t_2) 
 \int^{t_2}_y dt_3\, 
 \int^p_q \big( d\xi \, \p_\xi f^I{}_J(t_3,\xi)+  d\bar \xi \, \p_{\bar \xi} f^I{}_J(t_3,\xi) \big)
  \notag \\
  &= - \pi \delta^I_J \bar \Gamma^M(p,q) \Gamma^L{}_{KM}(x,y;z) + \delta^I_J \big( \cI_1^L{}_K(x,y;p,q;z) + \cI_2^L{}_K(x,y;p,q;z)\big)
  \notag
 \end{align}
 where the first term in the last line reduces to a polylogarithm of section \ref{sec:3.DHS},
\begin{align}
\Gamma^L{}_{KM}(x,y;z) &= \Gamma(b_L a^K a^M;x,y;z)    \label{Gsec.31} \\ 
&= \int^x_y dt_1 \, f^L{}_K(t_1,z) \int^{t_1}_y \bom_M(t_2) - \pi 
\int^x_y \bar \bom^L(t_1) \int^{t_1}_y \bom_K(t_2)\int^{t_2}_y \bom_M(t_3) \notag
 \end{align}
However, the two additional integrals $\cI_1^L{}_K$, $\cI_2^L{}_K$ in the last line of (\ref{Gsec.27}) require further simplifications
before they can be identified with multi-variable polylogarithms that occur in (\ref{Gsec.01}),
\begin{align}
\cI_1^L{}_K(x,y;p,q;z)  &= \int^x_y dt_1 \, f^L{}_K(t_1,z) \int^{t_1}_y dt_2 \int^p_q d\xi \, \p_\xi \p_{t_2} \cG(t_2,\xi)
 \label{Gsec.32} \\
&= \int^x_y dt_1 \, f^L{}_K(t_1,z)   \int^p_q d\xi \, \big( \p_\xi  \cG(\xi,t_1) -  \p_\xi  \cG(\xi,y) \big)
\notag \\
&\quad + \pi \Gamma_M(p,q) \int^x_y dt_1 \, f^L{}_K(t_1,z)  \int^{t_1}_y \bar \bom^M(t_2)
\notag \\
\cI_2^L{}_K(x,y;p,q;z) &=  - \pi     \int^x_y \bar \bom^L(t_1) \int^{t_1}_y   \bom_K(t_2) 
 \int^{t_2}_y dt_3\, 
 \int^p_q  d\xi \,  \p_\xi \p_{t_3} \cG(t_3,\xi) \notag \\
 &=   - \pi     \int^x_y \bar \bom^L(t_1) \int^{t_1}_y   \bom_K(t_2) 
 \int^p_q  d\xi \,  \big( \p_\xi \cG(\xi,t_2) -  \p_\xi \cG(\xi,y) \big) \notag \\
 &\quad - \pi^2 \Gamma_M(p,q)   \int^x_y \bar \bom^L(t_1) \int^{t_1}_y   \bom_K(t_2) \int^{t_2}_y \bar \bom^M(t_3) \notag
\end{align}
Both cases necessitate the \textit{$t_i$-reduction} of a product
$ f^L{}_K(t_1,z)   \p_\xi  \cG(\xi,t_1) $ or $\bom_K(t_2)  \p_\xi \cG(\xi,t_2)$.
In case of $\cI_2^L{}_K$, this is resolved through the weight-one interchange identity (\ref{preften.08})
\begin{align}
& - \pi     \int^x_y \bar \bom^L(t_1) \int^{t_1}_y   \bom_K(t_2) 
 \int^p_q  d\xi \,   \p_\xi \cG(\xi,t_2) 
\notag \\
 &=  \pi     \int^x_y \bar \bom^L(t_1) \int^p_q  d\xi  \int^{t_1}_y  \,
 \Big\{
 \omega_K(\xi) \p_{t_2 }\cG(t_2,\xi) - \omega_M(\xi) \p_{t_2}\Phi^M{}_K(t_2) 
 - \omega_M(t_2) \p_{\xi}\Phi^M{}_K(\xi) 
 \Big\} \notag \\
 &= \pi  \, \bigg\{
 {-} \Gamma_M(p,q)   \int^x_y \bar \bom^L(t_1) \int^{t_1}_y dt_2 \, \p_{t_2}\Phi^M{}_K(t_2) 
 - \int^p_q d\xi \, \p_{\xi}\Phi^M{}_K(\xi) 
 \int^x_y \bar \bom^L(t_1) \int^{t_1}_y \bom_M(t_2) \notag \\
&\quad\quad \quad \quad + \int^p_q  \bom_K(\xi) \int^x_y \bar \bom^L(t_1) 
\int^{t_1}_y dt_2 \, \p_{t_2}\cG(t_2,\xi)
\bigg\}  \label{Gsec.34}
\end{align}
The last term is still incompatible with the fibration bases occurring
in the multi-variable polylogarithms of section \ref{sec:Jmv} and will be 
seen to cancel later on.

\sm

We shall proceed to simplifying the integral~$\cI_1^L{}_K$ where
the tensorial Fay identity (\ref{hf.22}) is needed to
\textit{$t_1$-reduce} the bilinear $ f^L{}_K(t_1,z)   \p_\xi  \cG(\xi,t_1) $
in the integrand of (\ref{Gsec.32}). In this way, the most challenging 
contribution to $\cI_1^L{}_K$ takes the form,
\begin{align}
&\int^x_y dt_1 \, f^L{}_K(t_1,z)   \int^p_q \! d\xi \, \p_\xi  \cG(\xi,t_1)
= \int^x_y dt    \int^p_q \! d\xi \, \Big\{
\p_\xi \Phi^L{}_M(\xi) f^M{}_K(t,z) + f^L{}_M(t,\xi) f^M{}_K(\xi,z) \notag \\
&\quad  \quad\quad \quad \! \! - f^L{}_M(\xi,z) f^M{}_K(t,z)+ \omega_M(\xi) f^{LM}{}_K(t,\xi) + \omega_M(t) f^{ML}{}_K(\xi,z) + \omega_M(\xi) f^{ML}{}_K(t,z) 
\Big\} \notag \\
&= \int^p_q \! d\xi \, \p_\xi \cG(\xi,z) \int^x_y dt \, f^L{}_K(t,z) + \Gamma_M(x,y) \int^p_q \! d\xi \, f^{ML}{}_K(\xi,z)
+ \Gamma_M(p,q) \int^x_y dt \, f^{ML}{}_K(t,z) \notag \\
&\quad\quad 
+  \int^p_q \! d\xi \,f^M{}_K(\xi,z) \int^x_y dt \, f^{L}{}_M(t,\xi)
+  \int^p_q  \bom_M(\xi) \int^x_y dt \, f^{LM}{}_K(t,\xi)
 \label{Gsec.33}
\end{align}
The last line features two terms $ \int^x_y dt \, f^{L}{}_M(t,\xi)$ and  $ \int^x_y dt \, f^{LM}{}_K(t,\xi)$ 
which are not yet in a suitable fibration basis for integration over $\xi$. The former has already been
simplified in (\ref{Gsec.19}), and the latter requires a separate computation along the lines of section \ref{sec:clo.1},
\begin{align}
\int^x_y dt \, f^{LM}{}_K(t,\xi) &= \int^x_y dt \, f^{LM}{}_K(t,q)  - \delta^M_K \int^x_y dt \int^\xi_q \big(
d\eta \, \p_\eta \p_t \cG^L(t,\eta) + d\bar \eta \, \p_{\bar \eta} \p_t \cG^L(t,\eta)
\big) \notag \\
&=  \int^x_y dt \, f^{LM}{}_K(t,q)+ \delta^M_K 
\int^\xi_q d\eta \, \big( \p_\eta \cG^L(\eta,x) -  \p_\eta \cG^L(\eta,y) \big) \notag \\
&\quad + \pi \delta^M_K \, \bigg\{ 
 \int^\xi_q \bar \bom^R(\eta) \int^x_y dt \, f^L{}_R(t,\eta) 
+ \int^x_y  \bar \bom^R(t) \int^\xi_q d \eta\, f^L{}_R(\eta,t) 
\bigg\} 
\label{Gsec.35}
\end{align}
For both integrals $ \int^x_y dt \, f^L{}_R(t,\eta) $ and $ \int^\xi_q d \eta\, f^L{}_R(\eta,t) $
in the last line, we perform another change of fibration basis via (\ref{Gsec.19}). The 
latter then produces a term $-\delta^L_R \int^t_y ds \, \p_s \cG(s,\xi)$ which -- upon
integration against $ \int^x_y  \bar \bom^R(t)$ and $\int^p_q \bom_M(\xi)$ in the last lines 
of (\ref{Gsec.35}) and (\ref{Gsec.33}) --
cancels the term $ \int^p_q  \bom_K(\xi) \int^x_y \bar \bom^L(t_1) 
\int^{t_1}_y dt_2 \, \p_{t_2}\cG(t_2,\xi)$ from the simplification of $\cI_2^L{}_K$ in (\ref{Gsec.34}).

\sm

As a result of the above manipulations, the sum over the integrals $\cI_1^L{}_K$ 
and $\cI_2^L{}_K$ in (\ref{Gsec.32}) is expressible in terms of homotopy-invariant 
multi-variable polylogarithms in (\ref{Gsec.01}) with all $p$-dependence in the upper integration limit:
\begin{align}
&\cI_1^L{}_K(x,y;p,q;z)  + \cI_2^L{}_K(x,y;p,q;z)  = \Gamma(a^K b_L c_y;p,q;x;y)
+ \pi \bar \Gamma^L(x,y) \Gamma(a^K c_y;p,q;z;y) \notag \\
&\quad
+ \pi \bar \Gamma^M(x,y) \Gamma(a^K b_L a^M;p,q;z)
- \pi^2 \Gamma_M(p,q) \Gamma_K(x,y) \bar \Gamma^{LM}(x,y)
+ \pi^2 \Gamma_{KM}(p,q) \bar \Gamma^{LM}(x,y)
\notag \\
&\quad
+ \Gamma_M(x,y) \Gamma(a^K b_L b_M;p,q;z)
+ \Gamma_M(p,q) \Gamma(a^K b_L b_M;x,y;z)
+ \Gamma_M(p,q) \Gamma(b_L b_M a^K;x,y;q)
\notag \\
&\quad
+ \delta^L_K \Gamma(c_y;p,q;z;y) \Gamma(c_z;x,y;q;z)
- \Gamma(b_L a^M;x,y;q) \Gamma(a^K b_M;p,q;y)
\label{Gsec.39}
\end{align}
Together with (\ref{Gsec.27}), this prepares the combination of
multi-variable polylogarithms in (\ref{Gsec.26}) for integration over $p$
and illustrates the closure of $\Gamma(\mw;x,y;p;p_1,\cdots, p_n)$ under taking primitives in $p$
in a non-trivial case that relies on a tensorial Fay identity.

\newpage

\section{Coincident limits of Fay identities}
\label{sec:5}

In this section, we shall investigate the coincident limits of the modular tensors $\partial_x \cG^{I_1 \cdots I_r} (x,y)$ and $f^{I_1 \cdots I_r}{}_J(x,y)$  as $y \to x$,  as well as the coincident limits of the Fay identities constructed in section \ref{sec:fay} for  three points $x,y,z$, as  $z\rightarrow x$ or $z \rightarrow y$. In particular, we will show that the coincident limit of the modular tensors $\partial_x \cG^{I_1\cdots I_r}(x,y) $ produces constant modular tensors $\Nf^{ P_1\cdots P_s}$ of various  ranks $s\leq r{+}1$ that restrict to (almost) holomorphic Eisenstein series at genus one.
We will also \textit{$x$-reduce} products $f^{ \overrightarrow{I} } {}_J(x,y) f^{ \overrightarrow{P} J} {}_K(y,x) $  of $f$-tensors which share both points $x$ and $y$. Hence, primitives with respect to the shared point $x$ can be constructed in the same function space of higher-genus polylogarithms~\cite{DHS:2023} as in the  case of products (\ref{4.ff.99}) or (\ref{rewr.31}) of $f$-tensors with a single  point shared by an arbitrary number of factors.

\subsection{Coincident limits of genus-one Fay identities}
\label{sec:cl.0}

A crucial step in this section is to generalize the coincident limit of the Kronecker-Eisenstein coefficients on the torus,
\beq
\lim_{y \rightarrow x} f^{(r)}(x{-}y) = -{\rm G}_r
\, , \ \ \ \ r \geq 3
\label{csec.01}
\eeq
to arbitrary genus. Our normalization for the holomorphic Eisenstein series ${\rm G}_r$ is as follows, 
\beq
{\rm G}_r = \sum_{m,n \in \mathbb Z^2 \atop{(m,n) \neq (0,0)}} \frac{1}{(m\tau {+}n)^r}  
\, , \ \ \ \ r \geq 3
\label{csec.02}
\eeq
Recall that the restriction of both equations to $r\geq 3$ is required by two types of subtleties.  First, the short-distance behavior of $f^{(2)}(z)$ features a contribution $\frac{\pi(z{-}\bar z)}{z\Im \tau}$ whose limit as $z\rightarrow 0$ depends on the direction along which the limit is taken. Second, while the double sums in (\ref{csec.02}) are absolutely convergent for $r>2$ they are only conditionally convergent at smaller values  $r\leq 2$. The holomorphic quasi modular Eisenstein series ${\rm G}_2$ may be defined using the Eisenstein summation prescription and is related to the  
almost holomorphic modular completion $\widehat {\rm G}_2$ by the second relation below (see for example \cite{Zagier123} and \cite{DHoker:2022dxx}),
\beq
{\rm G}_2 = \lim_{M\rightarrow \infty} \sum^M_{m=-M}  \lim_{N\rightarrow \infty}  \sum^N_{n=-N}  \frac{\delta_{(m,n)\neq (0,0)}}{(m\tau {+}n)^2}  \, ,
\hskip1in
 \widehat {\rm G}_2= {\rm G}_2- \frac{\pi}{\Im \tau}
\label{csec.03}
\eeq
where the notation $\delta_{(m,n)\neq (0,0)}$ instructs us to drop the summand with $(m,n)= (0,0)$.
Alternatively, the modular version $\widehat {\rm G}_2$ of weight-two Eisenstein series arises in the limit,
\beq
\lim_{y \rightarrow x}\bigg( f^{(2)}(x{-}y) + \frac{\pi }{ \Im \tau}  \frac{ (\bar x{-} \bar y) }{x{-}y} \bigg) = - \widehat {\rm G}_2
\label{csec.04}
\eeq
which is well-defined by the subtraction of the problematic term $\sim \frac{\bar z}{z}$ of $f^{(2)}(z)$.
We will generalize the coincident limits (\ref{csec.01}) and (\ref{csec.04}) to arbitrary genus and encounter
modular tensors $\Nf^{ P_1\cdots P_s}$ with a higher-genus analogue of the integral representations 
of (almost) holomorphic Eisenstein series \cite{Lerche:1987qk}, 
\begin{align}
{\rm G}_r &= \bigg( \prod_{j=1}^r \int_{\Sigma} \frac{d^2 x_j}{\Im \tau} \bigg) \, \partial_{x_1} \cG(x_{1}, x_2) 
 \partial_{x_2}   \cG(x_{2}, x_3) \cdots  
 \partial_{x_{r-1}}  \cG(x_{r-1}, x_r)  \partial_{x_r}  \cG(x_{r}, x_1) \, , \ \ \ \ r\geq 3
\notag \\
 \widehat {\rm G}_2 &=    \int_{\Sigma} \frac{d^2 x_1}{\Im \tau}
 \int_{\Sigma} \frac{d^2 x_2}{\Im \tau}\,   \Big(
\partial_{x_1} \cG(x_{1}, x_2) 
 \partial_{x_2}   \cG(x_{2}, x_1)  -  \partial_{x_1}  
 \partial_{x_2} \cG(x_{1}, x_2) \Big)
 \label{csec.05}
\end{align}
in terms of the Arakelov Green function $\cG(x,y)$ on the torus defined in (\ref{agft}). In contrast to
the vanishing of Eisenstein series ${\rm G}_{2\ell+1}$ at odd weight, their higher-genus
counterparts $\Nf^{ P_1\cdots P_s}$ turn out to be non-trivial also  at odd rank $s \in 2\mathbb N {+}1$.

\sm

In view of the relations (\ref{csec.01}) and (\ref{csec.04}) between Kronecker-Eisenstein coefficients and (almost) holomorphic Eisenstein series, the coincident limit $y\rightarrow x$ of the genus-one Fay identity (\ref{exfay.16}) takes the form  (see Appendix A of \cite{Gerken:2019cxz} or section 6.3 of \cite{Gerken:2020aju}),
\begin{align}
  \label{coin.67} 
 f^{(r)}(z)  f^{(s)}(z) 
 &=
\mbinom{r{+}s}{ r} f^{(r+s)}(z)   
- \sum_{\ell=4}^{r} \mbinom{r{+}s{-}1{-}\ell }{s{-}1} \, {\rm G}_\ell \, f^{(r+s-\ell)}(z) 
  \no \\ & \quad 
 + (-1)^{s} {\rm G}_{r+s}  
 - \sum_{\ell=4}^{s} \mbinom{r{+}s{-}1{-}\ell }{r{-}1} \, {\rm G}_\ell \, f^{(r+s-\ell)}(z)  
  \notag \\
  &\quad
 - \mbinom{r{+}s{-}2}{r{-}1} \Big[  \partial_z f^{(r+s-1)}(z) + \widehat {\rm G}_2 \, f^{(r+s-2)}(z) \Big] 
\end{align}
with at most one $z$-dependent factor in each term on the right side. For instance, the coincident limit
of the weight-two Fay identity (\ref{gen1.03}) akin to partial fraction gives rise
to the following identity involving double poles,
\beq
\big(  f^{(1)}(z) \big)^2 = 2 f^{(2)}(z) - \partial_z f^{(1)}(z) -  \widehat {\rm G}_2
 \label{csec.06}
\eeq
The main results of this section will be the generalizations of (\ref{coin.67}) to arbitrary genus in (\ref{3.coin.0}) and (\ref{coin.75}) below which produces an \textit{$x$-reduced} form for  the product  $f^{ \overrightarrow{I} } {}_J(x,y) f^{ \overrightarrow{P} J} {}_K(y,x) $ in terms of $f$-tensors, their derivatives in the second point and constant modular tensors $\Nf^{ P_1\cdots P_s}$.

\subsection{Higher-genus coincident limits at weight two}
\label{sec:cl.1}

We start by generalizing the coincident limits of (\ref{csec.04}) and (\ref{csec.06}) of weight two and genus one to  arbitrary genus. This may be achieved by organizing the  coincident limit $z\to y$ of the modular scalar three-point Fay 
identity (\ref{3.a.2}) at higher genus by grouping together those terms whose limit is immediate and those terms whose limit is not,
\begin{align}
\lim_{z\to y} & \Big [  \Big ( \p_x \cG(x,y)  -  \p_x \cG(x,z) \Big ) \p_y \cG(y,z) - \om_I (x) \p_y \cG^I(y,z) \Big ]
\no \\ 
&\quad = 
 \om_I(y) \p_x \cG^I(x,y) - \p_x \p_y \cG_2(x,y) -  \p_x \cG(x,y) \p_y \cG(y,x) 
  \label{csec.11}
 \end{align}
Here and below, the short-distance behavior $\p_y \cG(y,z) \sim (z{-}y)^{-1}$ introduces derivatives of the accompanying functions of $y$ and $z$. The difference of $\p_x \cG(x,y)  $ and $ \p_x \cG(x,z)$ inside the limit of the first line can be converted to a derivative of a meromorphic function using the relations (\ref{stringgf}) and (\ref{A.GG}) between the Arakelov Green function $\cG(x,y)$ and the prime form $E(x,y)$, which in turn is  defined in (\ref{defpform}),
\begin{align}
\p_x \cG(x,y) - \p_x \cG(x,z) & =  - \p_x \ln { E(x,y) \over E(x,z)} + 2 \pi i \om_I(x)  \left ( \Im \int ^y_z \bom^I \right )
\label{1.inta}
\end{align}
Grouping terms according to tensorial modular properties proves the following Lemma.

{\lem
\label{7.lem:7}
The coincident limit of the modular scalar three-point Fay  identity (\ref{3.a.2}) at arbitrary genus is given by, 
\begin{align}
\om_I(x) \cC^I(y)
& = 
\p_x \cG(x,y) \p_y \cG(y,x) 
-  \p_x \p_y \cG(x,y) 
+ \p_x \p_y \cG_2(x,y)   - \om_I(y) \p_x \cG^I(x,y)  
 \label{csec.13}
 \end{align}
with the following well-defined limit, 
\begin{align}
\cC^I(y) &=  \lim_{z\to y} \left [  \p_y \cG^I(y,z) +  {  \pi  \over z-y} \int ^y _z \bar \bom^I  \right ]
 \label{csec.12}
 \end{align}}

Similar to the limit of the genus-one term $(\bar x{-}\bar y)/(x{-}y)$ in (\ref{csec.04}), the limit of the second term inside the brackets of  (\ref{csec.12}) by itself would depend  on the direction in which the points $z$ and $y$ approach one another. However, the combination with  $ \p_y \cG^I(y,z)$  leads to a well-defined $(1,0)$-form $\cC^I(y)$ limit. In order to see this, we note that the right side of (\ref{csec.13}) is manifestly single-valued in $x$ and $y$ so that  $\cC^I(y)$ must be single-valued in~$y$. Furthermore,  one verifies that the right side is holomorphic in $x$, as the left side is. Integrating against $\bar \om^I(x)$ and discarding total derivatives of the non-singular and single-valued combination $ \p_y \cG_2(x,y)   - \om_I(y)  \cG^I(x,y) $ gives the following integral representation,
\beq
\cC^I(y) =  
\int _\Sigma d^2 x \, \bar \om^I(x) \, \Big ( \p_x \cG(x,y) \p_y \cG(y,x) - \p_x \p_y \cG(x,y)  \Big )
 \label{csec.17}
 \eeq
The double poles of the terms inside the parentheses on the right side cancel one  another, so that the integral is absolutely convergent and produces a well-defined $(1,0)$-form in $y$.  To obtain a more tractable expression, we evaluate the $\pby$ using the Laplace equations (\ref{AG.05}) for the Arakelov Green function, and express the result in terms of the $\Phi$-tensor,
\bea
\pby \cC^I(y) = - \pi \bar \om^M(y) \p_y \Phi ^I{}_M(y)
= \pby \p_y \Phi ^{MI}{}_M(y)
 \label{csec.19}
 \eea 
Therefore, the combination $\cC^I(y) - \p _y \Phi ^{MI}{}_M(y)$ is holomorphic and single-valued in $y$, so it can be expanded in terms of holomorphic Abelian differentials, i.e.\ we have,
 \bea
 \label{4.coin2}
\cC^I(y) =  \p_y \Phi ^{MI}{}_M(y) +  \om_M(y) \Nf^{M I}
\eea
for a $y$-independent tensor $\Nf^{IJ}$. Upon insertion into (\ref{csec.17}), we  obtain an integral representation for $\Nf^{IJ}$ by integrating against $\bar \om^J(y)$,
\begin{align}
\Nf^{IJ} & =  \int _\Sigma d^2 y \, \bar \om^J(y) \cC^I(y)
 \label{csec.20} \\ 
& =  
\int _\Sigma d^2 x \, \int _\Sigma d^2 y \, \bar \om^I(x) \, \bar \om^J(y) \Big ( \p_x \cG(x,y) \p_y \cG(y,x) - \p_x \p_y \cG(x,y)  \Big )
\notag
\end{align}
This integral is absolutely convergent as the double poles of the terms inside the 
parentheses cancel one another. By the manifest symmetry of the second line in $I,J$,
we deduce that $\Nf^{IJ} = \Nf^{JI} $. Comparison with the integral representation
(\ref{csec.05}) of the almost holomorphic Eisenstein series $ \widehat {\rm G}_2$
 identifies the following restriction to genus one, 
\beq
\Nf^{IJ}\big|_{h=1} = \widehat {\rm G}_2
 \label{csec.00}
\eeq
and shows that the coincident limit (\ref{4.coin2}) at arbitrary genus restricts to the coincident limit (\ref{csec.04}) of $f^{(2)}$ at genus one (we recall that  $\Phi ^{I_1\cdots I_r}{}_J(y) |_{h=1} = 0$ by their vanishing trace~(\ref{Gtrcless})).

\subsubsection{Coincident limit of the tensorial weight-two Fay identity}
\label{sec:cl.1.2}

Based on the simplified coincident limit (\ref{4.coin2}), the
higher-genus Fay identity in (\ref{csec.13}) may be recast as follows, 
\begin{align}
0 & =  
\p_x \cG(x,y) \p_y \cG(y,x) - \p_x \p_y  \cG (x,y) + \p_x \p_y \cG_2(x,y) 
 \label{csec.21} \\
 &\quad
  - \Nf^{IJ} \om_I(x)  \om_J(y)  
 - \om_I(y) \p_x \cG^I(x,y)  
-\om_I(x)  \p_y \Phi ^{MI}{}_M(y)
\notag
\end{align}
The symmetry of the right side in $x,y$ is manifest in the first four terms. 
The symmetry of the remaining terms $ - \om_I(y) \p_x \cG^I(x,y)  
-\om_I(x)  \p_y \Phi ^{MI}{}_M(y)$ under $I\leftrightarrow J$ can be established
from the corollary $ \om_J(y) f^{MJ}{}_M(x,y) -\om_J(x)  f ^{MJ}{}_M(y,x) =0$
of the weight-two interchange identity in (\ref{ften.09}) 
whose last two terms cancel upon contraction with $\delta_I^J$.

\sm

The coincident limit (\ref{csec.21}) of the scalar Fay identity
at weight two can be unified with the traceless component (\ref{1stGphi})
of the tensorial weight-two Fay identity (\ref{hf.22}) to the compact~form,
\beq
f^I{}_J(x,y) f^J{}_K(y,x) = \delta^I_K \p_x \p_y \cG(x,y) - \omega_J(y) f^{I \shuffle J}{}_K(x,y) - \omega_J(x) 
\cF^{JI}{}_K(y)
 \label{csec.26} 
 \eeq
where the last term is given by,
\begin{align}
\cF^{JI}{}_K(y)  &=  \lim_{z\rightarrow y}
\bigg[ f^{JI}{}_K(y,z) + 
{  \pi  \delta^I_K \over z-y} \int ^z _y \bar \bom^J
 \bigg] 
= \p_y \Phi^{JI}{}_K(y) - \delta^I_K \cC^J(y) 
\notag \\
 &=   \p_y \Phi^{JI}{}_K(y) - \delta^I_K \p_y \Phi^{MJ}{}_M(y) -  \delta^I_K \Nf^{JM} \omega_M(y)
  \label{csec.27}  
 \end{align}
One can view (\ref{csec.21}) and (\ref{csec.26}) as the simplest 
 examples of \textit{$x$-reductions} that involve derivatives 
$\p_y f^{\vI M}{}_J(x,y) = - \delta^M_J \p_x \p_y \cG^{\vI}(x,y)$ of $f$-tensors, 
or $d_y \boldsymbol{f}^{\vI M}{}_J(x,y)$ in the notation of section \ref{sec:3-x}.
At genus $h=1$, the Fay identity (\ref{csec.26}) reduces to (\ref{csec.06})
in view of $f^{I_1\cdots I_r}{}_J(x,y) \rightarrow f^{(r)}(x{-}y)$ as well as 
$\cF^{JI}{}_K(y) \rightarrow - \widehat {\rm G}_2$ and 
$ \p_x \p_y \cG(x,y) \rightarrow  \p_x  f^{(1)}(x{-}y)$.

\subsection{Coincident limits of higher-weight $f$-tensor}
\label{sec:cl.2}

Starting from weight three, one can evaluate the coincident
limits $z\rightarrow y$ of the modular tensors $f^{\overrightarrow{I}}{}_J(y,z)$ or
 $\p_y\cG^{P_1 \cdots P_s}(y,z)$ in a more direct way.
 For rank $s\geq 2$, we introduce the shorthand,
\beq
 \cC^{P_1 \cdots P_s}(y) =  \lim_{z\to y}   \p_y \cG^{P_1 \cdots P_s}(y,z) 
 \label{csec.31}
 \eeq 
for the coincident limit at weight $s{+}1$ without the need for any addition of 
anti-holomorphic Abelian integrals as in (\ref{csec.12}).
The goal of this section is to establish both the well-definedness 
and the explicit form of the limits (\ref{csec.31}) through a recursive strategy.
 
\subsubsection{Coincident limit of weight-three tensors}
\label{sec:cl.2.1}
 
We shall first illustrate the recursive computation of the limits (\ref{csec.31})
through the weight-three example at rank $s=2$. The first step is to combine 
the anti-holomorphic derivatives (\ref{ften.93}) of the  $\p_y\cG^{IJ}(y,z)$
tensors in both variables to obtain,
\begin{align}
\partial_{\bar y} \lim_{z \rightarrow y}\partial_y \cG^{IJ}(y,z) &= 
 \pi  \lim_{z \rightarrow y} \big[ \bar \omega^J(z)  \p_y \cG^{I}(y,z) 
-  \bar \omega^I(y)  \p_y \cG^{J}(y,z)  \big]
- \pi \bar \om^M(y) \partial_y \Phi^{IJ}{}_M(y)
\label{coin.02}  
\end{align}
The limits of the individual terms on the right side are
ill-defined since they are lacking the anti-holomorphic Abelian integrals
of the expressions for $\cC^I(y)$ in (\ref{csec.12}).
However, the combination in the square bracket of (\ref{coin.02})
conspires to yield a well-defined limit,
\bea
\label{coin.34}
&&
\lim_{z \rightarrow y} \big[ \bar \omega^J(z)   \p_y\cG^{I}(y,z) 
-  \bar \omega^I(y) \p_y \cG^{J}(y,z)  \big] 
\no \\ && \hskip 0.4in =  
\bar \omega^J(z)  \cC^{I}(y) -  \bar \omega^I(y)  \cC^{J}(y)  
+ \pi \lim_{z \rightarrow y}  \frac{1}{y{-}z} \bigg\{
 \bar \omega^J(z) \int^y_z \bar \bom^I
 -  \bar \omega^I(y) \int^y_z \bar \bom^J
\bigg\} 
\notag\\ && \hskip 0.4in =
 \bar \omega^J(z)  \cC^{I}(y) -  \bar \omega^I(y)  \cC^{J}(y) 
\eea
since the curly bracket in second line of (\ref{coin.34}) vanishes with $(\bar y{-}\bar z)^2$.
Hence, the anti-holomorphic derivative (\ref{coin.02}) of the limit (\ref{csec.31}) at rank $s=2$ 
is well-defined,
\beq
\p_{\bar y} \cC^{IJ}(y)  =  \pi \big[ \bar \omega^J(z)  \cC^{I}(y) -  \bar \omega^I(y)  \cC^{J}(y) \big]
- \pi \bar \om^M(y) \partial_y \Phi^{IJ}{}_M(y) 
\label{coin.35}
\eeq
With the expression (\ref{4.coin2}) for $\cC^{I}(y)$
in terms of $\Phi$ and rank-two $\Nf$ tensors, one can readily integrate (\ref{coin.35}), 
\begin{align}
 \cC^{IJ}(y) &= 
\partial_y \Phi^{IMJ}{}_M(y)
- \partial_y \Phi^{JMI}{}_M(y)
+ \partial_y \Phi^{MIJ}{}_M(y)  \label{cl.09} \\
&\quad +
 \partial_y \Phi^{I}{}_M(y) \Nf^{MJ}
- \partial_y \Phi^{J}{}_M(y) \Nf^{MI}
+ \omega_M(y) \Nf^{MIJ}
\notag
\end{align}
which introduces a rank-three tensor $ \Nf^{MIJ}$ independent on $y$. 
Upon integration against $\bar \omega^M(y)$ and discarding the
total derivatives of the single-valued $\Phi$ tensors, we arrive at
the following integral representation of the new tensor $\Nf^{MIJ}$ in (\ref{cl.09}),
\begin{align}
\Nf^{MIJ} &= \int_\Sigma d^2 y \, \bar \omega^M(y)  \cC^{IJ}(y) 
= \int_\Sigma d^2 y \, \bar \omega^M(y) \lim_{z\rightarrow y} \p_y \cG^{IJ}(y,z) 
\label{coin.37} \\
&= \int_\Sigma d^2 y \, \bar \omega^M(y) \lim_{z\rightarrow y}  \int_\Sigma d^2 x \, \p_y\cG(y,x) \bar \omega^I(x) \p_x \cG^{J}(x,z) \notag
 \\
&= \int_\Sigma d^2 y \, \bar \omega^M(y)  \int_\Sigma d^2 x \,  \bar \omega^I(x)
\int_\Sigma d^2 w \, \bar \omega^J(w)\,  \p_y\cG(y,x)  \p_x\cG(x,w) \p_w  \cG(w,y)\notag
\end{align}
In passing to the second and third line, we have inserted the recursive definitions
of $\p_y \cG^{IJ}(y,z) $ and $ \p_x \cG^{J}(x,z)$ as convolutions of lower-rank $\cG$-tensors. The third line of
(\ref{coin.37}) manifests the cyclic symmetry $\Nf^{MIJ} =\Nf^{IJM}$, and integration by parts with respect to
all of $x,y,w$ furthermore reveals the reflection property $\Nf^{IJM} = - \Nf^{MJI}$.  Moreover, the
integrals in (\ref{coin.37}) are absolutely convergent which establishes that not only
the anti-holomorphic derivative (\ref{coin.02}) but also the limit (\ref{csec.31})
defining $ \cC^{IJ}(y) $ itself is well-defined.

\subsubsection{Recursion for higher-weight coincident limits}
\label{sec:cl.2.2}

The steps of section \ref{sec:cl.2.1} in the rank-two case can be repeated to show that the limits $ \cC^{P_1 \cdots P_s}(y) $  in (\ref{csec.31}) are well-defined at arbitrary rank $s\geq 2$. The inductive step in showing this has two steps. In the first step, the anti-holomorphic derivatives (\ref{ften.93}) are used to establish the relation,
\begin{align}
\partial_{\bar y} \lim_{z \rightarrow y}\partial_y \cG^{P_1 \cdots P_s}(y,z) &=  \pi  \lim_{z \rightarrow y} \big[ \bar \omega^{P_s}(y) \partial_y \cG^{P_1 \cdots P_{s-1}}(y,z) 
-  \bar \omega^{P_1}(y) \partial_y \cG^{P_2 \cdots P_s}(y,z)  \big]
\notag \\
&\quad
- \pi \bar \om^M(y) \partial_y \Phi^{P_1 \cdots P_s}{}_M(y) \label{coin.03}
\end{align} 
which amounts to the following recursion relation, 
\beq
\partial_{\bar y} \cC^{P_1\cdots P_s}(y) =
\pi \bar \omega^{P_s}(y) \cC^{P_1 \cdots P_{s-1}}(y) 
-  \pi \bar \omega^{P_1}(y) \cC^{P_2 \cdots P_s}(y)
 - \pi \bar \om^M(y) \partial_y \Phi^{P_1 \cdots P_s}{}_M(y)
\label{coin.51}
\eeq
since both $\cC$-tensors on the right side have lower rank $s{-}1$ than the one on the left side.
If we assume $s\geq 3$ here (see section \ref{sec:cl.2.1} for the $s=2$ case),
then the limits on the right side are individually well-defined by the inductive hypothesis.

\sm

Each step of integrating (\ref{coin.51}) introduces a new modular rank-$(s{+}1)$ tensor $\Nf^{MP_1  \cdots P_s}$
\begin{align}
\cC^{P_1\cdots P_s}(y) &= \int d^2 x \, \p_y \cG(y,x)
 \big[  \bar \omega^{P_1}(y) \cC^{P_2 \cdots P_s}(y) 
 - \bar \omega^{P_s}(y) \cC^{P_1 \cdots P_{s-1}}(y)  \big] \notag \\
&\quad
 +  \partial_y \Phi^{M P_1 \cdots P_s}{}_M(y)
 + \omega_M(y) \Nf^{MP_1  \cdots P_s}
\label{coin.52}
\end{align}
representing the holomorphic piece in the kernel of $\partial_{\bar y}$. Upon integration
against $\bar \omega^{M}(y)$ as in (\ref{coin.37}), one arrives at integral representations
\begin{align}
\Nf^{MP_1  \cdots P_s} &= \int_\Sigma d^2 y \, \bar \omega^M(y)  \cC^{P_1  \cdots P_s}(y) 
= \int_\Sigma d^2 y \, \bar \omega^M(y) \lim_{z\rightarrow y} \p_y \cG^{P_1  \cdots P_s}(y,z) 
  \label{precoin.53}
\end{align}
which can be simplified by the recursive definition of 
$\p_y \cG^{P_1  \cdots P_s}(y,z) $ as an iterated convolution as in (\ref{coin.37}) for $r\geq 3$,
\begin{align}
\Nf^{I_1 I_2 \cdots I_r} &= \bigg( \prod_{j=1}^r \int_\Sigma d^2 x_j \,  \bar \omega^{I_j}(x_j) \bigg)
 \p_{x_1}\cG(x_1,x_2)   \p_{x_2}\cG(x_2,x_2) 
 \cdots
  \p_{x_{r-1}}\cG(x_{r-1},x_r)   \p_{x_r}\cG(x_r,x_1)
  \label{coin.53}
\end{align}
These expressions for the tensors $\Nf^{I_1  \cdots I_r}$ imply their
dihedral symmetry under permutations of the indices $I_j$,
\begin{align}
\Nf^{I_1 I_2 \cdots I_r} = \Nf^{I_2  \cdots I_r I_1} \, , \ \ \ \ \ \
\Nf^{I_1I_2  \cdots I_r} =(-1)^r \Nf^{I_r  \cdots I_2 I_1}
  \label{coin.54}
\end{align}
where the alternating sign under the reflection
$I_1I_2  \cdots I_r \rightarrow I_r  \cdots I_2 I_1$ stems from the
total of $r$ integrations by parts in the derivation via (\ref{coin.53}).

\sm

Finally, the absolute convergence of the integrals in (\ref{coin.53}) for any $r\geq 2$
implies that not only the anti-holomorphic derivatives (\ref{coin.51}) but also the
limits $\cC^{I_1\cdots I_{r-1}}(y)$ themselves are well-defined if their lower-rank counterparts are.
This completes the inductive proof that the limits (\ref{csec.31}) are well-defined,
where the cancellation in (\ref{coin.34}) can be bypassed within the 
inductive step once the well-defined limit (\ref{cl.09}) at $s=2$ is taken as a base case.

\subsubsection{Explicit higher-weight coincident limits}
\label{sec:cl.2.3}

The inductive proof of the previous subsection led to the integral representation
(\ref{coin.53}) for $\Nf$-tensors of rank $\geq 3$ (see (\ref{csec.20}) for the
extra term in the integrand of the rank-two case) as well as the recursion relation (\ref{coin.52})
that relates $\cC^{P_1\cdots P_s}(y)$ to convolutions of its lower-rank analogues.
An explicit solution of this recursion (\ref{coin.52}) is presented in the following theorem:

{\thm
\label{thm:clim}
The coincident limits $ \cC^{I_1 \cdots I_r}(y) =  \lim_{z\to y}   \p_y \cG^{I_1 \cdots I_r}(y,z) $
at rank $r\geq 2$ are given by,
\begin{align}
\cC^{I_1 \cdots I_r}(y) & =  \omega_M(y)  \Nf^{M I_1 I_{2}\cdots I_r}
+ \p_y \Phi^{M I_1 \cdots I_r}{}_M(y)
 \label{explcs}\\
&\quad +
\sum_{1 \leq p \leq q \atop{(p,q) \neq (1,r)}}^r
(-1)^{r-q} \big[
\p_y \Phi^{ I_1 I_2 \cdots I_{p-1}\shuffle I_r I_{r-1}\cdots I_{q+1}  }{}_M(y) \Nf^{M I_p I_{p+1}\cdots I_q}
\notag \\
&\quad  \quad  \quad  \quad  \quad \quad  \quad \quad 
+ \p_y \Phi^{(I_1 I_2 \cdots I_{p-1}\shuffle I_r I_{r-1}\cdots I_{q+1} )M  I_p I_{p+1}\cdots I_q}{}_M(y)
\big]
\notag
\end{align}}
The proof of the theorem proceeds in two steps. In a first step, we
demonstrate by induction in $r$ that the expression (\ref{explcs}) obeys the differential equation
(\ref{coin.51}) relating ${\cal C}$ of different rank. This is readily accomplished by means of the variant
($p \neq 1$ and $q\neq r$),
\begin{align}
\p_{\bar y }\p_y \Phi^{(I_1 I_2 \cdots I_{p-1}\shuffle I_r I_{r-1}\cdots I_{q+1} ) \overrightarrow{Q} }{}_M(y) 
&= -\pi \bar \omega^{I_1}(y) \p_y \Phi^{(I_2 \cdots I_{p-1}\shuffle I_r I_{r-1}\cdots I_{q+1} ) \overrightarrow{Q} }{}_M(y)  \label{varshuf} \\
&\quad
-\pi \bar \omega^{I_r}(y) \p_y \Phi^{(I_1 I_2 \cdots I_{p-1}\shuffle I_{r-1}\cdots I_{q+1} ) \overrightarrow{Q} }{}_M(y)  \notag
\end{align}
of (\ref{prf.02}) for shuffle products of $f$-tensors which
casts the $\bar y$ derivative of (\ref{explcs}) into the form,
\begin{align}
\p_{\bar y }\cC^{I_1 \cdots I_r}(y) &=  
- \pi \bar \omega^M(y)
 \p_y \Phi^{I_1 \cdots I_r}{}_M(y)
  \label{varshuf.01} \\
&\quad - \pi 
\sum_{1 \leq p \leq q \atop{(p,q) \neq (1,r)}}^r
(-1)^{r-q} \Big[ \delta_{p\neq 1}
\bar \omega^{I_1}(y) \p_y \Phi^{ I_2 \cdots I_{p-1}\shuffle I_r I_{r-1}\cdots I_{q+1} }{}_M(y) \Nf^{M I_p I_{p+1}\cdots I_q}
\notag \\
&\quad  \quad  \quad  \quad  \quad \quad  \quad \quad  +
\delta_{q\neq r} \bar \omega^{I_r}(y)  \p_y \Phi^{ I_1 I_2 \cdots I_{p-1}\shuffle I_{r-1}\cdots I_{q+1} }{}_M(y) \Nf^{M I_p I_{p+1}\cdots I_q}
\Big. \notag \\
&\quad  \quad  \quad  \quad  \quad \quad  \quad \quad 
+ \delta_{p\neq 1} \bar \omega^{I_1}(y)  \p_y \Phi^{(I_2 \cdots I_{p-1}\shuffle I_r I_{r-1}\cdots I_{q+1} )M  I_p I_{p+1}\cdots I_q}{}_M(y) \notag \\
&\quad  \quad  \quad  \quad  \quad \quad  \quad \quad 
+\delta_{q\neq r} \bar \omega^{I_r}(y)  \p_y \Phi^{(I_1 I_2 \cdots I_{p-1}\shuffle I_{r-1}\cdots I_{q+1} )M  I_p I_{p+1}\cdots I_q}{}_M(y)
\Big] \notag
\end{align}
The notation $\delta_{p\neq 1}$ and $\delta_{q\neq r}$ indicates that the respective
terms are absent for $p=1$ and $q=r$, and we set $ \p_y \Phi^\emptyset{}_M(y) = \omega_M(y)$ 
in the $(p,q)=(2,r)$ contribution to the second line as well as the $(p,q)=(1,r{-}1)$ contribution to the third line.
The first term on the right side of (\ref{varshuf.01}) accounts for the last term in the aspired 
differential equation (\ref{coin.51}). The remaining terms match the target expression
since the coefficients of $\bar \omega^{I_1}(y)$ in the second \& fourth line 
and the coefficients 
of $\bar \omega^{I_r}(y)$ in the third \& fifth line match $-\pi \cC^{I_2 \cdots I_r}(y)$  
and $\pi \cC^{I_1 \cdots I_{r-1}}(y)  $, respectively, by
the inductive hypothesis. 

\sm

The second step of the proof is to show that the identity (\ref{explcs})
is not off by a holomorphic term in $y$. This can be verified by integration against 
$\bar \omega^J(y)$ where only the first term on the right side of (\ref{explcs})
contributes and yields $\Nf^{J I_1\cdots I_r}$. This matches the 
$\bar \omega^J(y)$ integral of the left side by the first step of (\ref{precoin.53}),
completing the proof of (\ref{explcs}).

\sm

With the decomposition (\ref{fphig.1}) of the $f$-tensors, the expressions (\ref{explcs}) for
$ \cC^{I_1 \cdots I_r}(y)$ make the coincident limit
\beq
f^{I_1 \cdots I_r J}{}_K(y,y) = \partial_y \Phi^{I_1 \cdots I_r J}{}_K(y) - \delta^J_K \cC^{I_1 \cdots I_r}(y) \, , \ \ \ \ r\geq 2
\label{sumclim}
\eeq
of $f$-tensors fully explicit, see (\ref{csec.27}) for the more subtle case with $r=1$ where
$f^{I_1J}{}_K(y,z) $ itself does not have a well-defined $z \rightarrow y$ limit.
At genus one, comparison of the integral representations 
(\ref{coin.53}) and (\ref{csec.05}) implies that the
modular tensors $\Nf^{P_1 \cdots P_s}$ at rank $s \geq 3$ 
reduce to holomorphic Eisenstein series,
\beq
\Nf^{P_1 \cdots P_s}  \big|_{h=1} = \left\{ \begin{array}{cl} 
 \big. {\rm G}_{s} &:\, s\geq 3 \\
\big. \widehat  {\rm G}_{2} &:\, s=2
\end{array} \right.
\eeq
where we have incorporated the earlier rank-two result (\ref{csec.00})
involving the almost holomorphic Eisenstein series $ \widehat  {\rm G}_{2}$.
Together with the vanishing genus-one restrictions of the $\Phi$-tensors, this implies
\begin{align}
\cC^{I_1 \cdots I_r}(y)\big|_{h=1} =  {\rm G}_{r+1}\, , \ \ \ \ 
f^{I_1 \cdots I_r J}{}_K(y,y) \big|_{h=1}  = - {\rm G}_{r+1}  \, , \ \ \ \ r\geq 2
\end{align}
whereas (\ref{4.coin2}) and (\ref{csec.00}) identify 
$\cC^{I_1}(y)\big|_{h=1} =\widehat  {\rm G}_{2}$ in the weight-two case.
In summary, the coincident limits (\ref{sumclim}) of higher-genus $f^{I_1 \cdots I_r J}{}_K(y,z)$-tensors
are considerably richer than their genus-one counterparts in view of the
hierarchy of modular tensors $\Nf^{P_1 \cdots P_s}$ of rank
$2\leq s \leq r{+}1$ in the expansion (\ref{explcs}) of $\cC^{I_1 \cdots I_r}(y)$.

\subsection{Coincident limit of Fay identities for arbitrary weight}
\label{sec:cl.3}

We are now ready to state and prove the coincident limit as $z \to y$ of the three-point Fay identity established in Theorem \ref{3.thm:8}. The main result of this section is  Theorem \ref{4.thm:1} for arbitrary rank, the proof of which is relegated to Appendix \ref{appB.5}.

\sm

The starting point is formula (\ref{4.ff.99}) for the Fay identity for three points of Theorem~\ref{3.thm:8}. As we take the limit $z \to y$, the left side converges to $f^{\vI} {}_J(x,y) \, f^{\vP J} {}_L(y,x)$, which is  to be \textit{$x$-reduced}. On the right side, all terms admit regular limits as $z \to y$, except for the following two cases: terms with singularities $f^{I}{}_J(y,z)= \delta^I_J (y{-}z)^{-1}+ {\rm reg}$ and terms that are affected by the direction dependent $z \to y$ limit of $f^{I_1 I_2}{}_J(y,z)$ discussed below (\ref{csec.12}).
\begin{enumerate}
\itemsep=-0.03in
\item 
When $s=0$, the term on the first line of (\ref{4.ff.99}), and the $k=r, \, r{-}1$ terms on the second line have singular or direction-dependent limits. 
\item 
When $s=1$, the limits of the term on the first line of (\ref{4.ff.99}), the terms $(k,\ell) = (r,0), \, (r,1), \, (r{-}1,0)$ on the second line, and the $\ell=1$ term on the third line are singular or direction dependent.
\item 
When $s \geq 2$, the terms $(k,\ell) = (r,0), \, (r,1), \, (r{-}1,0)$ on the second line of (\ref{4.ff.99}) and the $\ell=1, \, 2$ terms on the third line have singular or direction-dependent limits.  
\end{enumerate}
As the $z \to y$ limit of the left side of (\ref{4.ff.99})  is convergent in all cases, so must the limit of the combined right side be. Indeed, all the singularities tabulated above for each case, combine and cancel one another to produce  well-defined limits as $z \to y$. The net result of these finite limits may be expressed in terms of the following tensor functions of a single variable,
\bea
\label{4.cF}
\cF ^{I_1 \cdots I_r}{}_J(y) = 
\begin{cases} 
  f^{I_1 \cdots I_r}{}_J(y,y) &:\, r \geq 3 \cr 
  \p_y \Phi ^{I_1 I_2}{}_J(y) - \delta ^{I_2} _J \, \cC^{I_1}(y) &:\, r=2 \cr 
  \p_y \Phi ^{I_1}{}_J(y) &:\,  r=1 \cr 
 \om_J(y) &:\, r=0
\end{cases}
\eea
where the tensor function $\cC^I(y)$ was defined in (\ref{csec.12}) and evaluated in (\ref{4.coin2}). The coincident limit of the three-point Fay identity (\ref{4.ff.99}) then results in the following theorem.
{\thm 
\label{4.thm:1}
The coincident limits of the three-point Fay identities  allows us to \textit{$x$-reduce} the contracted product
$f^{\vI} {}_J(x,y) \, f^{\vP J} {}_K(y,x)$ with multi-indices $\vI= I_1\cdots I_r$ and $\vP=P_1 \cdots P_s$ of length $r\geq 1$ and $s\geq 0$,
\bea
\label{3.coin.0}
f^{\vI} {}_J(x,y) \, f^{\vP J} {}_K(y,x)
& = &
f^\vI {}_J(x,y) \, \cF ^{\vP  J} {}_K(y) - (-)^s \p_y f^{(\vPt  \, \shuffle \, I_1 \cdots I_{r-1}) I_r} {}_K(x,y) 
\no \\ && 
- \sum_{k=0}^r  \sum_{\ell=0}^s (-)^{\ell-s} 
f^{P_s \cdots P_{\ell+1} \shuffle I_1 \cdots I_k}{}_J(x,y) \,  \cF^{P_1 \cdots P_\ell J I_{k+1} \cdots I_r}{}_K(y) 
\no \\ && 
- \sum_{\ell=0}^s (-)^{\ell-s} \cF^{P_1 \cdots P_\ell}{}_J(y) \,  f^{P_s \cdots P_{\ell+1}J \, \shuffle \, \vI}{}_K(x,y) 
\eea
Alternatively, the rightmost term of the first line may be re-expressed using,
\bea
\label{4.coin}
 \p_y f^{(\vPt  \, \shuffle \, I_1 \cdots I_{r-1}) I_r} {}_K(x,y)  
= - \delta^{I_r}_K \p_x \p_y \cG^{\vPt  \, \shuffle \, I_1 \cdots I_{r-1}}(x,y)
\eea}
The proof of the theorem is relegated to Appendix \ref{appB.5}, and the adaptation 
of (\ref{3.coin.0}) to the three cases $s=0$, $s=1$ and $s\geq 2$ can be found in (\ref{adapta})
to (\ref{adaptc}). The
derivation of uncontracted Fay identities from contracted ones in section \ref{s:fay.6.7}
straightforwardly carries over to the coincident limit $z \rightarrow y$, leading to,
\bea
f^{\overrightarrow{I}}{}_K(x,y)  f^{\overrightarrow{P}Q}{}_L(y,x) 
& = & \delta^Q_L f^{\overrightarrow{I}}{}_J(x,y)   f^{\overrightarrow{P} J}{}_K(y,x)  
\label{ucon.12} 
 \\ &&
+ f^{\overrightarrow{I}}{}_K(x,y)  f^{\overrightarrow{P}Q}{}_L(y,a) 
-  \delta^Q_L f^{\overrightarrow{I}}{}_J(x,y)    f^{\overrightarrow{P} J}{}_K(y,a)
\notag
\eea
with an arbitrary point $a\in\Sigma$. Moreover, the coincident
Fay identities (\ref{3.coin.0}) can be used to extend the reduction of products of
$f^{\overrightarrow{R} }{}_{M}(z,y)$ and an arbitrary number 
of $f^{\overrightarrow{P_j}}{}_{K_j}(x_j,z)$ 
in section \ref{s:fay.6.8} to situations with $x_j \rightarrow y$.

\sm

An alternative representation of the coincident Fay identity (\ref{3.coin.0}) is stated in the theorem below, whose proof is given in Appendix \ref{prfcoin}.
{\thm 
\label{finthm}
The \textit{$x$-reduction (\ref{3.coin.0})} for the contracted product 
$f^{\vI} {}_J(x,y) \, f^{\vP J} {}_K(y,x)$  is equivalent to,
\begin{align}
&f^{ \overrightarrow{I}}{}_J(x,y) f^{\overrightarrow{P} J}{}_K(y,x) 
=  - (-1)^s \om_J(y) f^{   \overrightarrow{I}\shuffle \overleftarrow{P}  J }{}_K(x,y) 
 \label{coin.75} \\
&\quad\quad \! \!   
+(-1)^s \delta^{I_r} _K \partial_x \partial_y 
    \cG^{I_1 \cdots I_{r-1}   \shuffle \overleftarrow{P} }(x,y) 
 +(-1)^s f^{ I_1 \cdots I_{r-1} \shuffle \overleftarrow{P} }{}_J(x,y)
 \big[
  \delta^{I_r}_K \cC^J(y) - \p_y \Phi^{ J  I_r  }{}_K(y) 
 \big]
\notag \\
&\quad\quad   \!  \! 
-  \sum_{k =0}^{r-1}  \sum_{\ell=0}^s  \delta_{(k,\ell)\neq (r-1,0)}
 (-1)^{s-\ell} f^{ I_1 \cdots I_k \shuffle P_s \cdots P_{\ell+1} }{}_J(x,y)
 f^{ P_1 \cdots P_\ell J I_{k+1} \cdots I_r  }{}_K(y,y)
\notag  \\
&\quad\quad \!  \! 
- \sum_{\ell=1}^s (-1)^{s-\ell } \big[ 
f^{P_1\cdots P_\ell}{}_J(y,a_\ell ) f^{ P_s\cdots P_{\ell +1}J \shuffle \overrightarrow{I} }{}_K(x,y) 
\,{-}\, f^{P_1\cdots P_{\ell-1}J}{}_K(y,a_\ell ) f^{P_s\cdots P_\ell  \shuffle \overrightarrow{I} }{}_J(x,y)
\big] 
\notag
\end{align}
with arbitrary points $a_1,\cdots,a_s \in \Sigma$.}

\subsubsection{Comments on Theorem \ref{finthm}}

The alternative form (\ref{coin.75}) of the coincident Fay identity manifests the cancellation of the
first term $f^\vI {}_J(x,y) \, \cF ^{\vP  J} {}_K(y)$ on the right side of (\ref{3.coin.0}) and the reduction of 
several ${\cal F}^{P_1\cdots P_\ell}{}_J(y)$ in its second and third line
to their $\p_y \Phi^{P_1\cdots P_\ell}{}_J(y)$ parts. Moreover, (\ref{coin.75})
 is a more suitable starting point to make contact with 
the coincident Fay identity (\ref{coin.67}) at genus one.
The shuffle products of $f$-tensors reduce to multiples
of $f^{(r)}$ with the binomial coefficients in (\ref{binom})
 upon restriction to genus one.\footnote{One has to
shift $s \rightarrow s{+}1$ in (\ref{coin.67}) to account for the
extra upper index $J$ besides $\overrightarrow{P}=P_1\cdots P_s$ in the
factor $f^{\overrightarrow{P} J}{}_K(y,x) $ on the left side of (\ref{coin.75}).} 
The first term on the right side of (\ref{coin.75}) reproduces the first term
$\sim f^{(r+s)}$ on the right side of (\ref{coin.67}). The
last term of (\ref{coin.67}) originates from the second line of (\ref{coin.75}).
The third line of (\ref{coin.75}) produces the term $\sim {\rm G}_{r+s}$ in (\ref{coin.67})
from the extremal term $(k,\ell)=(0,s)$ and the sums over $\ell$ in (\ref{coin.67}) 
from the remaining summands. The last line of (\ref{coin.75}) drops out
at $h=1$ since both of the $a_\ell$-dependent $f$-tensors reduce to their
respective $\p \Phi$ parts.

\newpage

\section{Meromorphic Fay identities}
\label{sec:7}

In this section, we spell out counterparts of the identities of  sections \ref{sec:3} to \ref{sec:5} among non-meromorphic and single-valued $f$-tensors for meromorphic but multi-valued integration kernels on surfaces of arbitrary genus.  More specifically, we propose interchange identities that we shall prove and Fay identities that we shall conjecture for the meromorphic expansion coefficients $g^{I_1\cdots I_r}{}_J(x,y)$. Remarkably, the expressions for both the interchange and Fay identities that we find in terms  of the Enriquez kernels will  match with the analogous identities for DHS kernels  $f^{I_1\cdots I_r}{}_J(x,y)$ as carbon copies of one another  upon the formal substitution $f \rightarrow g $. We also investigate the coincident limits $y \rightarrow x$ of the Enriquez kernels which introduce
meromorphic analogues of the modular tensors $\Nf^{P_1\cdots P_{r}}$ seen
in the coincident limits of $f^{I_1\cdots I_r}{}_J(x,y)$ in the previous section.

\sm 

At genus one, the Fay identities of Kronecker-Eisenstein kernels $f^{(k)}(z)$
and $g^{(k)}(z)$ play a two-fold crucial role for integration on the torus: first, 
for a general proof that elliptic polylogarithms close
under taking primitives \cite{BrownLevin, Enriquez:2023}; second, for an explicit
derivation of differential and algebraic identities among elliptic polylogarithms
\cite{BrownLevin, Broedel:2014vla, Broedel:2017kkb},
elliptic multiple zeta values \cite{Enriquez:Emzv, Broedel:2015hia, Matthes:thesis, Mafra:2019xms} and 
modular graph forms \cite{Gerken:2019cxz, Gerken:2020aju, Gerken:2020xte} (as a reformulation of
the holomorphic subgraph reduction developed earlier in \cite{DHoker:2016mwo,DHoker:2016quv, Gerken:2018zcy}). 
By analogy with this impact of genus-one Fay identities, the interchange and Fay identities of this work are expected  to crucially feed into derivations and classifications of relations among configuration-space 
periods at arbitrary genus. 

\sm

By extending the interchange  and Fay identities among higher-genus $f$-tensors to their meromorphic counterparts from the Enriquez connection, the results of this section  pave the way for the study of iterated integrals
of $g^{I_1\cdots I_r}{}_J(x,y)$ including the hyper-elliptic polylogarithms
of \cite{Baune:2024}. In particular, the subsequent identities
will be applied to change fibration bases of the meromorphic polylogarithms
of \cite{Baune:2024} similar to those in section \ref{sec:clo} for non-meromorphic polylogarithms.

\subsection{Basics of the Enriquez coefficients}
\label{sec:7.1}

Enriquez introduced meromorphic flat connections, on the universal cover of an arbitrary compact Riemann surface, which have at most simple poles at the marked points and prescribed monodromies \cite{Enriquez:2011}.
Expanding the two-variable case of the Enriquez connection in certain non-commutative generators gives rise to 
merormophic integration kernels $g^{I_1\cdots I_r}{}_J(x,y)$ which are uniquely defined through their functional identities \cite{Enriquez:2011}.

{\thm[Enriquez \cite{Enriquez:2011}]
\label{thm:en}
There exists a unique family of differentials, denoted by $\omega^{I_1\cdots I_r}{}_J(x,y)$ 
in Enriquez's work and normalized with additional powers of $-2\pi i$ in this work,
\beq
g^{I_1\cdots I_r}{}_J(x,y) = (-2\pi i)^r \omega^{I_1\cdots I_r}{}_J(x,y) \, , \ \ \ \ \ \ 
r\geq 0
\label{Enorm}
\eeq
depending on two points $x,y$ in the universal covering space
of a compact Riemann surface~$\Sigma$ of genus $h$ and its complex-structure moduli
such that
\begin{enumerate}
\itemsep=0in
\item $g^{I_1\cdots I_r}{}_J(x,y)$ are $(1,0)$-forms in $x$ and scalars in $y$;
\item $g^{I_1\cdots I_r}{}_J(x,y)$ are meromorphic in all variables;
\item the $r=0$ instance is given by the Abelian differentials, $g^{\emptyset}{}_J(x,y) = \omega_J(x)$;
\item the monodromies of $g^{I_1\cdots I_r}{}_J(x,y)$ in $x,y$ vanish around the
$\mA^L$-cycles and obey the following recursion around the $\mB_L$-cycles,\footnote{The $ \mB$ monodromy in $y$ in the second line of (\ref{ome.02}) is derived in part b) of Lemma 9 of \cite{Enriquez:2011}.}
\begin{align}
\label{ome.02} 
g^{I_1 \cdots I_r}{}_J(x {+} \mB_L,y) & = 
g^{I_1 \cdots I_r}{}_J(x,y) 
+ \sum_{k=1}^r { (-2\pi i)^k \over k!} \, \delta ^{I_1 \cdots I_k} _L \, g^{I_{k+1} \cdots I_r}{}_J(x,y)
 \\
g^{I_1 \cdots I_r}{}_J(x ,y{+} \mB_L) & =  
g^{I_1 \cdots I_r}{}_J(x,y) 
+\delta ^{I_r}_J \sum_{k=1}^r { (2\pi i)^k \over k !} \, g^{I_1 \cdots I_{r-k}} {}_L(x,y) \, \delta^{I_{r-k+1} \cdots I_{r-1}}_L \notag
\end{align}
where the generalized Kronecker-deltas are given by
$\delta^{I_{1} I_2 \cdots I_k}_L=\delta^{I_{1}}_L \delta^{I_{2}}_L \cdots \delta^{ I_k}_L$;
\item given a (simply connected) fundamental domain 
 $\Sigma_f$ for the surface $\Sigma$ with $x,y \in \Sigma_f$, 
 all of $g^{I_1\cdots I_r}{}_J(x,y)$ with $r\neq 1$ are regular as $y\rightarrow x$;
the $r=1$ instance, however, exhibits a simple pole with the following residue,
\beq
g^I{}_J(x,y) =   \frac{\delta^I_J}{x{-}y} + {\rm reg}
\label{ome.01}
\eeq
\item at $y\neq x$, the only poles of $g^{I_1 \cdots I_r}{}_J(x,y)$ at $r\geq 1$
are those mandated by their monodromy relations and the pole of $g^I{}_J(x,y) $.
\end{enumerate}}
As detailed in section 8 of the published  version of  \cite{Enriquez:2011}, the  Enriquez kernels for genus one coincide with the meromorphic Kronecker-Eisenstein coefficients defined by (\ref{exfay.22}),
\beq
g^{I_1\cdots I_r}{}_J(x,y) \big|_{h=1} = g^{(r)}(x{-}y)
\label{ome.03}
\eeq

\subsubsection{Decomposition into trace and traceless components}

The $g^{I_1\cdots I_r}{}_J(x,y)$ obey a direct analogue of the decomposition (\ref{fphig.1})
of the $f$-tensors: Their dependence on $y$ is concentrated in the trace $\delta^{I_r}_J$
with respect to the last two indices \cite{Enriquez:2011}, 
\bea
\label{ome.04}
g^{I_1\cdots I_r}{}_J(x,y)  & = & \varpi^{I_1\cdots I_r}{}_J(x) - \delta^{I_r}_J \mcG^{I_1\cdots I_{r-1}}(x,y)
\no \\
\varpi^{I_1\cdots I_s J}{}_J(x) & = & 0
\eea
Therefore, the traceless part $\varpi^{I_1\cdots I_r}{}_J(x)$ solely depends on the point $x$, and this dependence is  holomorphic on $x$ in the universal covering space of $\Sigma$. Its $\mB$ monodromies follow from the traceless projection of the first line in (\ref{ome.02}) with respect to $I_r,J$,
\begin{align}
\varpi^{I_1\cdots I_r}{}_J(x{+}\mB_L) &= \varpi^{I_1\cdots I_r}{}_J(x)
+ \sum_{k=1}^{r-1} \frac{(-2\pi i)^k}{k!} \delta^{I_1\cdots I_k}_L   \varpi^{I_{k+1}\cdots I_r}{}_J(x) \notag \\
&\quad
+ \frac{(-2\pi i)^r}{r!} \bigg( \delta^{I_1\cdots I_r}_L  \omega_J(x)
- \frac{1}{h} \delta^{I_r}_J \delta^{I_1\cdots I_{r-1}}_L  \omega_L(x) \bigg)
\label{trcless}
\end{align}
By the tracelessness of $\varpi^{I_1\cdots I_{r}}{}_J(x)$, their restrictions to genus one vanish,
and the trace of (\ref{ome.03}) relates the $\mcG^{I_1\cdots I_{r-1}}(x,y)$ to the meromorphic Kronecker-Eisenstein kernels
\beq
\varpi^{I_1\cdots I_r}{}_J(x) \big|_{h=1} = 0
\, , \ \ \ \ \ \
\chi^{I_1\cdots I_s}{}(x,y) \big|_{h=1} = -  g^{(s+1)}(x{-}y)
\label{ochi.03}
\eeq
The pole structure of $g^{I_1\cdots I_{r}}{}_J(x,y)$ in items 5.\ and 6.\ of Theorem \ref{thm:en} readily translates into the $y\rightarrow x$ behavior,
\beq
\chi(x,y) = \frac{1}{y{-}x}+ {\rm reg}
\label{polechi}
\eeq
whereas all of $\chi^{I_1 \ldots I_s}(x,y) $ with $s\geq 1$ are regular as $y\rightarrow x$. Moreover, the
only poles of $\chi^{I_1 \ldots I_s}(x,y) $ with $s\geq 0$  at $y\neq x$ are those mandated by (\ref{polechi}) and the monodromies,
\begin{align}
\chi^{I_1\cdots I_{s}}(x{+}\mB_L,y) &= \chi^{I_1\cdots I_{s}}(x,y) 
+ \sum_{k=1}^s \frac{(-2\pi i)^k}{k!} \delta^{I_1 \cdots I_k}_L \chi^{I_{k+1} \cdots I_s}(x,y)
-\frac{(-2\pi i)^{s+1}}{(s{+}1)! h } \delta^{I_1\cdots I_s}_L \omega_L(x)
\notag \\
\chi^{I_1\cdots I_{s}}(x,y{+}\mB_L) &= \chi^{I_1\cdots I_{s}}(x,y) 
- \sum_{k=0}^s \frac{(2\pi i)^{s-k+1}}{(s{-}k{+}1)!} \, g^{I_1 \cdots I_k}{}_L(x,y) \delta_L^{I_{k+1}\cdots I_s}
\label{mclim.01}
\end{align}
Explicit expressions for  $g^{I_1\cdots I_r}{}_J(x,y)$ at $h\geq 2$ remain somewhat cumbersome to exhibit explicitly at this time, though recent work \cite{Baune:2024} offers formulas in the local Schottky  parametrization  of moduli space that lend themselves to numerical evaluation for genus two.

\subsection{Meromorphic interchange identities}
\label{sec:7.2}

In this section, we produce the meromorphic counterparts of the interchange identities for $f$-tensors in section \ref{sec:3} which will allow us to \textit{$x$-reduce}  expressions of the type $\omega_M(x)g^{I_1\cdots I_r}{}_J(y,x)$. The definition of \textit{$z$-reduced} in the meromorphic case is analogous to but simpler than the one given for the non-meromorphic case in section \ref{sec:3-x}. For mutually distinct points $z_1, \cdots, z_N$, the exterior algebra generated by the differential forms,
\bea
\bom_I(z_i), \qquad\qquad g ^{I_1 \cdots I_r}{}_J(z_i, z_j)dz_i, \qquad\qquad 
\p_{z_j} g ^{I_1 \cdots I_r}{}_J(z_i, z_j)dz_i\wedge dz_j
\label{angens}
\eea
will be denoted $\tilde \cA_N$. It is manifestly closed under addition, the wedge product, and application of the Dolbeault differentials $ \p_i = d z_j \p _{z _j}$ and  $\bar \p _j = d\bar z_j \p _{\bar z_j}$, the latter since the forms are all meromorphic and the points $z_i$ are mutually distinct. An arbitrary element of $\tilde \cA_N$ is defined to be \textit{$z_i$-reduced}, for a given value of $i$, if it is a linear combination of $z_i$-independent terms and those generators of $\tilde \cA_N$ that depend on $z_i$ with coefficients that are independent of $z_i$. In short, \textit{$z_i$-reduced} products of the generators (\ref{angens}) feature no more than one $z_i$-dependent factor besides $dz_i$. 

\sm

The simplest example occurs at rank $r=1$ where the meromorphic analogue of the basic interchange identity (\ref{ften.08a}) with a contracted index reads,
\bea
 \omega_M(x) g^M{}_J(y,x) + \omega_M(y) g^M{}_J(x,y)    = 0
\label{ome.06}
\eea
Through the decomposition (\ref{ome.04}) of the Enriquez kernels, this
translates into the following meromorphic counterpart of (\ref{preften.08}),
\bea
\omega_M(x)   \varpi^M{}_J(y) + \omega_M(y)   \varpi^M{}_J(x)    
 -   \omega_J(x)   \mcG(y,x)  -   \omega_J(y)   \mcG(x,y)   = 0
  \label{ome.07}
\eea
and illustrates that the traceless component $\varpi^M{}_J(x)$ of $g^M{}_J(x,y)$ compensates for the lack of translation invariance in the trace component $\chi(x,y)$ at genus $h\geq 2$. For arbitrary rank $r\geq 0$, we obtain the  meromorphic analogue of the non-meromorphic contracted interchange identities of Theorem  \ref{intlemma}.

{\thm
\label{mintlemma}
The differentials $\mQ^{I_1 \cdots I_r}{}_J(x,y) $ defined by,
\bea
 \label{ome.08}
\mQ^{I_1 \cdots I_r}{}_J(x,y) & =&  
\omega_M(x) \, g^{I_1 \cdots I_r M}{}_J(y,x)
+ (-1)^r \omega_M(y) \, g^{I_r \cdots I_1 M}{}_J(x,y)
\\ && 
+ \sum_{k = 1}^r (-1)^{k+r}
 \Big [
\varpi^{I_1 \cdots I_k}{}_M(y) \, \varpi ^{I_r \cdots I_{k + 1} M}{}_J(x)
- \varpi^{I_r \cdots I_{k} }{}_M(x) \, \varpi^{I_1 \cdots I_{k-1} M}{}_J(y) 
 \Big ]
 \notag
\eea
satisfy the following properties:
\begin{enumerate}
\itemsep=0in
\item they are $(1,0)$ forms in $x,y$ and obey the symmetry 
          $\mQ^{I_1 \cdots I_r}{}_J(x,y) = (-)^r \, \mQ^{I_r \cdots I_1}{}_J(y,x) $;
\item have vanishing $\mA$ monodromy in $x$ and $y$, and their $\mB$ monodromy in $y$ are given by,
\bea
\label{8.thm.2}
\mQ^{I_1 \cdots I_r}{}_J(x,y{+} \mB_L ) = 
\mQ^{I_1 \cdots I_r}{}_J(x,y)  + \sum _{k=1}^r 
{(-2\pi i)^k \over k!} \,
 \delta ^{I_1 \cdots I_k}_L \, \mQ^{I_{k+1} \cdots I_r}{}_J(x,y) 
\eea
\item they are holomorphic in $x$ and $y$ for all $r \geq 0$;
\item as a consequence, they  vanish identically for all $r \geq 0$,
\beq
\mQ^{I_1 \cdots I_r}{}_J(x,y) =0 
\label{pisarezero}
\eeq
\end{enumerate}
}

We note that the last line in (\ref{ome.08}) may alternatively be expressed solely in terms of the differentials $g^{I_1 \cdots I_r}{}_J(x,y)$, leading to the following representation,
\bea
 \label{ome.088}
 \mQ^{I_1 \cdots I_r}{}_J(x,y) & = & 
\omega_M(x) g^{I_1 \cdots I_r M}{}_J(y,x) + (-1)^r \omega_M(y) g^{I_r \cdots I_1 M}{}_J(x,y)
 \no \\ &&
 + \sum_{k = 1}^r (-1)^{k+r}
 \big[
g^{I_1 \cdots I_k}{}_M(y,a_k) g^{I_r \cdots I_{k + 1} M}{}_J(x,b_k)
\no \\ && \hskip 1in
- g^{I_1 \cdots I_{k-1} M}{}_J(y,a_k) g^{I_r \cdots I_{k} }{}_M(x,b_k)
 \big]
\eea
where $a_1,\cdots, a_r, b_1,\cdots, b_r$ are arbitrary points  in the universal cover of $\Sigma$. Solving the vanishing of $\mQ^{I_1 \cdots I_r}{}_J(x,y)$ in (\ref{ome.08}) for the first term on the right side provides the
\textit{$x$-reduced} form of $\omega_M(x) g^{I_1 \cdots I_r M}{}_J(y,x)$. 
The proof of Theorem \ref{mintlemma} is relegated to Appendix \ref{app:EZ}.

\subsubsection{Uncontracted interchange identities}
\label{sec:7.2.1}

Since the decompositions (\ref{rewr.1}) literally carry over to $f \rightarrow g$
as well as $\partial \cG \rightarrow \mcG$ and $\partial \Phi \rightarrow \varpi$, we obtain a meromorphic version of the uncontracted interchange identities (\ref{rewr.3}) that \textit{$x$-reduce} $ \omega_J(x)  g^{\overrightarrow{I} L}{}_K(y,x) $ with free indices $J,K,L$ and $\overrightarrow{I}= I_1\cdots I_r$.

{\cor The meromorphic un-contracted  interchange identities takes the form,
\begin{align}
 \omega_J(x)  g^{\overrightarrow{I} L}{}_K(y,x) 
&=  -  (-1)^r \omega_J(y)  g^{ \overleftarrow{I}  L}{}_K(x,y) +  \omega_J(x) g^{\overrightarrow{I} L}{}_K(y,a) - \delta^L_K \omega_M(x)   g^{\overrightarrow{I} M}{}_J(y,a)  \notag \\
&\quad
+ (-1)^r \omega_J(y)  g^{  \overleftarrow{I} L}{}_K(x,b) 
 - (-1)^r  \delta^L_K \omega_M(y) g^{ \overleftarrow{I} M}{}_J(x,b)  \notag \\
 &\quad+ \delta^L_K 
\sum_{k = 1}^r (-1)^{k+r}
 \big[
g^{I_1 \cdots I_{k-1} M}{}_J(y,a_k) g^{I_r \cdots I_{k } }{}_M(x,b_k) \notag \\
 &\quad\quad\quad\quad\quad\quad\quad\ 
 - g^{I_1 \cdots I_k}{}_M(y,a_k) g^{I_r \cdots I_{k + 1} M}{}_J(x,b_k)
 \big]
 \label{ome.10} 
\end{align}
with $\vI= I_1\cdots I_r$ and two additional arbitrary points $a,b$ in the universal cover of $\Sigma$.}

\subsubsection{Swapping identities}
\label{sec:7.2.2}

As another important two-point identity among Enriquez kernels, we shall here
introduce the meromorphic analogue of the symmetry property,
\beq
\partial_x\partial_y \cG^{I_1\cdots I_r}(x,y)  =  (-)^r \partial_x\partial_y  \cG^{I_r \cdots I_1}(y,x)
\label{swid.01}
\eeq
which straightforwardly follows from differentiating (\ref{Gtrcless}) with respect to $x$ and $y$.
In contrast to their single-valued counterparts
$\partial_x \cG^{I_1\cdots I_r}(x,y) $, the trace components $\chi^{I_1\cdots I_r}(x,y) $ of the Enriquez
kernels in (\ref{ome.04}) do not feature an exposed holomorphic derivative in $x$. Still, 
the trace component $\partial \cG \rightarrow \chi$ of the substitution rule $f \rightarrow g$
converts (\ref{swid.01}) to a valid identity stated
in the following theorem.
{\thm
\label{swid.02}
The $(1,0)$ forms $\cU^{I_1 \cdots I_r}(x,y) $  in $x$ and $y$ defined by,
\beq
\cU^{I_1 \cdots I_r}(x,y) = \p_y \chi^{I_1 \cdots I_r }(x,y) - (-)^r \p_x \chi^{I_r \cdots I_1}(y,x)
\label{swid.03}
\eeq 
vanish identically for any $r\geq 0$
\beq
\cU^{I_1 \cdots I_r}(x,y) =  0
\label{swid.04}
\eeq}
The proof of the theorem can be found in Appendix \ref{app:more}. It relies on the
$\mathfrak B$ monodromies in (\ref{mclim.01}) as well as the
cancellation of poles $\p_y \chi(x,y) = -\frac{1}{(x{-}y)^2}+ {\rm reg}$ from
$\cU(x,y) $.

\sm

By the decomposition (\ref{ome.04}), the $y$ derivatives of the Enriquez kernels
$g^{I_1\cdots I_r K}{}_J(x,y)$ are concentrated in the trace with respect to $K$ and $J$, i.e.
\beq
\p_y g^{I_1\cdots I_r K}{}_J(x,y) = -\delta^{K}_J  \p_y   \chi^{I_1\cdots I_r}(x,y)
\label{swid.05}
\eeq
and one arrives at the following corollary of Theorem \ref{swid.02}
{\cor Derivatives of the Enriquez kernels obey the swapping identities for all $r\geq 0$
\label{swid.06}
\beq
 \p_y g^{I_1 \cdots I_r K}{}_J(x,y) = (-)^r \p_x g^{I_r \cdots I_1 K}{}_J(y,x)
 \label{swid.06eq}
\eeq}
The swapping identities will play a key role in the change of fibration bases 
for the meromorphic polylogarithms in section \ref{sec:7.9}.

\subsection{Meromorphic Fay identities}
\label{sec:7.3}

As a meromorphic analogue of the simplest tensorial Fay identity (\ref{hf.22}) of
the $f$-tensors, the Enriquez kernels $g^I{}_J$ and $g^{I_1 I_2}{}_J$
are proposed to obey
\begin{align}
& g^M{}_J(x,y) g^J{}_K(y,z) + g^M{}_J(y,x) g^J{}_K(x,z) - g^M{}_J(x,z) g^J{}_K(y,z) \notag  \\
&\quad + \omega_J(x) g^{MJ}{}_K(y,x) 
+ \omega_J(y) g^{JM}{}_K(x,z) +  \omega_J(x) g^{JM}{}_K(y,z) = 0
\label{ome.00}
\end{align}
Through the decomposition (\ref{ome.04}) of the Enriquez kernels,
the trace with respect to $M,K$ implies the following
meromorphic counterpart of (\ref{3.a.2})
\begin{align}
0 & =    \mcG(x,y)  \mcG(y,z) +  \mcG(y,x)   \mcG(x,z) -   \mcG(x,z) \mcG(y,z)
\notag \\
&\quad - \om_I (x)   \mcG^I(y,z) - \om_I(y)   \mcG^I(x,z) +  \mcG_2(x,y)
\label{ome.11}
 \end{align}
where the symmetric function
$\chi_2(x,y)=\chi_2(y,x)$ can be viewed
as the meromorphic analogue of
$\p_x\p_y \cG_2(x,y)$ in (\ref{defgns}) and can be rewritten in
analogy with (\ref{papag2}),
  \begin{align}
 h\,  \mcG_2(x,y)  &=  \om_M(x) g^{IM}{}_I(y,x) +   \varpi ^J{}_I(x)   \varpi^I{}_J(y)  \notag \\
 &=  \om_M(y) g^{IM}{}_I(x,y) +   \varpi ^J{}_I(x)   \varpi^I{}_J(y) 
\label{ome.12}
\end{align} 
The traceless part of (\ref{ome.00}) in turn yields the meromorphic analogue of (\ref{1stGphi})
\begin{align}
  \mcG(y,x)  \varpi^M{}_K(x) &= -  \mcG(x,y)  \varpi^M{}_K(y)
+ \omega_J(x)  \varpi^{JM}{}_K(y)
+ \omega_J(y)  \varpi^{JM}{}_K(x) \notag\\
&\quad + \omega_J(x) g^{MJ}{}_K(y,x) 
- \tfrac{1}{h} \delta^M_K  \omega_J(x) g^{LJ}{}_L(y,x) \notag\\
&\quad +   \varpi^M{}_J(y)  \varpi^J{}_K(x)
- \tfrac{1}{h} \delta^M_K  \varpi^L{}_J(y)  \varpi^J{}_L(x)
\label{ome.99}
\end{align}
The products $g^M{}_J(y,x) g^J{}_K(x,z)$, $\mcG(y,x)   \mcG(x,z) $ and $  \mcG(y,x)  \varpi^M{}_K(x)$
may be \textit{$x$-reduced} with the help of equations (\ref{ome.00}), (\ref{ome.11}) and (\ref{ome.99}).  Alternatively, (\ref{ome.00}) and (\ref{ome.11}) can be used to \textit{$z$-reduce} the products  $g^M{}_J(x,z) g^J{}_K(y,z)$ and  $ \mcG(x,z) \mcG(y,z)$.

\subsubsection{Contracted meromorphic Fay identities}
\label{sec:7.3.1}

More generally, the contracted Fay identities (\ref{exfay.15})
and (\ref{4.ff.99}) at arbitrary weight $\geq 2$ 
are proposed to carry over to the Enriquez kernels as follows:
{\conj
\label{mfaysscal} The contracted product 
$g^{\overrightarrow{P}M}{}_J(y,z) g^{\overrightarrow{I}J}{}_K(x,z)$
for multi-indices $\vI= I_1\cdots I_r$ and $\vP= P_1\cdots P_s$, which is a $(1,0)$ form in $x,y$ and a scalar in $z$, 
can be \textit{$z$-reduced} in terms of Enriquez integration kernels as follows,
\begin{align}
&g^{\overrightarrow{P}M}{}_J(y,z) g^{\overrightarrow{I}J}{}_K(x,z)= (-1)^{s}  \omega_J(y) g^{\overrightarrow{I} M \overleftarrow{P} J}{}_K(x,y)   \label{ome.13} \\
&\quad 
+ g^{\overrightarrow{I} M}{}_J(x,y) g^{\overrightarrow{P}J}{}_K(y,z) 
+\sum_{k = 0}^r   g^{I_1 \cdots I_k}{}_J(x,y) g^{(\overrightarrow{P} \shuffle J I_{k+1}\cdots I_r )M}{}_K(y,z) 
 \notag \\
&\quad + g^{\overrightarrow{P}M}{}_J(y,x) g^{\overrightarrow{I}J}{}_K(x,z)
+\sum_{\ell=0}^s g^{P_1 \cdots P_\ell }{}_J(y,x) g^{(\overrightarrow{I} \shuffle J  P_{\ell+1} \cdots P_s)M}{}_K(x,z)
\notag \\
&\quad 
+ \sum_{\ell=1}^s (-1)^{s-\ell} 
\big( g^{P_1 \cdots   P_\ell}{}_J(y,a_\ell) g^{\overrightarrow{I} M P_s\cdots P_{\ell+1} J}{}_K(x,b_\ell)
-  g^{P_1 \cdots P_{\ell-1} J}{}_K(y,a_\ell) g^{\overrightarrow{I} M P_s\cdots  P_\ell}{}_J(x,b_\ell) \big)
\notag
\end{align}
with arbitrary points $a_1,\cdots, a_s, b_1,\cdots, b_s$ in the universal cover of $\Sigma$.}

\sm 
In Appendix \ref{appprfs}, the analogous Fay identities in Theorem \ref{3.thm:7} on the \textit{$z$-reduction} of the scalar $f^{\overrightarrow{P}M}{}_J(y,z) f^{\overrightarrow{I}J}{}_K(x,z)$ are shown to imply Theorem \ref{3.thm:8} for the \textit{$x$-reduction} of $f^{\overrightarrow{I}} {}_J(x,z)  f^{\overrightarrow{P} J}{}_K(y,x)$. This proof only relies on identities of $f$-tensors that apply in identical form for the Enriquez kernels, for instance that the dependence of both
$f^{I_1\cdots I_r}{}_J(x,y) $ and $ g^{I_1\cdots I_r}{}_J(x,y) $ on $y$ is concentrated in the trace $\delta^{I_r}_J$, see (\ref{fphig.1}) and (\ref{ome.04}). Accordingly, the following meromorphic version of Theorem \ref{3.thm:8} is a corollary of Conjecture~\ref{mfaysscal}:
{\conj 
\label{mfaysform} The contracted product $g^{\overrightarrow{I}} {}_J(x,z)   g^{\overrightarrow{P} J}{}_K(y,x)$, which is a $(1,0)$-form in the repeated point $x$, may be \textit{$x$-reduced} as follows  in terms of Enriquez integration kernels,
\begin{align}
g^{\overrightarrow{I}} {}_J & (x,z)   g^{\overrightarrow{P} J}{}_K(y,x) =
g^{\overrightarrow{I}} {}_J(x,z)   g^{\overrightarrow{P} J}{}_K(y,z)   \label{ome.14}  \\
&\quad
- \sum_{\ell=0}^s (-1)^{\ell-s} \sum_{k = 0}^r g^{ P_s \cdots P_{\ell+1} \shuffle I_1 \cdots I_k } {}_J(x,y) g^{P_1\cdots  P_\ell J I_{k + 1} \cdots I_r}{}_K(y,z)
\notag\\
&\quad
-  \sum_{\ell=0}^s (-1)^{\ell-s} g^{P_1 \cdots P_\ell}{}_J(y,z) \big[  g^{(P_s\cdots P_{\ell+1} \shuffle \overrightarrow{I} )J} {}_K(x,y) +  g^{(P_s\cdots P_{\ell+1} J \shuffle  I_1 \cdots I_{r-1} ) I_r} {}_L(x,z) \big] \notag\end{align}}

Even though the analogous Fay identities (\ref{exfay.15})
and (\ref{4.ff.99}) among DHS kernels are related to (\ref{ome.13}) and (\ref{ome.14})
by the substitution $g^{\overrightarrow{I}} {}_J(u,v)
\leftrightarrow f^{\overrightarrow{I}} {}_J(u,v)$ in each term, our proofs of 
Theorems \ref{3.thm:7} and \ref{3.thm:8} do not readily carry over to the meromorphic case:
First, consistency of DHS-kernel identities under antiholomorphic $x,y,z$ derivatives 
as in step 1 of section \ref{prfsec.1} translates into
the behaviour of Enriquez-kernel identities under the $\mB$ monodromies (\ref{ome.02}). The latter involve different powers
of $2\pi i$ and are from a combinatorial perspective more difficult to control than the antiholomorphic derivatives 
(\ref{ften.02}), (\ref{ften.03}) of DHS kernels. Second, our proofs of DHS-kernel
identities rely on demonstrating that the vanishing quantities integrate to zero, see step 2 in section \ref{prfsec.1}; the
meromorphic counterpart of this step is to demonstrate consistency of (\ref{ome.13}) and (\ref{ome.14}) with $\mA$-periods in $x,y$ which -- in contrast to $\int_\Sigma d^2 z\, \bar \omega^K(z) f^{I_1\cdots I_r} {}_J(z,x) = 0$ for $r\geq 1$ -- do
not vanish for $g^{I_1\cdots I_r} {}_J(z,x)$ with $r\notin 2\mathbb N{+}1$ \cite{Enriquez:2011} and thereby complicate computations.
Still, alternative proofs of Conjectures \ref{mfaysscal} and \ref{mfaysform} were advanced in \cite{Baune:2024ber, paper.II} based on different methods. Moreover, appendix A of \cite{DHoker:2025dhv} proves all cases of (\ref{ome.13}) and (\ref{ome.14})
with $\overrightarrow{P} = \emptyset$ by adapting the steps of section \ref{prfsec.1} in proving DHS-kernel identities 
to the meromorphic case.

\subsubsection{Uncontracted meromorphic Fay identities}
\label{sec:7.3.2}

The derivation of uncontracted Fay identities among $f$-tensors from contracted
ones in section \ref{sec:4.9} is solely based on the decomposition (\ref{fphig.1}) into
traces and traceless parts which carries over to the Enriquez kernels as 
seen in (\ref{ome.04}). Accordingly, the uncontracted Fay identities (\ref{rewr.8}) hold 
in identical form for $f \rightarrow g$,
\begin{align}
& g^{\overrightarrow{P}M}{}_K(y,z)  g^{\overrightarrow{I}Q}{}_L(x,z) =
 \delta^Q_L g^{\overrightarrow{P}M}{}_J(y,z)  g^{\overrightarrow{I} J}{}_K(x,z) \label{ome.17} \\
 &\quad
   +  g^{\overrightarrow{P}M}{}_K(y,z)  g^{\overrightarrow{I}Q}{}_L(x,a)
  -  \delta^Q_L g^{\overrightarrow{P}M}{}_J(y,z)    g^{\overrightarrow{I} J}{}_K(x,a)  
\notag
\end{align}
where Conjecture \ref{mfaysscal} may be used to \textit{$z$-reduce}  the first term
on the right side. Similarly, the uncontracted Fay identities (\ref{rewr.12})
have the direct meromorphic analogue
\begin{align}
&g^{\overrightarrow{I}}{}_K(x,z)  g^{\overrightarrow{P}Q}{}_L(y,x) 
= \delta^Q_L g^{\overrightarrow{I}}{}_J(x,z)   g^{\overrightarrow{P} J}{}_K(y,x)  
\label{ome.18} \\
&\quad + g^{\overrightarrow{I}}{}_K(x,z)  g^{\overrightarrow{P}Q}{}_L(y,a) 
-  \delta^Q_L g^{\overrightarrow{I}}{}_J(x,z)    g^{\overrightarrow{P} J}{}_K(y,a)
\notag
\end{align}
where Conjecture \ref{mfaysform} may be used to \textit{$x$-reduce} the first term on the right side.

\sm

In conclusion, the meromorphic uncontracted Fay identities
(\ref{ome.17}) and (\ref{ome.18}) ensure~that the elimination of repeated points
$z$ and $x$ in $g^{\overrightarrow{P}M}{}_K(y,z)  g^{\overrightarrow{I}Q}{}_L(x,z)$ 
and $g^{\overrightarrow{I}}{}_K(x,z)  g^{\overrightarrow{P}Q}{}_L(y,x) $
does not rely on the contracted index $J$ in (\ref{ome.13}) and (\ref{ome.14}), respectively.

\subsubsection{Iterated meromorphic Fay identities}
\label{sec:7.3.3}

We have seen in section \ref{s:fay.6.8} that iterative use of uncontracted Fay identities among
$f$-tensors \textit{$z$-reduces} products of $f^{\overrightarrow{R} }{}_{M}(z,y)$ and an arbitrary number 
of $f^{\overrightarrow{P_j}}{}_{K_j}(x_j,z)$. The conjectures in this section imply that the meromorphic uncontracted Fay  identities (\ref{ome.17}) and (\ref{ome.18}) take the same form as those of the $f$-tensors in (\ref{rewr.8}) and (\ref{rewr.12}). As a consequence, the reduction algorithm of section \ref{s:fay.6.8} should
carry over to arbitrary products $\prod_{j=1}^N g^{\overrightarrow{P_j}}{}_{K_j}(x_j,z)$ 
of Enriquez kernels after applying the substitution rule $f\rightarrow g$ to the $C$- 
and $D$-tensors in (\ref{rewr.32}).

\sm

In conclusion, the above meromorphic Fay identities involving three points are sufficient 
to \textit{$z$-reduce} arbitrary products of Enriquez kernels of $(1,0)$-form 
degree $\leq 1$ in $z$.  The algorithmic reduction of such products will be 
important to derive identities among iterated integrals of the Enriquez kernels.

\subsection{Meromorphic coincident limits}
\label{sec:7.4}

This section is dedicated to the coincident limits $z \rightarrow y$
of Enriquez kernels $g^{I_1\cdots I_r}{}_J(y,z)$ and their
corollaries for Fay identities. We provide evidence that the results
of section \ref{sec:5} on coincident limits of $f$-tensors -- in particular
the modular tensors $\Nf^{I_1\ldots I_r}$ that do not depend on any point -- have a direct
meromorphic counterpart.

\subsubsection{Coincident limits of Enriquez kernels}
\label{sec:7.4.1}

While the coincident limits of $f$-tensors were studied based on
anti-holomorphic derivatives, our analysis of their meromorphic
counterpart $g^{I_1\cdots I_r}{}_J(y,z)$ at $z \rightarrow y$
relies on monodromies. With the monodromies (\ref{mclim.01}) of the
trace components $\chi^{I_1\cdots I_{r}}(y,z)$ of the Enriquez kernels at hand,
it is straightforward to determine simultaneous monodromies
as both of $y,z$ are moved around the cycle $\mB_L$, e.g. %
\begin{align}
\chi^I(y{+}\mB_L,z{+}\mB_L) &= \chi^I(y,z)
-2\pi i  \varpi^I{}_L(y)
+ \frac{(2  \pi i)^2}{2} \bigg( 1 - \frac{1}{h} \bigg) \delta^I_L \omega_L(y)
\label{mclim.02} \\
\chi^{IJ}(y{+}\mB_L,z{+}\mB_L) &= \chi^{IJ}(y,z)
+ 2\pi i \big(  \delta^{J}_L \chi^{ I}(y,z)  -  \delta^{I}_L \chi^{ J}(y,z) - \varpi^{I J}{}_L(y) \big)
\notag \\
&\quad + \frac{(2  \pi i)^2}{2} \bigg(   \delta^{I}_L \varpi^{J}{}_L(y) - \frac{1}{2}  \delta^{J}_L \varpi^{I}{}_L(y)\bigg) 
+ \frac{(2\pi i)^3} {3!} \bigg( \frac{1}{h} - 1  \bigg) \delta^{IJ}_L \omega_L(y) \notag
\end{align}
Upon comparison with (\ref{trcless}), the simultaneous
$\mB$ monodromy $\chi^I(y{+}\mB_L,z{+}\mB_L) - \chi^I(y,z)$ is found to be identical with
that of $\varpi^{M I}{}_M(y)$. In the coincident limit $z \rightarrow y$, the
difference $\chi^I(y,y) - \varpi^{M I}{}_M(y)$ is therefore a single-valued and
holomorphic $(1,0)$-form in $y$ which can thus be expanded in $\omega_M(y)$,
\beq
\chi^I(y,y) =  \varpi^{M I}{}_M(y) +
\omega_M(y)  \mN^{MI}
\label{mclim.03}
\eeq
for some $y$-independent $ \mN^{MI}$. The restriction of $g^{I_1\cdots I_r}{}_J(x,y)$ to meromorphic Kronecker-Eisenstein kernels at genus one, see (\ref{ome.03}), together with the coincident limits,
\beq
\lim_{y \rightarrow x}
g^{(r)}(x{-}y)= -{\rm G}_r \, , \ \ \ \ \ \ r\geq 2
\label{ytoxgg}
\eeq
implies that
the quantity $ \mN^{MI}$ in (\ref{mclim.03}) can be viewed as a higher-genus uplift of the
quasi-modular holomorphic Eisenstein series~(\ref{csec.03}),
\beq
\mN^{IJ} \big|_{h=1} = {\rm G}_2
\label{aclim.04}
\eeq
The right side of (\ref{mclim.03}) is obtained from
the non-meromorphic identity (\ref{4.coin2}) through the formal
substitution rule $\p \Phi \rightarrow  \varpi$ and $\Nf^{MI}
\rightarrow \mN^{MI}$. However, the coincident
limit of $\chi^I(y,z)$ leading to the left side of (\ref{mclim.03}) does not necessitate any analogue of
the subtraction in (\ref{csec.12}) prior to the limit $z\rightarrow y$ of $\p_y {\cal G}^I(y,z)$.
It is tempting to apply the same substitution rules to the limits $z\rightarrow y$
of $\p_y {\cal G}^{I_1 \cdots I_r}(y,z)$ at higher rank $r\geq 2$ in section \ref{sec:cl.2}.
Indeed, substituting $\p \Phi \rightarrow  \varpi$ and $\Nf^{MI}
\rightarrow \mN^{MI}$ into the expression (\ref{cl.09}) for
$\lim_{z\rightarrow y}\p_y {\cal G}^{IJ}(y,z)$ completely captures the
monodromy (\ref{mclim.02}) of
$\chi^{IJ}(y{+}\mB_L,z{+}\mB_L) - \chi^{IJ}(y,z)$
in the limit $z \rightarrow y$: The first five terms on the right side of
\begin{align}
\chi^{IJ}(y,y) &=
\varpi^{IMJ}{}_M(y)
-  \varpi^{JMI}{}_M(y)
+ \varpi^{MIJ}{}_M(y)  \notag \\
&\quad +
\varpi^{I}{}_M(y) \mN^{MJ}
- \varpi^{J}{}_M(y) \mN^{MI} 
+ \omega_M(y) \mN^{MIJ}
\label{holo.09}
\end{align}
fully capture the monodromies of the left side in $y$,
which introduces yet another $y$-independent meromorphic
function $\mN^{MIJ}$ in the last term. The matching of the
monodromies on both sides relies on the symmetry
$\mN^{IM} = \mN^{MI}$ of the meromorphic functions in (\ref{mclim.03})
which we shall justify in the discussion below (\ref{holo.12}).

\sm

More generally, we expect the following meromorphic analogue of
the $z\rightarrow y$ limit (\ref{explcs}) of $\p_y {\cal G}^{I_1 \cdots I_r}(y,z)$:
{\conj
\label{cj.mero}
The coincident limits of the $\delta^J_K$ trace components
$\chi^{I_1 \cdots I_r}(y,z)$ of the Enriquez kernels
$g^{I_1 \cdots I_r J}{}_K(y,z)$ in (\ref{ome.04}) with $r\geq 1$ are given by,
\begin{align}
\chi^{I_1 \cdots I_r}(y,y) & =  \omega_M(y) \,  \mN^{M I_1 I_{2}\cdots I_r}
+  \varpi^{M I_1 \cdots I_r}{}_M(y)
\label{mclim.04}\\
&\quad +
\sum_{1 \leq p \leq q \atop{(p,q) \neq (1,r)}}^r
(-1)^{r-q} \bigg[
\varpi^{I_1 I_2 \cdots I_{p-1}\shuffle I_r I_{r-1}\cdots I_{q+1} }{}_M(y) \, 
\mN^{M I_p I_{p+1}\cdots I_q} 
\notag \\
&\quad  \quad  \quad  \quad  \quad \quad  \quad \quad  \;
+ \varpi^{(I_1 I_2 \cdots I_{p-1}\shuffle I_r I_{r-1}\cdots I_{q+1} )M  I_p I_{p+1}\cdots I_q}{}_M(y)
\bigg]
\notag
\end{align}
provided that the meromorphic $y$-independent quantities $\mN$ obey cyclic symmetry,
\beq
\mN^{I_1 I_2 \cdots I_r} = \mN^{I_2  \cdots I_r I_1} 
 \label{holo.10}
\eeq}
Assuming the cyclic symmetries (\ref{holo.10}) at lower rank, we have verified (\ref{mclim.04}) by comparing monodromies on both sides up to and including rank four. In case the cyclic symmetry in (\ref{holo.10}) fails at some ranks $r\geq 3$, then the right side of (\ref{mclim.04}) needs to be augmented by counterterms involving at least one factor of $\mN^{I_1 I_2 \cdots I_s} {-} \mN^{I_2  \cdots I_s I_1}$ in each term to match the $\mB$ monodromies. 
By the restriction (\ref{ochi.03}) of the components $\varpi$ and $\chi$ of the Enriquez kernels to genus one
and their coincident limits (\ref{ytoxgg}), we recover holomorphic Eisenstein series of modular weight $r\geq 3$ from the
genus-one instance of  (\ref{holo.09}) and (\ref{mclim.04}),
\beq
\mN^{I_1 I_2 \cdots I_r}  \big|_{h=1} = {\rm G}_r
 \label{holo.11}
\eeq
We leave it as two important open problems to find $\mA$-cycle-integral
or theta-function representations for $\mN^{I_1 I_2 \cdots I_r}$
and to determine their modular properties.

\sm

Note that the cyclic symmetry (\ref{holo.10}) of the meromorphic quantities $\mN^{I_1 I_2 \cdots I_r}$
is expected to extend to the full dihedral group:

 {\conj
\label{alsorefl}
The meromorphic quantities $\mN^{I_1 I_2 \cdots I_r}$ 
with $r\geq 2$ exhibit alternating parity under reflection
$I_1 I_2 \cdots I_r \rightarrow I_r \cdots I_2 I_1$ of their indices:
\beq
\mN^{I_1I_2  \cdots I_r} =(-1)^r \mN^{I_r  \cdots I_2 I_1}
 \label{holo.refl}
\eeq}

It is worth highlighting why the DHS counterparts $\widehat \mN^{I_1  \cdots I_r}$ of 
$\mN^{I_1  \cdots I_r}$ admit a simple proof of their dihedral symmetries (\ref{coin.54}) which
cannot be readily adapted to coincident Enriquez kernels: The dihedral properties of $\widehat \mN^{I_1  \cdots I_r}$
are manifest from their integral representation (\ref{coin.53}) which in turn follows from the surface integral
of the coincident DHS kernels (\ref{explcs}) over $y$. One cannot derive meromorphic integral representations
of $\mN^{I_1  \cdots I_r}$ with the same ease: By the non-zero $\mA$-periods of generic Enriquez kernels 
on the right side of (\ref{mclim.04}), its $\mA$-period does not result in expressions for $\mN^{I_1  \cdots I_r}$ with
transparent dihedral properties. Still, the proof of interchange and Fay identities of Enriquez kernels in \cite{paper.II}
identifies the reflection parity in Conjecture \ref{alsorefl} as a consequence of the conjectural cyclicity
$\mN^{I_1 I_2 \cdots I_r}  = \mN^{I_2  \cdots I_r I_1}$.

\subsubsection{Coincident limits of meromorphic Fay identities at weight two}
\label{sec:7.4.2}

With the above candidate expressions for the limits $z \rightarrow y$ of
$\chi^{I_1 \cdots I_r}(y,z)$, we shall now spell out the coincident limits of the
conjectural meromorphic Fay identities of section \ref{sec:7.3}.
In the same way as the pole $\p_x {\cal G}(x,y)=  (y{-}x)^{-1} + {\rm reg}$ in
non-meromorphic Fay identities introduced $y$-derivatives of $f^{I_1\cdots I_r}{}_J(x,y)$
into their coincident limits of section \ref{sec:cl.3}, the
simple pole $\chi(x,y)=  (y{-}x)^{-1} + {\rm reg}$
gives rise to contributions $\partial_y g^{I_1\cdots I_r}{}_J(x,y)$
to the subsequent formulas. As an additional simplifying feature of the
meromorphic setting, one does not encounter any analogues of the
Abelian integrals in (\ref{1.inta}) or the terms $\sim (\bar z{-}\bar y)$
in (\ref{3.coin.13}) which compensate for the ill-defined $z\rightarrow y$ limit
of $\p_y \cG^I(y,z)$.

\sm

The simplest example is the $z \rightarrow y$ limit of the meromorphic
Fay identity (\ref{ome.11}). Using the coincident limit (\ref{mclim.03})
of $\chi^I(y,z)$, we obtain
\begin{align}
0 & = 
\chi(x,y) \chi(y,x) -   \p_y  \chi(x,y)  + \chi_2(x,y)
 \label{holo.12} \\
&\quad  - \mN^{IJ}  \om_I(y) \om_J(x) 
- \om_I(y) \chi^I(x,y) 
-\om_I(x)  \varpi^{MI}{}_M(y)
\notag
\end{align}
as a meromorphic analogue of (\ref{csec.21}) with $\chi_2$ given by (\ref{ome.12}). By integrating $x$ and $y$ over
the $\mA^P$ and $\mA^Q$ cycles, one can deduce the symmetry property
$\mN^{IJ}=\mN^{JI}$ from (\ref{holo.12}):
\begin{itemize}
\item $\chi(x,y) \chi(y,x)$ and $ \chi_2(x,y) $ are symmetric under $x \leftrightarrow y$
by inspection, and so is $\p_y  \chi(x,y) = \p_x  \chi(y,x) $ by the swapping identity
in Theorem \ref{swid.02} at rank $r=0$; hence, the respective $\mA$-periods
$\oint_{\mA^P} dx \oint_{\mA^Q} dy $ of $ \chi_2(x,y) $ and the non-singular
combination $\chi(x,y) \chi(y,x) -   \p_y  \chi(x,y)$ are symmetric under
$P\leftrightarrow Q$;
\item the last two terms $ - \om_I(y) \chi^I(x,y)  -\om_I(x)  \varpi^{MI}{}_M(y)$ are also symmetric under 
$x \leftrightarrow y$ by the corollary $\omega_M(x) g^{IM}{}_I(y,x)= \omega_M(y) g^{IM}{}_I(x,y)$ of the
interchange identity (\ref{pisarezero}) and therefore have a $P\leftrightarrow Q$ symmetric integral against $\oint_{\mA^P} dx \oint_{\mA^Q} dy $;
\end{itemize}
By virtue of these observations, the computation of $\mN^{QP }=
\oint_{\mA^P} dx \oint_{\mA^Q} dy \, \mN^{IJ}  \om_I(y) \om_J(x) $
from (\ref{holo.12}) yields a symmetric function under $P\leftrightarrow Q$
as was used in the matching of $\mB$ monodromies in (\ref{holo.09}).

\sm

The scalar coincident Fay identity (\ref{holo.12}) and the
traceless two-tensor identity (\ref{ome.99}) can be combined to
the following meromorphic counterpart of (\ref{csec.26}) 
\beq
g^I{}_J(x,y) g^J{}_K(y,x) = \delta^I_K  \p_y \chi(x,y)   - \omega_J(y) g^{I \shuffle J}{}_K(x,y) - \omega_J(x)
g^{JI}{}_K(y,y)
\label{holo.13}
\eeq
which also follows from the $z\rightarrow y$ limit of (\ref{ome.00}).

\subsubsection{Coincident limits of meromorphic Fay identities at arbitrary weight}
\label{sec:7.4.3}

At higher weight, we shall use the compact notation,
\bea
\label{holo.14}
\cX^{I_1 \cdots I_r}{}_J(y) =
\begin{cases}
 g^{I_1 \cdots I_r}{}_J(y,y) &:\, r \geq 2 \cr
 \varpi^{I_1}{}_J (y) &:\,  r=1 \cr
\om_J(y) &:\, r=0
\end{cases}
\eea
analogous to (\ref{4.cF}). The trace part of 
$g^{I_1 \cdots I_r}{}_J(y,y) = \varpi^{I_1 \cdots I_r}{}_J(y)  - \delta^{I_r}_J \chi^{I_1 \cdots I_{r-1}}(y,y) $ is proposed to
admit the further decomposition via (\ref{mclim.04}), though the conjectures in this section would not be affected by tentative counterterms in (\ref{mclim.04}) involving $\mN^{I_1 I_2 \cdots I_s} {-} \mN^{I_2  \cdots I_s I_1}$. In the notation of (\ref{holo.14}), the meromorphic coincident Fay identity at arbitrary weight (which is conjectural since the underlying three-point Fay identities (\ref{ome.14}) are) takes the form,
{\conj
\label{me.coin}
The contracted product $g^{\vI} {}_J(x,y) \, g^{\vP J} {}_K(y,x)$ may be \textit{$x$-reduced} as follows,
\bea
\label{holo.15}
g^{\vI} {}_J(x,y) \, g^{\vP J} {}_K(y,x)
& = &
g^\vI {}_J(x,y) \, \cX^{\vP  J} {}_K(y) + (-)^s \delta^{I_r}_K   \p_y \chi^{\vPt  \, \shuffle \, I_1 \cdots I_{r-1}}(x,y)  
\no \\ &&
- \sum_{k=0}^r  \sum_{\ell=0}^s (-)^{\ell-s}
g^{P_s \cdots P_{\ell+1} \shuffle I_1 \cdots I_k}{}_J(x,y) \,  \cX^{P_1 \cdots P_\ell J I_{k+1} \cdots I_r}{}_K(y)
\no \\ &&
- \sum_{\ell=0}^s (-)^{\ell-s} \cX^{P_1 \cdots P_\ell}{}_J(y) \,  g^{P_s \cdots P_{\ell+1}J \, \shuffle \, \vI}{}_K(x,y)
\eea}
This contracted version of the meromorphic coincident Fay identity can be uplifted to an
uncontracted version through the meromorphic analogue of (\ref{ucon.12}):
\begin{align}
&g^{\overrightarrow{I}}{}_K(x,y)  g^{\overrightarrow{P}Q}{}_L(y,x)
= \delta^Q_L g^{\overrightarrow{I}}{}_J(x,y)   g^{\overrightarrow{P} J}{}_K(y,x) 
\label{holo.21} \\
&\quad + g^{\overrightarrow{I}}{}_K(x,y)  g^{\overrightarrow{P}Q}{}_L(y,a)
-  \delta^Q_L g^{\overrightarrow{I}}{}_J(x,y)   g^{\overrightarrow{P} J}{}_K(y,a)
\notag
\end{align}
with an arbitrary point $a$ on the universal cover of $\Sigma$.
For the single-valued analogue of the coincident Fay identities (\ref{holo.15}) in
Theorem \ref{4.thm:1}, the steps in Appendix \ref{prfcoin} lead to the
reformulation in Theorem \ref{finthm}. By adapting the computations of
Appendix \ref{prfcoin} to $f^{I_1 \cdots I_r}{}_J \rightarrow g^{I_1 \cdots I_r}{}_J $,
the meromorphic coincident Fay identities (\ref{holo.15}) can be shown to admit the following alternative form:
{\conj
\label{me.coin2}
The product $g^{\vI} {}_J(x,y) \, g^{\vP J} {}_K(y,x)$ may be alternatively \textit{$x$-reduced} via,
\begin{align}
&g^{ \overrightarrow{I}}{}_J(x,y) g^{\overrightarrow{P} J}{}_K(y,x)
=  - (-1)^s \om_J(y) g^{   \overrightarrow{I}\shuffle \overleftarrow{P}  J }{}_K(x,y)
+(-1)^s \delta^{I_r} _K   \partial_y
   \chi^{I_1 \cdots I_{r-1}   \shuffle \overleftarrow{P} }(x,y)  
\label{holo.22} \\
&\quad \quad 
-  \sum_{k =0}^{r-1}  \sum_{\ell=0}^s
(-1)^{s-\ell} g^{ I_1 \cdots I_k \shuffle P_s \cdots P_{\ell+1} }{}_J(x,y)
g^{ P_1 \cdots P_\ell J I_{k+1} \cdots I_r  }{}_K(y,y)
\notag  \\
&\quad \quad
- \sum_{\ell=1}^s (-1)^{s-\ell } \big[
g^{P_1\cdots P_\ell}{}_J(y,a_\ell ) g^{ P_s\cdots P_{\ell +1}J \shuffle \overrightarrow{I} }{}_K(x,y)
\,{-}\, g^{P_1\cdots P_{\ell-1}J}{}_K(y,a_\ell ) g^{P_s\cdots P_\ell  \shuffle \overrightarrow{I} }{}_J(x,y)
\big]
\notag
\end{align}
with arbitrary points $a_1,\cdots,a_s$ on the universal cover of $\Sigma$.}

In comparison to the non-meromorphic analogue (\ref{coin.75}) of (\ref{holo.22}), the
coincident limit $g^{  J I_r  }{}_K(y,y)$ in the $(k,\ell)=(r{-}1,0)$ term
of the second line is by itself well-defined and there is no need
to sidestep the ill-defined $z\rightarrow y$ limit of $f^{  J I_r  }{}_K(y,z)$
via $ \p_y \Phi^{ J  I_r  }{}_K(y) -  \delta^{I_r}_K \cC^J(y) $.

\subsection{Change of fibration basis for meromorphic polylogarithms}
\label{sec:7.9}

This section aims to provide an introduction to the implications of our identities  among Enriquez kernels for the corresponding iterated integrals. We shall focus on the meromorphic polylogarithms introduced in section 5.6 of \cite{Baune:2024} for hyperelliptic $\Sigma$ and extended here to arbitrary  Riemann surfaces of genus $h$,
\beq
 \tGamma{ \overrightarrow{I}_{\!\!1} &\overrightarrow{I}_{\!\!2} &\cdots &\overrightarrow{I}_{\!\!\ell} }{
J_1 &J_2 &\cdots &J_\ell }{ p_1 &p_2 &\cdots &p_\ell}{x,y} 
= 
\int^x_y dt \, g^{  \overrightarrow{I}_{\!\!1} }{}_{J_1} (t,p_1) \,
 \tGamma{  \overrightarrow{I}_{\!\!2} &\cdots &\overrightarrow{I}_{\!\!\ell} }{
 J_2 &\cdots &J_\ell }{ p_2 &\cdots &p_\ell}{t,y} \, , \ \ \ \ \ \
  \tGamma{ \emptyset }{ \emptyset }{ \emptyset }{x,y} = 1
  \label{gatw.01}
\eeq
This relation provides a definition of the polylogarithms, recursively in the length $\ell \geq 0$, and in the number of points $p_1,\cdots ,p_\ell$, as we shall now explain. For multi-indices $\overrightarrow{I}_{\!\!1}$, $\overrightarrow{I}_{\!\!2}$, $\cdots$, $\overrightarrow{I}_{\!\!\ell}$ involving $n_1$, $n_2$, $\cdots$, $n_\ell \geq0$ letters, the specialization of (\ref{gatw.01}) to genus one exactly
matches\footnote{We depart from the normalization conventions in section 5.6 of \cite{Baune:2024} by powers of $-2\pi i$ due to the relative factors in (\ref{Enorm}) between the integration kernels $g^{I_1\cdots I_r}{}_J(x,y)$ in (\ref{gatw.01}) and the $\omega^{I_1\cdots I_r}{}_J(x,y)$ employed in the hyperelliptic polylogarithms in the reference.} the formulation of elliptic polylogarithms via $g^{(n_i)}$-kernels \cite{Broedel:2017kkb} once the lower endpoint $y$ of the integration path of is fixed to the origin of the universal cover of the torus, 
\beq
 \tGamma{ \overrightarrow{I}_{\!\!1} &\overrightarrow{I}_{\!\!2} &\cdots &\overrightarrow{I}_{\!\!\ell} }{
J_1 &J_2 &\cdots &J_\ell }{ p_1 &p_2 &\cdots &p_\ell}{x,y} \, \Big |_{h=1 \atop{y=0}} = 
\tilde \Gamma\big( \smallmatrix n_1&n_2 &\cdots &n_\ell \\ p_1  &p_2 &\cdots &p_\ell  \endsmallmatrix ;x\big)
  \label{gatw.02}
\eeq
As is familiar from the elliptic polylogarithm $\tilde \Gamma\big( \smallmatrix 1 \\ a \endsmallmatrix ;x\big)$
with $a=0$ and $a=x$ \cite{Broedel:2014vla, Broedel:2018iwv}, the integral $\tGamma{ I }{ J }{ p}{x,y}$ 
over the singular integration kernel $g^I{}_J(t,p)= \frac{\delta^I_J}{t{-}p}+{\rm reg}$ exhibits endpoint divergences if $p=x$ or $p=y$ which require regularization, for instance via tangential base points \cite{Deligne:1989, Panzer:2015ida, Abreu:2022mfk}. At higher length $\ell \geq 2$, the meromorphic higher-genus polylogarithms (\ref{gatw.01}) are taken to be shuffle-regularized such that the treatment of endpoint divergences
is determined by the regularization prescription for $\tGamma{ I }{ J }{ p}{x,y}$.

\sm

Empty multi-indices $\overrightarrow{I}_{\!\!k} = \emptyset$ in (\ref{gatw.01}) refer to 
integration kernels $\omega_{J_k}(t)$ which do not depend on $p_k$, and we shall
then omit the third row of the respective column as in 
$\tGamma{  \cdots &\emptyset &\cdots   }{ \cdots &J_k &\cdots  }{  \cdots &  &\cdots  }{x,y}$.  
 If all the multi-indices $ \overrightarrow{I}_{\!\!k}$ 
 are empty, the polylogarithms (\ref{gatw.01}) reduce to iterated Abelian integrals 
 $\Gamma_{J_1\cdots J_r} ( x,y)$ obtained from the
 special case (\ref{Gsec.14}) of the generically non-meromorphic polylogarithms
 $\Gamma (\mw ;x,y;p)$  of \cite{DHS:2023}  for words $\mw$ in letters $a^{J_k}$ only,
 \beq
 \tGamma{ \emptyset &\emptyset &\cdots &\emptyset}{
J_1 &J_2 &\cdots &J_\ell }{   &  &\cdots & }{x,y}
 = \Gamma (a^{J_1} a^{J_2}\cdots a^{J_\ell} ;x,y;p)  = \Gamma_{J_1\cdots J_\ell} ( x,y)
 \label{gatw.03}
 \eeq
 The dependence of (\ref{gatw.01}) on the points $p_k$ is concentrated in the traces with respect to $J_k$ with the rightmost index of  $\overrightarrow{I}_{\!\!k} $, and the relation $\chi^{I_1\cdots I_s}(t,p) =  - \frac{1}{h}g^{I_1\cdots I_s K}{}_K(t,p)$ propagates as follows to the polylogarithms in (\ref{gatw.01}),
\beq
 \int^x_y dt \, \chi^{  \overrightarrow{I}_{\!\!1} }(t,p_1) \,
 \tGamma{  \overrightarrow{I}_{\!\!2} &\cdots &\overrightarrow{I}_{\!\!\ell} }{
 J_2 &\cdots &J_\ell }{ p_2 &\cdots &p_\ell}{t,y} 
=  - \frac{1}{h} \tGamma{ \overrightarrow{I}_{\!\!1}K &\overrightarrow{I}_{\!\!2} &\cdots &\overrightarrow{I}_{\!\!\ell} }{
K &J_2 &\cdots &J_\ell }{ p_1 &p_2 &\cdots &p_\ell}{x,y}  
 \label{gatw.04}
 \eeq
By the meromorphicity of the Enriquez kernels, the polylogarithms in (\ref{gatw.01})
are meromorphic in all points $x,y,p_1,\cdots,p_\ell$ in the universal cover of $\Sigma$ 
and in the moduli of the surface. Accordingly, total differentials reduce to the components
involving the holomorphic derivative, and one for instance simplifies $d_\xi $
to $d\xi \partial_{\xi}$ in,
\beq
 \tGamma{ \overrightarrow{I}_{\!\!1} &\cdots &\overrightarrow{I}_{\!\!k} &\cdots &\overrightarrow{I}_{\!\!\ell} }{
J_1 &\cdots &J_k &\cdots &J_\ell }{ p_1 &\cdots &p_k &\cdots &p_\ell}{x,y}
=  \tGamma{ \overrightarrow{I}_{\!\!1} &\cdots &\overrightarrow{I}_{\!\!k} &\cdots &\overrightarrow{I}_{\!\!\ell} }{
J_1 &\cdots &J_k &\cdots &J_\ell }{ p_1 &\cdots &q &\cdots &p_\ell}{x,y}
+ \int^{p_k}_q d\xi \,  \p_\xi \tGamma{ \overrightarrow{I}_{\!\!1} &\cdots &\overrightarrow{I}_{\!\!k} &\cdots &\overrightarrow{I}_{\!\!\ell} }{
J_1 &\cdots &J_k &\cdots &J_\ell }{ p_1 &\cdots &\xi &\cdots &p_\ell}{x,y}
 \label{gatw.05}
\eeq
In the remainder of this section, we shall combine (\ref{gatw.05}) with the meromorphic interchange and Fay identities of sections \ref{sec:7.2} and \ref{sec:7.3} to perform \textit{changes of 
fibration bases} for the meromorphic higher-genus polylogarithms of (\ref{gatw.01}). 
The existence of the change-of-fibration-basis identities is essential
for the closure of (\ref{gatw.01}) under integration as detailed in the case of non-meromorphic 
higher-genus polylogarithms in early section \ref{sec:clo}. The explicit form
of these identities to be derived below from (\ref{gatw.05}) at low length $\ell\leq 2$
is valuable for applications to Feynman integrals, string amplitudes or other 
situations in physics. The key ideas of the subsequent examples straightforwardly
apply to higher length $\ell$ and allow us to algorithmicially (in fact recursively in $\ell$) relegate the 
dependence of (\ref{gatw.01}) on an arbitrary point $p_k$ solely to the endpoint of the integration path as seen in (\ref{gatw.05}).

\sm 

In (\ref{gatw.05}) and throughout, we shall assume that $x,y,p_1,\cdots,p_\ell$ are pairwise distinct to avoid
endpoint divergences and to illustrate
the importance of Fay identities in a streamlined way. The subsequent methods to find primitives can be straightforwardly
extended to (shuffle-regularized) cases where some of $x,y,p_1,\cdots,p_\ell$ coincide, e.g.\ by adding up partial derivatives
in different variables when evaluating $d_\xi \,\tGamma{ \overrightarrow{I}_{\!\!1} &\cdots &\overrightarrow{I}_{\!\!k} &\cdots &\overrightarrow{I}_{\!\!\ell} }{
J_1 &\cdots &J_k &\cdots &J_\ell }{ p_1 &\cdots &\xi &\cdots &p_\ell}{x,\xi}$ as in \cite{Broedel:2014vla, Baune:2024ber}.

\subsubsection{Length one}
\label{sec:7.9.1}

At length one, the opening line (\ref{gatw.05}) leads us to the following
change of fibration basis,
\begin{align}
&\tGamma{ I_1 \cdots I_r }{J}{p}{x,y} -  \tGamma{ I_1 \cdots I_r }{J}{q}{x,y}
= \int^p_q d\xi \int^x_y dt\, \p_\xi  g^{I_1 \cdots I_r}{}_J(t,\xi) \notag \\
&\quad = -\delta^{I_r}_J \int^p_q d\xi \int^x_y dt\, \p_\xi  \chi^{I_1 \cdots I_{r-1}}(t,\xi) 
\no \\ & \quad
= (-)^r\delta^{I_r}_J \int^p_q d\xi \int^x_y dt\, \p_t  \chi^{I_{r-1} \cdots I_{1}}(\xi,t)  
\notag \\
&\quad =  (-)^r\delta^{I_r}_J \int^p_q d\xi \, \big(   \chi^{I_{r-1} \cdots I_{1}}(\xi,x)
-   \chi^{I_{r-1} \cdots I_{1}}(\xi,y) \big) \notag \\ 
&\quad = \frac{1}{h}  (-)^{r}\delta^{I_r}_J 
\,\bigg\{
 \tGamma{ I_{r-1}\cdots I_1 K }{ K }{ y }{p,q} -  \tGamma{ I_{r-1}\cdots I_1 K }{ K }{ x }{p,q}
\bigg\}
 \label{gatw.08}
\end{align}
From the first line to the second, we have used the fact that $\p_\xi  g^{I_1 \cdots I_r}{}_J(t,\xi) $ is proportional to $\delta ^{I_r}_J$ and in going to the third line we have applied the swapping identity in Theorem \ref{swid.02}.
The integrals over $\xi$ in the fourth line were lined up with the polylogarithms $\tilde \Gamma$
in the last line by means of the integration identity (\ref{gatw.04}) for the trace components
$\chi^{P_1 \cdots P_s}(x,y) $ of the Enriquez kernels.

\sm

One could have also carried out the change of fibration basis without
isolating the $\xi$-dependence of $\p_\xi g^{I_1 \cdots I_r}{}_J(t,\xi)$ through
the trace decomposition (\ref{ome.04}). With the alternative representation
(\ref{swid.06eq}) of the swapping identity, one arrives at,
\begin{align}
\tGamma{ I_1 \cdots I_r }{J}{p}{x,y} -  \tGamma{ I_1 \cdots I_r }{J}{q}{x,y}
= (-)^r \, \bigg\{
 \tGamma{ I_{r-1}\cdots I_1 I_r }{ J }{ y }{p,q} -  \tGamma{ I_{r-1}\cdots I_1 I_r }{ J }{ x }{p,q}
\bigg\}
 \label{gatw.09}
\end{align}
Equivalence to the earlier form (\ref{gatw.08}) of the change of fibration basis at length one 
follows from the fact that the traceless part of $\tGamma{ I_{r-1}\cdots I_1 I_r }{ J }{ y }{p,q}$ 
is independent on $y$. Note that the specialization of (\ref{gatw.09}) to genus one
and $y=q=0$ leads to well-known identities among elliptic polylogarithms (see section 2.2.2 of \cite{Broedel:2014vla} 
for derivations of closely related identities),
\beq
\tilde \Gamma\big( \smallmatrix r \\ p \endsmallmatrix ;x\big)
- \tilde \Gamma\big( \smallmatrix r \\ 0 \endsmallmatrix ;x\big)
= (-)^r\Big\{
\tilde \Gamma\big( \smallmatrix r \\ 0 \endsmallmatrix ;p\big)
- \tilde \Gamma\big( \smallmatrix r \\ x \endsmallmatrix ;p\big)
\Big\} + \delta_{r,1} i \pi
\label{g1spec}
\eeq
where the addition of $i\pi$ in the case of $r=1$ is due to the
simple pole of the integration kernel $g^{(1)}$ and tied to the specialization
$y=q$  (see e.g.\ \cite{Kaderli:2022qeu}). Within the identity (\ref{gatw.09}) at general
$h\geq 1$, however, the points $x,y,p,q$ are understood to be pairwise
distinct by the discussion before section \ref{sec:7.9.1}. If we furthermore require
the integration paths from $y$ to $x$ and from $q$ to $p$ to not intersect,\footnote{\label{2piinote}Intersecting
integration paths from $y$ to $x$ and from $q$ to $p$ with pairwise distinct $x,y,p,q$ would introduce 
additive offsets of $\pm 2\pi i$ into (\ref{gatw.09}). We are grateful to Carlos Rodriguez for 
valuable discussions and numerical checks of contributions $\pm 2 \pi i$ in (\ref{gatw.09})
or $\pm i \pi$ in (\ref{g1spec}), in particular for pointing out their universality for
arbitrary genus $h\geq 0$.} 
then there is no
analogue of $i\pi$ in (\ref{gatw.09}), even for the case of $r=1$ and $I_1=J$ where
the integration kernel acquires simple poles.
 
\subsubsection{Length two}
\label{sec:7.9.2}

The key steps in the length-one computation of (\ref{gatw.08}) generalize to carrying out the
change of fibration bases in higher-genus polylogarithms of arbitrary length $\ell\geq 2$.
We shall explicitly present the case of length $\ell=2$ with multi-indices
$\overrightarrow{I} = I_1 \cdots I_r$, $\overrightarrow{P}=P_1\cdots P_s$
and its reflection $\overleftarrow{P}=P_s\cdots P_1$, again starting from (\ref{gatw.05}),
\begin{align}
&\tGamma{ \overrightarrow{I} &\overleftarrow{P} M }{ K &R }{ z &p}{x,y} 
-  \tGamma{ \overrightarrow{I} &\overleftarrow{P} M }{ K &R }{ z &q}{x,y}
= \int^p_q d\xi \int^x_y dt_1\,  g^{\overrightarrow{I}  }{}_K(t_1,z)
\int^{t_1}_y dt_2 \,
 \p_\xi  g^{ \overleftarrow{P} M  }{}_R(t_2,\xi) \notag \\
&\quad = (-)^s  \int^p_q d\xi \int^x_y dt_1\,  g^{\overrightarrow{I}  }{}_K(t_1,z)
\, \big( g^{ \overrightarrow{P} M  }{}_R(\xi,t_1) - g^{ \overrightarrow{P} M  }{}_R(\xi,y) \big) 
\notag \\
&\quad = (-)^s \delta^M_R \int^p_q d\xi \int^x_y dt_1\,  g^{\overrightarrow{I}  }{}_J(t_1,z)
\, \big( g^{ \overrightarrow{P} J  }{}_K(\xi,t_1) - g^{ \overrightarrow{P} J }{}_K(\xi,y) \big) 
\notag \\
&\quad =  (-)^s \delta^M_R \bigg\{
 \int^p_q d\xi \int^x_y dt\,  g^{\overrightarrow{I}  }{}_J(t,z) g^{ \overrightarrow{P} J  }{}_K(\xi,t)
- \tGamma{ \overrightarrow{I}   }{ J }{ z}{x,y} 
\tGamma{ \overrightarrow{P} J }{ K }{ y}{p,q} 
\bigg\}
 \label{gatw.21}
\end{align}
We have applied the swapping identity $ \p_\xi  g^{ \overleftarrow{P} M  }{}_R(t_2,\xi) = (-)^s \p_{t_2}  g^{ \overrightarrow{P} M  }{}_R(\xi,t_2)$ in passing to the second line
and then used a relabeling of the trace-decomposition identity (\ref{ome.18})
to introduce contracted indices $J$ in the integrand. In this way, the
bilinear $g^{\overrightarrow{I}  }{}_J(t,z) g^{ \overrightarrow{P} J  }{}_K(\xi,t)$ in
the last line is amenable to the uncontracted Fay
identity in Conjecture \ref{mfaysform}. 

\sm

The points $x,y,z,p,q$ in (\ref{gatw.21}) are again taken to be pairwise distinct to avoid endpoint divergences,
and we furthermore assume the integration paths from $y$ to $x$ and from $q$ to~$p$ to not intersect.
One would otherwise encounter additive contributions involving $(2\pi i)^2$ or $2\pi i$ multiplying length-one
polylogarithms due to integration kernels with simple poles (i.e.\ in cases with $r=1$ or $s=0$),
see footnote \ref{2piinote}. The nesting of such contributions with powers of $2\pi i$ is a local effect
on the surface and follows the same combinatorial rules as the appearance of $2\pi i$ in
change-of-fibration-basis identities among polylogarithms (\ref{g0poly}) on the sphere, see for instance
\cite{Kaderli:2022qeu} for the parallels between genus zero and $h=1$.

\sm

Since the goal of this section is to arrive at 
a fibration basis with all the $p$-dependence in the integration limit,
it is essential to perform the integration over $t$ in (\ref{gatw.21}) prior to that over~$\xi$. This
is accomplished by $t$-reducing the last line of (\ref{gatw.21}) via (\ref{ome.14})
and expressing the $t$-integrals in terms of polylogarithms,
\begin{align}
\tGamma{ \overrightarrow{I} &\overleftarrow{P} M }{ K &R }{ z &p}{x,y} 
&-  \tGamma{ \overrightarrow{I} &\overleftarrow{P} M }{ K &R }{ z &q}{x,y}
= \delta^M_R \bigg\{
(-)^s 
\tGamma{ \overrightarrow{I} }{ J }{ z}{x,y} \bigg[
\tGamma{ \overrightarrow{P} J }{ K }{ z}{p,q} -  \tGamma{ \overrightarrow{P} J }{ K }{ y}{p,q}\bigg] \notag \\
&\quad \quad\quad - \sum_{\ell=0}^s(-)^\ell \sum_{k=0}^r \int^p_q d\xi \, 
g^{P_1\cdots P_\ell J I_{k+1} \cdots I_r}{}_K(\xi,z) 
\tGamma{ P_s\cdots P_{\ell+1} \shuffle I_1\cdots I_k }{ J }{ \xi}{x,y} \notag \\
&\quad\quad\quad   - \sum_{\ell=0}^s(-)^\ell \bigg[
\int^p_q d\xi \, 
g^{ P_1\cdots P_\ell }{}_J(\xi,z) 
\tGamma{ (P_s\cdots P_{\ell+1} \shuffle \overrightarrow{I}) J }{ K }{ \xi}{x,y}  \notag \\
&\quad \quad\quad\quad \quad\quad
+ \tGamma{  P_1 \cdots P_\ell }{ J }{ z }{p,q} 
 \tGamma{ (P_s\cdots P_{\ell+1}  J\shuffle I_1\cdots I_{r-1}) I_r }{ K }{z}{x,y} 
\bigg]
\bigg\}
 \label{gatw.22}
\end{align}
In order to express the leftover integrals over $\xi$ in terms of $\tilde \Gamma( \cdots ;p,q )$ 
in the aspired fibration basis, it remains to apply the length-one identity (\ref{gatw.09}) to rewrite both $\tGamma{ P_s\cdots P_{\ell+1} \shuffle I_1\cdots I_k }{ J }{ \xi}{x,y}$ and 
$\tGamma{ (P_s\cdots P_{\ell+1} \shuffle \overrightarrow{I}) J }{ K }{ \xi}{x,y} $ in the fibration basis of $ \tilde \Gamma(\cdots;\xi,q) $. With the recursion (\ref{recshf}) for the shuffle product, the leftover $\xi$-integrals take the form,
\begin{align}
\int^p_q d\xi \, &
g^{P_1\cdots P_\ell J I_{k+1} \cdots I_r}{}_K(\xi,z) 
\tGamma{ P_s\cdots P_{\ell+1} \shuffle I_1\cdots I_k }{ J }{ \xi}{x,y}   \label{gatw.23} \\
&=  \tGamma{ P_s\cdots P_{\ell+1} \shuffle I_1\cdots I_k }{ J }{ q}{x,y} 
\tGamma{ P_1\cdots P_\ell J I_{k+1} \cdots I_r }{ K }{ z}{p,q} 
\notag \\
&\quad + (-)^{k+s-\ell} \bigg\{ 
\tGamma{ P_1\cdots P_\ell J I_{k+1} \cdots I_r & ( P_{\ell+2} \cdots P_s \shuffle I_k\cdots I_1 )P_{\ell+1} }{  K&J }{ z &y }{p,q}  - (x\leftrightarrow y)
\bigg\} \notag \\
&\quad + (-)^{k+s-\ell} \bigg\{ 
 \tGamma{ P_1\cdots P_\ell J I_{k+1} \cdots I_r & ( P_{\ell+1} \cdots P_s \shuffle I_{k-1}\cdots I_1) I_k }{  K&J }{ z &y }{p,q}  - (x\leftrightarrow y)
\bigg\} \notag
\end{align}
as well as
\begin{align}
\int^p_q d\xi \, 
g^{ P_1\cdots P_\ell }{}_J(\xi,z) 
&\tGamma{ (P_s\cdots P_{\ell+1} \shuffle \overrightarrow{I}) J }{ K }{ \xi}{x,y}
= \tGamma{ (P_s\cdots P_{\ell+1} \shuffle \overrightarrow{I}) J }{ K }{ q}{x,y}
 \tGamma{ P_1\cdots P_\ell  }{ J }{ z }{p,q} \notag  \\
 &\quad \quad + (-)^{s-\ell+r+1} \bigg\{
 \tGamma{ P_1\cdots P_\ell & ( P_{\ell+1} \cdots P_s \shuffle \overleftarrow{I}) J }{ J &K }{ z &y}{p,q}
  - (x\leftrightarrow y)\bigg\}
 \label{gatw.24} 
\end{align}
The above computations express a generic length-two polylogarithm $ \tGamma{ \overrightarrow{I}_{\!\!1} &\overrightarrow{I}_{\!\!2} }{ J_1 &J_2 }{  p_1 &p_2 }{x,y} $ in the fibration basis of $ \tilde\Gamma(\cdots;p_2,q) $. The alternative fibration basis of  $ \tilde \Gamma( \cdots ;p_1,q) $ can be readily attained by employing the shuffle product, 
\beq
 \tGamma{ \overrightarrow{I}_{\!\!1} &\overrightarrow{I}_{\!\!2} }{ J_1 &J_2 }{  p_1 &p_2 }{x,y} = \tGamma{ \overrightarrow{I}_{\!\!1}   }{ J_1  }{  p_1   }{x,y}  \tGamma{ \overrightarrow{I}_{\!\!2} }{ J_2 }{   p_2 }{x,y}  - \tGamma{ \overrightarrow{I}_{\!\!2} &\overrightarrow{I}_{\!\!1} }{ J_2 &J_1 }{  p_2 &p_1 }{x,y} 
 \label{shfib}
 \eeq
and applying relabelings of the changes of fibration bases (\ref{gatw.09}) and (\ref{gatw.22}) at length $\ell=1$ and $\ell=2$ to the $p_1$-dependence on the right side.

\subsubsection{Arbitrary length}
\label{sec:7.9.3}

Similar to the non-meromorphic change-of-fibration-basis identities 
in sections \ref{sec:clo.2} and \ref{sec:clo.3}, our method (\ref{gatw.05}) to change fibration bases of
meromorphic polylogarithms is recursive in their length. The use of swapping
identities and meromorphic Fay identities in the $\ell=2$ computations of (\ref{gatw.21}) and (\ref{gatw.22})
generalizes to arbitrary length $\ell$ and 
necessitates change-of-fibration-basis identities at length $\leq \ell{-}1$ to perform the
$\xi$-integral on the right side of (\ref{gatw.05}). 
For instance, $ \tGamma{ \overrightarrow{I}_{\!\!1}  &\cdots &\overrightarrow{I}_{\!\!\ell} }{
J_1  &\cdots &J_\ell }{ p_1  &\cdots &p_\ell}{x,y} $ can be brought into the fibration basis of 
$\tilde\Gamma(\cdots;p_\ell,q)$ as follows:
\begin{itemize}
\item straightforwardly adapt the steps of (\ref{gatw.21}) to perform the innermost integral, resulting in contributions of the form $g^{\overrightarrow{I}_{\! \! \ell-1}  }{}_J(t_{\ell-1},p_{\ell-1}) g^{ \overrightarrow{P} J  }{}_K(\xi,t_{\ell-1})$ to the leftover integrand;
\item \textit{$t_{\ell-1}$-reduce} this bilinear and integrate the
outcome of Fay identities over $t_1,\cdots,t_{\ell-1}$ in terms of $ \tGamma{ \overrightarrow{I}_{\!\!1}  &\cdots &\overrightarrow{I}_{\!\!\ell-2} &\overrightarrow{K} }{
J_1  &\cdots &J_{\ell-2} &L }{ p_1  &\cdots &p_{\ell-2} &p_{\ell-1}}{x,y} $ and $ \tGamma{ \overrightarrow{I}_{\!\!1}  &\cdots &\overrightarrow{I}_{\!\!\ell-2} &\overrightarrow{K} }{
J_1  &\cdots &J_{\ell-2} &L }{ p_1  &\cdots &p_{\ell-2} &\xi}{x,y} $ of length $\ell{-}1$ as in (\ref{gatw.22});
\item assuming that change-of-fibration-basis identities at length $\ell{-}1$ are available, express all the
$ \tGamma{ \overrightarrow{I}_{\!\!1}  &\cdots &\overrightarrow{I}_{\!\!\ell-2} &\overrightarrow{K} }{
J_1  &\cdots &J_{\ell-2} &L }{ p_1  &\cdots &p_{\ell-2} &\xi}{x,y} $ of the previous step
in the fibration basis of $ \tilde \Gamma(\cdots; \xi,q) $ to perform the $\xi$-integral as in (\ref{gatw.23}) and (\ref{gatw.24}).
\end{itemize}
The rewriting of $ \tGamma{ \overrightarrow{I}_{\!\!1} &\cdots &\overrightarrow{I}_{\!\!k} &\cdots &\overrightarrow{I}_{\!\!\ell} }{
J_1 &\cdots &J_k &\cdots &J_\ell }{ p_1 &\cdots &p_k &\cdots &p_\ell}{x,y}$ at arbitrary $k \leq  \ell$ in the fibration basis of  $ \tilde \Gamma(\cdots; p_\ell,q) $ can be reduced to the previous 
case of $k=\ell$ by moving the column of $p_k$ to the rightmost position via shuffle identities similar to (\ref{shfib}) and importing relations at lower length.

\newpage

\section{Conclusions and further directions}
\label{sec:conc}

As the main result of this work, we have generalized the Fay identities among the Kronecker-Eisenstein integration kernels in the elliptic polylogarithms of Brown and Levin \cite{BrownLevin} to compact Riemann surfaces of arbitrary genus. Our higher-genus Fay identities are bilinear relations among the single-valued tensorial integration kernels $f^{I_1\cdots I_r}{}_J(x,y)$ that furnish the backbone of the flat connection used to generate the polylogarithms in \cite{DHS:2023}.
Bilinears of the schematic form $f(y,z)f(x,z)$ and $f(x,z) f(y,x)$ with arbitrary
collections of tensor indices are rewritten in (\ref{exfay.15}) and (\ref{4.ff.99}) without
repeated appearance of the points $z$ and $x$, respectively. These Fay identities among
single-valued but non-meromorphic kernels $f^{I_1\cdots I_r}{}_J(x,y)$ are proposed to
apply in identical form to the meromorphic but multi-valued kernels $g^{I_1\cdots I_r}{}_J(x,y)$
introduced by Enriquez \cite{Enriquez:2011}.

\sm

Already at genus one, Fay identities among Kronecker-Eisenstein kernels are crucial to demonstrate the closure of elliptic polylogarithms under taking primitives \cite{BrownLevin} and to develop concrete integration algorithms \cite{Broedel:2014vla}. 
Similarly, the higher-genus Fay identities in this work are essential to change fibration bases and to determine primitives
of the polylogarithms in \cite{DHS:2023} when they are multiplied by more than one $f$-tensor, i.e.\ necessary conditions for closure under integration.
In fact, Fay identities involving three points on the surface suffice to integrate products of $f^{I_1\cdots I_r}{}_J(x,y)$ kernels involving an arbitrary number of points (as is familiar from genus one). By the meromorphic Fay identities in this work, the same algorithms and arguments for closure under integration apply to iterated integrals of the multi-valued Enriquez kernels $g^{I_1\cdots I_r}{}_J(x,y)$ including the hyperelliptic polylogarithms of~\cite{Baune:2024}.

\sm

The coincident limit of the single-valued Fay identities introduces modular tensors $\Nf^{I_1\cdots I_r}$ of all ranks $r\geq 2$ that do not depend on marked points and reduce to (almost) holomorphic Eisenstein series upon specialization to genus one. By analogy with the role of coincident Fay identities at genus one for the differential equations \cite{Gerken:2019cxz} of modular graph forms \cite{DHoker:2015wxz, DHoker:2016mwo}, the modular tensors $\Nf^{I_1\cdots I_r}$ are expected to govern the differential relations among modular graph tensors at arbitrary genus \cite{DHoker:2020uid}. Our coincident Fay identities among Enriquez kernels similarly relate the $y \rightarrow x$ limit of $g^{I_1\cdots I_r}{}_J(x,y)$ to certain meromorphic functions on Torelli space that should prominently feature in the differential structure of higher-genus analogues of elliptic multiple zeta values \cite{Enriquez:Emzv}.

\sm

In applications to string scattering amplitudes, the Fay identities in this work will be a driving force for bootstrap approaches to their moduli-space integrand and the integrations over the moduli in the low-energy expansion. This can for instance be anticipated from the simplifications of fermionic correlation functions -- cyclic products of so-called Szeg\"o kernels -- in terms of $f$-tensors \cite{DHoker:2023khh}. Our Fay identities will facilitate the identification of simplifying amplitude structures from the conspiracy of these fermionic correlators with bosonic ones.

\newpage

\appendix

\section{The prime form and the Arakelov Green function}
\label{appdefs}

The purpose of this appendix is to collect some basic results on $\tet$-functions, the prime form, the Arakelov Green function, and some aspects of their relation with string amplitudes. For a more detailed and systematic exposition, we refer to  \cite{DHoker:2022dxx, DHoker:1988pdl}.

\sm

The \textit{Siegel upper half space} of rank $h$ is the set of $h \times h$ symmetric matrices with complex entries and positive definite imaginary part,
\bea
\cH_h = \Big \{ \Omega \in \CC^{h \times h} \hbox{ such that } \Omega ^t = \Omega, \, \Im (\Omega) >0 \Big \}
\eea
The space $\cH_h$ may also be given as the K\"ahler coset $\cH_h = Sp(2h, \RR)/U(h)$ where $Sp(2h,\RR)$ is the group of matrices $M \in Sp(2h,\RR)$ defined by,
\bea
M = \left ( \begin{matrix} A & B \cr C & D \end{matrix} \right )\, ,
\hskip 0.6in 
M^t \mJ M = \mJ \, ,
\hskip 0.6in
\mJ = \left ( \begin{matrix} 0 & -I \cr I & 0 \end{matrix} \right )
\eea
and $A,B,C,D$ are real $h \times h$ matrices. The space $\cH_1$ is the Poincar\'e upper half plane.

\sm

\textit{Riemann $\tet$-functions} are holomorphic functions of $ (\zeta, \Omega) \in \CC^h \times \cH_h$ that may be augmented by characteristics $\delta = [\delta ', \delta ''] $ with $\delta', \delta '' \in \CC^h$, and are defined as follows,
\bea
\tet [\delta ] (\zeta |\Omega) = \sum_{n \in \ZZ^h} \exp \Big ( i \pi (n+\delta') ^t \Omega (n+\delta') + 2 \pi i (n+\delta')^t (\zeta +\delta'') \Big )
\eea
Half-integer characteristics $\delta ', \delta '' \in (\ZZ/2\ZZ)^h$ are either even or odd depending on whether the integer $4(\delta')^t \delta ''$ is even or odd or, equivalently, whether $\tet[\delta](\zeta|\Omega)$ is even or odd in $\zeta$. 

\sm

The homology group $H_1 (\Sigma, \ZZ)$ of a compact Riemann surface $\Sigma$  (which is connected by definition) of genus $h$ is isomorphic to $\ZZ ^{2h}$ and may be generated by a canonical homology basis of cycles $\mA^I$ and $\mB_J$ with intersection pairing $\mJ (\mA^I, \mA^J) =  \mJ(\mB_I, \mB_J)=0$ and $\mJ (\mA^I, \mB_J) = \delta ^I_J $ where $I,J = 1, \cdots, h$. The dual cohomology group $H^1(\Sigma, \ZZ)$ may be generated by $h$ holomorphic Abelian differentials $\bom_J$ which are canonically normalized on $\mA^I$-cycles as in (\ref{bsec.01}). The modular group $Sp(2h, \ZZ)$ transforms a canonical homology basis into a canonical homology basis. Its transformation laws were summarized in section \ref{sec:2.1}.

\sm

The period matrix $\Omega$ of a Riemann surface $\Sigma$ of genus $h$, defined in (\ref{bsec.01}), is an element of the corresponding  Siegel upper half space $\cH_h$ for all genera $h$. However, the converse does not hold globally for genus $h \geq 2$ and does not hold even locally for  genus $ h \geq 4$ where it gives rise to the Schottky problem. The subspace of $\cH_h$ whose elements correspond to the period matrix of a compact Riemann surface  is referred to as \textit{Torelli space} and denoted $\cT_h$. Equivalently, one may define $\cT_h$ as the moduli space of Riemann surfaces of genus $h$ endowed with a choice of canonical homology basis.

\subsection{The prime form}

For any odd half-integer characteristics $\nu$, one defines a  holomorphic $(1,0)$ form,
\bea
\om_\nu (x) = \om_I(x)  \, \p^I \tet [\nu] (0|\Omega) \, , \hskip 1in \p^I \tet [\nu] (0|\Omega) = { \p \over \p \zeta _I} \tet [\nu] (\zeta|\Omega) \Big |_{\zeta=0} 
\eea
whose $2(h-1)$ zeros are all double zeros as a consequence of the Riemann vanishing theorem. As a result, its square root $h_\nu(x) $ is a holomorphic $(\half, 0)$ form on $\Sigma$, namely it is a spinor with spin structure $\nu$. The \textit{prime form }
is defined as follows,
\bea
E(x,y) = { \tet [\nu] ( \zeta |\Omega) \over h_\nu (x) \, h_\nu(y)} \, ,
\hskip 1in \zeta _I = \int _y^x \bom _I
\label{defpform}
\eea
The prime form $E(x,y)$ is a holomorphic $(-\half, 0)$ form in $x$ and $y$, and is independent of the choice of $\nu$. While its $\mA$ monodromy is given by a factor of $\pm 1$, its $\mB$ monodromy  is non-trivial and renders $E(x,y)$ multiple-valued on $\Sigma$. Meromorphic Abelian differentials of the second and third kind, given by, 
\bea
\p_x \p_y \ln E(x,y)\, , \hskip 1in  \p_x \ln {E(x,u) \over E(x,v) }  
\eea 
respectively, are single-valued $(1,0)$ forms in $x$ and $y$, but the latter is multiple-valued  in $u,v$. For local complex coordinates $x,y$ parametrizing nearby points, the prime form satisfies,
\bea
E(x,y) = x-y + \cO\big((x-y)^3\big)
\eea
and vanishes nowhere else on $\Sigma$. Therefore $E(x,y)$ plays the role of a local difference function between points on a Riemann surface of arbitrary genus. As a result, we have,
\bea
\pbx \, \p_x \ln { E(x,y) \over E(x,z)} = \pi \delta (x,y) - \pi \delta (x,z)
\eea
The Riemann-Roch theorem precludes the existence of a single-valued meromorphic $(1,0)$ form with only a  single simple pole on a compact Riemann surface $\Sigma$, a fact familiar from electrostatics that one cannot place a single charge on a compact manifold. 

\subsection{The Arakelov Green function}

The Arakelov Green function $\cG(x,y)$ is a single-valued scalar Green function of $x,y \in \Sigma$, which is symmetric $\cG(y,x) = \cG(x,y)$ and is defined by the following equations, 
\bea
\label{A.1}
\pbx \p_x \cG(x,y) = - \pi \delta (x,y) + \pi \kappa(x)\, ,
\hskip 1in 
\int _\Sigma d^2x \, \kappa (x) \, \cG(x,y) =0
\eea
Here, $\kappa(x)$ is the pull-back to $\Sigma$ of the translation-invariant K\"ahler form on the Jacobian variety $J(\Sigma)= \CC^h/(\ZZ^h+ \Omega \ZZ^h)$, and is given in terms of the Abelian differentials $\om_I$ by,
\bea
\kappa (x) = {1 \over h} \, \om_I(x) \bar \om^I(x) \, , \hskip 1in \int _\Sigma d^2x \, \kappa(x) =1
\eea
As  a result, $\kappa$ is a modular invariant and conformally invariant volume form on $\Sigma$. Since the defining equations for the Arakelov Green function in (\ref{A.1}) are modular and conformally invariant, so is their unique solution $\cG(x,y)$.

\sm

An explicit construction of $\cG(x,y)$ may be given in terms of the so-called \textit{string Green function} $G(x,y) = G(y,x)$ defined by,
\bea
G(x,y) = - \log | E(x,y) |^2 + 2\pi \left ( \Im \int^x_y \bom_I \right ) \left (   \Im \int^x_y \bom^I \right )
\label{stringgf}
\eea
While the second term on the right side transforms as a scalar in $x,y$, the first term does not as it involves the logarithm of a differential form of non-zero weight. As a result, $G(x,y)$ transforms non-trivially under conformal transformations $x\to x', y \to y'$, 
\bea
G'(x',y') = G(x,y)+ u(x) + u(y)
\eea
for some function $u(x)$ which depends on $x$ and the conformal transformation $x \to x'$.  To properly define $G(x,y)$ and relate it to $\cG(x,y)$, we choose a (simply connected) fundamental domain $\Sigma_f$ for the Riemann surface~$\Sigma$, in terms of which the Arakelov Green function may be obtained by,
\bea
\label{A.GG}
\cG(x,y) = G(x,y) - \gamma (x) - \gamma (y) + \gamma_0
\eea
where $\gamma (x)$ and $\gamma _0$ are given by,
\bea
\gamma (x) = \int _{\Sigma_f} d^2 y \, \kappa (y) G(x,y) \, ,
\hskip 0.5in 
\gamma _0 = \int _{\Sigma_f} d^2 x \, \kappa (x) \gamma (x)
\eea
The Arakelov Green function, so obtained, is properly modular and conformally invariant.

\newpage

\section{Vanishing cyclic forms ${\cal V}^{(w)}_I$ at arbitrary genus}
\label{sec:6.3}

In this appendix, we propose a recursive construction of single-valued and holomorphic
$(1,0)$-forms ${\cal V}^{(w)}_I$ in $w{+}1$ points $x_1,\cdots,x_{w+1}$ on 
a higher-genus surface $\Sigma$ which
\begin{enumerate}
\itemsep=0in
\item are cyclically symmetric under $x_i \rightarrow x_{i +1}$ with $x_{w+2}=x_1$;
\item share the pole structure of the rational function (\ref{cycl0}) of $r=w{+}1$ points on the sphere;
\item generalize the vanishing elliptic $V_w(1,\cdots ,w{+}1)$ functions of (\ref{vids.01}) to arbitrary genus;
\item generalize the vanishing ${\cal V}^{(2)}_I(x_1,x_2,x_3)$ function in (\ref{calv2}) to higher multiplicity;
\item are solely expressed in terms of $f$-tensors and holomorphic Abelian differentials;
\item integrate to zero against $\prod_{j=1}^{w+1}\int_\Sigma d^2 x_j \,\bar \omega^{K_j}(j) $ and thus vanish by the earlier properties.
\end{enumerate}
By the cyclic invariance of ${\cal V}^{(w)}_{I}(1,2,\cdots,w{+}1) = {\cal V}^{(w)}_{I}(x_1,x_2,\cdots,x_{w{+}1})$,
one can fully characterize the subsequent construction of the aspired ${\cal V}^{(w)}_{I}$ functions
through those terms where the free vector index $I$ is carried by $(1,0)$-forms $f^{\overrightarrow{P}}{}_I(1,a)$ or $\omega_I(1)$ in $x_1$ as opposed to $x_{j\neq 1}$. Imposing the ${\cal V}^{(w)}_{I}$ functions to reproduce the vanishing elliptic $V_w(1,\cdots ,w{+}1)$ functions (\ref{vids.01}) upon restriction to genus one admits the two choices $a=2$ and $a=w{+}1$ for the second point of the characterizing $f^{\overrightarrow{P}}{}_I(1,a)$ factors.

\subsection{Examples at low weights}
\label{appvw.1}

 In fact, the construction of vanishing ${\cal V}^{(w)}_{I}$ functions for all values $w\leq 5$ -- and conjecturally for all higher $w\geq 6$ -- succeeds with two additional simplifying features 
\begin{itemize}
\itemsep=0in
\item there are no factors of $\omega_I(1)$ carrying the free vector index of ${\cal V}^{(w)}_{I}(1,2,\ldots,w{+}1)$, so that the free index $I$ is always carried by a factor of $f^{\overrightarrow{P}}{}_I(1,a)$ for $\overrightarrow{P} \not= \emptyset$;
\item all factors of $f^{\overrightarrow{P}}{}_I(1,a)$ in ${\cal V}^{(w)}_{I}(1,2,\ldots,w{+}1)$ have $a=2$ and not $a=w{+}1$
\end{itemize}
which are illustrated by the following examples:
\begin{align}
{\cal V}^{(1)}_{I}(1,2) &= \omega_J(2) f^J{}_I(1,2) + {\rm cycl}(1,2)
\label{van.01} \\
{\cal V}^{(2)}_{I}(1,2,3)
&= \omega_J(2) \omega_K(3) f^{KJ}{}_I(1,2) 
+ \omega_J(2) f^J{}_K(3,1) f^K{}_I(1,2)+ {\rm cycl}(1,2,3)
\notag \\
{\cal V}^{(3)}_{I}(1,2,3,4)
&= 
\omega_J(2) \omega_K(3) \omega_L(4) f^{LKJ}{}_I(1,2) 
+ \omega_J(2) f^J{}_K(3,4) \omega_L(4) f^{LK}{}_I(1,2) \notag \\
&\quad
+ \omega_L(2) \omega_J(3) f^J{}_K(4,1) f^{KL}{}_I(1,2)
+ \omega_J(2) f^J{}_K(3,4) f^K{}_L(4,1) f^L{}_I(1,2)\notag \\
&\quad
+ \omega_J(2) \omega_K(3) f^{KJ}{}_L(4,1) f^L{}_I(1,2)
 + {\rm cycl}(1,2,3,4) \notag
\notag
\end{align}
Note that the vanishing of ${\cal V}^{(1)}_{I}(1,2) $ is equivalent to the
weight-one interchange identity (\ref{ften.08a}), and that the vanishing of 
${\cal V}^{(2)}_{I}(1,2,3) $ can be verified through the tensorial Fay
identity (\ref{hf.22}) at weight two. One may suspect that the
vanishing of the higher-weight ${\cal V}^{(w)}_{I}$ in this section can be
deduced from a sequence of three-point Fay identities (\ref{exfay.15}) or (\ref{4.ff.99}),
and it would be interesting to demonstrate this at generic multiplicity. In absence of
a direct computation, we have proven the vanishing of 
${\cal V}^{(3)}_{I}$ in (\ref{van.01}) and ${\cal V}^{(4)}_{I},{\cal V}^{(5)}_{I}$ below
by checking
\begin{itemize}
\item integrating to zero: The expressions for ${\cal V}^{(w)}_{I}$ involve no more than $w$ factors of $f^{ \overrightarrow{P} }{}_J(a,b)$ without any cycles among the pairs $x_a,x_b$. Hence, for each term in ${\cal V}^{(w)}_{I}$, at least one of the points $x_c$ only enters through the $(1,0)$-form leg $x_a=x_c$ of a single $f^{ \overrightarrow{P} }{}_J(a,b)$ without any instance of $b=c$. Such a term is a total derivative of a single-valued function of $x_c$ and thus integrates to zero against $\int_\Sigma d^2x_c \, \bar \omega^K(c)$.
\item meromorphicity: The anti-holomorphic derivatives in all points vanish
for $ x_{j} \neq x_i$ as has been tested by computer algebra up to and including $w=5$.
\item absence of poles: The residues of the poles 
$(x_{j}{-}x_{j+1})^{-1}$ in ${\cal V}^{(w)}_{I}$ are given by the lower-weight function
${\cal V}^{(w-1)}_{I}(1,\cdots,j,j{+}2,\cdots,w{+}1)$. Since the latter 
satisfies the earlier vanishing conditions to the weights $w\leq 6$ we tested, 
the absence of poles in ${\cal V}^{(w)}_{I}$ is established by induction in $w$.
\end{itemize}
The restriction of our expressions for ${\cal V}^{(w\leq 5)}_{I}$ to genus one
via $f^{I_1\cdots I_r}{}_J(x,y) \rightarrow f^{(r)}(x{-}y)$ matches the expansion
(\ref{vids.01}) of the vanishing $V_w(1,2,\cdots,w{+}1)$ functions as one can
easily see from the unit coefficients on both sides.

\sm

The representation (\ref{van.01}) of the cyclic seeds can be lined
up with a recursively defined family of modular tensors, 
\begin{align}
\label{van.02}
P_I(1,2) & = \omega_I(1) 
 \\
P_I(1,2,3) &= \omega_K(1) f^K{}_I(2,3) = P_K(1,2) f^K{}_I(2,3) \notag \\
P_I(1,2,3,4) &= \omega_K(1) f^K{}_L(2,3) f^L{}_I(3,4)
+ \omega_K(1) \omega_L(2) f^{LK}{}_I(3,4)
\notag \\
&= P_L(1,2,3)  f^L{}_I(3,4) + P_K(1,2) P_L(2,3)  f^{LK}{}_I(3,4) \notag \\
P_I(1,2,3,4,5) &= P_L(1,2,3,4)  f^L{}_I(4,5) 
+ P_K(1,2) P_L(2,3) P_M(3,4)  f^{MLK}{}_I(4,5)
\notag \\
&\quad 
+ \big[ P_K(1,2,3) P_L(3,4) + P_K(1,2) P_L(2,3,4) \big] f^{LK}{}_I(4,5) 
\notag
\end{align}
in the sense that,
\begin{align}
{\cal V}^{(1)}_{I}(1,2) &= P_I(2,1,2) + {\rm cycl}(1,2)
\label{van.03} \\
{\cal V}^{(2)}_{I}(1,2,3)
&=  P_I(2,3,1,2) + {\rm cycl}(1,2,3)
\notag \\
{\cal V}^{(3)}_{I}(1,2,3,4)
&= 
P_I(2,3,4,1,2) 
 + {\rm cycl}(1,2,3,4) \notag
\end{align}
In the first line of (\ref{van.02}), we refer to an implicit adjacent leg in the notation $ \omega_I(1) = P_I(1,2)$
to preserve the weight $n{-}2$ of the other $P_I(1,2,\cdots,n)$.

\subsection{Higher-weight conjectures}
\label{appvw.2}

We shall next present a conjectural higher-multiplicity generalization
of the expressions (\ref{van.01}) or (\ref{van.03}) for ${\cal V}^{(w\leq 3)}_{I}$ by proposing
a recursive construction of higher-point tensor functions $P_I(a_1,a_2,\cdots,a_k)$ such that,
\begin{align}
{\cal V}^{(n-1)}_{I}(1,2,\cdots,n)
&= 
P_I(2,3,\cdots,n,1,2) 
 + {\rm cycl}(1,2,\cdots,n) 
 \label{van.04}
\end{align}
obeys the properties in the preamble of this appendix. Our proposal for an
all-multiplicity family of $P_I$ -- in particular the extension of the
recursion (\ref{van.02}) -- is most conveniently stated in the shorthand notation,
\begin{align}
\hat P_I(1,2,\cdots,n) 
&= P_I(1,2,\cdots,n,n{+}1) 
 \label{van.05}
\end{align}
where the last leg without $(1,0)$-form degree is kept implicit.\footnote{Given that the last point $n{+}1$
in the $f^{ \overrightarrow{P} }{}_I(n,n{+}1) $-tensors entering  $\hat P_I(1,2,\cdots,n) $ 
is excluded from the notation on the left side, the $\hat P_I$ can be thought of as being defined with respect to a cyclic reference ordering $1,2,\ldots,w$. As an exceptional property of the $n=1$ case, the notation $\hat P_I(1)=\omega_I(1)$ reflects all the variables of this tensor function.} The
recursion relations (\ref{van.02}) then take the more compact form,
\begin{align}
\label{van.06}
\hat P_I(1) &= \omega_I(1) 
\\
\hat P_I(1,2) & = \hat P_K(1) f^K{}_I(2,3) 
\no \\
\hat P_I(1,2,3) &=  \hat P_K(1,2)  f^K{}_I(3,4) +\hat  P_K(1) \hat P_L(2)  f^{LK}{}_I(3,4) \notag \\
\hat P_I(1,2,3,4) &=  \hat P_K(1,2,3)  f^K{}_I(4,5) 
+ \big[ \hat P_K(1,2) \hat P_L(3) + \hat P_K(1) \hat P_L(2,3) \big] f^{LK}{}_I(4,5) \notag \\
&\quad
+ \hat P_K(1)\hat P_L(2) \hat P_M(3)  f^{MLK}{}_I(4,5)
\notag
\end{align}
and we shall also spell out the next instance relevant for ${\cal V}^{(4)}_{I}(1,2,3,4,5)$:
\begin{align}
&\hat P_I(1,2,3,4,5) =  \hat P_K(1,2,3,4)  f^K{}_I(5,6)  \label{hatp5} \\
&\quad
+ \big[ \hat P_K(1,2,3) \hat P_L(4) 
+ \hat P_K(1,2) \hat P_L(3,4) 
+ \hat P_K(1) \hat P_L(2,3,4) \big] f^{LK}{}_I(5,6) \notag \\
&\quad
+ \big[ \hat P_K(1,2)\hat P_L(3) \hat P_M(4) 
+ \hat P_K(1)\hat P_L(2,3) \hat P_M(4) 
+  \hat P_K(1)\hat P_L(2) \hat P_M(3,4) \big] f^{MLK}{}_I(5,6) \notag \\
&\quad 
 + \hat P_K(1)\hat P_L(2) \hat P_M(3) \hat P_N(4)  f^{NMLK}{}_I(5,6)
\notag
\end{align}
We have verified the validity of (\ref{hatp5}) and a similar 16-term expression for 
$\hat P_I(1,2,\cdots,6) $ to reproduce the functions ${\cal V}^{(4)}_{I}$
and ${\cal V}^{(5)}_{I}$ with the  desired properties via (\ref{van.04}) and (\ref{van.05}).
The pattern exhibited by these low-weight formulas leads us to the following conjecture at arbitrary multiplicity. 

\sm

{\conj
The recursion relation for the seeds $\hat P_I$ of the cyclic forms ${\cal V}^{(n-1)}_{I}$
in the sense of (\ref{van.04}) and (\ref{van.05}) at arbitrary multiplicity is given by,
\begin{align}
&\hat P_I(1,2,\cdots,n) = \sum_{j=1}^{n-1} \ \sum_{A_1 A_2\cdots A_j = {\bf N}} 
  \hat P_{K_1}(A_1)
 \hat P_{K_2}(A_2)\cdots
 \hat P_{K_{j}}(A_j) f^{K_{j} \cdots K_2 K_1}{}_I(n,n{+}1)
 \label{van.07}
\end{align}
where ${\bf N}$ stands for the sequence ${\bf N}=1 \, 2 \, \cdots \, (n{-}1)$ and the sum over ${\bf N} = A_1 \cdots A_j$ stands for the sum over all deconcatenations of ${\bf N}$ into non-empty ordered sets $A_1, \cdots, A_j$ (see below for examples).   Decomposing the sum over $j$ into its individual terms gives the following  more explicit formula,
\begin{align}
\hat P_I(1,\cdots,n) = & \hat P_K(1,\cdots,n{-}1)  f^K{}_I(n,n{+}1) 
 + \sum_{AB={\bf N}} \! \hat P_{K_1}(A)  \hat P_{K_2}(B)  f^{K_2 K_1}{}_I(n,n{+}1) 
\notag \\ 
&+ \sum_{ABC = {\bf N}} \hat P_{K_1}(A)  \hat P_{K_2}(B) \hat P_{K_3}(C)  f^{K_3 K_2 K_1}{}_I(n,n{+}1) 
\no \\ & + \cdots
\notag \\
 & + 
 \hat P_{K_1}(1)
 \hat P_{K_2}(2)\cdots
 \hat P_{K_{n-1}}(n{-}1) f^{K_{n-1} \cdots K_2 K_1}{}_I(n,n{+}1)
 \label{van.07expl}
\end{align}
For example, the sum $ \sum_{AB={\bf N}}$ runs over all deconcatenations of ${\bf N} = 1\, 2 \, \cdots \, (n{-}1)$ into two disjoint and non-empty ordered sequences $A=1\, 2\, \cdots \, i$ and $B=(i{+}1)\, \cdots \, (n{-}1)$ for $i=1, \cdots , n{-}2$, while the sum  $\sum_{ABC = {\bf N}}$  runs over $A=1\, 2\, \cdots\,  i$, $ B=(i{+}1)\, \cdots\,  j$, and $C=(j{+}1)\, \cdots \, (n{-}1)$ with $1\leq i < j \leq n{-}2$.
}

\sm

\sm 

We have verified that the recursive construction (\ref{van.04}),
(\ref{van.05}) and (\ref{van.07}) yields vanishing ${\cal V}^{(w)}_{I}$-functions up to and including
$w=5$ by the analysis of holomorphicity and vanishing surface integrals as detailed in section \ref{appvw.1}.
The validity of the construction at arbitrary weights $w\geq 6$ is expected but conjectural.

\newpage

\section{Proofs of the main lemmas and theorems}
\label{appprfs}

In this appendix, we collect the proofs of Lemma \ref{5.lem:1}, Lemma \ref{5.lem:6}, and  Theorems  \ref{3.thm:7},  \ref{3.thm:8},  \ref{4.thm:1},  \ref{finthm}, \ref{mintlemma} and \ref{swid.02}. The lemmas and 
theorems were stated in the main body of the paper, but their proofs are too lengthy to be 
given there in any detail.

\subsection{Proof of Lemma \ref{5.lem:1}}
\label{appB.0}

To prove Lemma \ref{5.lem:1}, we proceed as follows. The combination $\cS^{\vI | \vP }{}_K(x,y,z)$ defined in (\ref{5.lem:1a}) is a $(1,0)$ form in both $x$ and $y$ and  a scalar in $z$. Following the
two steps in section \ref{prfsec.1}, we shall first prove that it is holomorphic in $x,y,z$ so that it must be independent of $z$, and a linear combination of $\om_A(x) \om_B(y) $ with coefficients that are independent of $x$ and $y$. Then, by showing that its integral against $\bar \om^A(x) \bar \om^B(y)$ vanishes, we establish the Lemma.

\sm

To show that $\cS ^{\vI | \vP }{}_K(x,y,z)$ is holomorphic in $x,y,z$, we begin by noticing its symmetry,
\bea
\label{B.1.sym}
\cS ^{\vI | \vP  }{}_K(x,y,z) = \cS ^{\vP | \vI  }{}_K(y,x,z) 
\eea
and evaluate its $\bar \p$ derivatives with respect to $x,y,z$, 
\begin{align}
\label{B.1.a}
\pbx \cS ^{\vI | \vP  }{}_K(x,y,z)   &=  
- \pi \bar \om^{I_1} (x) \, \cS ^{I_2 \cdots I_r | \vP } {}_K(x,y,z) 
&  r & \geq 1
\no \\
\pby \cS ^{\vI | \vP  }{}_K(x,y,z)   &=  
- \pi \bar \om^{P_1} (y) \, \cS ^{ \vI | P_2 \cdots P_s } {}_K(x,y,z) 
& s & \geq 1
\no \\
\pbz \cS ^{\vI | \vP }{}_K(x,y,z)   &=  
\, \pi  \, \bar \om^J (z) \, \cS ^{I_1 \cdots I_{r-1}  | P } {}_J(x,y,z) \, \delta ^{I_r} _K
\no \\ 
&\quad 
+\pi  \, \bar \om^J (z) \, \cS ^{\vI | P_1 \cdots P_{s-1}  } {}_J(x,y,z) \, \delta ^{P_s} _K
&  r,s & \geq 1
\end{align}
When $r=0$, the corresponding equations become,
\begin{align}
\label{B.1.b}
\pbx \cS ^{\emptyset | \vP  }{}_K(x,y,z)   &=  0
\no \\
\pby \cS ^{\emptyset | \vP  }{}_K(x,y,z)  & = 
- \pi \, \bar \om^{P_1} (y) \, \cS ^{ \emptyset  | P_2 \cdots P_s } {}_K(x,y,z) 
& s & \geq 1
\no \\
\pbz \cS ^{\emptyset | \vP }{}_K(x,y,z)  & = 
\, \pi  \, \bar \om^J (z) \, \cS ^{ \emptyset  | P_1 \cdots P_{s-1}  } {}_J(x,y,z) \, \delta ^{P_s} _K
& s & \geq 1
\end{align}
while for $s=0$ we have,
\begin{align}
\label{B.1.c}
\pbx \cS ^{\vI | \emptyset  }{}_K(x,y,z)   &=  
- \pi \, \bar \om^{I_1} (x) \, \cS ^{I_2 \cdots I_r | \emptyset } {}_K(x,y,z) 
&  r & \geq 1
\no \\
\pby \cS ^{\vI | \emptyset  }{}_K(x,y,z)  & =  0
\no \\
\pbz \cS ^{\vI | \emptyset }{}_K(x,y,z)   &=  
\, \pi  \, \bar \om^J (z) \, \cS ^{I_1 \cdots I_{r-1}  | \emptyset } {}_J(x,y,z) \, \delta ^{I_r} _K
&  r & \geq 1
\end{align}
Finally, for $r=s=0$, all $z$-dependence cancels out and the function reduces to,
\bea
\cS ^{\emptyset | \emptyset  }{}_K(x,y,z)   = \om_J(x) f^J{}_K(y,x) + \om_J(y) f^J{}_K(x,y) 
\eea
which vanishes identically by the basic interchange identity given in (\ref{ften.08a}).

\sm

We shall now proceed with a proof  by induction on $n=r{+}s$ for $r,s \geq 0$.  For given $n \geq 1$ we assume that 
$\cS ^{\vI' | \vP'  }{}_K(x,y,z)$ vanishes for all pairs $(r',s')$ such that $r'{+}s' \leq n{-}1$ and $\vI ' = I_1 \cdots I_{r'}$ and ${\vP}' = P_1 \cdots P_{s'}$. From the structure of the differential equations in (\ref{B.1.a}), (\ref{B.1.b}), and (\ref{B.1.c}), it follows that the $\bar \p$ derivatives in $x,y,z$ of all $\cS ^{\vI | \vP  }{}_K(x,y,z)$ with $r{+}s=n$ then vanish, so that $\cS ^{\vI| \vP  }{}_K(x,y,z)$ is independent of $z$ and is a holomorphic $(1,0)$ form in both $x$ and $y$. Integrating against $\bar \om^A(x)$ and $\bar \om^B(y)$, we see by inspection of the defining equation (\ref{5.lem:1a}) that the integral in $y$ of the first and third lines vanishes and that the integral in $x$ of the second  and fourth lines vanishes (recall that the $f$-tensors are total derivatives in their first argument of single-valued functions). Therefore, we have,
\bea
  \int _\Sigma d^2 x\,\bar \om^A(x) \int _\Sigma d^2y \, \bar \om^B(y) \, \cS ^{\vI | \vP  }{}_K(x,y,z) =0
  \eea
Since $\cS ^{\vI | \vP  }{}_K(x,y,z)$ is a holomorphic $(1,0)$ form in $x$ and $y$, which is independent of $z$, as was already established earlier, it follows  that $\cS ^{\vI | \vP  }{}_K(x,y,z)=0$ for all $r{+}s=n$. This completes the proof by induction on $n$ of Lemma \ref{5.lem:1}.

\subsection{Proof of Theorem \ref{3.thm:7}}
\label{appB.1}

To prove Theorem \ref{3.thm:7}, we begin by recasting the relation (\ref{exfay.15}) in terms of the following sum, in which the variables $x$ and $z$ occur only through a single factor in each term,
\bea
\cY ^{ \vI |\vP | M}{}_K (x,y,z) =
f^{\vI M}{}_J(x,y) f^{\vP J}{}_K (y,z)  
+ \sum_{k=0}^r f^{I_1 \cdots I_k} {}_J(x,y) f^{(\vP \shuffle J I_{k+1} \cdots I_r)M}{}_K(y,z)
\qquad
\label{defY}
\eea
This sum is precisely the second line in (\ref{exfay.15}).  Inspection of (\ref{exfay.15}) reveals that 
Theorem \ref{3.thm:7} may be expressed as the vanishing of the combination  $\cR  ^{\vI | \vP |M}{}_K(x,y,z) $ defined as follows,
\bea
\label{B.sigma}
\cR ^{\vI | \vP |M}{}_K(x,y,z) 
& = & 
\cY ^{ \vI |\vP | M}{}_K (x,y,z) + \cY ^{ \vP |\vI | M}{}_K (y,x,z) 
\no \\ &&
- f^{\vP M }{}_J(y,z) f^{\vI J}{}_K(x,z) + \cZ^{\vI | \vP | M}{}_K(x,y)
\eea
where $\cZ^{\vI | \vP | M}{}_K(x,y)$ is given by, 
\bea
\cZ^{\vI | \vP | M}{}_K(x,y) & = & 
 (-)^s \om_J(y) f^{\vI M \vPt J}{}_K(x,y)
\no \\ &&
+ \sum_{\ell=1}^s (-)^{s-\ell} \Big ( f^{P_1 \cdots P_\ell}{}_J(y,b_\ell) f^{\vI M P_s \cdots P_{\ell+1} J}{}_K(x,a_\ell) 
\no \\ && \hskip 0.9in
- f^{P_1 \cdots P_{\ell-1}J }{}_K(y,b_\ell) f^{\vI M P_s \cdots P_\ell }{}_J(x,a_\ell) \Big )
\label{defZ}
\eea
To prove the vanishing of $\cR ^{\vI | \vP |M}{}_K(x,y,z) $ we shall first show below  that it is holomorphic in $x,y,z$. Since $\cR ^{\vI | \vP |M}{}_K(x,y,z) $ is a scalar in $z$ it must then be independent of $z$, and since it is a $(1,0)$ form in $x$ and $y$, it must admit a decomposition into  a linear combination of $\om_A(x) \om_B(y)$ with coefficients that are independent of $x$ and $y$. Second, we shall show that its integral against $\bar \om^A (x) \bar \om^B(y)$ vanishes,
\bea
\int _\Sigma d^2x \, \bar \om^A (x) \int _\Sigma d^2y \, \bar \om^B (y) \, \cR ^{\vI | \vP |M}{}_K(x,y,z)  =0
\eea
The combination of these two results then implies the vanishing of $\cR ^{\vI | \vP |M}{}_K(x,y,z) $, thereby completing the proof of  Theorem \ref{3.thm:7}. 

\sm

To show holomorphicity of $\cR $, we begin by evaluating the $\bar \p$ derivatives of $\cY$ in (\ref{defY}),
\bea
\pbx \cY ^{ \vI|\vP | M}{}_K (x,y,z) & = & 
- \pi \, \bar \om^{I_1}(x) \, \cY^{I_2 \cdots I_r | \vP|M}{}_K(x,y,z)
+\pi \delta (x,y) f^{(\vP \shuffle \vI)M} {}_K(y,z)
\no \\
\pby \cY ^{ \vI |\vP | M}{}_K (x,y,z)  & = &
- \pi \, \bar \om ^{P_1} (y) \, \cY ^{ \vI | P_2 \cdots P_s |M}{}_K (x,y,z) 
- \pi \delta(y,x) f^{(\vP \shuffle \vI)M}{}_K(x,z)
\no \\
\pbz \cY ^{ \vI |\vP | M}{}_K (x,y,z)  & = &
\pi \,  \bar \om^L(z) \Big ( f^{\vI M}{}_K(x,y) f^{\vP} {}_L(y,z) 
+  \om_J(x) f^{\vP \shuffle J \vI}{}_L(y,z) \, \delta ^M_K \Big )
\no \\&&
+ \pi \, \delta ^M_K \, \bar \om^L(z) \sum_{\ell=1}^r f^{I_1 \cdots I_\ell}{}_J(x,y) f^{\vP \shuffle J I_{\ell+1} \cdots I_r}{}_L(y,z)
\eea
and the $\bar \p$ derivatives of $\cZ$ in (\ref{defZ}),
\bea
\pbx \cZ^{\vI|\vP|M}{}_K(x,y) & = & - \pi \bar \om^{I_1} (x) \cZ^{I_2 \cdots I_r  | \vP|M} {}_K(x,y)
\no \\
\pby \cZ^{\vI|\vP|M}{}_K(x,y) & = & - \pi \bar \om^{P_1} (y) \, \cZ^{\vI | P_2 \cdots P_s |M} {}_K(x,y) \hskip 1in s \geq 2
\no \\
\pby \cZ^{\vI| P|M}{}_K(x,y)  & = & - \pi \bar \om^{P} (y) \, \om_J(y) \, f^{\vI M J }{}_K(x,y) \hskip 1in s= 1
\eea
Combining these results with the definition of $\cR ^{\vI | \vP |M}{}_K(x,y,z) $, we obtain the following formulas for its $\bar \p$ derivatives,
\bea
\label{B.8}
\pbx \cR ^{\vI|\vP|M}{}_K(x,y,z) & = & - \pi \, \bar \om^{I_1}(x) \, \cR ^{I_2 \cdots I_r |\vP|M}{}_K(x,y,z) 
\no \\
\pby \cR ^{\vI|\vP|M}{}_K(x,y,z) & = & - \pi \, \bar \om^{P_1}(y) \, \cR^{\vI | P_2 \cdots P_s |M}{}_K(x,y,z) 
\no \\
\pbz \cR ^{\vI|\vP|M}{}_K(x,y,z) & = & \pi  \, \bar \om^L(z) \, \cS ^{\vI | \vP  }{}_L(x,y,z) \, \delta ^M_K
\eea
where the combination $\cS ^{\vI | \vP }{}_L(x,y,z)$ is the one defined in (\ref{5.lem:1a}) of Lemma \ref{5.lem:1}.

\sm

Since we have already proven  $\cS ^{\vI | \vP  }{}_L(x,y,z)=0$ in Lemma \ref{5.lem:1}, it follows from the third line in (\ref{B.8}) that $\cR ^{\vI |\vP|M}{}_K(x,y,z) $ is a holomorphic scalar on $\Sigma$, and thus independent of~$z$. Next, we proceed by induction on the sum $r{+}s$ as advocated in section \ref{prfsec.1}. We have already established that the case $r=s=0$ corresponds to the weight-two identity (\ref{hf.22}), which was proven earlier. Now let us assume that we have established the vanishing of $\cR ^{\vI |\vP |M}{}_K(x,y,z) $ for all $r,s \geq 0$ such that $r{+}s\leq n$. The first two equations in (\ref{B.8}) then imply that the combinations $\cR ^{\vI |\vP|M}{}_K(x,y,z) $ are holomorphic $(1,0)$ forms  in $x$ and $y$ for  all $r,s$ such that $r{+}s=n{+}1$. To further establish that the combination  vanishes, we show that its integral against $\bar \om^A(x) \bar \om^B(y)$ vanishes. To see this, note that the integral of $\cZ^{\vI | \vP | M}{}_K(x,y)$ against $\bar \om^A(x)$ vanishes term by term. Similarly, the first term on the second line in (\ref{B.sigma}) integrates to zero against $\bar \om^A(x)$. Finally, $\cY^{\vI | \vP | M}{}_K(x,y,z)$ integrates to zero against $\bar \om^B(y)$, so that all terms in (\ref{B.sigma}) individually integrate to zero against the combined $\bar \om^A(x) \bar \om^B(y)$. Therefore $\cZ^{\vI | \vP | M}{}_K(x,y)=0$ for all $r{+}s=n{+}1$, thus completing our proof by induction on $r{+}s$.

\subsection{Proof of Lemma \ref{5.lem:6}}
\label{appB.2}

The starting point for the proof of Lemma \ref{5.lem:6} is again Lemma \ref{5.lem:1} which states the vanishing of $\cS ^{\vI | \vP}{}_K(x,y,z)$ in (\ref{5.lem:1a}). This result  implies that the contracted product $ f^{\vI} {}_J(x,z)   f^{\vP J}{}_K(y,x) $, which is a $(1,0)$ form in the repeated point $x$, may be expressed as follows, 
\bea
\label{3.obv}
 f^{\vI} {}_J(x,z)   f^{\vP J}{}_K(y,x) & = & 
 f^{\vI} {}_J(x,z)  f^{\vP J}{}_K(y,z) 
- f^{\vP}{}_J(y,z) \Big ( f^{\vI J}{}_K(x,y) - f^{\vI J}{}_K(x,z) \Big ) 
\no \\ &&
 - \sum_{k=0}^r f^{I_1 \cdots I_k } {}_J(x,y) f^{\vP \shuffle JI_{k+1} \cdots I_r }{}_K(y,z) 
\no \\ &&
- \sum_{\ell=0}^s f^{P_1 \cdots P_\ell}{}_J(y,x) f^{\vI \shuffle J P_{\ell+1} \cdots P_s} {}_K(x,z)
\eea
Recall that the essence of Lemma \ref{5.lem:6} is to \textit{$x$-reduce} the left side. Clearly the sum over $\ell$ 
on the last line is not \textit{$x$-reduced}, and the contracted index $J$ enters the factors of
$f(y,x)$ and $f(x,z)$ in different positions as compared to the left side. But the position of $J$ may be rearranged
to be of the same form as on the left side by using the following identity, 
\bea 
f^{P_1 \cdots P_\ell}{}_J(y,x)  f^{\vI \shuffle J P_{\ell+1} \cdots P_s} {}_K(x,z) 
& = & 
  f^{\vI \shuffle  P_{\ell} \cdots P_s} {}_J(x,z)  f^{P_1 \cdots P_{\ell-1} J}{}_K (y,x) 
\label{c3id} \\ && 
  - f^{\vI \shuffle  P_{\ell} \cdots P_s} {}_J(x,z)  \p_y \Phi ^{P_1 \cdots P_{\ell-1} J}{}_K (y) 
  \no \\ &&
+ f^{\vI \shuffle J P_{\ell+1} \cdots P_s} {}_K(x,z) \p_y \Phi ^{P_1 \cdots P_\ell}{}_J(y) 
\no
\qquad
\eea
Note that all the extra terms that are produced by this rearrangement on the last two lines of (\ref{c3id}) are properly \textit{$x$-reduced}. Using the above result, we may  now write a new equivalent version of (\ref{3.obv}) in which all terms with products of two $x$ dependent $f$-tensors have the same tensorial structure, 
\bea
\label{3.obv4}
 f^{\vI} {}_J(x,z)   f^{\vP J}{}_K(y,x) & = & 
\Lambda ^{\vI | \vP  }{}_K(x,y,z) 
- \sum_{\ell=1}^s   f^{\vI \shuffle  P_{\ell} \cdots P_s} {}_J(x,z)  f^{P_1 \cdots P_{\ell-1} J}{}_K (y,x)
\qquad
\eea
where $\Lambda ^{\vI | \vP  }{}_K(x,y,z) $ is \textit{$x$-reduced} by construction and defined as follows, 
\begin{align}
\label{3.obv7}
\Lambda ^{\vI | \vP }{}_K  (x,y,z) &= 
 f^{\vI} {}_J(x,z)  f^{\vP J}{}_K(y,z) 
- f^{\vP}{}_J(y,z) \Big ( f^{\vI J}{}_K(x,y) - f^{\vI J}{}_K(x,z) \Big ) 
\no \\ &\quad
- \om_J(y) f^{\vI \shuffle J \vP}{}_K(x,z)  - \sum_{k=0}^r f^{I_1 \cdots I_k } {}_J(x,y) f^{\vP \shuffle JI_{k+1} \cdots I_r }{}_K(y,z) 
\no \\ &\quad
+ \sum_{\ell=1}^s \Big (  f^{\vI \shuffle  P_{\ell} \cdots P_s} {}_J(x,z)  \p_y \Phi ^{P_1 \cdots P_{\ell-1} J}{}_K (y) 
\no \\ &\quad \hskip 0.6in 
-  f^{\vI \shuffle J P_{\ell+1} \cdots P_s} {}_K(x,z) \p_y \Phi ^{P_1 \cdots P_\ell}{}_J(y) \Big )
\end{align}
It is straightforward to rearrange the last line in (\ref{3.obv7}) into the form presented in (\ref{3.obv9}) of Lemma \ref{5.lem:6} by using the following relations, 
\bea
\p_y \Phi ^{P_1 \cdots P_{\ell-1} J}{}_K (y) & = & 
f ^{P_1 \cdots P_{\ell-1} J}{}_K (y) +  \p_y \cG^{P_1 \cdots P_{\ell-1}} (y,a_\ell) \, \delta ^J_K
\no \\ 
\p_y \Phi ^{P_1 \cdots P_\ell}{}_J(y) & = &
f ^{P_1 \cdots P_\ell}{}_J(y) + \p_y \cG^{P_1 \cdots P_{\ell-1}} (y,a_\ell) \, \delta ^{P_\ell}_J
\eea
and observing that the terms involving $\p_y \cG^{P_1 \cdots P_{\ell-1}} (y,a_\ell)$ 
cancel for arbitrary values of $a_\ell$.

\subsubsection{Inverting equation (\ref{3.obv4})}

To prove the relation (\ref{recxx.1}) of Lemma \ref{5.lem:6}, it remains to \textit{invert} the relation (\ref{3.obv4}) and express $ f^{\vI} {}_J(x,z)   f^{\vP J}{}_K(y,x) $ solely in terms of $\Lambda ^{\vI' | \vP' }{}_K  (x,y,z)$
for various combinations $\vI'$ and $\vP'$ of $\vI$ and $\vP$. To do so, we recast (\ref{3.obv4}) by moving the sum over $\ell$ from the right side to the left side of the equation, and then including the term on the left as the $\ell=0$ contribution to the sum, so as to obtain,
\bea
\Lambda ^{\vI | \vP  }{}_K(x,y,z) = 
\sum_{k=0}^s   f^{\vI \shuffle  P_{k+1} \cdots P_s} {}_J(x,z)  f^{P_1 \cdots P_k J}{}_K (y,x)
\eea
using the convention $\vI \shuffle  P_{k+1} \cdots P_s = \vI$ for the case $k=s$. We now use this formula to evaluate the right side of (\ref{recxx.1}) as follows,
\begin{align}
\sum_{\ell =0}^s  (-)^{s-\ell} \,& \Lambda ^{\vI \shuffle P_s \cdots P_{\ell+1} | P_1 \cdots P_\ell }{}_K(x,y,z)
\label{3.th.6} \\ 
&\quad \hspace{-1.5cm} = 
\sum_{\ell =0}^s (-)^{s-\ell} \sum_{k=0}^\ell   f^{(\vI \shuffle P_s \cdots P_{\ell+1}) \shuffle  P_{k+1} \cdots P_\ell} {}_J(x,z)  f^{P_1 \cdots P_k J}{}_K (y,x)
\no \\ 
&\quad  \hspace{-1.5cm}=
\sum_{k=0}^s  f^{P_1 \cdots P_k J}{}_K (y,x) \sum_{\ell =k}^s (-)^{s-\ell}   f^{(\vI \shuffle P_s \cdots P_{\ell+1}) \shuffle  P_{k+1} \cdots P_\ell} {}_J(x,z)  
\no \\ 
&\quad\hspace{-1.5cm}  =
f^{\vP J}{}_K(y,x) f^{\vI}{}_J(x,z) + 
\sum_{k=0}^{s-1}  \cB_k^{\vI|P_{k+1} \cdots P_s}{}_J (x,z)  f^{P_1 \cdots P_k J}{}_K (y,x) 
\notag
\end{align}
where we denote the coefficients as follows for $0 \leq k \leq s-1$, 
\bea
\label{3.th.7}
\cB_k^{\vI|P_{k+1} \cdots P_s}{}_J (x,z) = \sum_{\ell =k}^s (-)^{s-\ell}   f^{(\vI \shuffle P_s \cdots P_{\ell+1}) \shuffle  P_{k+1} \cdots P_\ell} {}_J(x,z)  
\eea
The first term on the right side of the last line of (\ref{3.th.6}) is precisely the left side of (\ref{recxx.1}). Thus, to prove Lemma \ref{5.lem:6}, it will suffice to prove that the sum over $k$ on the last line of (\ref{3.th.6}) vanishes. Since all dependence on $y$ is concentrated in the coefficients $f^{P_1 \cdots P_k J}{}_K (y,x)$, and these functions are linearly independent of one another for different values of $k$, the coefficient functions $\cB_k^{\vI|P_{k+1} \cdots P_s}{}_J (x,z) $ should vanish for each value of $k$ in the range $0 \leq k \leq s{-}1$. This is indeed the case as we shall now prove.

\subsubsection{Vanishing of $\cB_k$ in (\ref{3.th.7})}

We begin by using the associativity of the shuffle product to rewrite (\ref{3.th.7}) as follows, 
\bea
\label{3.th.7a}
\cB_k^{\vI|P_{k+1} \cdots P_s}{}_J (x,z) = 
\sum_{\ell =k}^s (-)^{s-\ell}   f^{\vI \shuffle (P_s \cdots P_{\ell+1} \shuffle  P_{k+1} \cdots P_\ell) } {}_J(x,z)  
\eea
Henceforth we shall drop the argument $(x,z)$ which is common to all functions below. For  $k = s{-}1$ we clearly have  $
\cB_{s-1}^{\vI|P_s}{}_J  =0$.  Henceforth we set $0 \leq k \leq s{-}2$.  Next, we identify all the contributions for which the last index in a given shuffle $ (P_s \cdots P_{\ell+1} \shuffle  P_{k+1} \cdots P_\ell) $ is $P_m$ for $m$ in the range $k{+}1 \leq m \leq s$.  For each value of $m$ only two ``terms" in the sum will contribute (of course, each ``term" is really a sum of shuffles). To this end we use the following decomposition formula, 
\begin{align}
f^{\vI \shuffle (P_s \cdots P_{\ell+1} \shuffle  P_{k+1} \cdots P_\ell)} {}_J
& = 
f^{\vI \shuffle   ( ( P_s \cdots P_{\ell+1} \shuffle P_{k+1} \cdots P_{\ell-1} ) P_\ell  ) } {}_J
\notag \\ &\quad
+ f^{\vI \shuffle ( ( P_s \cdots P_{\ell+2} \shuffle P_{k+1} \cdots P_\ell ) P_{\ell+1} ) } {}_J
\end{align}
Substituting this decomposition into (\ref{3.th.7a}), and changing summation variables in the second sum $\ell{+}1 \to \ell$ gives,
\begin{align}
\cB_k^{\vI|P_{k+1} \cdots P_s}{}_J  & =  
 (-)^{s-k} f^{\vI \shuffle P_s \cdots P_{k+1} } {}_J 
+ \sum_{\ell =k+1}^{s-1}  (-)^{s-\ell}   
f^{\vI \shuffle  \big \{ ( P_s \cdots P_{\ell+1} \shuffle P_{k+1} \cdots P_{\ell-1} ) P_\ell  \big \} } {}_J 
\notag \\ &\quad
+ f^{\vI \shuffle  P_{k+1} \cdots P_s} {}_J 
- \sum_{\ell =k+2}^{s}  (-)^{s-\ell}   
f^{\vI \shuffle \big \{ ( P_s \cdots P_{\ell+1} \shuffle P_{k+1} \cdots P_{\ell-1} ) P_{\ell} \big \} } {}_J
\label{3.th.12}
\end{align}
Except for the contribution $\ell=k{+}1$ in the first sum and $\ell=s$ in the second sum, all other terms in the two sums cancel one another. The remaining contributions are readily seen to cancel the first terms in the two lines on the right side of (\ref{3.th.12}). This completes the proof of Lemma \ref{5.lem:6}.

\subsection{Proof of Theorem \ref{3.thm:8}}
\label{appB.2a}

To prove equation (\ref{4.ff.99}) of  Theorem \ref{3.thm:8}, we start from the expression (\ref{recxx.1}) of Lemma~\ref{5.lem:6}, which we repeat here for convenience,
\bea
\label{3.th.5}
 f^{\vI} {}_J(x,z)   f^{\vP J}{}_K(y,x) & = & 
 \sum_{\ell =0}^s (-)^{s-\ell} \, \Lambda ^{\vI \shuffle P_s \cdots P_{\ell+1} | P_1 \cdots P_\ell }{}_K(x,y,z)
\eea
Our first step is a reorganization of the expression (\ref{3.obv9}) for $\Lambda ^{\vI | \vP }{}_K(x,y,z)$ where the first terms $ f^{\vI } {}_J(x,z)  f^{\vP J}{}_K(y,z) $  and $- \om_J(y) f^{\vI \shuffle J \vP }{}_K(x,z) $ in the first and second line are absorbed into  extensions of the sums on the last line of (\ref{3.obv9}) to
$\ell=s{+}1$ and $\ell=0$, respectively. Upon setting the arbitrary points in
(\ref{3.obv9}) to $a_\ell = z$, we arrive at the decomposition,
\begin{align}
\label{3.th.4}
\Lambda ^{\vI | \vP  }{}_K(x,y,z)  &=  \sum_{j=1}^5 \Lambda_{j} ^{\vI | \vP  }{}_K(x,y,z)
\end{align}
in terms of the following shorthands,
\begin{align}
 \Lambda_{1} ^{\vI | \vP  }{}_K(x,y,z)&= - f^{\vP}{}_J(y,z)  f^{\vI J}{}_K(x,y) 
\no \\ 
 \Lambda_{2}^{\vI | \vP  }{}_K(x,y,z)
&=
- \sum_{k=0}^r f^{I_1 \cdots I_k } {}_J(x,y) f^{\vP \shuffle JI_{k+1} \cdots I_r }{}_K(y,z) 
\no \\  
\Lambda_{3} ^{\vI | \vP  }{}_K(x,y,z)&=
 \sum_{m=0}^s  f^{\vI \shuffle  P_{m+1} \cdots P_s} {}_J(x,z)  f ^{P_1 \cdots P_m J}{}_K (y,z) 
\no \\ 
 \Lambda_{4}^{\vI | \vP  }{}_K(x,y,z)
&=  f^{\vP}{}_J(y,z) \, f^{\vI J}{}_K(x,z)
\no \\  
\Lambda_{5} ^{\vI | \vP}{}_K(x,y,z) &=
- \sum_{m=0}^s  f ^{P_1 \cdots P_m}{}_J(y,z)  \, f^{\vI \shuffle J P_{m+1} \cdots P_s} {}_K(x,z) 
\label{3atoe}
\end{align}
The main task of this proof is to obtain the complete right side of (\ref{4.ff.99}) from the sums 
$\sum _{\ell=0} ^ s (-)^{s-\ell} \Lambda_j^{\vI \shuffle P_s \cdots P_{\ell+1} | P_1 \cdots P_\ell}{}_K (x,y,z) $ in (\ref{recxx.1}) over the individual contributions in (\ref{3atoe}) with $j=1,2,\cdots,5$. However, before doing so in section \ref{B.2a.2}, we shall first establish several combinatorial identities to rearrange the iterated shuffle products from the sums over $\ell$.

\subsubsection{Combinatorial lemmas}
\label{B.2a.1}

The following Lemma \ref{preitsh.01} on alternating sums of shuffle products and its Corollaries \ref{itsh.01}, \ref{itsh.09} and \ref{itsh.13} will be instrumental in simplifying the sums in (\ref{3.th.4}) and  (\ref{3atoe}).
{\lem
\label{preitsh.01}
For multi-indices $\overrightarrow{P} = P_1 \cdots P_s$
of length $s$, the alternating sum,
\beq
\Upsilon(\overrightarrow{P})
= \sum_{\ell =0}^s (-1)^{s-\ell} P_1  \cdots P_\ell \shuffle P_s  \ldots P_{\ell+1} 
\label{itsh.02}
\eeq
vanishes for any non-empty $\overrightarrow{P}$ and otherwise yields the neutral element
$\emptyset$ of shuffle multiplication,
\beq
\Upsilon(\overrightarrow{P}) = \left\{
\begin{array}{cl}
0 &\hbox{ for } \ \overrightarrow{P} \neq \emptyset \\
\emptyset &\hbox{ for } \ \overrightarrow{P} = \emptyset
 \end{array}
\right.
\label{itsh.03}
\eeq}
The proof of Lemma \ref{preitsh.01} proceeds by first evaluating the cases with $s=0$ and $s=1$,
\begin{align}
\Upsilon(\emptyset) &=  \emptyset \shuffle  \emptyset  =  \emptyset 
\label{itsh.07} \\
\Upsilon(P_1) &=  {-} \emptyset \shuffle P_1 + P_1 \shuffle \emptyset
  = 0
\notag
\end{align}
The vanishing of $\Upsilon(P_1\cdots P_s)$ for arbitrary $s\geq 1$ is then proven by  induction on $s$, starting with $\Upsilon(P_1)$ for $s=1$ in (\ref{itsh.07}). Assuming that  
$\Upsilon(\overrightarrow{R})=0$ for all multi-indices $\overrightarrow{R}$ of length $s{-}1$, 
the recursive definition (\ref{recshf}) of the shuffle product simplifies the length-$s$ case to,
\begin{align}
\Upsilon(\overrightarrow{P})
&= P_1  \bigg ( \sum_{\ell =1}^s (-1)^{s-\ell}  P_2 \cdots P_\ell \shuffle P_s \ldots P_{\ell+1}  
\bigg )\notag \\
&\quad + P_s  \bigg ( \sum_{\ell =0}^{s-1} (-1)^{s-\ell} P_1 \cdots P_\ell \shuffle  P_{s-1} \ldots P_{\ell+1}   \bigg )
 \notag \\
&= P_1\Upsilon(P_2\cdots P_s ) -  P_s\Upsilon(P_1\cdots P_{s-1}) = 0
\label{itsh.08}
\end{align}
since both of $\Upsilon(P_2\cdots P_s)$ and $\Upsilon(P_1\cdots P_{s-1})$ vanish by the inductive hypothesis. Together with the case of $\overrightarrow{P}=\emptyset$ in (\ref{itsh.07}),
this concludes the proof of Lemma~\ref{preitsh.01}.

{\cor
\label{itsh.01}
Upon shuffle multiplication with an arbitrary multi-index $\overrightarrow{Q} = Q_1\cdots Q_t$ of length $t$,
the combination
\beq
\Upsilon(\overrightarrow{P}, \overrightarrow{Q}) = \Upsilon(\overrightarrow{P}) \shuffle  \overrightarrow{Q}
= \sum_{\ell =0}^s (-1)^{s-\ell} P_1  \cdots P_\ell \shuffle P_s  \ldots P_{\ell+1} \shuffle \overrightarrow{Q}
\label{shitsh.02}
\eeq
simplifies to
\beq
\Upsilon(\overrightarrow{P}, \overrightarrow{Q}) = \left\{
\begin{array}{cl}
0 &\hbox{ for } \ \overrightarrow{P} \neq \emptyset \\
\overrightarrow{Q} &\hbox{ for } \ \overrightarrow{P} = \emptyset
 \end{array}
\right.
\label{shitsh.03}
\eeq}
Corollary \ref{itsh.01} is a simple consequence of (\ref{itsh.03}) and $\emptyset$ being the neutral element of shuffle multiplication.

{\cor
\label{itsh.09}
The combination of shuffles,
\beq
\Xi( P_1 P_2 \cdots P_m, J,\overrightarrow{Q}) =
\sum_{\ell=0}^m (-)^{m-\ell} P_1  \cdots P_\ell \, \shuffle \, J \left ( P_m  \cdots P_{\ell+1} \, \shuffle \, \overrightarrow{Q} \right )
\label{itsh.10}
\eeq
admits the following simplified representation,
\beq
\Xi( P_1 P_2 \cdots P_m, J,\overrightarrow{Q}) = P_1 P_2 \cdots P_m J  \overrightarrow{Q}
\label{itsh.11}
\eeq}
The proof is again most conveniently performed via induction in the length $m$ of the first entry. In the base
case at $m=0$, we evidently have,
\bea
\Xi( \emptyset, J,\overrightarrow{Q}) 
= \emptyset \shuffle J( \emptyset \shuffle \overrightarrow{Q})= J  \overrightarrow{Q}
\eea 
For $m\geq 1$, we apply the  recursion (\ref{recshf}) for the shuffle product to (\ref{itsh.10}),
\begin{align}
\Xi( P_1  \cdots P_m, J,\overrightarrow{Q}) &=
P_1 \bigg (  \sum_{\ell=1}^m (-)^{m-\ell} P_2 \cdots P_\ell \, \shuffle \, J \left ( P_m  \cdots P_{\ell+1} \, \shuffle \, \overrightarrow{Q}  \right ) \bigg )
\notag \\
&\quad + J  \biggl ( \sum_{\ell=0}^m (-)^{m-\ell} P_1 \cdots P_\ell \, \shuffle \,  \left ( P_m \cdots P_{\ell+1} \, \shuffle \, \overrightarrow{Q}  \right ) \bigg )
\notag \\
&= P_1 \Xi( P_2 \cdots P_m, J,\overrightarrow{Q}) + J \Upsilon( P_1 \cdots P_m, \overrightarrow{Q})
\label{itsh.12}
\end{align}
The second term $\Upsilon( P_1 \cdots P_m, \overrightarrow{Q})$ of the third line vanishes as a consequence of Corollary \ref{itsh.01} (the first entry of $\Upsilon$ is non-empty for $m\geq 1$). The first term of the third line may be simplified on the basis of the inductive hypothesis
$\Xi( P_2 \cdots P_m, J,\overrightarrow{Q}) = P_2 \cdots P_m J  \overrightarrow{Q}$
for words of length $m{-}1$ in the first entry, which gives (\ref{itsh.11}) and proves Corollary \ref{itsh.09}. 

{\cor
\label{itsh.13}
The combination of shuffles,
\beq
\Theta(P_1 \cdots P_s, J,  \overrightarrow{I}) =
\sum_{\ell=0}^s (-)^{s-\ell} P_s  \cdots P_{\ell+1} \, \shuffle \, J P_{1}  \cdots P_\ell \, \shuffle \,\overrightarrow{I}
\label{itsh.14}
\eeq
admits the following simplified representation,
\beq
\Theta(P_1 \cdots P_s, J,  \overrightarrow{I}) 
= (-)^{s} P_s  \cdots P_{1} J \, \shuffle \,\overrightarrow{I}
\label{itsh.15}
\eeq
}
This time, the proof relies on induction in the combined length $r{+}s$
of the multi-indices $P_1 \cdots P_s$ and $\overrightarrow{I} = I_1  \cdots I_r$:
After checking the base case $\Theta(\emptyset, J, \overrightarrow{I})
= \emptyset  \shuffle J \shuffle \overrightarrow{I}
=J \shuffle \overrightarrow{I}$ 
at $s=0$, we apply (\ref{recshf}) in the inductive step at $s\geq 1$ and arbitrary $r$:
\begin{align}
\Theta(P_1 \cdots P_s, J,  \overrightarrow{I}) &=
J  \bigg ( \sum_{\ell=0}^s (-)^{s-\ell} P_s  \cdots P_{\ell+1} \, \shuffle \,  P_{1} \cdots P_\ell  \, \shuffle \,\overrightarrow{I} \bigg )  \\
&\quad +P_s \bigg (
\sum_{\ell=0}^{s-1} (-)^{s-\ell}  P_{s-1} \cdots P_{\ell+1} \, \shuffle \, J P_{1} \cdots P_\ell \, \shuffle \,\overrightarrow{I} 
\bigg ) 
\notag \\
&\quad +I_1 \bigg (
\sum_{\ell=0}^s (-)^{s-\ell} P_s  \cdots P_{\ell+1} \, \shuffle \, J P_{1} \cdots P_\ell \, \shuffle \,I_2 \cdots I_r \bigg )
\notag \\
&= J \Upsilon(P_1\cdots P_s, \overrightarrow{I}) - P_s 
\Theta(P_1 \cdots P_{s-1}, J, \overrightarrow{I}) 
+ I_1 \Theta(P_1 \cdots P_{s}, J, I_2\cdots I_r) 
\notag \\
&= ( -)^s P_s \big(P_{s-1}\cdots P_1 J \, \shuffle \,\overrightarrow{I} \big)
+  ( -)^s   I_1 \big(P_{s}\cdots P_1 J \, \shuffle \, I_2\cdots I_r\big)
\label{itsh.16}
\no
\end{align}
In passing to the last line, we have used Corollary \ref{itsh.01} to
set $ \Upsilon(P_1\cdots P_s,  \overrightarrow{I}) =0$ (using $s\geq 1$)
and the inductive hypothesis $ \Theta(P_1\cdots P_{s-1}, J,  \overrightarrow{I}) 
= (-1)^{s-1} P_{s-1}\cdots P_1 J  \, \shuffle \,\overrightarrow{I} $
as well as $ \Theta(P_1 \cdots P_{s}, J, I_2\cdots I_r) = (-1)^s
P_{s}\cdots P_1 J \, \shuffle \, I_2\cdots I_r$, both of which have reduced overall length $r{+}s{-}1$
of the first and last entry. The result of (\ref{itsh.15}) is obtained based on the recursion (\ref{recshf}) in reverse order.

\subsubsection{Application to the proof of Theorem \ref{3.thm:8}}
\label{B.2a.2}

Equipped with combinatorial identities of section \ref{B.2a.1}, we can
now proceed to performing the sums $\sum _{\ell=0} ^ s (-)^{s-\ell} \Lambda_j^{\vI \shuffle P_s \cdots P_{\ell+1} | P_1 \cdots P_\ell}{}_K (x,y,z) $ as in (\ref{recxx.1}) 
over the $\Lambda_{j=1,2,\cdots,5}$ in (\ref{3atoe}).

\sm

\noindent
$\boldsymbol{j=1\!:}$ ~ The summation over the terms $ \Lambda_{1} ^{\vI | \vP  }{}_K(x,y,z)$ in the first line of (\ref{3atoe}) straightforwardly produces the sum on the third line of (\ref{4.ff.99}) corresponding to the first term inside the parentheses of the summand.

\sm

\noindent
$\boldsymbol{j=2\!:}$ ~  We begin with the summation of the terms 
$ \Lambda_{2} ^{\vI | \vP   }{}_K(x,y,z)$ produced by the second line of (\ref{3atoe}). To this end we write out this contribution more explicitly as follows,
\beq
\Lambda_2^{\vI \shuffle P_s \cdots P_{\ell+1} | P_1 \cdots P_\ell}{}_K (x,y,z)
=
- \sum_{k=0}^r \sum_{m=\ell} ^s f^{P_s \cdots P_{m+1}  \shuffle I_1 \cdots I_k}{}_J(x,y) f^{P_1 \cdots P_\ell \shuffle J  ( P_m \cdots P_{\ell+1} \shuffle \hat I_k)}{}_K(y,z)
\eeq
where we shall use the abbreviation $\hat I_k = I_{k+1} \cdots I_r$ throughout this appendix. 
Its contribution to the right side of (\ref{4.ff.99}) is given by the sum over $\ell$, 
\bea
&&
\sum _{\ell=0} ^ s (-)^{s-\ell} \Lambda_2^{\vI \shuffle P_s \cdots P_{\ell+1} | P_1 \cdots P_\ell}{}_K (x,y,z) 
 \label{moretract}\\ && \hskip 0.3in =
- \sum_{k=0}^r \sum_{m=0} ^s (-)^{s-m} f^{P_s \cdots P_{m+1}  \shuffle I_1 \cdots I_k}{}_J(x,y) 
\sum_{\ell=0}^m (-)^{m-\ell}  f^{P_1 \cdots P_\ell \shuffle J  ( P_m \cdots P_{\ell+1} \shuffle \hat I_k)}{}_K(y,z)
\no
\eea
where we have swapped the summations over $\ell$ and $m$. The sum over $\ell$ 
realizes the combination $\Xi$ in (\ref{itsh.10}) at $  \overrightarrow{Q} = \hat I_k$ 
which we shall simplify via Corollary \ref{itsh.09},
\begin{align}
\sum_{\ell=0}^m (-)^{m-\ell}  f^{P_1 \cdots P_\ell \shuffle J  ( P_m \cdots P_{\ell+1} \shuffle \hat I_k)}{}_K(y,z)
&=  f^{\Xi(P_1 \cdots P_m,J, \hat I_k)}{}_K(y,z) \\
& =   f^{P_1 \cdots P_m J \hat I_k}{}_K(y,z) =  f^{P_1 \cdots P_m J I_{k+1}\ldots I_r}{}_K(y,z)
\notag
\end{align}
As a consequence, (\ref{moretract}) takes the more tractable form,
\begin{align}
&\sum _{\ell=0} ^ s (-)^{s-\ell} \Lambda_2^{\vI \shuffle P_s \cdots P_{\ell+1} | P_1 \cdots P_\ell}{}_K (x,y,z) 
 \\ 
 &\quad =
- \sum_{k=0}^r \sum_{m=0} ^s (-)^{s-m} f^{P_s \cdots P_{m+1}  \shuffle I_1 \cdots I_k}{}_J(x,y) 
f^{P_1 \cdots P_m  J  I_{k+1} \cdots I_r }{}_K(y,z)
\notag
\end{align}
Changing summation variables from $m$ to $\ell$ in the above formula precisely produces the double sum on the second line of (\ref{4.ff.99}). 

\sm

\noindent
$\boldsymbol{j=3\!:}$ ~ The summation of the terms
$ \Lambda_{3} ^{\vI | \vP  }{}_K(x,y,z)$ in the third line of (\ref{3atoe})
gives the first term in (\ref{4.ff.99}). To show this, we collect the sum over $\ell$  as follows,
\begin{align}
&\sum_{\ell =0}^s (-)^{s-\ell} \, \Lambda_3^{\vI \shuffle P_s \cdots P_{\ell+1} | P_1 \cdots P_\ell }{}_K(x,y,z)  \label{3.cor.9} \\
&\quad =
\sum_{m=0} ^s f^{P_1 \cdots P_m J}{}_K(y,z)  
\sum_{\ell=m}^s (-)^{s-\ell}  f^{ (\vI \, \shuffle \, P_s \cdots P_{\ell+1} ) \, \shuffle \, P_{m+1} \cdots P_\ell}{}_J(x,z)  \notag
\end{align}
Using associativity of the shuffle product, we may drop the parentheses in the superscript of the second factor. For each value of $m$, the sum over $\ell$ can be
lined up with the alternating combination $\Upsilon$ defined in (\ref{shitsh.02})
with $P_{m+1}\cdots P_s$ in the place of $P_{1}\cdots P_s$,
\begin{align}
\sum_{\ell=0}^s (-)^{s-\ell}  f^{ (\vI \, \shuffle \, P_s \cdots P_{\ell+1} ) \, \shuffle \, P_{m+1} \cdots P_\ell}{}_J(x,z) 
&=  
f^{  \Upsilon(P_{m+1} P_{m+2}\cdots P_s, \vI)  }{}_J(x,z)  \notag \\
&=  \delta_{m,s} f^{ \vI  }{}_J(x,z)
\label{itsh.41}
\end{align}
where we have used the vanishing of $\Upsilon$ with a non-empty first entry
established in Corollary~\ref{itsh.01}. Hence, the first sum in (\ref{3.cor.9}) collapses
 to the term $m=s$, resulting in 
\beq
\sum_{\ell =0}^s (-)^{s-\ell} \, \Lambda_3^{\vI \shuffle P_s \cdots P_{\ell+1} | P_1 \cdots P_\ell }{}_K(x,y,z)  
=   f^{\overrightarrow{P} J}{}_K(y,z) f^{\overrightarrow{I}} {}_J(x,z) 
\eeq
which gives precisely the first term in (\ref{4.ff.99}).

\sm

\noindent
$\boldsymbol{j=4,5\!:}$ ~
 The summation over $\ell$ of the last two terms
$ \Lambda_{4} ^{\vI | \vP  }{}_K(x,y,z)$ 
and $ \Lambda_{5} ^{\vI | \vP }{}_K(x,y,z)$ in (\ref{3atoe}) 
 produces the sum on the third line of (\ref{4.ff.99}) involving the second term inside the parentheses of the summand. To show this, we begin with the summation of the last term, 
 \begin{align}
 &\sum_{\ell =0}^s (-)^{s-\ell} \, \Lambda_5^{\vI \shuffle P_s \cdots P_{\ell+1} | P_1 \cdots P_\ell }{}_K(x,y,z)
\label{3.cor.8}
\\
&\quad =
- \sum_{\ell=0}^s (-)^{s-\ell} \sum_{m=0}^\ell f^{P_1 \cdots P_m}{}_J(y,z) f^{(\vI \, \shuffle \, P_s \cdots P_{\ell+1} ) \, 
\shuffle \, J P_{m+1} \cdots P_\ell}{}_K(x,z)
\no \\ 
&\quad =
- \sum_{m=0}^s  f^{P_1 \cdots P_m}{}_J(y,z) \sum_{\ell=m}^s (-)^{s-\ell}  f^{(\vI \, \shuffle \, P_s \cdots P_{\ell+1} ) \, 
\shuffle \, J P_{m+1} \cdots P_\ell}{}_K(x,z)
\notag
\end{align}
The sum over $\ell$ 
realizes the combination $\Theta$ in (\ref{itsh.14}) with $P_{m+1} P_{m+2}\cdots P_s$
in the place of $P_1P_2\cdots P_s$ which we shall simplify via Corollary \ref{itsh.13}:
\begin{align}
\sum_{\ell=m}^s (-)^{s-\ell}  f^{(\vI \, \shuffle \, P_s \cdots P_{\ell+1} ) \, 
\shuffle \, J P_{m+1} \cdots P_\ell}{}_K(x,z) &= f^{\Theta(P_{m+1}\cdots P_s,J,\vI)
}{}_K(x,z)  \\
&=  (-1)^{s-m} f^{ P_s P_{s-1}\cdots P_{m+1}  \shuffle \vI
}{}_K(x,z)
\notag
\end{align}
Thus, (\ref{3.cor.8}) becomes, 
\beq
\label{3.cor.7}
\sum_{\ell =0}^s (-)^{s-\ell} \, \Lambda_5^{\vI \shuffle P_s \cdots P_{\ell+1} | P_1 \cdots P_\ell }{}_K(x,y,z)
=
- \sum_{m=0}^s  (-)^{s-m} f^{P_1 \cdots P_m}{}_J(y,z)  f^{\vI \, \shuffle \, P_s \cdots P_{m+1} J  }{}_K(x,z)
\eeq
Assembling this result with the summation of the terms 
$ \Lambda_{4} ^{\vI | \vP }{}_K(x,y,z)$ in (\ref{3atoe}) and
renaming the summation variable $m$ in (\ref{3.cor.7}) to $\ell$, we have,
\begin{align}
&\sum_{\ell =0}^s (-)^{s-\ell} \, 
\big[
\Lambda_4^{\vI \shuffle P_s \cdots P_{\ell+1} | P_1 \cdots P_\ell  }{}_K(x,y,z)
+ \Lambda_5^{\vI \shuffle P_s \cdots P_{\ell+1} | P_1 \cdots P_\ell  }{}_K(x,y,z)
\big]
\notag \\
&\quad =
- \sum_{\ell=0}^s  (-)^{s-\ell} f^{P_1 \cdots P_\ell }{}_J(y,z) \left (   f^{\vI \, \shuffle \, P_s \cdots P_{\ell +1} J  }{}_K(x,z)
- f^{(\vI \, \shuffle \, P_s \cdots P_{\ell+1} ) J}{}_K(x,z)  \right )
\no \\ 
&\quad =
- \sum_{\ell=0}^s  (-)^{s-\ell} f^{P_1 \cdots P_\ell }{}_J(y,z)  
f^{( P_s \cdots P_{\ell+1} J  \, \shuffle \, I_1 \cdots I_{r-1} ) I_r}{}_K(x,z) 
\label{3.cor.6}
\end{align}
where the simplification in passing to the last line was carried out using the basic property of the shuffle product,
\bea
 f^{\vI \, \shuffle \, P_s \cdots P_{\ell +1} J  }{}_K(x,z)
= f^{(\vI \, \shuffle \, P_s \cdots P_{\ell+1} ) J}{}_K(x,z) 
+ f^{(I_1 \cdots I_{r-1}  \, \shuffle \, P_s \cdots P_{\ell+1} J ) I_r}{}_K(x,z) 
\eea
followed by the commutativity $I_1 \cdots I_{r-1}  \, \shuffle \, P_s \cdots P_{\ell+1} J 
=P_s \cdots P_{\ell+1} J  \, \shuffle \,   I_1 \cdots I_{r-1}$. In this form, the last line of
(\ref{3.cor.6}) is readily seen to match the sum on the third line of (\ref{4.ff.99}) corresponding to the second term inside the parentheses of the summand. 

\sm

In summary, the combination of the five terms in (\ref{3atoe}) reproduces the complete
right side of (\ref{4.ff.99}) -- its first line via $\Lambda_3$, its second line via $\Lambda_2$
and its third line via $\Lambda_1$ (first term inside the parenthesis) and $\Lambda_{4,5}$ (second term inside the parenthesis).

\subsection{Proof of Theorem \ref{4.thm:1}}
\label{appB.5}

In the proof of Theorem \ref{4.thm:1}, it is helpful to treat the cases $s=0$, $s=1$ and $s \geq 2$ separately, as the structure of the singularities and direction dependent limits that occur in these three cases is significantly different. In slight abuse of terminology, we shall refer to both the simple pole $f^J {}_K(y,z) = \delta^J_K (y{-}z)^{-1}+{\rm reg}$ and to the direction dependent $z \rightarrow y$ limit of $f^{JI} {}_K(y,z)$ 
(see section \ref{sec:cl.1}) as ``singular''.

\subsubsection{Proof for the case $s=0$}

For $s=0$ we have $\vP=\emptyset$ and the Fay identity in (\ref{4.ff.99}) simplifies as follows, 
\begin{align}
f^{\vI} {}_J(x,z) \, f^J {}_K(y,x)
& = 
f^\vI {}_J(x,z) \, f^J {}_K(y,z)
- \sum_{k=0}^r f^{I_1 \cdots I_k}{}_J(x,y)  f^{J I_{k+1} \cdots I_r}{}_K(y,z) 
\notag \\ &\quad 
- \om_J(y) \Big ( f^{\vI J}{}_K(x,y) 
+ f^{(J \shuffle I_1 \cdots I_{r-1}) I_r}{}_K(x,z)  \Big ) 
\label{3.coin.11}
\end{align}
The limit of any term involving $f^{I_1 \cdots I_r}{}_J(y,z)$ with $r\geq 3$ is regular and given by $\cF^{I_1 \cdots I_r}{}_J(y)$ defined in (\ref{4.cF}). The singular terms are as follows,
\bea
\hbox{sing}_{s=0}(x,y,z) = \Big ( f^\vI {}_J(x,z) - f^\vI {}_J(x,y) \Big )  f^J {}_K(y,z) - f^{I_1 \cdots I_{r-1}}{}_J(x,y) f^{J I_r}{}_K(y,z) 
\quad
\eea
The calculation of their limit may be organized as follows, using the fact that the contribution from the tensor $\Phi$ to $f$ cancels in the first term,  
\begin{align}
\label{3.coin.12}
\lim_{z \to y} \hbox{sing}_{s=0}(x,y,z) & = 
 \delta ^{I_r} _K \lim_{z \to y} \bigg [ \Big ( \p_x \cG ^{I_1 \cdots I_{r-1}} (x,z) - \p_x \cG ^{I_1 \cdots I_{r-1}} (x,y)  \Big )  \p_y \cG(y,z)
 \\ &\ \ \hskip 0.6in 
+ f^{I_1 \cdots I_{r-1}}{}_J(x,y) \p_y \cG ^J (y,z) \bigg ] - f^{I_1 \cdots I_{r-1}}{}_J(x,y) \p_y \Phi ^{J I_r}{}_K(y) 
\no
\end{align}
Here, only the $\p_y \cG(y,z)$ term in $f^J{}_K(y,z)$ contributes to a non-vanishing limit. The last term  arises as the finite limit from decomposing $f^{J I_r}{}_K(y,z)$. Expanding the difference inside the parentheses on the
first line of (\ref{3.coin.12}) to first order in $z{-}y$, we obtain, 
\bea
\label{3.coin.13}
&&
\p_x \cG ^{I_1 \cdots I_{r-1}} (x,z) - \p_x \cG ^{I_1 \cdots I_{r-1}} (x,y) 
\no \\ && \hskip 0.4in =
(z-y) \p_x \p_y \cG ^{I_1 \cdots I_{r-1}} (x,y) + (\bar z - \bar y) \p_x \pby \cG ^{I_1 \cdots I_{r-1}} (x,y)
\no \\ && \hskip 0.4in =
(z-y) \p_x \p_y \cG ^{I_1 \cdots I_{r-1}} (x,y) 
- \pi (\bar z - \bar y) \bar \om^J(y) f ^{I_1 \cdots I_{r-1}} {}_J (x,y)
\eea
Upon identifying the Abelian integral $(\bar z {-} \bar y) \bar \om^J(y)= \int^z_y \bar \bom^J+(\bar z {-} \bar y)^2$, the last term on the third line above combines with the $\p_y \cG ^J (y,z)$ on the second line of (\ref{3.coin.12}) to produce the well-defined limit $\cC^J(y)$ defined in (\ref{csec.12}). The latter combines with the last term  of (\ref{3.coin.12}) to produce the combination $\cF^{J I_r}{}_K(y)$ defined in (\ref{4.cF}). Therefore,  the limit in (\ref{3.coin.12}) becomes, 
\bea
\label{3.coin.14}
\lim_{z \to y} \hbox{sing}_{s=0} (x,y,z) & = & 
 \delta ^{I_r} _K \,  \p_x \p_y \cG ^{I_1 \cdots I_{r-1}} (x,y) 
 - f^{I_1 \cdots I_{r-1}}{}_J(x,y) \cF^{JI_r}{}_K(y) 
\eea
Combining this result with (\ref{4.coin}), we obtain the coincident limit for the case $s=0$, 
\bea
\label{4.coin.1}
f^{\vI} {}_J(x,y) \, f^J {}_K(y,x)
& = &
- \p_y f^{\vI}{}_K(x,y) - \om_J(y) f^{J \, \shuffle \, \vI}{}_K(x,y) 
\no \\ &&
- \sum_{k=0}^{r-1} f^{I_1 \cdots I_k}{}_J(x,y) \cF^{J I_{k+1} \cdots I_r}{}_K(y)
\eea
 Comparing with (\ref{3.coin.0}) for the case of $s=0$, we find agreement upon using the fact that the $k=r$ term in (\ref{3.coin.0}) cancels the first term on the first line of the right side of (\ref{3.coin.0}). This concludes the proof of Theorem  \ref{4.thm:1} for the case $s=0$.

\subsubsection{Proof for the case $s=1$}

The proof of Theorem \ref{4.thm:1} for the case $s=1$ proceeds analogously to the $s=0$ case.  The singular terms as $z \rightarrow y$ are as follows,
\begin{align}
\hbox{sing}_{s=1} (x,y,z) & = 
f^J{}_K(y,z) f^{P \, \shuffle \, \vI}{}_J(x,y) + f^{J I_r}{}_K(y,z) f^{P \, \shuffle \, I_1 \cdots I_{r-1}}{}_J(x,y)
\notag \\ &\quad 
- f^P{}_J(y,z) \Big ( f^{\vI J }{}_K(x,y) + f^{(J \, \shuffle \, I_1 \cdots I_{r-1}) I_r}{}_K(x,z) \Big )
\end{align}
The calculation of their limit may be organized as follows, 
\begin{align}
\lim_{z \to y} \hbox{sing}_{s=1} (x,y,z) & =  
\p_y \Phi ^J{}_K(y) f^{P \, \shuffle \, \vI}{}_J(x,y) 
+ \p_y \Phi ^{J I_r}{}_K(y) f^{P \, \shuffle \, I_1 \cdots I_{r-1}}{}_J(x,y)
\notag \\ &\quad
- \p_y \Phi ^P{}_J(y)  f^{J \, \shuffle \, \vI}{}_K(x,y) 
- \delta ^{I_r} _K \lim _{z \to y} \bigg [  \p_y \cG^J(y,z)  f^{P \, \shuffle \, I_1 \cdots I_{r-1}}{}_J(x,y)
\notag \\ &\quad  \hskip 0.3in 
+ \p_y \cG(y,z) \Big ( \p_x \cG^{P \, \shuffle \, I_1 \cdots I_{r-1}}(x,z) - \p_x \cG ^{P \, \shuffle \, I_1 \cdots I_{r-1}}(x,y) \Big )
\bigg ]
\end{align}
Expanding the difference on the last line to first order in $z-y$, as in (\ref{3.coin.13}), produces a double derivative term in $x,y$ and a term that combines with $\p_y \cG^J(y,z)$ to produce $\cC^J(y)$ which  combines with $\p_y \Phi ^{J I_r}{}_K(y)$ to produce $\cF^{J I_r}{}_K(y)$. Collecting all contributions, we obtain, 
\begin{align}
\lim_{z \to y} \hbox{sing}_{s=1} (x,y,z) 
& =  
\cF ^J{}_K(y) f^{P \, \shuffle \, \vI}{}_J(x,y) 
+ \cF ^{J I_r}{}_K(y) f^{P \, \shuffle \, I_1 \cdots I_{r-1}}{}_J(x,y)
\notag \\ &\quad
- \cF ^P{}_J(y)  f^{J \, \shuffle \, \vI}{}_K(x,y) 
+ \p_y f^{(P \, \shuffle \, I_1 \cdots I_{r-1} ) I_r}{}_K(x,y)
\end{align}
Combining the above limit of the singular terms with the limits of the regular terms then produces (\ref{3.coin.0}) and proves Theorem \ref{4.thm:1} for the case $s=1$.

\subsubsection{Proof for the case $s\geq 2$}

For $s\geq 2$, the term on the first line on the right side of (\ref{4.ff.99}) admits a regular limit. The terms $(k, \ell) = (r,0), \, (r, 1), \, (r{-}1,0)$ on the second line of (\ref{4.ff.99}) and the terms $\ell=1, \, 2$ on the third line of (\ref{4.ff.99}) are singular, and are given by,
\begin{align}
(-)^s \hbox{sing}_{s \geq 2}(x,y,z) & = 
- f^{\vPt \, \shuffle \, \vI}{}_J(x,y) f^J{}_K(y,z) 
- f^{\vPt \, \shuffle \, I_1 \cdots I_{r-1}}{}_J(x,y) f^{J I_r}{}_K(y,z) 
\label{coutright} \\ &\quad
+ f^{P_s \cdots P_2 \, \shuffle \, \vI}{}_J(x,y) f^{P_1 J}{}_K(y,z)
- f^{P_1 P_2}{}_J(y,z) f^{P_s \cdots P_3 J \, \shuffle \, \vI} {}_K(x,y)
\no \\ &\quad
+ f^{P_1}{}_J(y,z) \Big ( f^{(P_s \cdots P_2 \, \shuffle \, \vI)J}{}_K(x,y) 
+ f^{(P_s \cdots P_2 J \, \shuffle \, I_1 \cdots I_{r-1} )I_r }{}_K(x,z) \Big ) 
\qquad
\no
\end{align}
Decomposing each $f(y,z)$-tensor into its $\p_y \Phi(y)$ and $\p_y \cG(y,z)$ parts, we obtain,
\begin{align}
(-)^s \hbox{sing}_{s\geq 2}(x,y,z) & =  
- f^{\vPt \, \shuffle \, \vI}{}_J(x,y) \p_y \Phi ^J{}_K(y) 
- f^{\vPt \, \shuffle \, I_1 \cdots I_{r-1}}{}_J(x,y) \p_y \Phi ^{J I_r}{}_K(y) 
\notag\\ &\quad
+ f^{P_s \cdots P_2 \, \shuffle \, \vI}{}_J(x,y) \p_y \Phi ^{P_1 J}{}_K(y)
- \p_y \Phi ^{P_1 P_2}{}_J(y) f^{P_s \cdots P_3 J \, \shuffle \, \vI} {}_K(x,y)
\no \\ &\quad
+ \p_y \Phi^{P_1}{}_J(y) \Big ( f^{(P_s \cdots P_2 \, \shuffle \, \vI)J}{}_K(x,y) 
+ f^{(P_s \cdots P_2 J \, \shuffle \, I_1 \cdots I_{r-1} )I_r }{}_K(x,z) \Big ) 
\no \\ &\quad
+ \delta ^{I_r}_K \, \p_y \cG (y,z) \Big ( \p_x \cG^{\vPt \, \shuffle \, I_1 \cdots I_{r-1} }(x,z)
- \p_x \cG^{\vPt \, \shuffle \, I_1 \cdots I_{r-1} }{}(x,y) \Big ) 
\no \\ &\quad
+ f^{\vPt \, \shuffle \, I_1 \cdots I_{r-1}}{}_J(x,y) \p_y \cG ^J(y,z) \delta ^{I_r }_K 
\end{align}
Note that terms proportional to $\p_y \cG ^{P_1}(y,z)$, which arise at intermediate steps from the second line of (\ref{coutright}), cancel one another outright. Taking the limit, we obtain, 
\begin{align}
(-)^s \lim _{z \to y} \hbox{sing}_{s \geq 2}(x,y,z) & = 
- f^{\vPt \, \shuffle \, \vI}{}_J(x,y) \p_y \Phi ^J{}_K(y) 
- f^{\vPt \, \shuffle \, I_1 \cdots I_{r-1}}{}_J(x,y) \cF ^{J I_r}{}_K(y) 
 \\ &\quad
+ f^{P_s \cdots P_2 \, \shuffle \, \vI}{}_J(x,y) \p_y \Phi ^{P_1 J}{}_K(y)
- \p_y \Phi ^{P_1 P_2}{}_J(y) f^{P_s \cdots P_3 J \, \shuffle \, \vI} {}_K(x,y)
\no \\ &\quad
+ \p_y \Phi^{P_1}{}_J(y) f^{P_s \cdots P_2 J \, \shuffle \, \vI}{}_K(x,y) 
+ \delta ^{I_r}_K \, \p_x \p_y  \cG^{\vPt \, \shuffle \, I_1 \cdots I_{r-1} }(x,y)
\qquad
\no
\end{align}
The $\p_y \Phi$ factors on the second line may be promoted into the corresponding $\cF$ factors  because the differences cancel between the two terms. Thus, the final result for the limit may be expressed as follows, 
\begin{align}
(-)^s \lim _{z \to y} \hbox{sing}_{s \geq 2}(x,y,z) & =  
- f^{\vPt \, \shuffle \, \vI}{}_J(x,y) \cF ^J{}_K(y) 
- f^{\vPt \, \shuffle \, I_1 \cdots I_{r-1}}{}_J(x,y) \cF ^{J I_r}{}_K(y) 
\label{sings2} \\ &\quad
+ f^{P_s \cdots P_2 \, \shuffle \, \vI}{}_J(x,y) \cF ^{P_1 J}{}_K(y)
- \cF ^{P_1 P_2}{}_J(y) f^{P_s \cdots P_3 J \, \shuffle \, \vI} {}_K(x,y)
\no \\ &\quad
+ \cF^{P_1}{}_J(y) f^{P_s \cdots P_2 J \, \shuffle \, \vI}{}_K(x,y) 
+ \delta ^{I_r}_K \, \p_x \p_y  \cG^{\vPt \, \shuffle \, I_1 \cdots I_{r-1} }(x,y)
\qquad
\no
\end{align}
While the terms corresponding to $(k, \ell) = (r,0), \, (r, 1), \, (r{-}1,0)$ on the second line of (\ref{4.ff.99}) and the terms $\ell=1, \, 2$ on the third line of (\ref{4.ff.99}) were combined in the calculation of the limit of the singular terms, we see that
on the right side of (\ref{sings2}), the first term provides the term $(k, \ell) = (r,0)$ in the double sum of (\ref{3.coin.0}), the second term provides its $(k,\ell)=(r{-}1,0)$ term, and the third term provides its $(k,\ell)=(r,1)$ term. The fourth and fifth terms of (\ref{sings2}) provide the $\ell=2$ and $\ell=1$ terms in the single sum on the third line of (\ref{3.coin.0}). Finally, the last term of (\ref{sings2}) is identified as the second term on the right side of (\ref{3.coin.0}) via (\ref{4.coin}). Thus all terms in (\ref{3.coin.0}) are properly produced in the limit for $s \geq 2$. This concludes the proof of Theorem~\ref{4.thm:1} for the case $s \geq 2$ and thus for all cases. 

\subsection{Proof of Theorem \ref{finthm}}
\label{prfcoin}

We shall here prove the equivalence of the two representations
(\ref{3.coin.0}) and (\ref{coin.75}) of the coincident Fay identities
at arbitrary rank, weight and genus. As a first step, we specialize the general
identity (\ref{3.coin.0}) for $f^{ \overrightarrow{I} } {}_J(x,y) \, f^{ \overrightarrow{P} J} {}_K(y,x)$
 to the three cases of $\overrightarrow{P}=P_1 P_2\cdots P_s$ with $s=0$, $s=1$ or 
$s\geq 2$ and adapt the tensor functions $\cF ^{I_1 \cdots I_r}{}_J(y) $ in (\ref{4.cF}) to each~term:
\begin{itemize}
\itemsep=0in
\item $s=0$: The first term $f^\vI {}_J(x,y) \, \cF ^{J} {}_K(y) $ on the right side of (\ref{3.coin.0})
(with $\vI= I_1\cdots I_r$) readily cancels the $(k,\ell)=(r,0)$ term in the second line, and we are left with,
 \begin{align}
 \label{adapta}
f^{ \overrightarrow{I} } {}_J(x,y) \, f^{ J} {}_K(y,x) 
&= \delta^{I_r}_K \p_x \p_y  \cG^{ I_1 \cdots I_{r-1}  }(x,y)   -  \omega_J(y) f^{ \overrightarrow{I}  \shuffle J  }{}_K(x,y) 
 \\ & \quad 
+ f^{ I_1 \cdots I_{r-1} }{}_J(x,y) \big[ \delta^{I_r}_K {\cal C}^J(y) - \p_y \Phi^{J I_r}{}_K(y) \big] 
\notag \\
&\quad
\! - \sum_{k=0}^{r-2} f^{I_1\cdots I_k}{}_J(x,y) f^{J I_{k+1}\cdots I_r}{}_K(y,y)
\no
\end{align}
\item $s=1$: The first term $f^\vI {}_J(x,y) \, \cF ^{P J} {}_K(y) $ on the right side of (\ref{3.coin.0}) readily cancels the $(k,\ell)=(r,1)$ term in the second line, resulting in, 
 \begin{align}
&f^{ \overrightarrow{I} } {}_J(x,y) \, f^{P J} {}_K(y,x) = - \delta^{I_r}_K \p_x \p_y  \cG^{P \, \shuffle \, I_1 \cdots I_{r-1}   }(x,y) 
+ \omega_J(y) f^{\overrightarrow{I} \shuffle PJ}{}_K(x,y)
  \notag \\
&\quad\quad\quad\quad + \sum_{k=0}^{r-2} f^{P \, \shuffle \, I_1\cdots I_k}{}_J(x,y) f^{J I_{k+1}\cdots I_r}{}_K(y,y) 
- \sum_{k=0}^{r-1} f^{I_1\cdots I_k}{}_J(x,y) f^{PJ I_{k+1}\cdots I_r}{}_K(y,y) \notag \\
&\quad \quad\quad\quad
+  f^{P \, \shuffle \, I_1 \cdots I_{r-1} }{}_J(x,y)  \big[ \p_y \Phi^{J I_r}{}_K(y) 
 - \delta^{I_r}_K   {\cal C}^J(y) \big] \notag \\
 &\quad\quad\quad\quad
+   f^{P \, \shuffle \, \overrightarrow{I} }{}_J(x,y) \p_y \Phi^J{}_K(y)
-  \p_y \Phi^P{}_J(y)  f^{J \, \shuffle \, \overrightarrow{I} }{}_K(x,y) 
  \label{adaptb} 
\end{align}
\item $s\geq 2$: After isolating all cases of
$\cF ^{I_1 \cdots I_r}{}_J(y) $ with $r\leq 2$ which
depart from the expression $f^{I_1 \cdots I_r}{}_J(y,y) $ at generic rank
from the sums in (\ref{3.coin.0}), we have,
 \begin{align}
&f^{ \overrightarrow{I} } {}_J(x,y) \, f^{ \overrightarrow{P} J} {}_K(y,x) =
f^{\overrightarrow{I}} {}_J(x,y) \, f^{\overrightarrow{P}  J} {}_K(y,y) 
+  (-1)^s \delta^{I_r}_K \p_x \p_y  \cG^{ I_1 \cdots I_{r-1} \, \shuffle \,  \overleftarrow{P}  }(x,y)
\notag  \\
&\quad \quad + (-1)^s f^{ \overleftarrow{P} \shuffle I_1 \cdots I_{r-1} }{}_J(x,y)
\big[  \delta^{I_r}_K \cC^J(y) - \partial_y \Phi^{J I_r}{}_K(y)  \big] 
\notag  \\
&\quad \quad 
- \sum_{\ell=0}^s (-1)^{\ell-s} \sum_{ {k =0} }^r 
\delta_{k - \ell  \leq r- 2}
 f^{  I_1 \cdots I_k   \shuffle  P_s \cdots P_{\ell+1} }{}_J(x,y)  f^{P_1 \cdots P_\ell J I_{k +1} \cdots I_r}{}_K(y,y) 
\notag  \\
&\quad \quad 
- (-1)^{s} \omega_J(y) 
f^{  \overrightarrow{I}\shuffle \overleftarrow{P}  J }{}_K(x,y) 
- \sum_{\ell =3 }^s (-)^{\ell-s} f^{P_1 \cdots P_\ell}{}_J(y,y) 
f^{  \overrightarrow{I} \shuffle P_s \cdots P_{\ell+1} J  }{}_K(x,y) 
\notag  \\
&\quad \quad 
 - (-)^s  \Big (   f^{P_s \cdots P_1 \, \shuffle \,  \overrightarrow{I} }{}_J(x,y) \p_y \Phi^J {}_K (y) 
- \p_y \Phi ^{P_1}{}_J  (y)  f^{P_s \cdots P_2 J \, \shuffle \, \overrightarrow{I} }{}_K(x,y)  \Big )
\notag  \\
&\quad \quad 
 + (-)^s  \Big (   f^{P_s \cdots P_2 \, \shuffle \,  \overrightarrow{I} }{}_J(x,y) \p_y \Phi^{P_1J} {}_K (y) 
- \p_y \Phi ^{P_1 P_2}{}_J  (y)  f^{P_s \cdots P_3 J \, \shuffle \, \overrightarrow{I} }{}_K(x,y)  \Big )
\label{adaptc}
\end{align}
where the $\cC^I(y)$ from the $(k,\ell)=(r,1)$ term in the second line and
the $\ell=2$ term in the third line of (\ref{3.coin.0}) have already been cancelled.
The symbol $\delta_{k - \ell  \leq r- 2}$ in the third line of (\ref{adaptc}) excludes
the terms $(k,\ell) \in \{ (r,0),\, (r{-}1,0),\, (r,1) \}$ from the double sum over $k$ and $\ell$,
ensuring that $ f^{P_1 \cdots P_\ell J I_{k +1} \cdots I_r}{}_K(y,y) $ has at least three upper indices.
\end{itemize}
In the cases $s=0$ and $s=1$, the specializations (\ref{adapta}) and (\ref{adaptb})
of (\ref{3.coin.0})
straightforwardly line up with the corresponding $s=0$ and $s=1$ cases of (\ref{coin.75}).
Hence, the leftover task is to show agreement of (\ref{coin.75}) with the rewritten form
(\ref{adaptc}) of (\ref{3.coin.0}) at $s\geq 2$.

\sm

Several terms of (\ref{adaptc}) and (\ref{coin.75}) are easily seen to match:
\begin{itemize}
\item $ (-1)^s \delta^{I_r}_K \p_x \p_y  \cG^{ I_1 \cdots I_{r-1} \, \shuffle \,  \overleftarrow{P}  }(x,y)$
and $- (-1)^{s} \omega_J(y) 
f^{  \overrightarrow{I}\shuffle \overleftarrow{P}  J }{}_K(x,y) $

as well as $(-1)^s f^{ \overleftarrow{P} \shuffle I_1 \cdots I_{r-1} }{}_J(x,y)
\big[  \delta^{I_r}_K \cC^J(y) - \partial_y \Phi^{J I_r}{}_K(y)  \big] $;

\item the last two lines of (\ref{adaptc}) match the terms $\ell =1,2$ in
the last line of (\ref{coin.75}).
\end{itemize}
After taking these matches into account, it remains to verify that
\begin{align}
&f^{\overrightarrow{I}} {}_J(x,y) \, f^{\overrightarrow{P}  J} {}_K(y,y) 
- \sum_{\ell =3 }^s (-)^{\ell-s} f^{P_1 \cdots P_\ell}{}_J(y,y) 
f^{  \overrightarrow{I} \shuffle P_s \cdots P_{\ell+1} J  }{}_K(x,y) 
\label{laststep}  \\
&\quad \quad 
- \sum_{\ell=0}^s (-1)^{\ell-s} \sum_{ {k =0} }^r 
\delta_{k - \ell  \leq r- 2}
 f^{  I_1 \cdots I_k   \shuffle  P_s \cdots P_{\ell+1} }{}_J(x,y)  f^{P_1 \cdots P_\ell J I_{k +1} \cdots I_r}{}_K(y,y) 
\notag  \\
&\quad = -  \sum_{k =0}^{r-1}  \sum_{\ell=0}^s  \delta_{(k,\ell)\neq (r-1,0)}
 (-1)^{s-\ell} f^{ I_1 \cdots I_k \shuffle P_s \cdots P_{\ell+1} }{}_J(x,y)
 f^{ P_1 \cdots P_\ell J I_{k+1} \cdots I_r  }{}_K(y,y)
\notag  \\
&\quad  \quad 
- \sum_{\ell=3}^s (-1)^{s-\ell } \big[ 
f^{P_1\cdots P_\ell}{}_J(y,a_\ell ) f^{ P_s\cdots P_{\ell +1}J \shuffle \overrightarrow{I} }{}_K(x,y) 
\,{-}\, f^{P_1\cdots P_{\ell-1}J}{}_K(y,a_\ell ) f^{P_s\cdots P_\ell  \shuffle \overrightarrow{I} }{}_J(x,y)
\big] 
\notag
\end{align}
For this purpose, we rearrange the double sum in the second line of the left side 
according to $\sum_{\ell=0}^s  \sum_{k =0}^r  \delta_{k - \ell  \leq r- 2}
= \sum_{\ell=0}^s  \sum_{k =0}^{r-1} \delta_{k - \ell \neq r- 1}
+ \sum_{\ell=2}^s \delta_{r=k}$, leading to
\begin{align}
&- \sum_{\ell=0}^s (-1)^{\ell-s} \sum_{ {k =0} }^r 
\delta_{k - \ell  \leq r- 2}
 f^{  I_1 \cdots I_k   \shuffle  P_s \cdots P_{\ell+1} }{}_J(x,y)  f^{P_1 \cdots P_\ell J I_{k +1} \cdots I_r}{}_K(y,y) 
\label{laststep.2} \\ 
& = - \sum_{\ell=0}^s (-1)^{\ell-s} \sum_{k =0}^{r-1 }
\delta_{k - \ell \neq r- 1}
 f^{  I_1 \cdots I_k   \shuffle  P_s \cdots P_{\ell+1} }{}_J(x,y)  f^{P_1 \cdots P_\ell J I_{k +1} \cdots I_r}{}_K(y,y) \notag \\
 &\quad
 - \sum_{\ell=2}^{s-1} (-1)^{\ell-s} 
  f^{ \overrightarrow{I} \shuffle  P_s \cdots P_{\ell+1} }{}_J(x,y)  f^{P_1 \cdots P_\ell J }{}_K(y,y)
  -   f^{ \overrightarrow{I}  }{}_J(x,y)  f^{P_1 \cdots P_s J }{}_K(y,y)
 \notag
\end{align}
In the last line, we have exposed the last term of the sum $\sum_{\ell=2}^s$
which cancels the first term on the left side of (\ref{laststep}).
Since the middle line of (\ref{laststep.2}) matches the first line on the right
side of (\ref{laststep}), the last step is to check that
\begin{align}
& \sum_{\ell =3 }^s (-)^{\ell-s} f^{P_1 \cdots P_\ell}{}_J(y,y) 
f^{  \overrightarrow{I} \shuffle P_s \cdots P_{\ell+1} J  }{}_K(x,y) 
+  \sum_{\ell=2}^{s-1} (-1)^{\ell-s} 
  f^{ \overrightarrow{I} \shuffle  P_s \cdots P_{\ell+1} }{}_J(x,y)  f^{P_1 \cdots P_\ell J }{}_K(y,y)
 \notag \\
 &=  \sum_{\ell=3}^s (-1)^{s-\ell } \big[ 
f^{P_1\cdots P_\ell}{}_J(y,a_\ell ) f^{ P_s\cdots P_{\ell +1}J \shuffle \overrightarrow{I} }{}_K(x,y) 
\,{-}\, f^{P_1\cdots P_{\ell-1}J}{}_K(y,a_\ell ) f^{P_s\cdots P_\ell  \shuffle \overrightarrow{I} }{}_J(x,y)
\big] 
\end{align}
This is the case since the $\partial \cG$ terms of the $f$-tensors with $y$ as their first argument separately cancel on both sides, and the $\partial \Phi$ contributions are seen to match after shifting the summation variable
of the second term on the left side to $\ell{+}1 = m \in \{ 3,4,\cdots,s\}$. We have thus demonstrated
(\ref{laststep}) which concludes the proof of this appendix that
(\ref{3.coin.0}) is equivalent to~(\ref{coin.75}).

\subsection{Proof of Theorem \ref{mintlemma}}
\label{app:EZ}

The subsequent proof of Theorem \ref{mintlemma} is most conveniently performed 
in the original normalization convention $\omega^{I_1\cdots I_r}{}_J(x,y)$
of the Enriquez kernels \cite{Enriquez:2011} related to the $g^{I_1\cdots I_r}{}_J(x,y)$ 
in this work via (\ref{Enorm}).
The decomposition (\ref{ome.04}) then takes the form
\beq
\omega^{I_1\cdots I_r}{}_J(x,y)  =  \tvarpi^{I_1\cdots I_r}{}_J(x) 
- \delta^{I_r}_J \tilde \chi^{I_1\cdots I_{r-1}}(x,y) 
\, , \ \ \ \ \ \ \tvarpi^{I_1 \cdots I_{r-1}J}{}_J(x) = 0
\label{appe.04}
\eeq
with rescaled components 
\begin{align}
\tvarpi^{I_1 \cdots I_r}{}_J(x)  &=  (-2\pi i)^{-r} \varpi^{I_1 \cdots I_r}{}_J(x) 
\notag\\
\tilde \chi^{I_1 \cdots I_s}(x,y)  &=  (-2\pi i)^{-s-1} \chi^{I_1 \cdots I_s}(x,y) 
\end{align}
In this way, we can take advantage of the simplified monodromies of the
Enriquez kernels $\omega^{I_1\cdots I_r}{}_J(x,y)$ in demonstrating the vanishing of,
\begin{align}
 &\mQ^{I_1 \cdots I_r}{}_J(x,y)  =  (-2\pi i)^r \, \bigg\{
\omega_M(x) \, \omega^{I_1 \cdots I_r M}{}_J(y,x)
+ (-1)^r \omega_M(y) \, \omega^{I_r \cdots I_1 M}{}_J(x,y)
\label{redefpi}\\ &\quad\quad \quad
+ \sum_{k = 1}^r (-1)^{k+r}
 \Big [
\tvarpi^{I_1 \cdots I_k}{}_M(y) \, \tvarpi^{I_r \cdots I_{k + 1} M}{}_J(x)
- \tvarpi^{I_r \cdots I_{k} }{}_M(x) \,\tvarpi^{I_1 \cdots I_{k-1} M}{}_J(y) 
 \Big ]  \, \bigg\}\notag
\end{align}
claimed in Theorem \ref{mintlemma}.
To prove the theorem, we note that it is straightforward to verify item 1.

\sm

To prove item 2 we note that the $\mA$ monodromy of $\mQ^{I_1 \cdots I_r}{}_J(x,y)$ vanishes since the $\mA$ monodromy of  $\om^{I_1 \cdots I_r}{}_J(x,y)$ vanishes for all $r \geq 0$, thus establishing the first part of item~2. The heart of the theorem is the proof of the $\mB$ mondromy formula in (\ref{8.thm.2}).  The monodromies around $\mB$-cycles of $\om^{I_1 \cdots I_r}{}_J(x,y)$, given in (\ref{ome.02}), may be expressed as follows,
\bea
\omega ^{I_1 \cdots I_r}{}_J(x {+} \mB_L,y) & = & 
\omega ^{I_1 \cdots I_r}{}_J(x,y) + \Delta ^{(x)} _{B_L} \, \omega ^{I_1 \cdots I_r}{}_J(x,y)
\no \\
\omega ^{I_1 \cdots I_r}{}_J(x ,y{+} \mB_L) & = & 
\omega ^{I_1 \cdots I_r}{}_J(x,y) + \Delta ^{(y)} _{B_L} \, \omega ^{I_1 \cdots I_r}{}_J(x,y)
\eea
where the monodromy shifts are given by,
\bea
\label{9.a.8}
\Delta ^{(x)} _{\mB_L} \, \omega ^{I_1 \cdots I_r}{}_J(x,y) & = & 
\sum_{k=1}^r { 1 \over k!} \, \delta ^{I_1 \cdots I_k} _L \, \om^{I_{k+1} \cdots I_r}{}_J(x,y)
\no \\
\Delta ^{(y)} _{\mB_L} \, \omega ^{I_1 \cdots I_r}{}_J(x,y) & = &
\delta ^{I_r}_J \sum_{k=1}^r { (-)^k \over k!} \, \omega ^{I_1 \cdots I_{r-k}} {}_L (x,y) \, \delta^{I_{r-k+1} \cdots I_{r-1}}_L
\eea
Throughout, we shall extend the definition to include $\omega ^{I_1 \cdots I_r}{}_J(x,y) |_{r=0} = \om_J(x)$ which is a single-valued holomorphic Abelian differential. We shall also need the $\mB$ monodromy of the traceless part $\tvarpi^{I_1 \cdots I_r}{}_J(x)$ in (\ref{appe.04}), which may be readily deduced from the first equation in (\ref{9.a.8}) and can be found in (\ref{9.b.5}) below.

\sm

In view of the symmetry stated in item 1, the monodromies in $x$ and $y$ are equivalent to one another. The combinatorics of the calculation of the $\mB$ monodromy of $\mQ^{I_1 \cdots I_k}{}_J(x,y)$ will be simpler in the variable $y$ than in $x$, and we begin by computing the monodromy in $y$ of the four contributions in (\ref{redefpi}). The first two terms on the right side of (\ref{redefpi}) involve,
\bea
\label{9.b.4}
\Delta ^{(y)} _{\mB_L} \,  \om^{I_1 \cdots I_r M}{}_J(y,x) & = &
\sum_{k=1}^r { 1 \over k!} \, \delta ^{I_1 \cdots I_k}_L  \, \om ^{I_{k+1} \cdots I_r M}{}_J(y,x)
+ { \delta ^{I_1 \cdots I_r M}_L \over (r{+}1)!} \, \om_J(y)
 \\
\Delta ^{(y)} _{\mB_L} \,  \om^{I_r \cdots I_1 M}{}_J(x,y) & = &
\delta ^M _J \sum_{k=1}^r { (-)^k \over k !} \, \delta ^{I_1 \cdots I_{k-1}}_L  \, \om ^{I_r \cdots I_k}{}_L(x,y)
- (-)^r  \delta ^M_J \, { \delta ^{I_1 \cdots I_r }_L \over (r{+}1)!} \, \om_L(x) 
\no
\eea
One verifies that the contributions from the terms with denominators $(r{+}1)!$ cancel one another in the sum of these two terms that enters into (\ref{redefpi}). The $y$-dependent parts of the summands in (\ref{redefpi}) transform as follows,
\begin{align}
\label{9.b.5}
\Delta ^{(y)} _{\mB_L} \, \tvarpi^{I_1 \cdots I_{\ell-1}  M}{}_J(y) & = 
\sum_{n=1}^{\ell-1} { 1 \over n  !} \,  \delta ^{I_1 \cdots I_n}_L \, \tvarpi^{I_{n+1} \cdots I_{\ell-1} M}{}_J(y)
+ { 1 \over \ell \, !} \delta ^{I_1 \cdots I_{\ell-1}}_L \Big ( \delta ^M_L \, \om_J(y) 
-{ 1 \over  h}  \, \delta ^M_J \, \om_L(y) \Big )
\no \\
\Delta ^{(y)} _{\mB_L} \, \tvarpi^{I_1 \cdots I_{\ell} }{}_M(y) & = 
\sum_{n=1}^{\ell-1} { 1 \over n !} \,  \delta ^{I_1 \cdots I_n}_L \, \tvarpi^{I_{n+1} \cdots I_{\ell} }{}_M(y)
+  { 1 \over \ell !} \,  \delta ^{I_1 \cdots I_\ell}_L \, \omega_M(y)
-{ 1 \over  h \, \ell \, !} \, \delta ^{I_1 \cdots I_{\ell-1}}_L \, \delta ^{I_\ell} _M \, \om_L(y) 
\end{align}
Adding the contributions from these two terms in (\ref{redefpi}), one verifies that the contributions with denominators $h$ cancel one another.  As a result, the sum over $k$ in (\ref{redefpi}) evaluates to,
\bea
\label{9.b.6}
&&
\Delta ^{(y)} _{\mB_L} 
\sum_{k=1}^r (-)^{r+ k} \,  \Big (  \tvarpi^{I_1 \cdots I_k }{}_M(y)  \, \tvarpi^{I_r \cdots I_{k+1} M} {}_J (x)
-  \tvarpi^{I_r \cdots I_k }{}_M(x)  \,  \tvarpi^{I_1 \cdots I_{k-1} M} {}_J (y)   \Big )
\no \\ && \hskip 0.6in =
- \sum_{n=1}^{r-1} { 1 \over n!} \, \delta ^{I_1 \cdots I_n}_L 
\sum_{k=n+1}^r (-)^{r+k} \,\tvarpi^{I_r \cdots I_k}{}_M(x) \, \tvarpi^{I_{n+1} \cdots I_{k-1} M}{}_J(y)
\no \\ && \hskip 0.8in 
- \sum_{k=1}^r  { (-)^{r+k} \over k !} \, \tvarpi^{I_r \cdots I_k}{}_M(x) \, 
\delta ^{I_1 \cdots I_{k-1} M}_L \,  \, \om_J(y) 
\no \\ && \hskip 0.8in 
+ \sum_{n=1}^r { 1 \over n!} \, \delta ^{I_1 \cdots I_n}_L 
\sum_{k=n}^r (-)^{r+k} \, \tvarpi^{I_r \cdots I_{k+1} M}{}_J(x) \, \tvarpi^{I_{n+1} \cdots I_k}{}_M(y)
\eea
The remaining terms are as follows,
\bea
\frac{ \Delta ^{(y)} _{\mB_L} \mQ ^{I_1 \cdots I_r}{}_J(x,y) }{(-2\pi i)^r}
& = & 
\sum_{n=1}^r { 1 \over n!} \delta ^{I_1 \cdots I_n}_L \bigg [
\om_M(y) \, \om^{I_{n+1} \cdots I_r M}{}_J(x,y)
 \\ && \hskip 0.9in 
- \sum_{k=n+1}^r (-)^{r+k} \, \tvarpi^{I_r \cdots I_k}{}_M(y) \, \tvarpi^{I_{n+1} \cdots I_{k-1} M}{}_J(x)
\no \\ && \hskip 0.9in 
+ \sum_{k=n} ^r (-)^{r+k} \, \tvarpi^{I_r \cdots I_{k+1}M}{}_J(y) \, \tvarpi^{I_{n+1} \cdots I_k}{}_M(x) \bigg ]
\no \\ &&
+ (-)^r \om_J(x) \sum_{k=1}^r { (-)^k \over k !} \delta ^{I_1 \cdots I_{k-1}} _L 
\Big ( \om^{I_r \cdots I_k}{}_L(y,x) - \tvarpi^{I_r \cdots I_k}{}_L(y) \Big )
\no
\eea
The terms inside the square bracket almost make up $\mQ^{I_{n+1} \cdots I_r}{}_J(x,y)$.  Accounting for the difference,  we obtain after some simplifications, 
\bea
\label{9.b.9}
\Delta ^{(y)} _{\mB_L} \mQ^{I_1 \cdots I_r}{}_J(x,y)
& = & 
\sum_{n=1}^r {   (-2\pi i)^n \over n!} \delta ^{I_1 \cdots I_n}_L    \mQ^{I_{n+1} \cdots I_r}{}_J(x,y)
 \\ &&
+  (-2\pi i )^r  \sum_{n=1}^r { (-)^{r+n} \over n!} \bigg [ 
\om_J(x) \delta ^{I_1 \cdots I_{n-1}}_L \Big ( \om^{I_r \cdots I_n}{}_L(y,x) - \tvarpi^{I_r \cdots I_n}{}_L(y)  \Big ) 
\no \\ && \hskip 0.9in \hskip 0.2in 
- \om_M(x) \delta ^{I_1 \cdots I_{n}}_L \Big ( \om^{I_r \cdots I_{n+1} M}{}_J(y,x) - \tvarpi^{I_r \cdots I_{n+1}M }{}_J(y)  \Big ) \bigg ]
\no
\eea
Using the definition of the traces $\tilde \chi$ and the traceless parts $\tvarpi$ in  (\ref{appe.04}), the terms in the parentheses may be simplified as follows,
\bea
\om^{I_k \cdots I_n}{}_L(y,x) - \tvarpi^{I_k \cdots I_n}{}_L(y) & = & 
 - \delta ^{I_n} _L \,  \tilde \chi^{I_k \cdots I_{n+1} }(y,x)  
\no \\
\om^{I_k \cdots I_{n+1} M}{}_J(y,x) - \tvarpi^{I_k \cdots I_{n+1}M }{}_J(y) & = & 
 - \delta ^M_J \,   \tilde \chi^{I_k \cdots I_{n+1} }(y,x) 
\eea
It is readily verified that the second and third lines in (\ref{9.b.9}) precisely cancel one another, thereby completing the proof of item 2 of Theorem  \ref{mintlemma}. 

\sm

To prove item 3, namely holomorphicity in $x$, we notice that the second line in (\ref{redefpi}) is by itself holomorphic since $\tvarpi(x)$ is. The first line is automatically holomorphic in $x$ for $r \geq 1$ since its ingredients are individually holomorphic, while holomorphicity  for $r=0$ follows from the fact that the pole at $x=y$ manifestly cancels between the two terms. 

\sm

To prove item 4, we make use of the items 1, 2 and 3 established earlier.  In particular, we use the relations between the monodromy of $\mQ$ and the fact that $\mQ$ is holomorphic in $x,y$. Cutting the Riemann surface $\Sigma$ along a set of canonical homology cycles and decomposing the boundary of the resulting fundamental domain as follows, 
\bea
\mC =  \bigcup _{K=1} ^h  \, \mA^K \, \mB_K \, (\mA^K)^{-1} \, \mB_K^{-1} 
\label{fdomain}
\eea 
we use the holomorphicity of $\mQ$ to conclude that, by Cauchy's theorem in absence of poles,
\bea
\oint _\mC dx \, \mQ ^{I_1 \cdots I_r}{}_J (x,y) =0
\eea
The contributions from the integrals over $\mB_K$ and $\mB_K^{-1}$ cancel one another in view of the invariance of $\mQ$ under the $\mA^K$ transformation that maps $\mB_K$ to $\mB_K^{-1}$, and we are left with,
\bea
\sum_{K} \oint _{\mA^K} dx \, \Big (  \mQ^{I_1 \cdots I_r}{}_J (x{+}\mB_K, y) 
-  \mQ^{I_1 \cdots I_r}{}_J (x, y) \Big ) = 0
\eea
Using the monodromy relation established in item 2, this becomes, 
\bea
\sum _{k=1}^r {  (-2\pi i)^k \over k!}  \,  \sum_{K} \, \delta ^{I_1 \cdots I_k }_K 
\oint _{\mA^K} dx \,   \mQ  ^{I_{k+1} \cdots I_r} {}_J (x,y)   =  0 
\eea
For $r=1$, only the value $k=1$ contributes and the relation becomes,
\bea
\sum_K \delta ^{I_1}_K \oint _{\mA^K} dx \, \mQ _J (x,y) = \oint _{\mA^{I_1}} dx \, \mQ _J(x,y) = 0
\eea
Since $\mQ_J(x,y)$ is a single-valued holomorphic $(1,0)$ form in $x$ the above equation implies that $\mQ _J(x,y)=0$.  
By induction on the value of $r$, the integrals over $\mQ  ^{I_1 \cdots I_r} {}_J (x,y) $ vanish at arbitrary rank $r\geq 1$,
\bea
\oint _{\mA^K} dx \, \mQ  ^{I_1 \cdots I_r} {}_J (x,y) = 0
\eea
Since $ \mQ ^{I_1 \cdots I_r}{}_J(x,y)$ is a single-valued holomorphic $(1,0)$ form in $x$, it must vanish identically. 
This completes the proof of Theorem \ref{mintlemma}.

\subsection{Proof of Theorem \ref{swid.02}}
\label{app:more}

The first step in proving the vanishing of the combinations $\cU^{I_1 \cdots I_r}(x,y)$ in (\ref{swid.03})
is to evaluate their monodromies in $x$ and $y$. The $\mA$ monodromies in both points vanish
since those of $\chi^{I_1 \cdots I_r }(x,y) $ do, and the $\mB$ monodromies can be assembled from
the following consequence of the monodromies in (\ref{mclim.01}),
\bea
\p_y \chi^{I_1 \cdots I_r }(x{+}{\mB_L},y) 
& = & 
\p_y \chi^{I_1 \cdots I_r }(x,y) +
\sum _{k=1}^r { (-2\pi i)^k \over k!} \, \delta ^{I_1 \cdots I_k}_L \, \p_y \chi^{I_{k+1} \cdots I_r}(x,y)
\no \\
 \p_x \chi^{I_r \cdots I_1}(y,x{+}{\mB_L}) 
& = &   \p_x \chi^{I_r \cdots I_1}(y,x) +
\sum _{k=1}^r { (2\pi i)^k \over k!} \, \delta ^{I_1 \cdots I_{k-1}}_L \,  \p_x \chi^{I_r \cdots I_k }(y,x)
\eea
The resulting expression for the $\mB$ monodromy of $\cU^{I_1 \cdots I_r}(x,y)$ in $x$
allows us to recombine the terms of schematic form $ \p_y \chi(x,y)$ and 
$\p_x \chi(y,x)$ to lower-rank instances of (\ref{swid.03}),
\bea
 \cU^{I_1 \cdots I_r}(x{+}{\mB_L},y) 
= \cU^{I_1 \cdots I_r}(x,y) + \sum_{k=1}^r { (-2\pi i)^k \over k!} \, \delta ^{I_1 \cdots I_k}_L \, \cU^{I_{k+1} \cdots I_r}(x,y)
\label{umondr.01}
\eea
The swapping identity $\cU^{I_1 \cdots I_r}(x,y) = - (-)^r \cU^{I_r \cdots I_1}(y,x) $
which is evident from (\ref{swid.03}) similarly organizes the $\mB$ monodromy 
of $\cU^{I_1 \cdots I_r}(x,y)$ in $y$ into lower-rank combinations $\cU$.

\sm

The second step in proving Theorem \ref{swid.02} is to demonstrate holomorphicity
of $\cU^{I_1 \cdots I_r}(x,y) $ in $x,y$. According to the discussion around (\ref{polechi}), the only
poles as $y \rightarrow x$ occur in $\chi^{I_1 \cdots I_r }(x,y) $ at rank $r=0$. Hence, all of
$\p_y\chi^{I_1 \cdots I_r }(x,y) $ and thus $ \cU^{I_1 \cdots I_r}(x,y) $ at $r\geq1$ are non-singular as $y \rightarrow x$. At rank $r=0$ in turn, we have a double pole in,
\beq
\p_y\chi(x,y) = - \frac{1}{(x{-}y)^2} + {\rm reg}
\eeq
where simple poles are absent since $\chi(x,y) $ does not exhibit any logarithmic terms $\log(x{-}y)$.
Nevertheless, these double poles cancel out from the combination,
\beq
 \cU(x,y) = \p_y\chi(x,y)  - \p_x\chi(y,x) 
\eeq
in (\ref{swid.03}) and we have established holomorphicity of $ \cU^{I_1 \cdots I_r}(x,y) $ as
$y \rightarrow x$ at any rank $r\geq 0$. The monodromies (\ref{umondr.01}) then imply that
$ \cU^{I_1 \cdots I_r}(x,y) $ are regular when $x$ and $y$ are in distinct fundamental domains of $\Sigma$.

\sm

In order to conclude the proof of  Theorem \ref{swid.02}, we apply 
Cauchy's theorem in the absence of poles (as established in the previous step),
\begin{align}
0 = \oint _\mC dx \, \cU^{I_1 \cdots I_r}(x,y) 
&=
\sum_{K} \oint _{\mA^K} dx \, \Big (  \cU^{I_1 \cdots I_r}(x{+}\mB_K, y) 
-  \cU^{I_1 \cdots I_r}(x, y) \Big ) 
\notag \\
&= \sum_{k=1}^r { (-2\pi i)^k \over k!} \, \delta ^{I_1 \cdots I_k}_L \sum_{K} \oint _{\mA^K} dx \, 
 \cU^{I_{k+1} \cdots I_r}(x,y)
 \label{ointc}
\end{align}
using the decomposition of the boundary $\mC$ of the fundamental domain for $\Sigma$ in (\ref{fdomain})
and the $\mB$ monodromies (\ref{umondr.01}) in passing to the second and third line, respectively.
At rank  $r=1$, (\ref{ointc}) specializes to $ \oint _{\mA^K} \cU(x,y) =0$ which, together with
holomorphicity and single-valuedness of $\cU(x,y)$, implies the vanishing $\cU(x,y)=0$.
By induction on the value on $r$, one can then successively show that the analogous
$\mA^K$ periods vanish at arbitrary rank,
\beq
 \oint _{\mA^K} dx \, 
 \cU^{I_{1} \cdots I_r}(x,y) = 0
\eeq
which, together with its holomorphicity and single-valuedness, implies the vanishing of the respective $ \cU^{I_{1} \cdots I_r}(x,y) $.

\newpage

\section{Recursive construction of Fay identities}
\label{app.cons}

This appendix introduces two constructive methods to derive Fay identities
via iterated convolutions of the tensorial weight-two identity (\ref{hf.22}).
In this way, the non-constructive proofs of Theorems \ref{3.thm:7} and \ref{3.thm:8}
in Appendices \ref{appB.1} and \ref{appB.2a} are complemented by a recursive construction
that was initially used to propose the general form of the Fay identities (\ref{exfay.15})
and (\ref{4.ff.99}) before they were rigorously established.

\sm

The first method, which is described in sections \ref{s:fay.3.1} and \ref{s:fay.3.1.b},
is the more general one but suffers from inefficiencies in certain cases specified below.
The second method, which is described in sections \ref{s:fay.3.2} and \ref{s:fay.3.2.b},
is presented in less generality but offers a targeted fix for the shortcoming of 
the first method.

\subsection{A first method applied to weight three}
\label{s:fay.3.1}

In order to illustrate the first method to generate Fay identities at increasing weight,
we evaluate the auxiliary integral,
\beq
H^{IM}{}_K(x,y,z) = \int_\Sigma d^2 u \, \bar \omega^I(u) f^M{}_J(x,u) 
f^J{}_L(y,u)  f^L{}_K(u,z) 
 \label{mths.00} 
\eeq
in two different ways. In both cases, we will use the following convolution identities
which are straightforward consequences of (\ref{3.Phi}), (\ref{fphig.2}) and (\ref{ften.06}),
\begin{align}
\int_\Sigma d^2 u \, \bar \omega^I(u) f^{ \overrightarrow{Q} R}{}_K(x,u) f^{ \overrightarrow{P} }{}_J(u,a) &= - \delta^R_K f^{\overrightarrow{Q} I \overrightarrow{P} }{}_J(x,a) \label{mths.01} \\
\int_\Sigma d^2 u \, \bar \omega^I(u) f^{ \overrightarrow{Q} R}{}_K(x,u) \omega_J(u) &=
\delta^I_J f^{\overrightarrow{Q} R  }{}_K(x,b) 
 - \delta^R_K f^{\overrightarrow{Q} I  }{}_J(x,b) 
\notag
\end{align}
where $\overrightarrow{P} \neq \emptyset$, and
the second line involves an arbitrary point $b \in \Sigma$.
\begin{itemize}
\item[(i)] bring the first two factors $ f^M{}_J(x,u) f^J{}_L(y,u) $
in the integrand of (\ref{mths.00}) into a \textit{$u$-reduced} form (see section \ref{sec:3-x})
using (\ref{hf.22}) and then integrate term by term via (\ref{mths.01})
\begin{align}
H^{IM}{}_K(x,y,z) &= \int_\Sigma d^2 u \, \bar \omega^I(u) \Big(
f^{M}{}_{ J}( y, x)  f^{J}{}_{ L}( x, u) +
f^{M }{}_{J}( x, y) f^{J}{}_{ L}( y, u)   \label{mths.02} \\
 &\quad + \omega_J(x) f^{J M}{}_{ L}( y, u)  + \omega_J(y) f^{JM}{}_{L}( x, u) + 
\omega_J(x) f^{M J }{}_{L}(y, x) 
\Big) f^L{}_K(u,z)  \notag \\
&= - f^M{}_J(y,x) f^{IJ}{}_K(x,z)
 - f^M{}_J(x,y) f^{IJ}{}_K(y,z) \notag \\
 &\quad
  - \omega_J(x) f^{JIM}{}_K(y,z)
    - \omega_J(y) f^{JIM}{}_K(x,z)
\notag
\end{align}
\item[(ii)]
bring the last two factors $f^J{}_L(y,u)  f^L{}_K(u,z) $ in the integrand of (\ref{mths.00})
 into a \textit{$u$-reduced} form using (\ref{hf.22}) and then integrate term by term via (\ref{mths.01})
\begin{align}
H^{IM}{}_K(x,y,z) &= \int_\Sigma d^2 u \, \bar \omega^I(u) f^M{}_J(x,u) 
\Big (
f^{J}{}_{ L}( y, z) f^{L}{}_{K}( u, z) - f^{J}{}_{L}( u, y) f^{L}{}_{K}( y, z) \notag\\
&\quad   -  \omega_L(y) f^{LJ}{}_{K}( u, z)
  - \omega_L(y) f^{J L}{}_{K}( u, y) - \omega_L(u) f^{LJ}{}_{K}( y, z) 
\Big ) \label{mths.03} \\
&=  - f^M{}_J(y,z) f^{IJ}{}_K(x,z) + f^{IM}{}_J(x,y)  f^J{}_K(y,z) 
+ \omega_J(y) f^{IMJ}{}_K(x,y) \notag \\
&\quad + \omega_J(y)f^{IJM}{}_K(x,z)
+ f^I{}_J(x,b) f^{JM}{}_K(y,z)  -  f^M{}_J(x,b) f^{IJ}{}_K(y,z) 
\notag
\end{align}
\end{itemize}
Equating the two representations of $H^{IM}{}_K(x,y,z)$ obtained from (i) and (ii)
yields a weight-three Fay identity, based on the weight-two input (\ref{hf.22}). 
Setting the arbitrary point in (\ref{mths.03}) to $b=y$ cancels the
second term in the third line of (\ref{mths.02}), and we arrive at
\begin{align}
&- f^M{}_J(y,x) f^{IJ}{}_K(x,z)
- \omega_J(x) f^{JIM}{}_K(y,z)
- \omega_J(y) f^{JIM}{}_K(x,z)
\notag\\
&= -  f^M{}_J(y,z) f^{IJ}{}_K(x,z) + f^{IM}{}_J(x,y)  f^J{}_K(y,z) 
+ f^I{}_J(x,y) f^{JM}{}_K(y,z)  \notag \\
&\quad
+ \omega_J(y) f^{IMJ}{}_K(x,y)  
+ \omega_J(y)f^{IJM}{}_K(x,z)
\label{mths.04}
\end{align}
Solving this identity for $f^M{}_J(y,z) f^{IJ}{}_K(x,z) $ then reproduces the first example
in (\ref{explf34}).

\subsection{A first method applied to higher weight}
\label{s:fay.3.1.b}

The key idea in the above derivation carries over to arbitrary weight:
Adapt the above methods (i) and (ii) to the higher-weight
generalization of the weight-three integral in (\ref{mths.00})
\beq
{\cal H}^{I,\overrightarrow{M} ,\overrightarrow{P}, \overrightarrow{Q}}{}_K(x,y,z) = \int_\Sigma d^2 u \, \bar \omega^I(u) f^{\overrightarrow{M}}{}_J(x,u) 
f^{\overrightarrow{P}J}{}_L(y,u)  f^{ \overrightarrow{Q} L}{}_K(u,z) 
 \label{mths.06} 
\eeq
For each choice of the multi-indices $\overrightarrow{M} ,\overrightarrow{P}$
and $\overrightarrow{Q}$, one can derive a higher-weight
Fay identity from the auxiliary integral (\ref{mths.06}) by equating two methods of
reducing the number of $u$-dependent factors in the integrand:
\begin{itemize}
\item[(i)] bring the first two factors $f^{\overrightarrow{M}}{}_J(x,u) 
f^{\overrightarrow{P}J}{}_L(y,u)  $ into a \textit{$u$-reduced} form using lower-weight identities for repeated scalar points, see section \ref{sec:fay.2}
\item[(ii)] bring the last two factors $f^{\overrightarrow{P}J}{}_L(y,u)  
f^{ \overrightarrow{Q} L}{}_K(u,z) $ into a \textit{$u$-reduced} form using lower-weight identities for repeated one-form points, see section \ref{sec:fay.3}
\end{itemize}
In both cases, the integration over $u$ can be performed term by term via (\ref{mths.01})
after applying the Fay identities of (i) and (ii).

\sm

We shall now illustrate to what extent the weight-three Fay
identity (\ref{mths.04}) gives access to weight-four Fay identities. While (\ref{mths.00})
is the only weight-three integral amenable to the method of this section, there
are three weight-four instances of (\ref{mths.06}):
\begin{align*}
{\cal H}^{I,M , \emptyset,Q}{}_K(x,y,z) & ~ \leftrightarrow\;  \textrm{need Fay for} \
f^{M}{}_J(x,u) 
f^{J}{}_L(y,u) \hskip 0.13in \& \
f^{J}{}_L(y,u)  f^{Q  L}{}_K(u,z)\\
{\cal H}^{I,M N ,\emptyset, \emptyset}{}_K(x,y,z) &~ \leftrightarrow\; \textrm{need Fay for}\
f^{MN}{}_J(x,u) 
f^{J}{}_L(y,u) \ \& \
f^{J}{}_L(y,u)  f^{  L}{}_K(u,z)\\
{\cal H}^{I,M ,P, \emptyset}{}_K(x,y,z) & ~ \leftrightarrow\; \textrm{need Fay  for}\
f^{M}{}_J(x,u) 
f^{PJ}{}_L(y,u) \hskip 0.08in \& \
f^{PJ}{}_L(y,u)  f^{  L}{}_K(u,z)
\end{align*}
Two of the required Fay identities (for $f^{M}{}_J(x,u) 
f^{J}{}_L(y,u) $ and for $f^{J}{}_L(y,u)  f^{  L}{}_K(u,z)$) are of weight two and again
boil down to relabelings of (\ref{hf.22}). The remaining four required Fay identities have weight three,
\begin{align}
({\rm a}) &\ \  f^{J}{}_L(y,u)  f^{Q  L}{}_K(u,z)
&({\rm c}) &\ \  f^{M}{}_J(x,u) 
f^{PJ}{}_L(y,u)
\notag \\
({\rm b}) &\ \  f^{MN}{}_J(x,u) 
f^{J}{}_L(y,u)
&({\rm d}) &\ \  f^{PJ}{}_L(y,u)  f^{  L}{}_K(u,z)
\label{abcd}
\end{align}
Both of (a) and (c) are available from relabelings of (\ref{mths.04}), namely (a) by solving for the unique term with two $x$-dependent factors and relabeling $x \rightarrow u$ and (c) by solving for the unique term with two $z$-dependent factors and relabeling $z \rightarrow u$. The Fay identity (b) can also be extracted from (\ref{mths.04}) by applying the matrix commutator identity, 
\bea
 f^M{}_J(y,z) f^{IJ}{}_K(x,z) 
& = &  f^J{}_K(y,z) f^{IM}{}_J(x,z) 
\no \\ &&
-  f^J{}_K(y,a) f^{IM}{}_J(x,b) 
+  f^M{}_J(y,a) f^{IJ}{}_K(x,b)  
\label{matcmid}
 \eea
with arbitrary $a,b \in \Sigma$ to the unique repeatedly $z$-dependent term and relabeling the first term
$  f^J{}_K(y,z) f^{IM}{}_J(x,z) $ on the right side of (\ref{matcmid})
to match the names of the indices and variables of the target 
$f^{MN}{}_J(x,u)  f^{J}{}_L(y,u)$ in (b).

\sm

Finally, the term $ f^{PJ}{}_L(y,u)  f^{  L}{}_K(u,z)$ in (d) 
may share the form degrees with the term $f^{J}{}_L(y,u)  f^{Q  L}{}_K(u,z)$ in (a) 
but crucially differs from its index structure. 
This can be seen from the fact that the bilinear terms in $\cG$
are given by $\delta^J_K \p_y \cG^P(y,u) \p_u \cG(u,z)$ for (d) and 
$\delta^J_K \p_y\cG(y,u) \p_u \cG^Q(u,z)$ for (a), where the repeated
point $u$ enters the weight-two factors $\p_y \cG^P(y,u) $ and $\p_u \cG^Q(u,z)$ with
a different form degree.
In principle, this could be fixed by inserting separate permutations of (\ref{mths.04})
under $x \leftrightarrow y$ into one another. However, we refrain from spelling out this cumbersome
workaround and take the complication in deriving the Fay identity for (d)
as a motivation to introduce a separate method in the next section.

\sm

The simplifications of auxiliary integrals
(\ref{mths.06}) at various weights $\leq 6$ benefitted from the
following method to modify the position of the
upper contracted index $J$ in the second factor of
$f^{\overrightarrow{I}}{}_J(a,x)  f^{\cdots J \cdots}{}_K(b,c)$.
Even though the techniques of section \ref{sec:4.9} allow us to
reformulate Fay identities without any index contractions, the
identity (\ref{conlem.23}) below offers convenient shortcuts
in the recursive construction of Fay identities. 
The key realization which drives the rerouting of contracted
indices is that the antisymmetrized combination on the left side~of,
\begin{align}
&f^{\overrightarrow{I}P}{}_J(a,x)  f^{\overrightarrow{A}J \overrightarrow{B}Q \overrightarrow{C} }{}_K(b,c) 
- f^{\overrightarrow{I}Q}{}_J(a,x)  f^{\overrightarrow{A}P \overrightarrow{B}J \overrightarrow{C} }{}_K(b,c) 
   \label{conlem.23}  \\
   &\quad = f^{\overrightarrow{I}P}{}_J(a,y)  f^{\overrightarrow{A}J \overrightarrow{B}Q \overrightarrow{C} }{}_K(b,c) 
- f^{\overrightarrow{I}Q}{}_J(a,y)  f^{\overrightarrow{A}P \overrightarrow{B}J \overrightarrow{C} }{}_K(b,c) 
   \notag
\end{align}
does not depend on the point $x$. That is why this expression is equated to its
relabeling  $x \rightarrow y$ in the second line. Applications to
Fay identities arise if one of $b$ or $c$ coincides with $x$ on the left side,
so the two terms on the right side no longer exhibit the repeated point $x$.
As a net effect of (\ref{conlem.23}) in these cases, terms with the upper contracted index
$J$ in the position of $f^{\overrightarrow{A}J \overrightarrow{B}Q \overrightarrow{C} }{}_K(b,c) $ are
traded for others with $J$ in an arbitrary different position
$ f^{\overrightarrow{A}P \overrightarrow{B}J \overrightarrow{C} }{}_K(b,c) $
(where some of $\overrightarrow{A}$, $\overrightarrow{B}$ or $\overrightarrow{C}$ may be empty).
For instance, (\ref{conlem.23}) translates the Fay identity for $f^M{}_J(y,z) f^{IJ}{}_K(x,z)$ in the first line of (\ref{explf34}) into,
\begin{align}
f^M{}_J(y,z) f^{JI}{}_K(x,z)  &=
f^M{}_J(y,x) f^{JI}{}_K(x,z)
 + f^M{}_J(x,y) f^{JI}{}_K(y,z)
+ f^{MI}{}_J(x,y) f^J{}_K(y,z)  \notag \\
&\quad + \omega_J(x) f^{J M I}{}_K(y,z)
  + \omega_J(y) f^{(M \shuffle J) I }{}_K(x,z)
 + \omega_J(y) f^{M I J}{}_K(x,y)
\end{align}
with $J\leftrightarrow I$ swapped in the second factor. 
In this way, lower-weight Fay identities relevant to the two evaluation strategies (i) and (ii) of
the auxiliary integral (\ref{mths.06}) can be brought into the most opportune form.

\subsection{A second method applied to weight three}
\label{s:fay.3.2}

We shall next present an independent method to construct Fay identities from convolutions of
lower-weight instances which makes use of the auxiliary identity for $\overrightarrow{P} \neq \emptyset$,
\begin{align}
\int_{\Sigma} d^2 x\, \bar \omega^I(x) \p_u \cG^{  \overrightarrow{Q}  }(u,x) f^{ \overrightarrow{P} }{}_J(x,a) &=
f^{  \overrightarrow{Q}  I \overrightarrow{P}  }{}_J(u,a) 
\notag \\
 \int_{\Sigma} d^2 x\, \bar \omega^I(x) \p_u \cG^{  \overrightarrow{Q}  }(u,x) \omega_J(x) &=
\p_u \Phi^{  \overrightarrow{Q}  I }{}_J(u) 
 \label{mths.11}
\end{align}
derived from the trace of (\ref{mths.01}) in $R,K$. As a first example, we apply (\ref{mths.11})
to perform the following integral in two different ways:
\begin{align}
\widehat H^{IM}{}_K(u,y,z) &= \int_\Sigma d^2 x \, \bar \omega^I(x)\p_u \cG(u,x) f^M{}_J(y,x) f^J{}_K(x,z) 
 \label{mths.12}
 \end{align}
\begin{itemize}
\item[(i)] bring the last factors $f^M{}_J(y,x) f^J{}_K(x,z) $ into an \textit{$x$-reduced} form
using the $x\leftrightarrow y$ image of the weight-two Fay identity (\ref{hf.22}),
\begin{align}
\widehat H^{IM}{}_K(u,y,z) &=  \int_\Sigma d^2 x \, \bar \omega^I(x)\p_u \cG(u,x)
\big(  f^M{}_J(y,z) f^J{}_K(x,z) -  f^M{}_J(x,y) f^J{}_K(y,z) \notag  \\
&\quad \quad \quad - \omega_J(y) f^{MJ}{}_K(x,y) 
- \omega_J(y) f^{JM}{}_K(x,z) -  \omega_J(x) f^{JM}{}_K(y,z)  \big)
\notag \\
&= f^M{}_J(y,z) f^{IJ}{}_K(u,z) 
-  f^{IM}{}_J(u,y) f^J{}_K(y,z)
- \omega_J(y) f^{IMJ}{}_K(u,y)  \notag  \\
&\quad \quad \quad 
- \omega_J(y) f^{IJM}{}_K(u,z) -   f^{JM}{}_K(y,z) \p_u \Phi^{I}{}_J(u)
 \label{mths.13}
\end{align}
\item[(ii)] bring the first and the last factor $\p_u \cG(u,x) f^J{}_K(x,z) $
into an \textit{$x$-reduced} form using (\ref{hf.22}) when the middle term contributes
$f^M{}_J(y,x) \rightarrow - \delta^M_J \p_y \cG(y,x)$,
\begin{align}
\widehat H^{IM}{}_K(u,y,z) &= \int_\Sigma d^2 x \, \bar \omega^I(x)\p_u \cG(u,x) \big(
\p_y \Phi^M{}_J(y) f^J{}_K(x,z) 
-  \p_y \cG(y,x)  f^M{}_K(x,z) 
\big)  \notag \\
&= \p_y \Phi^M{}_J(y) f^{IJ}{}_K(u,z)  
\notag \\
&\quad+ \int_\Sigma d^2 x \, \bar \omega^I(x)\p_y \cG(y,x) \big( f^M{}_J(u,x)  f^J{}_K(x,z) \, {-}\, \p_u \Phi^M{}_J(u)  f^J{}_K(x,z)   \big)
\notag \\
&= \p_y \Phi^M{}_J(y) f^{IJ}{}_K(u,z)  
\,{-}\, \p_u \Phi^M{}_J(u)  f^{IJ}{}_K(y,z) \, {+} \, \widehat H^{IM}{}_K(y,u,z)  \! \! 
 \label{mths.14}
\end{align}
In the first step, we have rewritten $-\p_u \cG(u,x) f^M{}_K(x,z) $ as
$ f^M{}_J(u,x)  f^J{}_K(x,z) - \p_u \Phi^M{}_J(u)  f^J{}_K(x,z) $.
In passing to the last line,
the $y\leftrightarrow u$ image $\widehat H^{IM}{}_K(y,u,z) $ of the integral
(\ref{mths.12}) has been identified from the factors
$\p_y \cG(y,x)f^M{}_J(u,x)  f^J{}_K(x,z)$ in the integrand.
\end{itemize}
When equating the two expressions for $\widehat H^{IM}{}_K(u,y,z) $ in (i) and (ii),
the $y\leftrightarrow u$ image $\widehat H^{IM}{}_K(y,u,z) $ in (\ref{mths.14}) is understood to be evaluated through
the five-term expressions on the right side of (\ref{mths.13}) with $y$ and $u$ interchanged:
\begin{align}
& f^M{}_J(y,z) f^{IJ}{}_K(u,z) 
-  f^{IM}{}_J(u,y) f^J{}_K(y,z)
- \omega_J(y) f^{IMJ}{}_K(u,y)  \notag  \\
&\quad \quad \quad 
- \omega_J(y) f^{IJM}{}_K(u,z) -   f^{JM}{}_K(y,z) \p_u \Phi^{I}{}_J(u)
\notag \\
&= \p_y \Phi^M{}_J(y) f^{IJ}{}_K(u,z)  
- \p_u \Phi^M{}_J(u)  f^{IJ}{}_K(y,z) 
\notag \\
&\quad 
+ f^M{}_J(u,z) f^{IJ}{}_K(y,z) 
-  f^{IM}{}_J(y,u) f^J{}_K(u,z)
- \omega_J(y) f^{IMJ}{}_K(y,u)  \notag  \\
&\quad \quad \quad 
- \omega_J(u) f^{IJM}{}_K(y,z) -   f^{JM}{}_K(u,z) \p_y \Phi^{I}{}_J(y)
 \label{mths.15}
\end{align}
One can simultaneously uplift
the $\p\Phi$-tensors on both sides to $f$-tensors
$\p_u \Phi^{I}{}_J(u) {\rightarrow} f^{I}{}_J(u,a)$, $ \p_u \Phi^M{}_J(u) \rightarrow f^M{}_J(u,a) $
and
$ \p_y \Phi^M{}_J(y)  \rightarrow f^M{}_J(y,b), \   \p_y \Phi^{I}{}_J(y) \rightarrow f^{I}{}_J(y,b) $ 
with arbitrary $a,b \in \Sigma$ since the corresponding $\p_u \cG(u,a)$
and $\p_y \cG(y,b)$ cancel.
Setting $a,b \rightarrow z$ leads to cancellations of four terms, and one
arrives at the following weight-three Fay identity with manifest antisymmetry
in $x\leftrightarrow y$,
\begin{align}
0 &= \omega_J(x) f^{IJM}{}_K(y,z) -  \omega_J(y) f^{IJM}{}_K(x,z)
\notag \\
&\quad +  \omega_J(x) f^{IMJ}{}_K(y,x) -  \omega_J(y) f^{IMJ}{}_K(x,y)  
 \notag \\
 &\quad  + f^{IM}{}_J(y,x) f^J{}_K(x,z)  - f^{IM}{}_J(x,y) f^J{}_K(y,z)  \notag \\
 &\quad
 + f^I{}_J(y,z) f^{JM}{}_K(x,z) -f^I{}_J(x,z) f^{JM}{}_K(y,z) 
\label{hf.21}
\end{align}
which is in fact equivalent to (\ref{ipexpl}). Once the third term $ \omega_J(x) f^{IMJ}{}_K(y,x) $ on the right side
is eliminated through the weight-three interchange identity of Theorem \ref{intlemma}, the fifth
one $f^{IM}{}_J(y,x) f^J{}_K(x,z) $ is the only instance of the repeated point $x$.
Upon solving for the fifth term $ f^{IM}{}_J(y,x) f^J{}_K(x,z)$
and relabeling points and indices,
the Fay identity (\ref{hf.21}) provides the missing \textit{$u$-reduced}
rewriting of the term (d) in (\ref{abcd}). The trace component $\delta^K_M$ of (\ref{hf.21})
eliminates the repeated appearance of $x$ in $\p_y \cG^{I}(y,x) \p_x \cG(x,z)$
whereas the the previous method provided the analogous elimination for
 $\p_y \cG(y,x) \p_x \cG^I(x,z)$ via (\ref{mths.04}), which has a different form degree
 of $x$ in the weight-two function.
This illustrates the synergy between the
method of section \ref{s:fay.3.1.b} and the 
alternative method of the present section that we shall
next generalize to higher weight.

\subsection{A second method applied to higher weight}
\label{s:fay.3.2.b}

The higher-weight generalization of the method in the previous section
\ref{s:fay.3.2} relies
on the following variant of the integral (\ref{mths.12}):
\begin{align}
\widehat {\cal H}^{I, \overrightarrow{Q}M }{}_K(u,y,z) &= \int_\Sigma d^2 x \, \bar \omega^I(x)\p_u \cG(u,x) f^{\overrightarrow{Q}M}{}_J(y,x) f^J{}_K(x,z) 
 \label{mths.21}
 \end{align}
 We deliberately introduce less multi-indices than in the earlier
 family ${\cal H}^{I,\overrightarrow{M} ,\overrightarrow{P}, \overrightarrow{Q}}{}_K(x,y,z)$ 
 of auxiliary integrals in (\ref{mths.06}) to demonstrate that 
 the present method is recursive in weight. The mechanism to derive new Fay identities
 is again to equate the evaluations of the integral (\ref{mths.21}) using two different
 lower-weight Fay identities
\begin{itemize}
\item[(i)] bring the last two factors $f^{\overrightarrow{Q}M}{}_J(y,x) f^J{}_K(x,z) $ 
into an \textit{$x$-reduced} form using Fay identities obtained from the same
procedure at lower weight (see below why this is possible) and integrate the resulting expression term by
term via (\ref{mths.11})
\item[(ii)] bring the product $\p_u \cG(u,x) f^J{}_K(x,z) $
of the first and the last factor into an \textit{$x$-reduced} form when the middle term contributes
$f^{\overrightarrow{Q}M}{}_J(y,x) \rightarrow - \delta^{M}_J \p_y \cG^{\overrightarrow{Q}}(y,x)$; 
regardless on the choice of $\overrightarrow{Q}$,
this solely requires the weight-two Fay identity (\ref{hf.22}):
\begin{align}
\! \! \! \widehat {\cal H}^{I, \overrightarrow{Q}M}{}_K(u,y,z) &= \! \int_\Sigma  \!d^2 x \, \bar \omega^I(x)\p_u \cG(u,x) \big(
\p_y \Phi^{ \overrightarrow{Q}M }{}_J(y) f^J{}_K(x,z) 
\, {-} \, \p_y \cG^{ \overrightarrow{Q} }(y,x)  f^{M}{}_K(x,z) 
\big)  \notag \\
&= \p_y \Phi^{ \overrightarrow{Q} M}{}_J(y) f^{IJ}{}_K(u,z)  
 \label{mths.22} \\
&\quad+\! \int_\Sigma  \! d^2 x \, \bar \omega^I(x)\p_y \cG^{  \overrightarrow{Q} }(y,x) \big( f^M{}_J(u,x)  f^J{}_K(x,z) \, {-}\, \p_u \Phi^M{}_J(u)  f^J{}_K(x,z)   \big)
\notag \\
&= \p_y \Phi^{ \overrightarrow{Q} M}{}_J(y) f^{IJ}{}_K(u,z)  
- \p_u \Phi^M{}_J(u)  f^{\overrightarrow{Q} IJ}{}_K(y,z)  \notag \\
&\quad + \! \int_\Sigma  \! d^2 x \, \bar \omega^I(x)\p_y \cG^{  \overrightarrow{Q} }(y,x) 
\big(
 f^M{}_J(u,z) f^J{}_K(x,z) -  f^M{}_J(x,u) f^J{}_K(u,z) \notag  \\
&\quad \quad \quad - \omega_J(u) f^{MJ}{}_K(x,u) 
- \omega_J(u) f^{JM}{}_K(x,z) -  \omega_J(x) f^{JM}{}_K(u,z) 
\big)
\notag
\end{align}
The rewriting $-\p_u \cG(u,x) f^M{}_K(x,z) = f^M{}_J(u,x)  f^J{}_K(x,z) - \p_u \Phi^M{}_J(u)  f^J{}_K(x,z) $
in the first step is identical to that in (\ref{mths.14}). In the last step, the weight-two
Fay identity (\ref{hf.22}) leads to an \textit{$x$-reduced} 
parenthesis multiplying $\p_y \cG^{  \overrightarrow{Q} }(y,x) $, so we
can perform the $x$-integral for each term via (\ref{mths.11}):
\begin{align}
 \widehat {\cal H}^{I, \overrightarrow{Q}M}{}_K&(u,y,z) = 
 \p_y \Phi^{ \overrightarrow{Q} M}{}_J(y) f^{IJ}{}_K(u,z)  
- \p_u \Phi^M{}_J(u)  f^{\overrightarrow{Q} IJ}{}_K(y,z)    \label{mths.23}  \\
& + f^M{}_J(u,z) f^{\overrightarrow{Q} I J}{}_K(y,z) -  f^{\overrightarrow{Q} I M}{}_J(y,u) f^J{}_K(u,z) 
\notag  \\
&  - \omega_J(u) f^{ \overrightarrow{Q} I MJ}{}_K(y,u)
- \omega_J(u) f^{\overrightarrow{Q} I JM}{}_K(y,z) -  
\p_y \Phi^{\overrightarrow{Q} I }{}_J(y) f^{JM}{}_K(u,z) 
\notag
\end{align}
\end{itemize} 
For a given multi-index $\overrightarrow{Q}=Q_1Q_2\cdots Q_r$ of length $r$,
the next step is to equate (\ref{mths.23}) with
the outcome of integrating the weight-$(r{+}2)$ Fay identity
for $f^{\overrightarrow{Q}M}{}_J(y,x) f^J{}_K(x,z)$ in~(i) against 
$\int_\Sigma d^2 x \, \bar \omega^I(x)\p_u \cG(u,x) $. We shall argue in
two steps why this is guaranteed to yield a Fay identity of weight $(r{+}3)$ that 
\textit{$u$-reduces} $ f^{\overrightarrow{Q} I M}{}_J(y,u) f^J{}_K(u,z) $
for arbitrary $\overrightarrow{Q}$:
First, none of the terms in the expression for (\ref{mths.21}) due to (i) can involve a
repeated appearance of $u$ by inspection of the right side of (\ref{mths.11}). 
Second, after applying the
interchange identities of Theorem \ref{intlemma} to the term
$\omega_J(u) f^{ \overrightarrow{Q} I MJ}{}_K(y,u)$ on the right side of
(\ref{mths.23}), the only repeatedly $u$-dependent term in (\ref{mths.23})
and hence the entire Fay identity due to (i) $=$ (ii) 
is $ f^{\overrightarrow{Q} I M}{}_J(y,u) f^J{}_K(u,z) $. By repeating
this derivation for different lengths~$r$ of $\overrightarrow{Q}=Q_1Q_2\cdots Q_r$,
one arrives at an all-weight family of Fay identities that eliminate the repeated point $x$ in
$ f^{\overrightarrow{I}  M}{}_J(y,x) f^J{}_K(x,z) $ for arbitrary $\overrightarrow{I}$.
  
  \sm
  
The discussion below (\ref{abcd}) identified the
Fay identity (d) for $ f^{IJ}{}_L(y,u) f^L{}_K(u,z) $ as a bottleneck in applying the
method of section \ref{s:fay.3.1.b} to all weight-four cases. Also for
higher-weight instances of the auxiliary integral (\ref{mths.06}), the required
Fay identities for $ f^{\overrightarrow{I}J}{}_L(y,u) f^L{}_K(u,z) $ 
turn out to be most difficult to derive from the method of section 
\ref{s:fay.3.1.b}. Hence, the recursive generation of Fay identities
for $ f^{\overrightarrow{I}J}{}_L(y,u) f^L{}_K(u,z) $ in the present
section complements the earlier method by resolving a 
major obstacle in its systematic higher-weight application.

\sm

Given a \textit{$u$-reduction} of
$ f^{\overrightarrow{P}J}{}_L(y,u) f^L{}_K(u,z) $
at fixed $\overrightarrow{P}$, the method of section
\ref{s:fay.3.1.b} was found  to recursively 
generate higher-weight identities for 
$ f^{\overrightarrow{P}J}{}_L(y,u) f^{Q_1\cdots Q_r L}{}_K(u,z) $ in several examples, adding indices 
to the second factor.
The underlying auxiliary integrals take the form of ${\cal H}^{I,M ,\overrightarrow{P}, 
Q_1\cdots Q_r}{}_K(x,y,z) $ at fixed $\overrightarrow{P}$ and increasing
length~$r$ of $\overrightarrow{Q}=Q_1\cdots Q_r$. One can see from the
following reasoning that the Fay identity for $ f^{M}{}_J(x,u) f^{\overrightarrow{P}J}{}_L(y,u) $ required in the
rewriting (i) of ${\cal H}^{I,M ,\overrightarrow{P}, 
Q_1\cdots Q_r}{}_K(x,y,z) $ integrals is readily available:
First, the sequence ${\cal H}^{I,M ,\emptyset, 
Q_1\cdots Q_r}{}_K(x,y,z) $ at empty $\overrightarrow{P}$
can be shown to generate Fay identities for $ f^{J}{}_L(y,u) f^{\overrightarrow{Q}L}{}_K(u,z) $
at all weights. Second, by the observation in
section \ref{sec:obs}, these Fay identities involving
one $f$-tensor of weight one can be solved for the
unique repeatedly $z$-dependent term
$ f^{J}{}_L(y,z) f^{\overrightarrow{Q}L}{}_K(u,z) $, see (\ref{Pempty}).
Third, relabelings of the indices and points yield the desired Fay identities for
$ f^{M}{}_J(x,u) f^{\overrightarrow{P}J}{}_L(y,u) $.

\sm

In conclusion, the methods of section 
\ref{s:fay.3.1.b} and the present one are found to be in fruitful symbiosis: 
The latter method is known to generate all-weight Fay identities
for $ f^{\overrightarrow{P}J}{}_L(y,u) f^L{}_K(u,z) $, and the former method
is believed to then deduce more general Fay identities
$ f^{\overrightarrow{P}J}{}_L(y,u) f^{\overrightarrow{Q} L}{}_K(u,z) $ with
additional multi-indices $\overrightarrow{Q}$. From the combination of
both methods, we generated a selection of Fay identities
at weight $\leq 6$ with various choices of $ \overrightarrow{P}$ and 
$\overrightarrow{Q}$ which turned out to be sufficient to
propose the general Fay identities (\ref{exfay.15})
and (\ref{4.ff.99}) as conjectures. In view of the proof of (\ref{exfay.15})
and (\ref{4.ff.99}) in Appendices \ref{appB.1} and \ref{appB.2a}, we did not attempt 
a rigorous investigation if the method of this appendix will eventually
construct the Fay identities for $f^{\overrightarrow{P}J}{}_L(y,u) f^{\overrightarrow{Q} L}{}_K(u,z)$
with arbitrary $ \overrightarrow{P}$ and $\overrightarrow{Q}$. 


\end{document}